\documentclass[aps,showpacs,prd]{revtex4-1}

\usepackage[ansinew]{inputenc}
\usepackage{amsmath,amssymb,amsfonts,amsthm}
\usepackage{mathrsfs}
\usepackage{color}
\usepackage{graphicx}
\usepackage{pstricks}
\usepackage{subfigure}
\usepackage[all]{xy}
\usepackage{multirow}

\begin{document}
\vspace*{3cm}
\title{Supersymmetric rotating black hole spacetime tested by geodesics}
\author{Valeria Diemer (n\'{e}e Kagramanova), Jutta Kunz}
\address{
Institut f\"ur Physik, Universit\"at Oldenburg,
D--26111 Oldenburg, Germany
}

\begin{abstract}
We present the complete analytical solution of the geodesics equations 
in the supersymmetric BMPV spacetime~\cite{Breckenridge:1996is}. 
We study systematically the properties of massive and massless 
test particle motion. 
We analyze the trajectories with analytical methods based on the 
theory of elliptic functions.  
Since the nature of the effective potential depends strongly 
on the rotation parameter $\omega$, one has to distinguish 
between the underrotating case,
the critical case and the overrotating case, 
as discussed by Gibbons and Herdeiro 
in their pioneering study~\cite{Gibbons:1999uv}.
We discuss various properties which distinguish this spacetime 
from the classical relativistic spacetimes like Schwarzschild, 
Reissner-Nordstr\"om, Kerr or Myers-Perry.
The overrotating BMPV spacetime allows, for instance,
for planetary bound orbits for massive and massless particles. 
We also address causality violation as
analyzed in~\cite{Gibbons:1999uv}.
\end{abstract}

\date{\today}

\begin{figure*}[b!]
\begin{center}
\includegraphics[width=6cm]{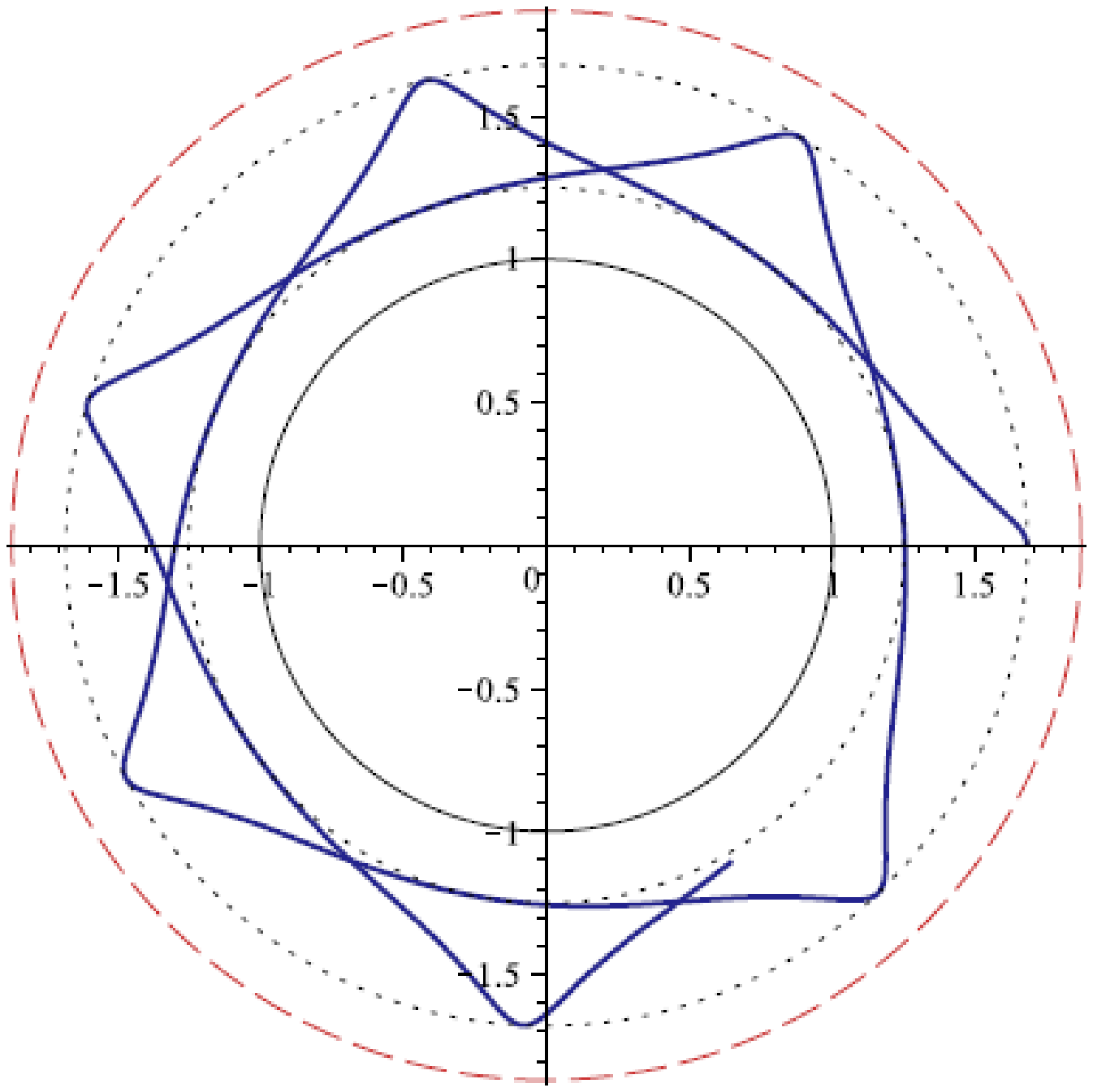}
\vspace*{4cm}
\end{center}
\end{figure*}

\pacs{04.20.Jb, 02.30.Hq}

\maketitle

\section{Introduction}

The Breckenridge-Myers-Peet-Vafa (BMPV) spacetime \cite{Breckenridge:1996is}
represents a fascinating solution of the bosonic sector of minimal supergravity
in five dimensions. 
It describes a family of charged rotating extremal black holes
with equal-magnitude angular momenta, % $|J_1|=|J_2|=J$,
that are associated with independent rotations in two orthogonal planes.

The BMPV solution has been analyzed in various respects.
At first interest focussed on the entropy of the
extremal black hole solutions. 
Here a microscopic derivation of the entropy led to perfect agreement 
with the classical value obtained from the horizon area of the black holes
$A=2 \pi^2 \sqrt{\mu^3-\omega^2}$,
where $\mu$ is a charge parameter and $\omega$ is the rotation parameter \cite{Breckenridge:1996is}.
Clearly, the entropy is largest in the static case,  
and vanishes in the critical case $\mu^3=\omega^2$.
For still faster rotation the radicant would become negative.

Gauntlett, Myers and Townsend \cite{Gauntlett:1998fz}
analyzed the BMPV spacetime further, pointing out that it describes
supersymmetric black hole solutions with a non-rotating horizon,
but finite angular momentum.
They argued that angular momentum can be stored in the gauge field,
and a negative fraction of the total angular momentum resides behind
the horizon, while
the effect of the rotation on the horizon is to make it squashed \cite{Gauntlett:1998fz}.

While Gauntlett, Myers and Townsend \cite{Gauntlett:1998fz} addressed already
the presence of closed timelike curves (CTSs) in the BMPV spacetime,
a thorough qualitative study of the geodesics and the possibility of time travel
was given by Gibbons and Herdeiro \cite{Gibbons:1999uv}.
Considering three cases for the BMPV spacetime,
Gibbons and Herdeiro pointed out, that while in the underrotating case
the CTCs are hidden behind the degenerate horizon,
the CTCs occur in the exterior region in the overrotating case.
Thus the BMPV spacetime contains naked time machines in this case.

Moreover, in the overrotating case the horizon becomes ill-defined,
since it becomes a timelike hypersurface, with the entropy
becoming naively imaginary 
\cite{Gibbons:1999uv,Herdeiro:2000ap,Herdeiro:2002ft,Cvetic:2005zi}.
This hypersurface is then referred to as {\sl pseudo-horizon}.
In fact, as shown by Gibbons and Herdeiro, this pseudo-horizon
cannot be traversed by particles or light following geodesics.
Thus the interior region of an overrotating BMPV spacetime
cannot be entered. Therefore the outer spacetime represents a {\sl repulson}.
The geodesics in the exterior region
between the pseudo-horizon and infinity are complete.

The repulson behavior of the overrotating BMPV spacetime
has been analyzed further by Herdeiro \cite{Herdeiro:2000ap}.
By studying the motion of charged testparticles in the spacetime
he realized that the repulson effect is still present.
Considering accelerated observers, however, he noted
that it could be possible to travel into the interior region
\cite{Herdeiro:2000ap}.

Herdeiro further showed, that when oxidising the 
overrotating $D=5$ BMPV solution to $D=10$
the causal anomalies are resolved.
Interestingly, here a relation between microscopic unitarity
and macroscopic causality emerged. The breakdown of causality
in the BMPV spacetime is associated with a breakdown of unitarity
in the super conformal field theory 
\cite{Herdeiro:2000ap,Herdeiro:2002ft} (see also \cite{Dyson:2006ia}).

The BMPV solution may be considered
as a subset of a more general family of solutions
found by Chong, Cvetic, L\"u and Pope \cite{Chong:2005hr}.
This more general family of solutions exhibits close similarities
and, in particular, a generic presence of CTCs.
However, it also contains further supersymmetric
black holes, where naked CTCs can be avoided, 
and in addition new topological solitons.

Along these lines 
Cvetic, Gibbons, L\"u and Pope \cite{Cvetic:2005zi}
gave a thorough analysis of further exact solutions of gauged supergravities
in four, five and seven dimensions.
Again, these solutions in general may possess CTCs, but interesting
regular black hole and soliton solutions are present as well.

Here we revisit the properties of the BMPV spacetime by constructing the
complete analytical solution of the geodesic equations 
in this spacetime. 
We systematically study the motion of massive and massless test particles
and analyze the trajectories with analytical methods 
based on the theory of elliptic functions.
We classify the possible orbits and present examples of these orbits
to illustrate the various types of motion.

Our paper is structured as follows. In section~\ref{section:metric} we introduce and discuss the metric of the BMPV spacetime. We present the Kretschmann scalar, which reveals the physical singularity at $r=0$ (in our coordinates). Subsequently, we here derive the equations of motion. 

In section~\ref{sec:beta} we discuss the properties of the $\vartheta$-motion and derive the solution of the $\vartheta$-equation. In section~\ref{sec:radial} we solve the radial equation in terms of the Weierstrass' $\wp$-function. We then discuss in detail the properties of the motion in terms of the effective potential in subsection~\ref{sec:pot}. We distinguish between the underrotating case, the overrotating case and the critical case, which are definded in terms of the values of the rotation parameter $\omega$. We then discuss the corresponding dynamics of massive and massless test particles for these cases. In subsection~\ref{sec:diag} we exemplify the properties of motion with parameteric diagrams from the radial polynomial.
 
In section~\ref{sec:varphi} and section~\ref{sec:psi} we solve the differential equations for the $\varphi$ and $\psi$ equations in terms of Weierstrass' functions. We solve the $t$ equation in section~\ref{sec:time}. In section~\ref{sec:ctc} we address causality. 
To illustrate our analytical solutions we present
in section~\ref{section:orbits} various types of trajectories.
We show two-dimensional orbits in the $\theta=\pi/2$ plane in
section~\ref{section:2dorbits}, and three-dimensional projections
of four-dimensional orbits in section~\ref{section:3dorbits}. 
In the last section we conclude.

\section{The metric and the equations of motion}~\label{section:metric}

\subsection{The metric}

The five dimensional metric describing the BMPV spacetime 
can be expressed as follows \cite{Breckenridge:1996is}
\begin{equation}
ds^2 = - \left( 1 - \frac{ \mu }{r^2} \right)^2 \left( dt - \frac{\mu\omega}{(r^2-\mu)} (\sin^2\vartheta d\varphi - \cos^2\vartheta d\psi ) \right)^2 + \left( 1 - \frac{ \mu }{r^2} \right)^{-2} {dr^2} + {r^2} \left( d\vartheta^2 + \sin^2\vartheta d\varphi^2 + \cos^2\vartheta d\psi^2 \right) \label{metric} \ .
\end{equation}
The parameter $\mu$ is related to the charge
and to the mass of these solutions,
while the parameter $\omega$ is related to their two equal magnitude angular momenta.
The coordinates $r$, $\vartheta$, $\varphi$, $\psi$ represent a spherical coordinate system,
where the angular coordinates
have the ranges $\vartheta\in [0, \frac{\pi}{2} ]$, $\varphi\in [0, 2\pi )$ and $\psi\in [0, 2\pi )$.

It is convenient to work with a normalized metric of the form
\begin{equation}
ds^2 = - \left( 1 - \frac{ 1 }{r^2} \right)^2 \left( dt - \frac{ \omega}{(r^2-1)} (\sin^2\vartheta d\varphi - \cos^2\vartheta d\psi ) \right)^2 + \left( 1 - \frac{ 1 }{r^2} \right)^{-2} {dr^2} + {r^2} \left( d\vartheta^2 + \sin^2\vartheta d\varphi^2 + \cos^2\vartheta d\psi^2 \right) \label{metric2} \ 
\end{equation}
 with dimensionsless coordinates and parameter
\begin{equation}
\frac{r}{\sqrt{\mu}} \rightarrow r \ , \,  \frac{t}{\sqrt{\mu}} \rightarrow t \ , \, \frac{\omega}{\sqrt{\mu}} \rightarrow \omega \ , \, \frac{ds}{\sqrt{\mu}} \rightarrow ds \ .
\end{equation}
Note, that in order to avoid a complicated notation we retain the same notation for the normalized coordinates and quantities.

The BMPV spacetime is a stationary asymptotically flat spacetime.
It has two hypersurfaces relevant for its physical interpretation in the following discussion.
The hypersurface where $g_{tt}$ vanishes looks like a non-rotating degenerate horizon
located at $r=1$. The second hypersurface is associated with the
causal properties of the spacetime. 
Representing the outer boundary of the region where CTCs arise,
it is located at $r_L=\omega^{1/3}$
and referred to as the velocity of light surface (VLS) 
\cite{Gibbons:1999uv,Herdeiro:2000ap,Herdeiro:2002ft,Cvetic:2005zi}.

For the proper description of the BMPV spacetime we need to distinguish the following three cases:
\begin{enumerate}
\item $\omega<1$: underrotating case
\item $\omega=1$: critical case
\item $\omega>1$: overrotating case
\end{enumerate}
In the underrotating case the BMPV spacetime describes extremal supersymmetric
black holes. Here indeed a degenerate horizon is  located at 
$r=1$, and the VLS is hidden behind the horizon.
Since CTCs arise only inside the VLS, the black hole spacetime outside the horizon
is free of CTCs.

In the overrotating case the surface $r=1$ becomes a timelike hypersurface.
Therefore this surface does not describe a horizon.
It is referred to as a {\sl pseudo-horizon},
whose area would be imaginary. 
In the overrotating case the VLS resides outside the surface $r=1$.
Thus the outer spacetime contains a naked time machine.
Since no geodesics can cross the pseudo-horizon, the spacetime 
represents a {\sl repulson}
\cite{Gibbons:1999uv,Herdeiro:2000ap,Herdeiro:2002ft,Cvetic:2005zi}.

In the critical case the surface $r=1$ has vanishing area.
Here the VLS coincides with this surface $r=1$. 
Thus there is no causality violation in the outer region $r>1$.

The Kretschmann scalar $\mathcal{K}=R^{\alpha \beta \gamma \sigma } R_{\alpha \beta \gamma \sigma}$, where $R^{\alpha \beta \gamma \sigma }$ are the contravariant components of the Riemann tensor, in the BMPV spacetime has the form
\begin{equation}
\mathcal{K} = \frac{(288 r^8+(-384 \omega^2-720) r^6+(1152 \omega^2+508) r^4-904 \omega^2 r^2+136\omega^4)}{r^{16}} \ ,
\end{equation}
which indicates that the spacetime has a physical point-like singularity at $r=0$.
At $r=1$ the Kretschmann scalar is finite.

Our detailed analytical study of the geodesics of neutral particles and light 
in the BMPV spacetime fully supports the 
previous analysis of  the properties of this spacetime.

\subsection{The Hamilton-Jacobi equation}

The Hamilton-Jacobi equation for neutral test particles is of the form (see e.g.~\cite{Misner})
\begin{equation}
-\frac{\partial S}{\partial \lambda} = \frac{1}{2} g^{\alpha \beta} \left( \frac{\partial S}{\partial x^\alpha} \right) \left( \frac{\partial S}{\partial x^\beta} \right) \label{eq:HJ} \ .
\end{equation}
We therefore need the non-vanishing inverse metric components $g^{\alpha \beta}$ given by
\begin{eqnarray}
&& g^{tt} = \frac{\omega^2-r^6}{r^2(r^2-1)^2} \ , \,\, g^{t\varphi} = \frac{\omega}{r^2(r^2-1)^2} \ , \,\, g^{t\psi} = - \frac{\omega}{r^2(r^2-1)^2} \nonumber \ , \\
&& g^{\varphi\varphi} = \frac{1}{r^2 \sin^2 \vartheta} \ , \,\, g^{\psi\psi} = \frac{1}{r^2 \cos^2 \vartheta}  \nonumber \ , \\ 
&& g^{rr} = \frac{ (r^2-1)^2 }{r^4} \ , \,\, g^{\vartheta \vartheta} = \frac{ 1 }{r^2} \ . 
\end{eqnarray}

We search for the solution $S$ of the equations~\eqref{eq:HJ} in the form:
\begin{equation}
S = \frac{1}{2}\delta \lambda - E t + \Phi \varphi + \Psi \psi + S_r(r) + S_\vartheta(\vartheta) \label{eq:S} \ ,
\end{equation}
where $E$ is the conserved energy of a test particle with a mass parameter $\delta$, and $\Phi$ and $\Psi$ are its conserved angular momenta. $\delta=0$ for massless test particles and $\delta=1$ for massive test particles. Since the metric components are functions of the coordinates $r$ and $\vartheta$, we have to separate the equations of motion with respect to these coordinates. $\lambda$ is an affine parameter.

%$\lambda$ is an affine parameter so normalized that $\displaystyle{\frac{d}{d\lambda}=\boldsymbol{p}}$ for the $4$--momentum $\boldsymbol{p}$.

We insert~\eqref{eq:S} into~\eqref{eq:HJ} and get
\begin{eqnarray} 
&& -r^2 \delta -\frac{(\omega^2 -r^6)E^2}{(r^2-1)^2} + 2 \frac{\omega E}{r^2-1} \left( \Phi - \Psi \right) - \frac{(r^2-1)^2}{r^2} S^2_r(r) \nonumber \\
&& = S^2_\vartheta(\vartheta) + \frac{\Phi^2}{\cos^2 \vartheta} + \frac{\Psi^2}{\cos^2 \vartheta}  \label{eq:HJ2} \ .
\end{eqnarray}

Since the left and right hand sides of the equation~\eqref{eq:HJ2} depend only on $r$ and $\vartheta$, respectively, we can equate both sides to a constant ${K}$, the separation constant. We obtain
\begin{eqnarray}
  S^2_\vartheta(\vartheta)  = {K} -   \frac{\Phi^2}{\sin^2 \vartheta} - \frac{\Psi^2}{\cos^2 \vartheta}    &  \equiv & \Theta            \label{eq:THETA} \ , \\
  \frac{(r^2-1)^4}{r^2} S^2_r(r)   =   2 \omega E ( \Phi - \Psi ) (r^2-1) - (\omega^2 -r^6)E^2 - ({K}+r^2\delta) (r^2-1)^2    & \equiv & R       \label{eq:R} \ , 
\end{eqnarray}
where we have introduced new functions $\Theta$ and $R$.

We can write the action $S$ (equation~\eqref{eq:S}) in the form
\begin{equation}
S = \frac{1}{2}\delta \lambda - E t + \Phi \varphi + \Psi \psi + \int_r{ \frac{r\sqrt{R}}{(r^2-1)^2} dr } + \int_\vartheta{ \sqrt{\Theta} d\vartheta } \label{eq:S2} \ .
\end{equation}

Following the standard procedure we differentiate equation~\eqref{eq:S2} with respect to the constants ${K}$, $\delta$, $\Phi$, $\Psi$ and $E$. The result is a constant which can be set zero. Combining the derived differential equations we get the Hamilton-Jacobi equations in the form
\begin{eqnarray}
&&\frac{rdr}{d\tau} = \sqrt{R} \ ,  \label{reqn1} \\
&& \frac{d\vartheta}{d\tau} = \sqrt{\Theta} \ , \label{varthetaeqn1} \\ 
&& \frac{d\varphi}{d\tau} = \frac{\omega E}{r^2-1} - \frac{\Phi}{\sin^2\vartheta}   \ , \label{varphieqn1} \\ 
&& \frac{d\psi}{d\tau} = - \frac{\omega E}{r^2-1} - \frac{\Psi}{\cos^2\vartheta}   \ , \label{psieqn1} \\ 
&& \frac{dt}{d\tau} = \frac{ \omega(\Phi-\Psi)(r^2-1) - E(\omega^2 - r^6) }{(r^2-1)^2} \label{teqn1} \ , 
\end{eqnarray}
where $\Theta$ and $R$ are given by~\eqref{eq:THETA} and~\eqref{eq:R}, and $\tau$ is a new affine parameter defined by~\cite{Mino:2003yg}
\begin{equation}
d\tau = \frac{d\lambda}{r^2} \ . \label{eq:tau}
\end{equation}

\section{Properties of the motion}~\label{section:motion}

\subsection{The $\vartheta$-equation}~\label{sec:beta}

\subsubsection{The restrictions from the $\vartheta$-equation}

Consider now the $\vartheta$-equation~\eqref{varthetaeqn1} with $\Theta$ defined in~\eqref{eq:THETA}
\begin{equation}
d\tau = \frac{d\vartheta}{ \sqrt{\Theta} } \ , \quad \Theta={K} -   \frac{\Phi^2}{\sin^2 \vartheta} - \frac{\Psi^2}{\cos^2 \vartheta}    \ . \label{varthetaeqn2}
\end{equation}

%To have real solutions of the angle $\vartheta$ we require $\Theta\geq 0$. This implies ${K}\geq 0$.

We introduce a new variable $\xi=\cos^2\vartheta$. The equation~\eqref{varthetaeqn2} reduces to
\begin{equation}
d\tau = - \frac{d\xi}{ 2 \sqrt{\Theta_\xi} } \ , \quad \Theta_\xi= -{K}\xi^2 + (K+\Psi^2-\Phi^2)\xi -\Psi^2  = \sum^2_{i=0}{b_i\xi^i}     \ , \label{xieqn1}
\end{equation}
with
\begin{equation}
b_2=-K \ , \quad b_1=K+\Psi^2-\Phi^2 \equiv {K}+{A}{B} \quad \mbox{and} \quad b_0=-\Psi^2\equiv -\frac{({A}+{B})^2}{4} \ , \label{b_coeffs} 
\end{equation}
where we introduced 
\begin{equation}
A= \Psi-\Phi \ , \quad B = \Psi+\Phi  \label{AB} \ .
\end{equation}

The discriminant $D_\xi$ of $\Theta_\xi$~\eqref{xieqn1} takes the form
\begin{equation}
D_\xi =  b_1^2 - 4 b_2 b_0  = ({K}-{A}^2)({K}-{B}^2) \label{Dxi} \ .
\end{equation}

The roots of the polynomial $\Theta_\xi$~\eqref{xieqn1} read
\begin{equation}
\xi_{1,2} = \frac{1}{2{K}} \left(  {K} +{A}{B} \mp \sqrt{ D_\xi } \right) \ , \label{zeros_xi}
\end{equation}

The discriminant $D_\beta$ must be positive or zero for the solutions~\eqref{zeros_xi} to be real. In~\eqref{Dxi} two cases are possible:
\begin{equation} {K} \geq {A}^2  \,\,\, \cup \,\,\, {K} \geq {B}^2  \label{condj1} \end{equation} 
or
\begin{equation} {K} < {A}^2  \,\,\ \cup  \,\,\, {K} < {B}^2 \label{condj2} \ . \end{equation}

For the upcoming analysis we keep in mind that since $\xi=\cos^2\beta$ the condition 
\begin{equation}
0 \leq \xi\leq 1 \, \label{xicond}
\end{equation} 
must be fulfilled. Under this condition we will see that only the case~\eqref{condj1} is relevant.

At first we consider ${K}>0$.

Let $0< m \leq 1$ and $0 < n \leq 1$ and ${A} = m \sqrt{{K}}$ and ${B} = n \sqrt{{K}}$. Inserting this into~\eqref{zeros_xi} we get
\begin{equation}
\xi_{1,2} = \frac{1}{2}(1+mn \mp \sqrt{(1-m^2)(1-n^2)}) \ . \label{zeros_xi2}
\end{equation}
Then considering the limits for $m\rightarrow 0$ and $m\rightarrow 1$ we obtain: 
\begin{eqnarray}
&& \lim_{m\rightarrow 0} \xi_{1,2} = \frac{1}{2}(1-\sqrt{1-n^2}) \ , \\
&& \lim_{m\rightarrow 1} \xi_{1,2} = \frac{1}{2}(1+n) \ .
\end{eqnarray}
If we take the limits for $n$ instead of $m$, we can simply replace $n$ by $m$ in the result above. 

Taking into account the conditions on $m$ and $n$, we observe that $\xi_{1,2}$ lie in the allowed region~\eqref{xicond}. 

If $m=n$ then 
\begin{equation}
\xi_{1} = m^2 \quad \text{and} \quad \xi_{2}=\frac{1}{2} \ . 
\end{equation}

Let now $0< m < 1$ and $0 < n < 1$ and ${A} = \frac{1}{m} \sqrt{K}$ and ${B} = \frac{1}{n} \sqrt{{K}}$ (${K}$ still positive). Inserting this into~\eqref{zeros_xi} we get
\begin{equation}
\xi_{1,2} = \frac{1+mn \mp \sqrt{(1-m^2)(1-n^2)}}{2mn} \ . \label{zeros_xi3}
\end{equation}
Taking the limit for $m\rightarrow 1$ yields:
\begin{equation}
\lim_{m\rightarrow 1} \xi_{1,2} = \frac{1}{2}(1+\frac{1}{n}) \ .
\end{equation}
Again, if we take the limit for $n$ instead of $m$, we can simply replace $n$ by $m$ in the result above. 

 Contrary to the case above, the condition~\eqref{xicond} is not fulfilled for $0<n<1$, since the values of $\xi$ become larger than one. 

If $m=n$ then 
\begin{equation}
\xi_{1} = 1 \quad \text{and} \quad \xi_{2}=\frac{1}{m^2} \ .
\end{equation}

In this case only $\xi_{1} = 1$ is an eligible root. Since both roots of the function $\Theta_\xi$ define the boundaries of the $\vartheta$--motion, they must satisfy~\eqref{xicond}.

If $K<0$ and fulfills the conditions~\eqref{condj2}, then with the substitution ${A} = \frac{1}{m} \sqrt{-K}$ and ${B} = \frac{1}{n} \sqrt{{-K}}$ where $0< m < 1$ and $0 < n < 1$ one can show that one zero is negative and the other is larger than one. Both cases do not satisfy the condition~\eqref{xicond}.

These observations show that ${K}$, ${A}$ and ${B}$ must satisfy the conditions~\eqref{condj1}. 
%or if $K<A^2$ $\cup$ $K<B^2$ then the ratio of $A^2$ and $B^2$ to $K$ must be equal which guaranties one possible zero $\xi=1$.

\subsubsection{$\vartheta(\tau)$-solution}

Thus, the highest coefficient $b_2 = -K$ in $\Theta_\xi$~\eqref{xieqn1} is negative. In this case the differential equation~\eqref{xieqn1} can be integrated as
\begin{equation}
\tau - \tau_0 =  \left(  \frac{1}{2\sqrt{-b_2} } \arcsin{  \frac{2b_2\xi + b_1}{\sqrt{D_\xi}}  } \right) \Biggl|^{\xi(\tau)}_{\xi_{0} }  \ . \label{betaeqn4}
\end{equation}

Here and later on the index $0$ means an initial value.

To find $\vartheta=\arccos(\pm\sqrt{\xi})$ as a function of $\tau$ we invert the solution~\eqref{betaeqn4}. Thus,
\begin{equation}
\vartheta (\tau) = \arccos\left( \pm\sqrt{ \frac{1}{2b_2} \left(  \sqrt{D_\xi} \sin{( 2\sqrt{-b_2}(\tau-\tau^\prime) )}  - b_1  \right) } \right) \ , \label{betaeqn5}
\end{equation}
where 
\begin{equation}
\tau^\prime =  \tau_0 - \frac{1}{2\sqrt{-b_2} } \arcsin{  \frac{2b_2\xi_0 + b_1}{\sqrt{D_\xi}}  } \ . \label{betaeqn6}
\end{equation}

\subsection{The radial equation}~\label{sec:radial}

The differential equation~\eqref{reqn1} contains a polynomial of order 6 in $r$ on the right hand side:
\begin{equation}
\left(\frac{dr}{d\tau}\right)^2 = \frac{1}{r^2} \sum^{3}_{i=0}{ a_i r^{2i} } \ , \label{reqn1_1}
\end{equation}
where 
\begin{eqnarray}
&& a_3 = E^2-\delta  \ , \quad a_2 =  2\delta  - K  \ , \nonumber \\
&& a_1 =   2 K - 2\omega E {A} - \delta   \ , \quad a_0 = - K+{A}^2 - (\omega E - {A} )^2 \label{req_coeff} \ .
\end{eqnarray}

%Taking into account the condition~\eqref{condj1} we notice that the coefficient $a_0$ from~\eqref{req_coeff} is negative:
%\begin{equation}
%a_0 = - 4(j^2-j^2_R) - (\omega E - 2{A})^2 \label{a0_coeff} \ .
%\end{equation}

Next, we introduce the new variable $x=r^2$. Then the polynomial on the right hand side in equation~\eqref{reqn1_1} becomes of order $3$:
\begin{equation}
\left(\frac{dx}{d\tau}\right)^2 = \sum^{3}_{i=0}{ 4 a_i x^{i} } \equiv P(x) \ , \label{reqn2}
\end{equation}
with the coefficients $a_i$ defined by~\eqref{req_coeff}.

To reduce the equation~\eqref{reqn2} to the Weierstrass form we use the transformation $x = \frac{1}{4a_3}\left(4y - \frac{4a_2}{3}\right)$. We get:
\begin{equation}
d\tau = \frac{dy}{\sqrt{P_3(y)}} \ , \quad \mbox{with} \quad P_3(y) = 4y^3 - g_2 y -g_3 \ , \label{reqn3}
\end{equation}
where
\begin{equation}
g_2=\frac{4a_2^2}{3} - 4a_1 a_3 \, , \qquad  g_3=\frac{a_1 a_2 a_3}{3} - 4 a_0 a_3^2 - \left(\frac{ 2 a_2 }{3} \right)^3 \ .
\end{equation}
The differential equation \eqref{reqn3} is of elliptic type and is solved by the Weierstra{\ss}' $\wp$--function~\cite{Markush}
\begin{equation}
y(\tau) = \wp\left(\tau - \tau^\prime; g_2, g_3\right) \ , \label{soly}
\end{equation}
where 
\begin{equation}
\tau^\prime=\tau_{ 0 }+\int^\infty_{y_{ 0 }}{\frac{dy}{\sqrt{4y^3-g_2y-g_3}}} \, \label{tauprime}
\end{equation}
with $y_{ 0 }=  a_3 r^2_{ 0 } + \frac{a_2}{3}$. $\tau_0$ and $r_0$ denote the initial values.

Then the solution of~\eqref{reqn2} acquires the form
\begin{equation}
r (\tau) = \sqrt{ \frac{1}{a_3} \left( \wp\left(\tau - \tau^\prime; g_2, g_3\right) - \frac{a_2}{3} \right) } \ . \label{solr}
\end{equation}
In~\eqref{solr} we choose the positive sign of the square root, since the singularity located at $r=0$ prohibits particles to reach negative radial values.

\subsubsection{ Properties of the motion. Effective potential }~\label{sec:pot}

The singularity is located at $r=0$, and the degenerate horizon 
or pseudo-horizon at $r=1$. 
For physical motion the values of $r$ (and $x=r^2$) must be real and positive therefore. 

We define the effective potential $V^\pm_{\rm eff}$ from~\eqref{reqn1_1} via
\begin{equation}
\left(\frac{dr}{d\tau}\right)^2 = r^4 \Delta_\omega (E-V^+_{\rm eff})(E-V^-_{\rm eff}) \ , \label{rpot}
\end{equation}
where $\Delta_\omega=1-\frac{\omega^2}{r^6}$. $\Delta_\omega=0$ corresponds
to the VLS.
The effective potential then reads
\begin{equation}
V^\pm_{\rm eff} = \frac{1}{r^4}\frac{\Delta}{\Delta_\omega} \left(  \omega {A} \pm \sqrt{ \Delta_{\rm eff} }  \right)  \,\,\, \text{with} \,\,\, \Delta_{\rm eff}=\omega^2 {A}^2 + r^6\Delta_\omega (K +r^2 \delta)  \ , \label{rpot1}
\end{equation}
where $\Delta=1-\frac{1}{r^2}$ and $\Delta=0$ describes the horizon or pseudo-horizon,
while ${A}=\Psi-\Phi$~\eqref{AB} is a combination of the angular momenta of the
test particle. 
A principal condition for physical $r$ values (real and positive) to exist is the positiveness of the RHS of~\eqref{rpot}. The counterpart of it will define forbidden regions for a test particle in the effective potential. The limit of the effective potential at infinity is defined by the test particle's mass parameter $\delta$:
\begin{equation}
\lim_{x\rightarrow\infty}{V^\pm_{\rm eff}} = \pm \sqrt{\delta}  \ . \label{rpot_limit}
\end{equation}

From the form of the potential~\eqref{rpot1} we recognize that the term $\Delta_\omega $ in the denominator together with the term $r^4$ lead to divergencies. Since $r=0$ is a physical singularity, we concentrate on the divergency caused by $\Delta_\omega \rightarrow 0$, i.e. $x\rightarrow \sqrt[3]{\omega^2}$ with $x=r^2$. Consider the Laurent series expansion of the potential~\eqref{rpot1} in the vicinity of $x=\sqrt[3]{\omega^2}$: 
\begin{equation}
V^\pm_{\rm eff} = \frac{1}{3} \frac{ (\omega {A} \pm \sqrt{\omega^2 {A}^2}) (\sqrt[3]{\omega^2} -1 ) }{ \sqrt[3]{\omega^4} (x - \sqrt[3]{\omega^2}) } + \, \mathrm{holomorphic \,\, part}  \ . \label{rpot2}
\end{equation}

We are interested in the coefficient at $(x-\sqrt[3]{\omega^2})^{-1}$, since the other terms are holomorphic. Analysing~\eqref{rpot2} we see that there are two factors which define the character of the potentials $V^{\pm}_{\rm eff}$~\eqref{rpot1}. The first one is the direction from which $x$ approaches the value $\sqrt[3]{\omega^2}$. In addition to this, the sign of the factor $\sqrt[3]{\omega^2} -1$, being a second factor, defines the final character of the potential and herewith the properties of the motion. In table~\ref{tab1} we show the asymptotic behaviour of the potential~\eqref{rpot1}.

\renewcommand{\arraystretch}{1.5}
\begin{table}[h]
\begin{center}
\begin{tabular}{c|c|c|c|}\cline{2-4}
& \multicolumn{3}{ |c| }{First case: ${A}>0$}  \\ \cline{2-4} 
                               & $x\rightarrow \sqrt[3]{\omega^2}$ & $\sqrt[3]{\omega^2}$ & limit \\ \hline\hline
\multicolumn{1}{ |c| }{\multirow{2}{*}{ $V^+_{\rm eff}$}}  & left      & \multirow{4}{*}{ $<1$ } & $\infty$  \\ \cline{2-2}\cline{4-4}
\multicolumn{1}{ |c|  }{}
                               & right     &                         & $-\infty$ \\ \cline{1-2}\cline{4-4}
\multicolumn{1}{ |c| }{\multirow{2}{*}{ $V^-_{\rm eff}$}}  & left      &                         & \multirow{2}{*}{ $w$ } \\ \cline{2-2}
\multicolumn{1}{ |c|  }{}
                               & right     &                         &  \\ \hline\hline
%%%%%%%%
\multicolumn{1}{ |c| }{\multirow{2}{*}{ $V^+_{\rm eff}$}}  & left      & \multirow{4}{*}{ $>1$ } & $-\infty$  \\ \cline{2-2}\cline{4-4}
\multicolumn{1}{ |c|  }{}
                               & right     &                         & $+\infty$ \\ \cline{1-2}\cline{4-4}
\multicolumn{1}{ |c| }{\multirow{2}{*}{ $V^-_{\rm eff}$}}  & left      &                         & \multirow{2}{*}{ $w$ } \\ \cline{2-2}
\multicolumn{1}{ |c|  }{}
                               & right     &                         &  \\ \hline\hline
\end{tabular}
\hspace{1cm}
\begin{tabular}{c|c|c|c|}\cline{2-4}
& \multicolumn{3}{ |c| }{Second case: ${A}<0$}  \\ \cline{2-4}
                               & $x\rightarrow \sqrt[3]{\omega^2}$ & $\sqrt[3]{\omega^2}$ & limit \\ \hline\hline
\multicolumn{1}{ |c| }{\multirow{2}{*}{ $V^+_{\rm eff}$}}  & left      & \multirow{4}{*}{ $<1$ } & \multirow{2}{*}{ $w$ } \\ \cline{2-2}
\multicolumn{1}{ |c|  }{}
                               & right     &                         &  \\ \cline{1-2}\cline{4-4} 
\multicolumn{1}{ |c| }{\multirow{2}{*}{ $V^-_{\rm eff}$}}  & left      &                         & $-\infty$  \\ \cline{2-2}\cline{4-4} 
\multicolumn{1}{ |c|  }{}
                               & right     &                         & $+\infty$ \\ \hline\hline
%%%%%%%%
\multicolumn{1}{ |c| }{\multirow{2}{*}{ $V^+_{\rm eff}$}}  & left      & \multirow{4}{*}{ $>1$ } & \multirow{2}{*}{ $w$ } \\ \cline{2-2}
\multicolumn{1}{ |c|  }{}
                               & right     &                         &  \\ \cline{1-2}\cline{4-4}
\multicolumn{1}{ |c| }{\multirow{2}{*}{ $V^-_{\rm eff}$}}  & left      &                         & $\infty$  \\ \cline{2-2}\cline{4-4} 
\multicolumn{1}{ |c|  }{}
                               & right     &                         & $-\infty$ \\ \hline\hline 
\end{tabular}
\caption{ Behaviour of the effective potential~\eqref{rpot1} with respect to the divergency at $\Delta_\omega =0$ (VLS at $x=\sqrt[3]{\omega^2}$) (cp. eq.~\eqref{rpot2}). Due to this a potential barrier forms which may either lie behind the degenerate horizon 
or in front of the pseudo-horizon,
 allowing for planetary bound orbits (see discussion in the text). Inversion of the sign of ${A}$ mirrors the $V^+_{\rm eff}$ and $V^-_{\rm eff}$ parts w.r.t.~the $x$-axis but does not change the general properties of the effective potential. Here $w=\frac{1}{2} \frac{1-\sqrt[3]{\omega^2}}{{A} \omega} \left( K + \delta \sqrt[3]{\omega^2} \right)$. When the minus (for ${A}>0$) or the plus part (for ${A}<0$) of the effective potential approaches $x=\sqrt[3]{\omega^2}$, which is a removable discontinuity in this case, it has a value $w$ there.  \label{tab1}}
\end{center}
\end{table}

\begin{figure*}[th!]
\begin{center}
\subfigure[][$\omega=0.7, {K}= 1$, ${A}=0.1 \sqrt{{K}}$]{\label{pot1}\includegraphics[width=7cm]{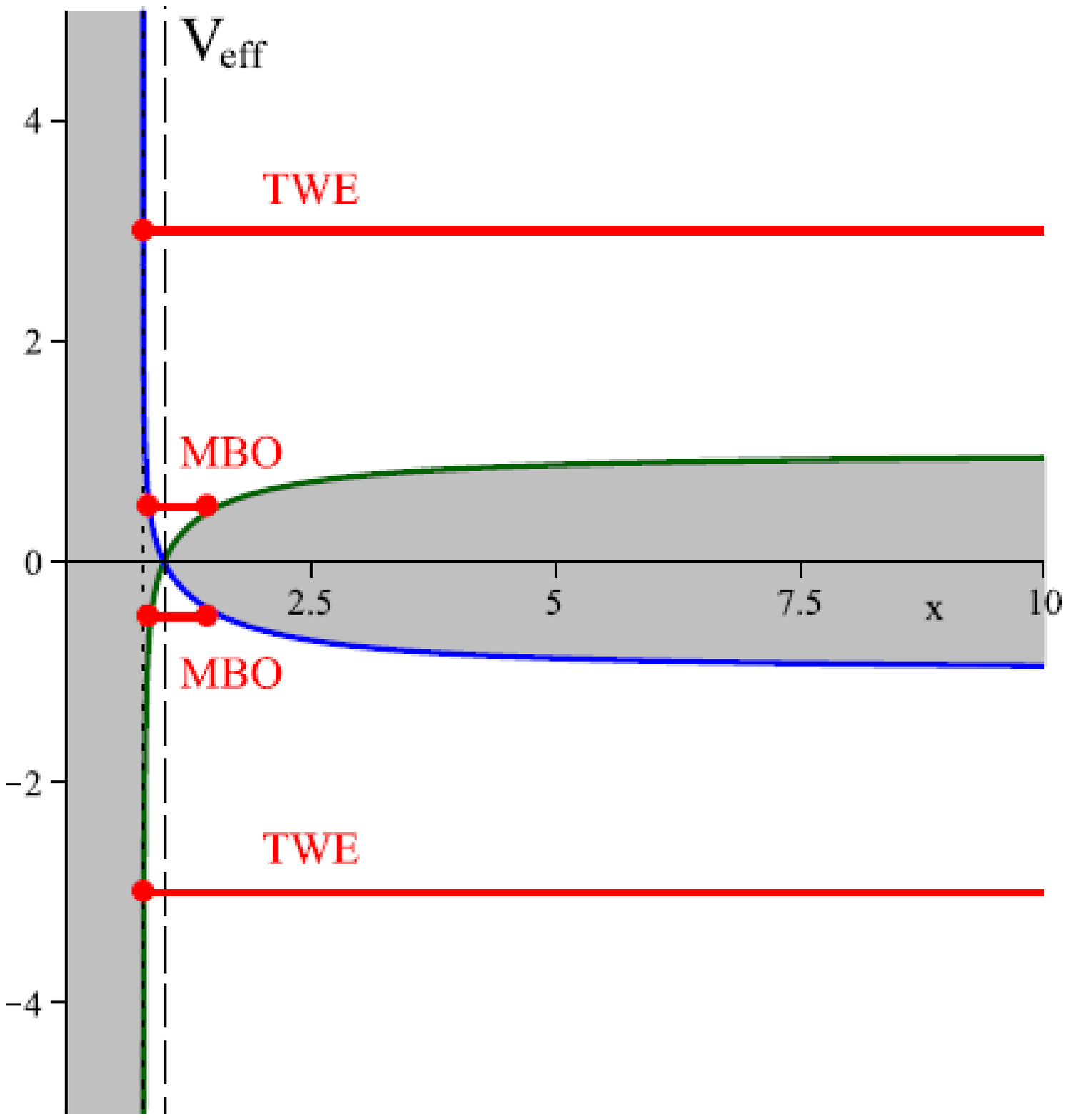}}
\subfigure[][$\omega=0.9, {K}= 36$, ${A}=0.1  \sqrt{{K}}$]{\label{pot2}\includegraphics[width=7cm]{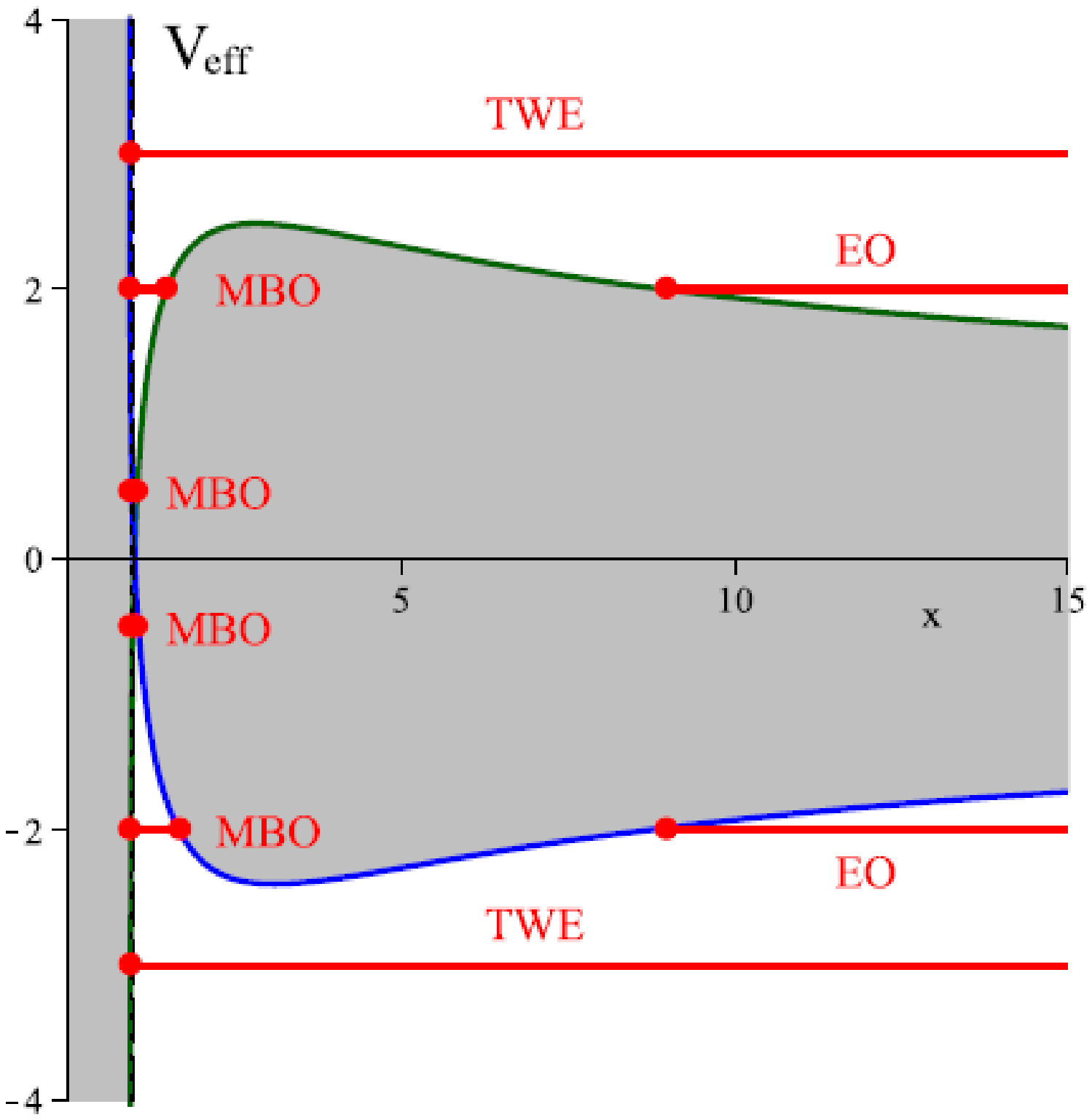}}
\subfigure[][$\omega=1.1, {K}= 10$, ${A}=-0.1  \sqrt{{K}}$; $|{A}|<|{A}^c|$]{\label{pot3}\includegraphics[width=7cm]{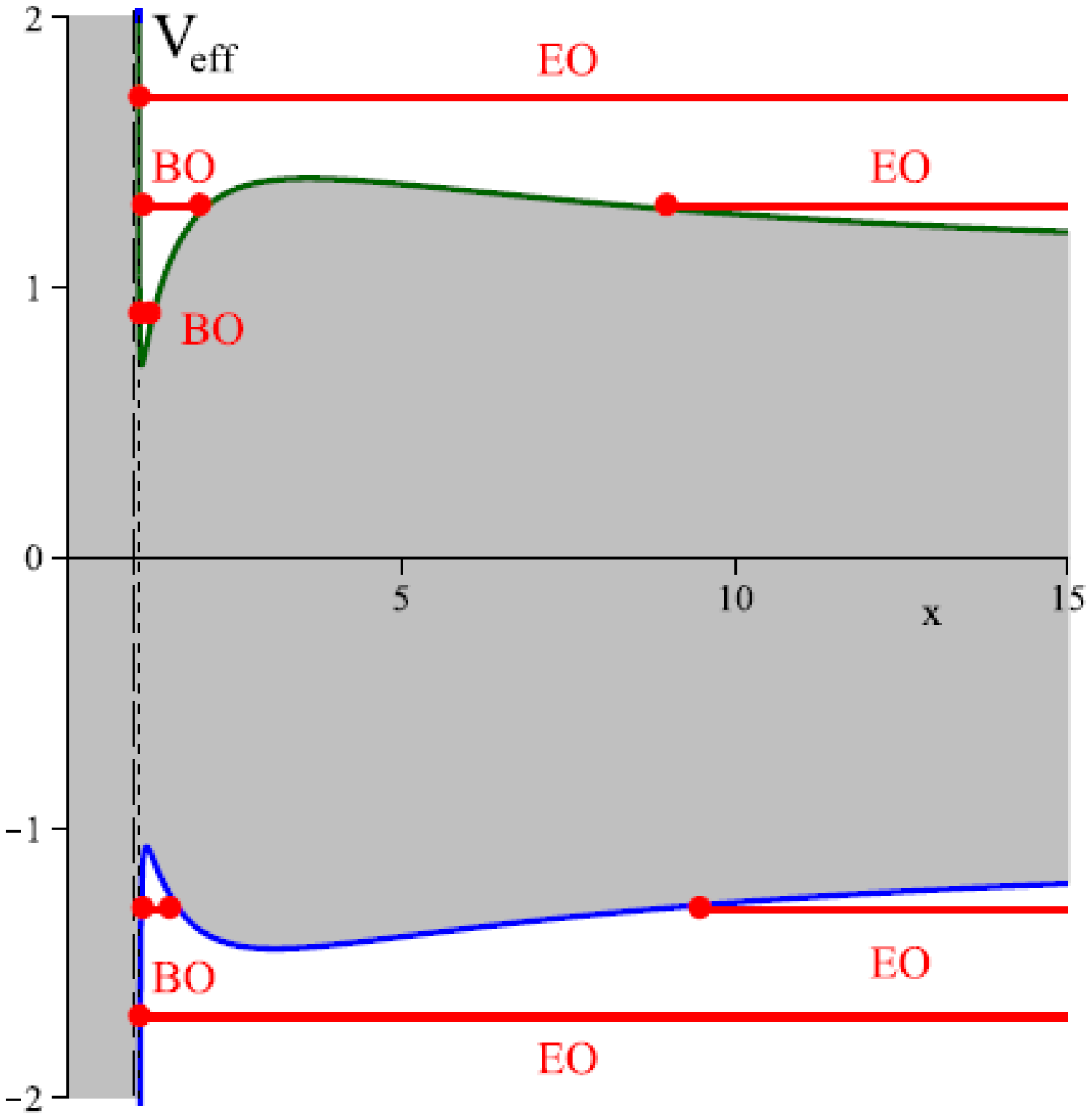}}
\subfigure[][$\omega=2.1, {K}= 0.16$, ${A}=-0.1  \sqrt{{K}}$; $|{A}|<|{A}^c|$]{\label{pot4}\includegraphics[width=7cm]{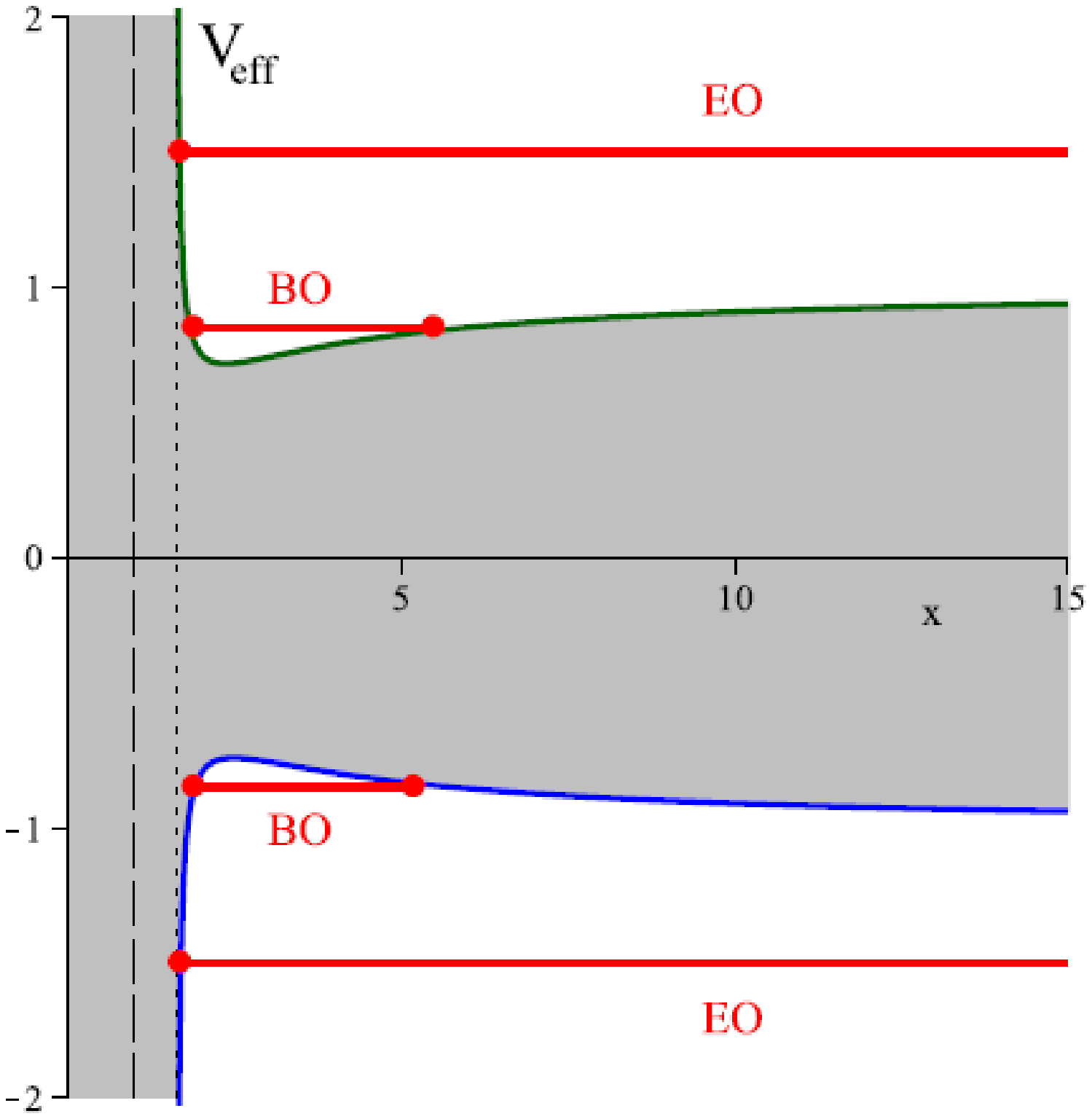}}
\end{center}
\caption{ Figures~\subref{pot1} and~\subref{pot2} show the effective potentials for $\omega<1$
(underrotating case) and $\boldsymbol{\delta=1}$. Figures~\subref{pot3} and~\subref{pot4} illustrate them for $\omega>1$ (overrotating case). The dashed line visualizes the horizon/pseudo-horizon at $x=1$ ($x=r^2$). The dotted line at $x=\sqrt[3]{\omega^2}$ (VLS) shows, where the effective potential~\eqref{rpot1} diverges (see the discussion in sec.~\ref{sec:pot}). The grey region indicates the forbidden regions where the RHS of~\eqref{rpot} becomes negative. For $\omega<1$ a planetary bound orbit is not possible (figs.~\subref{pot1} and~\subref{pot2}), since a test particle will necessarily cross the degenerate horizon located at $x=1$. This happens because the potential asymptotically approaches $x=\sqrt[3]{\omega^2}$, which is smaller than $1$ in this case (i.e., lies behind the horizon). In the wedge of the figures~\subref{pot1} and~\subref{pot2} for $E=0$ a circular orbit  at $x=1$ is possible. For $\omega>1$ the turning point of an orbit will lie in front of the pseudo-horizon, since in this case the value $x=\sqrt[3]{\omega^2}$ (VLS), which the potential asymptotically approaches, is larger than 1 (figs.~\subref{pot3} and~\subref{pot4}). Thus, for $\omega>1$ planetary bound orbits are possible. In the figs.~\subref{pot3} and~\subref{pot4} the value of ${A}$ is smaller than the critical value defined in the equation~\eqref{cond_jR}. Hence, the intersection point of the $V^\pm_{\rm eff}$ parts of the potential at $x=1$ belongs to the forbidden grey region. See also table~\ref{tab2} for a schematical representation of the orbit types. \label{fig:pots}}
\end{figure*}

Consider at first positive ${A}$. For $\omega<1$ the potential barrier extends all over the $V_{\rm eff}$-axis: when $x$ decreases (i.e. when $x\rightarrow \sqrt[3]{\omega^2}$ from the right) it stretches to $-\infty$ and for increasing $x$ (i.e. when $x\rightarrow \sqrt[3]{\omega^2}$ from the left) it stretches to $\infty$. The $V^+_{\rm eff}$ part of the effective potential is responsible for this effect, while $V^-_{\rm eff}$ has a finite limit. $V^\pm_{\rm eff}$ vanishes at $x=1$ ($\Delta=0$) which corresponds to the location of the horizon or pseudo-horizon. Both parts of the potential cross at this point. We show an example in the figure~\ref{fig:pots}\subref{pot1} and~\subref{pot2}. The grey region shows the forbidden regions where the RHS of~\eqref{rpot} becomes negative.

It is important to note here that the positive zeros of $\Delta_{\rm eff}$ and the vanishing of $\Delta_\omega$ define the discontinuities of the potential~\eqref{rpot1} relevant for the physical motion. In general, by Descartes' rule of signs, provided the condition~\eqref{condj1} is fulfilled, $\Delta_{\rm eff}$ has one positive zero (and at most $3$ negative zeros or $1$ negative and $2$ complex conjugate with negative real part). It means that the region of $V^\pm_{\rm eff}$ lying between the origin of the coordinates and this point contains complex values and is forbidden entirely. The region to the right of this point contains real values of the potentials $V^\pm_{\rm eff}$ and is generally allowed. The possible regions with physical motion will be finally determined by the positive or vanishing RHS of~\eqref{rpot}.

\begin{figure*}[th!]
\begin{center}
\subfigure[][$\omega=0.7, {K}= 1$, ${A}=\sqrt{{K}}$]{\label{pot1jR}\includegraphics[width=7cm]{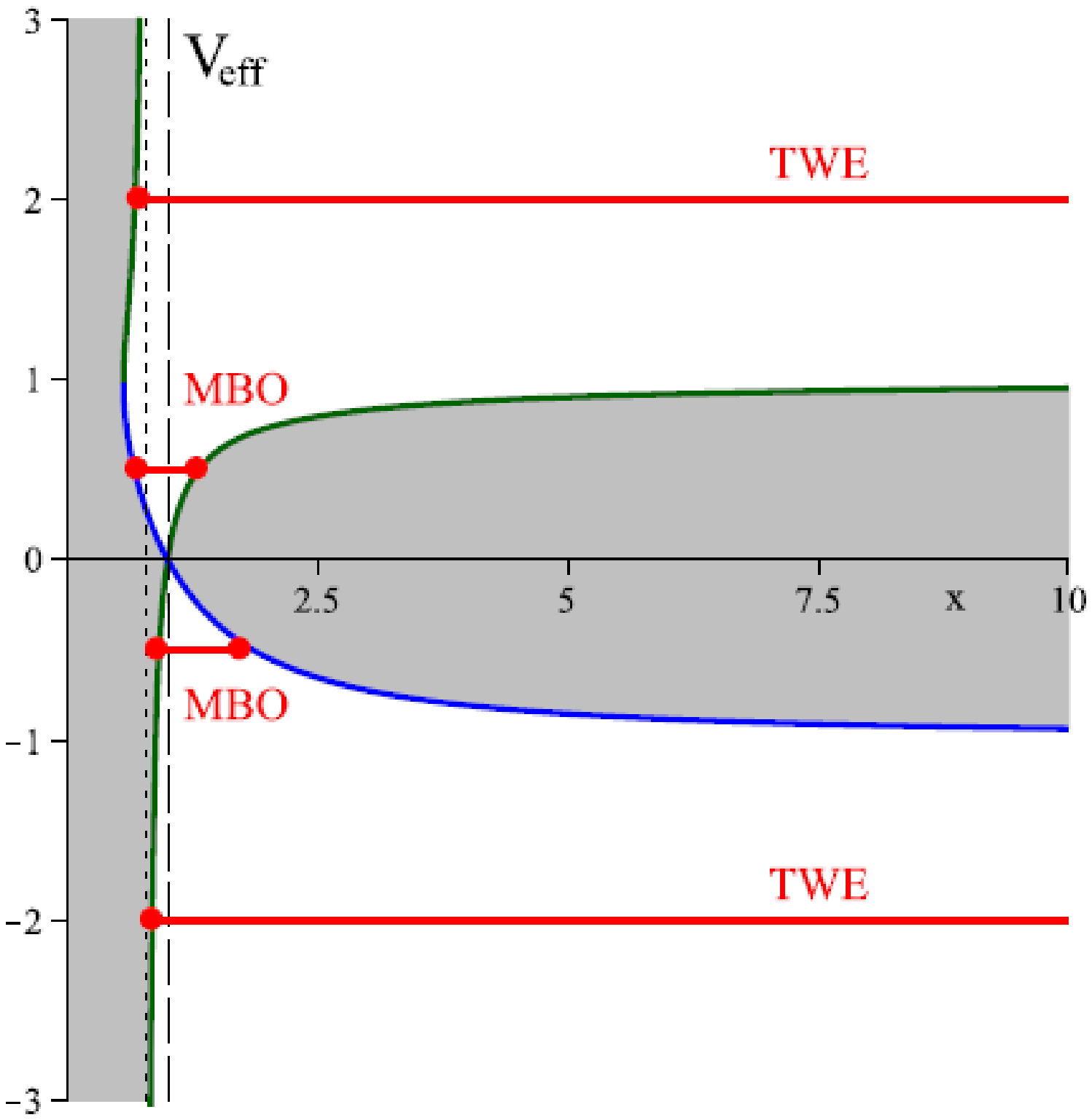}}
\subfigure[][$\omega=0.9, {K}= 36$, ${A}=\sqrt{{K}}$]{\label{pot2jR}\includegraphics[width=7cm]{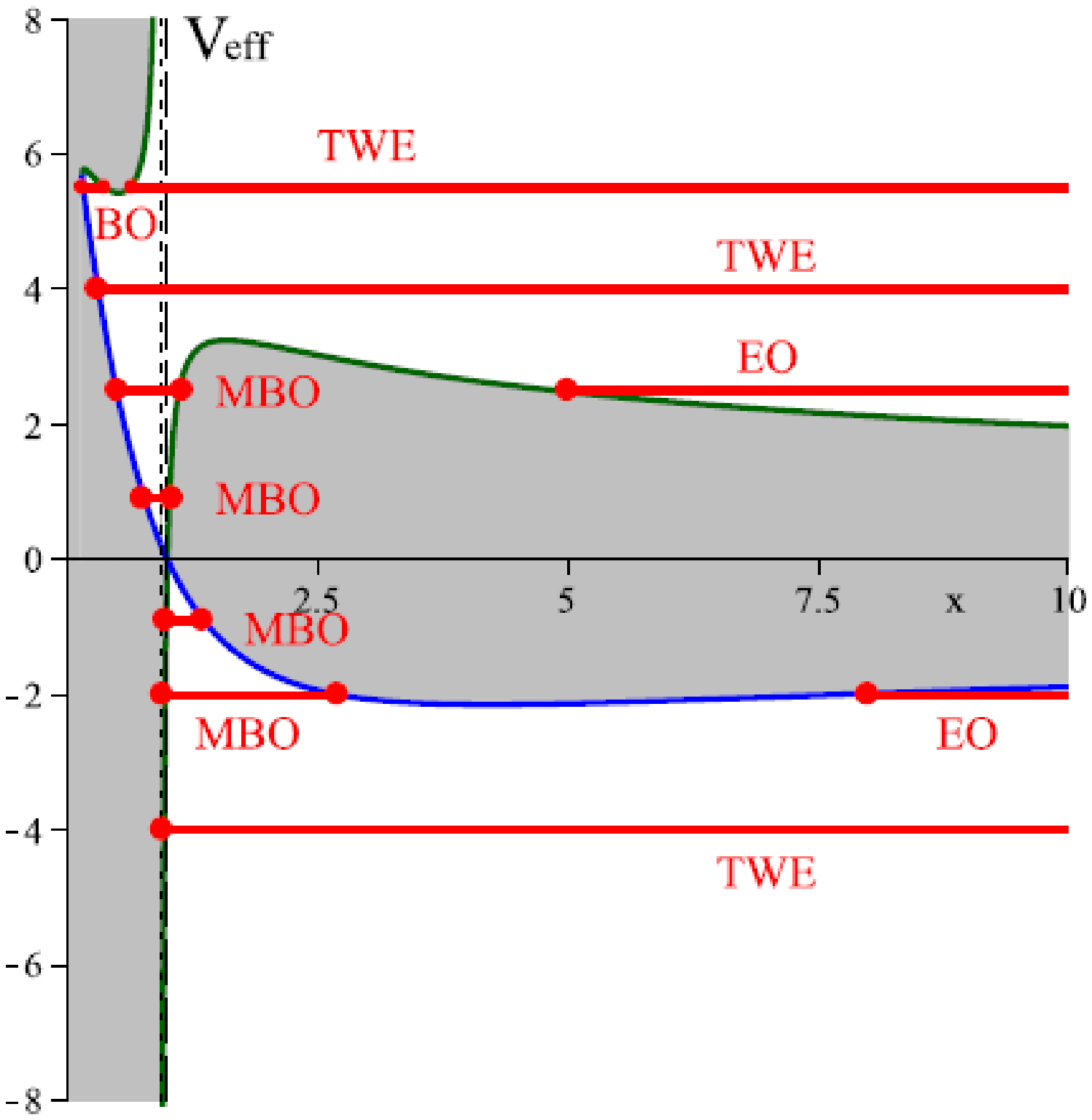}}
\subfigure[][$\omega=1.1, {K}= 10$, ${A}= \sqrt{{K}}$, $|{A}|>|{A}^c|$]{\label{pot3jR}\includegraphics[width=7cm]{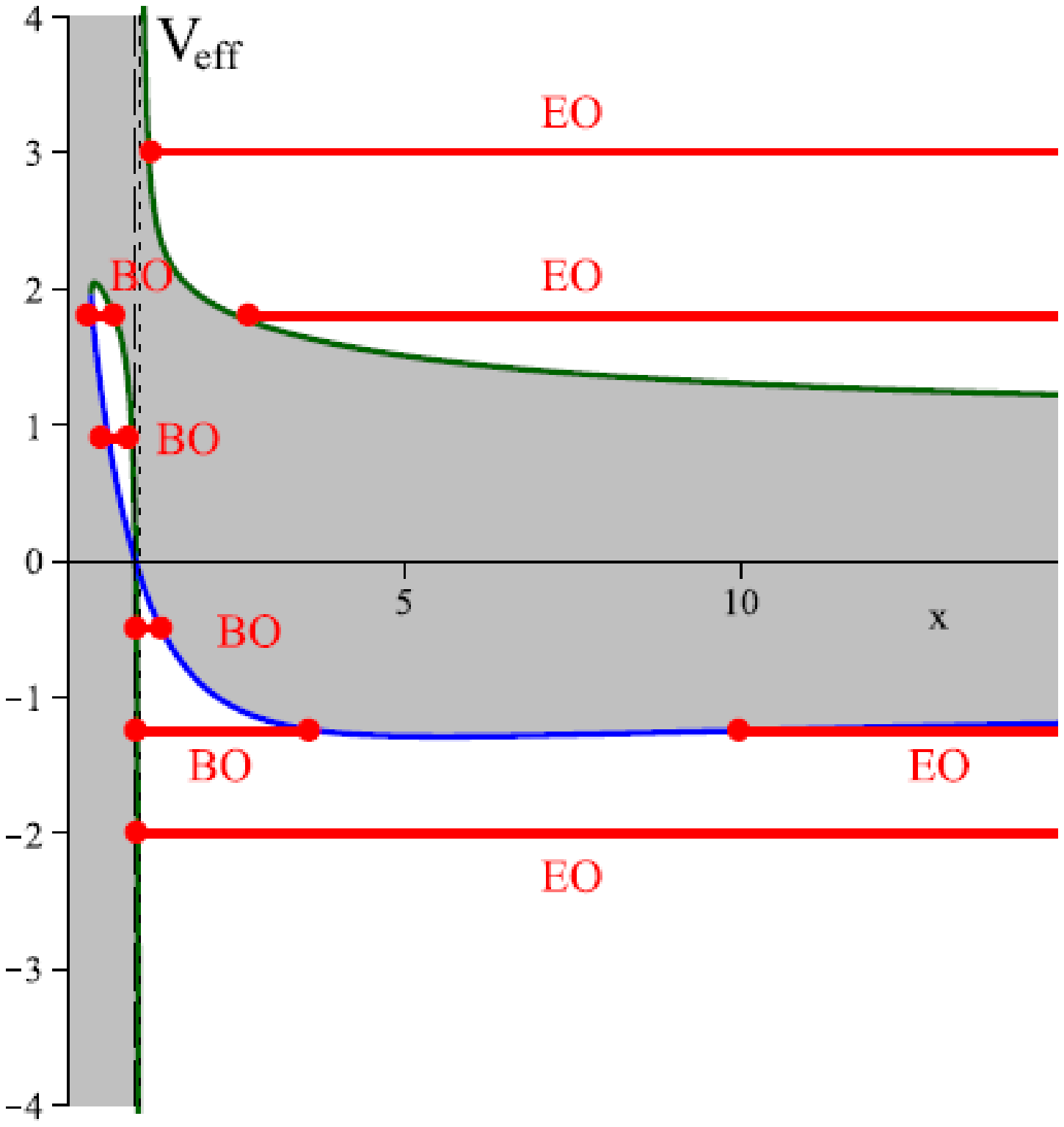}}
\subfigure[][$\omega=2.1, {K}= 400$, ${A}=- \sqrt{{K}}$, $|{A}|>|{A}^c|$]{\label{pot4jR}\includegraphics[width=7cm]{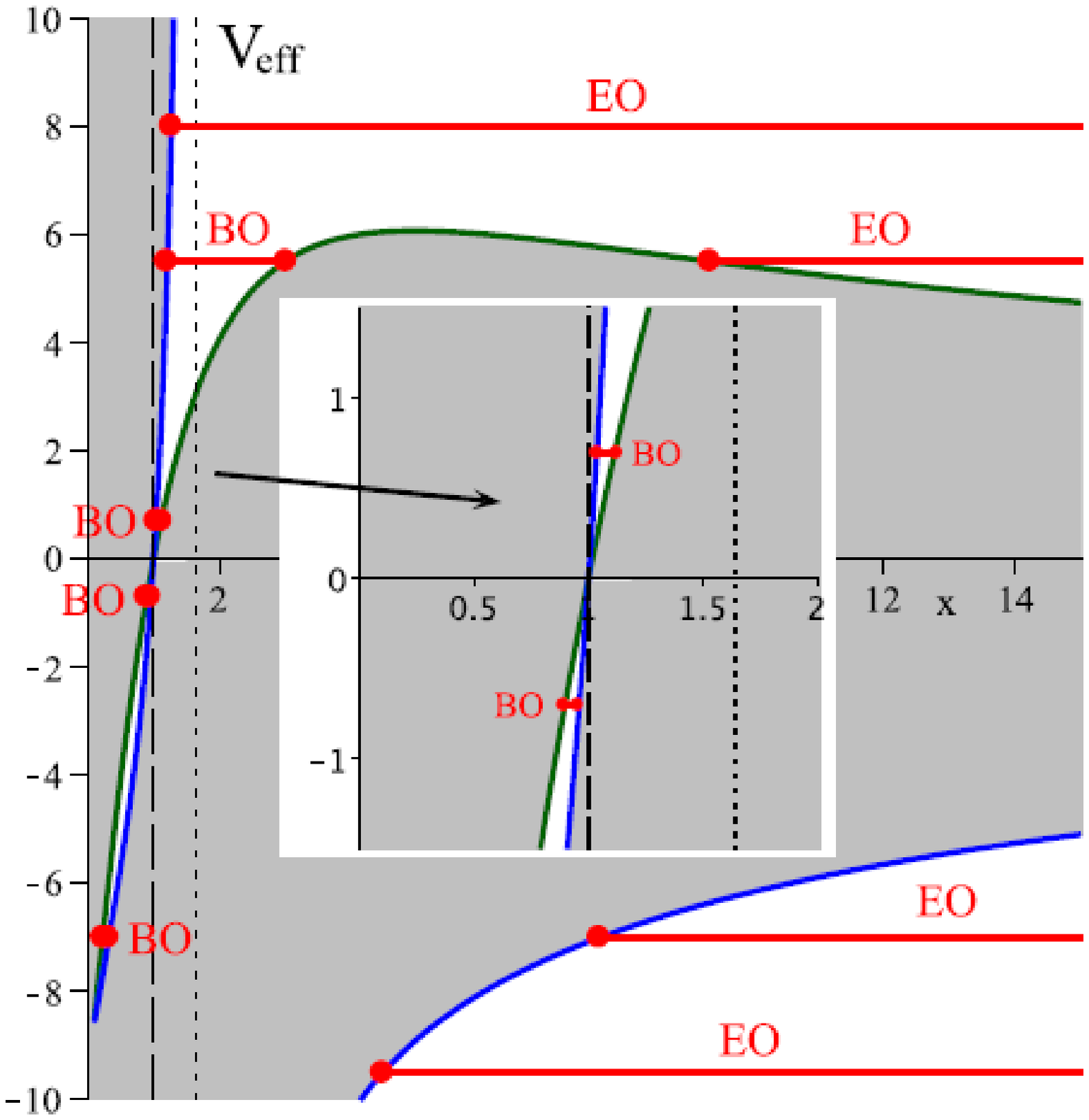}}
\end{center}
\caption{ Figures~\subref{pot1jR} and~\subref{pot2jR} show the effective potentials for $\omega<1$ (underrotating case) and $\boldsymbol{\delta=1}$. Figures~\subref{pot3jR} and~\subref{pot4jR} illustrate them for $\omega>1$ (overrotating case). The dashed line visualizes the horizon/pseudo-horizon at $x=1$ ($x=r^2$). The dotted line at $x=\sqrt[3]{\omega^2}$ (VLS) shows where the effective potential~\eqref{rpot1} diverges (see the discussion in sec.~\ref{sec:pot}). The grey region indicates the forbidden regions where the RHS of~\eqref{rpot} becomes negative. For $\omega<1$ bound orbits are possible (fig.~\subref{pot2jR}) for high values of ${K}$ and ${A}$. But they lie behind the degenerate horizon and are hidden for a remote observer. For $\omega>1$ bound orbits, where both turning points lie behind the pseudo-horizon, also exist. They belong to the white allowed region in a loop form located inside the forbidden grey region in figs.~\subref{pot3jR} and~\subref{pot4jR}. Here the value of ${K}$ is larger than the critical value ${K}^c$ defined in equation~\eqref{cond_j} (or with other words ${A}$ fulfills the inequality~\eqref{cond_jR}). In the wedges of the figures~\subref{pot3jR} and~\subref{pot4jR}  planetary bound orbits (BO) exist arbitrarily close to the pseudo-horizon,
with negative and positive energies, respectively, which are visible for a remote observer. For $E=0$ a circular orbit at $x=1$ is possible in all the pictures. See also table~\ref{tab2} for a schematical representation of the orbit types. \label{fig:potsjR}}
\end{figure*}

For $\omega>1$ the potential barrier also exists all over the $V_{\rm eff}$-axis. The asymptotic behaviour of the $V^\pm_{\rm eff}$ parts differs from the case with $\omega<1$: for decreasing $x$ (i.e. when $x\rightarrow \sqrt[3]{\omega^2}$ from the right) $V^+_{\rm eff}$ stretches to $+\infty$ and when $x$ increases (i.e. when $x\rightarrow \sqrt[3]{\omega^2}$ from the left) it goes to $-\infty$. The two parts of the potential do not always intersect at $x=1$ like in the previous case. It is possible that $x=1$ where the potential would become zero lies in the forbidden (grey) region as in the figs.~\ref{pot3} and~\subref{pot4}, or, like in the figs.~\ref{pot3jR} and~\subref{pot4jR}, an additional allowed region forms in the forbidden grey area. 

For negative ${A}$ the $V^\pm_{\rm eff}$-parts of the effective potential mirror w.r.t.~the $x$ axis. 

\paragraph{Investigation of $x=1$ traverse of $V^\pm_{\rm eff} $.}

To understand this behaviour of the effective potential consider the equations~\eqref{rpot} and~\eqref{rpot1}. When the potentials intersect, i.e. $V^+_{\rm eff}=V^-_{\rm eff}$, one of the intersection points is $x=1$. But for test particle motion to be possible, this point must lie in an allowed region. This means that $\Delta_{\rm eff}$ must fulfill the condition 
\begin{equation}
\Delta_{\rm eff}\geq 0 \ . \label{cond_Delta}
\end{equation}
Setting $x=r^2=1$ into $\Delta_{\rm eff}$ we get
\begin{equation} 
\Delta_{\rm eff} (x=1) = {A}^2 \omega^2 + (1-{\omega^2}) ( {K} + \delta)  \ . \label{rpot_Delta}
\end{equation}

With the condition~\eqref{cond_Delta} we get a restriction on the value of ${A}$ for which the potential~\eqref{rpot1} traverses $x=1$ (the equality sign in~\eqref{cond_Delta} we will consider below):
\begin{equation}
{A}^2 > \frac{\omega^2-1}{\omega^2}\left( {K} + \delta \right)={{A}^{c}}^2 \ , \label{cond_jR}
\end{equation}
where  ${A}^{c}$ implies a critical value. Choosing the values of ${K}$ and ${A}$ one has to take care of the condition ${K} \geq {A}^2$ from the inequality~\eqref{condj1}.

It is clear now why for $\omega<1$ the potentials always intersect at $x=1$: ${A}^2$ as a positive number is always larger than the negative RHS of~\eqref{cond_jR} and hence the condition~\eqref{cond_jR} is always fulfilled. 

The situation is different for $\omega>1$. In this case for ${A}^2 < {{A}^{c}}^2$ the point $x=1$ lies in the forbidden region as in figs.~\ref{pot3} and~\subref{pot4} and for ${A}^2 > {{A}^{c}}^2$ -- in the allowed region having the form of a loop as shown in figs.~\ref{pot3jR} and~\subref{pot4jR}.

For the special case when ${A}^2=K$ we infer a restriction on ${K}$ from~\eqref{cond_jR} of the form 
\begin{equation}
{K} > (\omega^2-1) \delta = {{K}^{c}} \ , \label{cond_j}
\end{equation} 
which defines a critical value of ${K}$ for ${A}^2=K$ which allows for an additional region for bound orbits to exist.

Consider now the case when 
\begin{equation}
{A}^2={{A}^{c}}^2=\frac{\omega^2-1}{\omega^2}\left( {K} + \delta \right)\ . \label{jR_special}
\end{equation} 
Then $\Delta_{\rm eff}$ reads
\begin{equation} 
\Delta_{\rm eff} =  (x-1)( \delta x^3 + ( \delta + {K} )x^2+( {K} + \delta )x+ {K} + \delta (1-\omega^2) )  \ . \label{rpot_Deltax}
\end{equation}
In this case $x=1$ is not only the intersection point of the plus and minus parts of the effective potentials but is also a discontinuity of $V^\pm_{\rm eff}$. With the Descartes' rule of signs we find that the second bracket in~\eqref{rpot_Deltax} has either 
\begin{enumerate}
\item no positive zeros if ${K} >\delta (\omega^2-1)$, 
\item or one positive zero if ${K} <\delta (\omega^2-1)$. 
\end{enumerate}

Inserting~\eqref{jR_special} into condition~\eqref{condj1} we get
\begin{equation}
{K}-{A}^2 \geq 0 \quad \Rightarrow \quad  {K} \geq (\omega^2-1) \delta \ .
\end{equation}

\begin{figure*}[th!]
\begin{center}
\includegraphics[width=7cm]{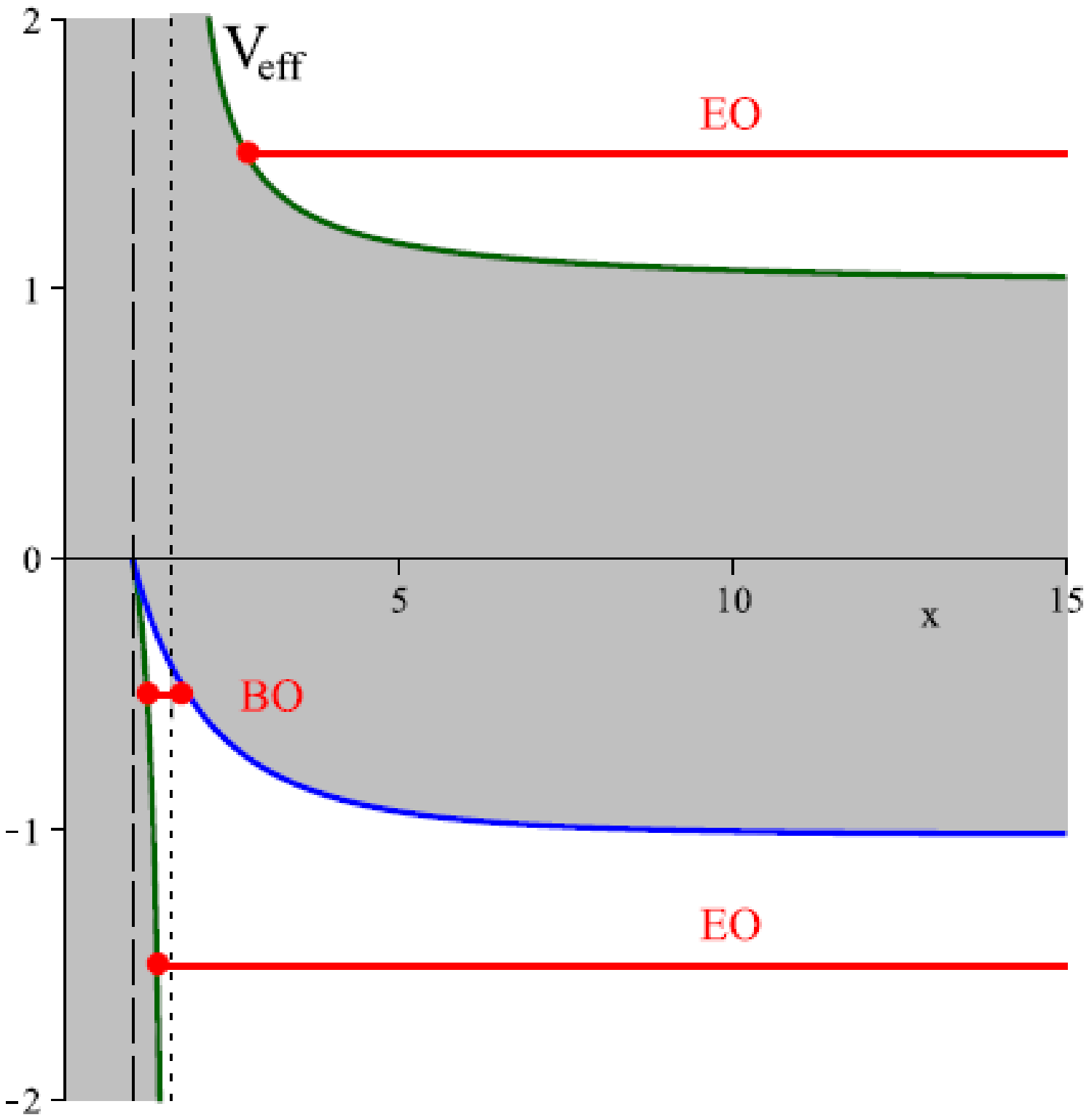}
\end{center}
\caption{In this effective potential with $\omega=2$ (overrotating case) and $K= 3.24$ for massive test particles ${A}$ is equal to the critical value ${{A}^{c}}^2=\frac{\omega^2-1}{\omega^2}\left( K + \delta \right)$ defined in equation~\eqref{jR_special} (cp. also the discussion close to the inequality~\eqref{cond_jR}). ${K}-{A}^2\geq 0$ according to the condition~\eqref{condj1}. The intersection point of the potentials $V^+_{\rm eff}$ and $V^-_{\rm eff}$ at $x=1$ is simultaneously a zero of $\Delta_{\rm eff}$ in~\eqref{rpot1} defining therewith one of the discontinuities of the effective potential~\eqref{rpot1} in the sense that to the left of this point $V^\pm_{\rm eff}$ has complex values which makes this region not suitable for physical motion. The region to the right is generally allowed. The exact regions of physical motion are finally determined by the positive or vanishing RHS of~\eqref{rpot}. The dashed line visualizes the pseudo-horizon at $x=1$ ($x=r^2$). The effective potential~\eqref{rpot1} approaches asymptotically $x=\sqrt[3]{\omega^2}$ (VLS) indicated by a dotted line. \label{pot:jRcrit}}
\end{figure*}

Thus, only the possibility $1$ above is relevant for physical motion. For ${K}=(\omega^2-1)\delta$ (in this case ${K}={K}^{c}$) the point $x=0$ is also a zero of $\Delta_{\rm eff}$ and coincides with the physical singularity. Recalling the discussion above we conclude that for ${A}^2={{A}^{c}}^2$ and ${K} \geq {A}^2$ the region to the left of the point $x=1$ is generally not allowed since $V^\pm_{\rm eff}$ has complex values there. An example of the corresponding effective potential is shown in fig.~\ref{pot:jRcrit}.

\subsubsection{Dynamics of massive test particles}

With this knowledge of the properties of the effective potential, we analyze now the radial motion by studying the polynomial $P(x)$~\eqref{reqn2} with the coefficients~\eqref{req_coeff}. The coefficient $a_0$ is shown to be negative, while the coefficient $a_3$ is positive or negative depending on whether $E^2 >\delta$ or $E^2<\delta$. Thus, only the sign of the coefficients $a_2$ and $a_1$ is not fixed. 

Using the Descartes' rule of signs we can determine the number of real positive solutions of the polynomial $P(x)$ in~\eqref{reqn2}. For $E^2<\delta$, i.e. $a_3<0$, at most $2$ real positive zeros are possible if $a_2>0$ for positive or negative $a_1$. If $a_2<0$ then for $a_1>0$ the number of real positive roots is $2$ and there are no positive solutions for $a_1<0$. Since we know that there is a potential barrier extending all over the $V_{\rm eff}$-axis, there must then be at least one positive root for any type of motion to exist. The last case would correspond to an energy value which lies in the forbidden region. This is for example the case for quite large $\omega$ and comparatively small ${K}$. For example, for this set of parameters all $a_i$ coefficients in the equation~\eqref{reqn2} are negative: $\delta=1$, $\omega=20$, ${K}=4$, ${A}=0.1 \sqrt{K}$ and an energy value of e.g. $E=0.99$. 

For $|E|<{1}$ for massive particles two positive turning points exist. This corresponds to 
\begin{enumerate}
\item for $\omega<1$ a many-world-bound orbit {\bf{MBO}} as in the figures~\ref{fig:pots}\subref{pot1} and~\subref{pot2},
\item for $\omega>1$ a planetary bound orbit {\bf{BO}} in the pictures~\ref{pot3} and~\subref{pot4}, or a bound orbit hidden for a remote observer behind the pseudo-horizon in the figures~\ref{pot3jR} and~\subref{pot4jR} or figure~\ref{pot:jRcrit}.
\end{enumerate}
Here the direction from which the potential approaches $\pm 1$ for $x\rightarrow \infty$ is important. Thus, in the pictures~\ref{pot1} and~\subref{pot4} the effective potential approaches $1$ from below (or $-1$ from above) and in the pictures~\ref{pot2} and~\subref{pot3} it approaches $1$ from above (or $-1$ from below).

For $|E|>{\delta}$, i.e. $a_3>0$, the number of positive roots is $3$ if $a_2<0$ and $a_1>0$. From the analysis of the effective potential we conclude that in this case both (planetary) bound and escape orbits are possible.  For other combinations of signs of the coefficients $a_2$ and $a_1$ the number of positive zeros is $1$ for $|E|>{\delta}$.

Thus, for $|E|>1$ for massive particles the orbit types are 
\begin{enumerate}
\item for $\omega>1$ a planetary bound orbit {\bf{BO}} and an escape orbit {\bf{EO}} or only an escape orbit which can be found in the figs.\ref{fig:pots}\subref{pot3} and~\subref{pot4}, or a bound orbit behind the pseudo-horizon and an escape orbit in the plots~\ref{pot3jR} and~\subref{pot4jR}
\item for $\omega<1$ a two-world escape {\bf{TWE}} shown in the plot~\ref{pot1} and a many-world-bound orbit and an escape orbit in the picture~\subref{pot2}. Also bound orbits in the inner region exist as illustrated in the figure~\ref{pot2jR}.
\end{enumerate}

The condition $|E|>{1}$ for massive particles under which bound orbits in this BMPV spacetime exist differs from the usual condition $|E|<{1}$ known from a large number of classical relativistic spacetimes such as Schwarzschild, Reissner-Nordström or Kerr.

Looking at the variation of the values of the parameters ${K}$ and ${A}$ over the plots~\ref{fig:pots},~\ref{fig:potsjR} and~\ref{pot:jRcrit} we see that not only the variation of the parameter $\omega$ influences the effective potential and correspondingly the types of orbits. Also the parameters ${K}$ and ${A}$ play a big role. Let us address this issue in detail. These parameters are not independent but related by the inequality~\eqref{condj1} which defines the maximal and the minimal values of ${A}$, namely $\sqrt{{K}}$ and $-\sqrt{{K}}$ respectively. 

In fig.~\ref{fig:potsjR}\subref{pot1jR} we have plotted the potential for the same $\omega$ ($\omega<1$, i.e., underrotating case) and ${K}$ as in the fig.~\ref{fig:pots}\subref{pot1} but with ${A}=\sqrt{{K}}$. We see that the difference is small: the potential bends slightly into the direction of the singularity. Compare now pictures~\ref{pot1jR} and~\ref{pot2jR}. If we just let ${A}$ grow, for example set it to its maximum value, we get a similar bending as before. Increasing ${K}$ instead leads to dramatical changes and the appearance of new orbit types. In fig.~\ref{pot2jR} behind the degenerate horizon a hook is formed which allows for a bound orbit {\bf{BO}}, that is not visible to an observer at infinity, together with a two-world-escape orbit {\bf{TWE}}. 
This feature is reminiscent of the bound orbits behind the event horizons 
in the Reissner-Nordström spacetime with highly charged test 
particles~\cite{Grunau:2010gd},
in the Kerr-Newmann spacetimes~\cite{Hackmann:2013pva},
or Myers-Perry spacetimes~\cite{Kagramanova:2012hw}.
Also the Kerr spacetime possesses this interesting 
property~\cite{Chandrasekhar83,Oneil}.

From the discussion in the previous paragraph we can expect that growing ${K}$ and ${A}$ would influence the form of the effective potential also for $\omega>1$ (overrotating case). Indeed, comparing figs.~\ref{pot3jR} and~\subref{pot4jR} with the figs.~\ref{pot3} and~\subref{pot4} we see that the potential forms a loop for large ${A}$. The loop is located in the positive part of the $V_{\rm eff}$-axis for ${A}>0$ and it is below the $x$-axis in the case of negative ${A}$. In this new region bound orbits are possible which are again hidden for a remote observer by the pseudo-horizon analogous to the previous case. The value of ${A}$ for which the loop forms is defined by the condition~\eqref{cond_jR}. The critical value ${A}^{c}$ given by~\eqref{jR_special} marks the beginning of the loop formation (see figure~\ref{pot:jRcrit}).

In the table~\ref{tab2} we summarize the results on the orbit types from the previous paragraphs. 

\renewcommand{\arraystretch}{1.5}
\begin{table}[t]
\begin{center}
\begin{tabular}{lcccl|c|c}\hline
type & region & + zeros & range of $x$ & orbit & $|E|$ & $\omega$ \\ \hline\hline
\.{B} & (\.{B}) & 2  & 
\begin{pspicture}(0,-0.2)(4.5,0.2)
\psline[linewidth=0.5pt]{->}(0,0)(4.5,0)
\psline[linewidth=0.5pt,linestyle=dashed](1,-0.2)(1,0.2)
\psline[linewidth=0.5pt,doubleline=true](0,-0.2)(0,0.2)
\psline[linewidth=1.2pt]{*-*}(0.5,0)( 1.7,0)
\end{pspicture}  
 & MBO &  \multirow{2}{*}{ $<1$ } & \multirow{5}{*}{ $<1$ }  \\
\.{B}$\rm _{E=0}$ & (\.{B}) & 2  & 
\begin{pspicture}(0,-0.2)(4.5,0.2)
\psline[linewidth=0.5pt]{->}(0,0)(4.5,0)
\psline[linewidth=0.5pt,linestyle=dashed](1,-0.2)(1,0.2)
\psline[linewidth=0.5pt,doubleline=true](0,-0.2)(0,0.2)
\psline[linewidth=1.2pt]{*-*}(1,0)(1,0)
\end{pspicture}  
 & MCO &  & \\  \cline{1-6}
%%%
\.{E} & (\.{E}) & 1  & 
\begin{pspicture}(0,-0.2)(4.5,0.2)
\psline[linewidth=0.5pt]{->}(0,0)(4.5,0)
\psline[linewidth=0.5pt,linestyle=dashed](1,-0.2)(1,0.2)
\psline[linewidth=0.5pt,doubleline=true](0,-0.2)(0,0.2)
\psline[linewidth=1.2pt]{*-}(0.5,0)( 4.5,0)
\end{pspicture}  
 & TWE & \multirow{3}{*}{ $>1$ } & \\
\.{B}E & (\.{B}E) & 3  & 
\begin{pspicture}(0,-0.2)(4.5,0.2)
\psline[linewidth=0.5pt]{->}(0,0)(4.5,0)
\psline[linewidth=0.5pt,linestyle=dashed](1,-0.2)(1,0.2)
\psline[linewidth=0.5pt,doubleline=true](0,-0.2)(0,0.2)
\psline[linewidth=1.2pt]{*-*}(0.5,0)( 1.7,0)
\psline[linewidth=1.2pt]{*-}(2.2,0)( 4.5,0)
\end{pspicture}  
 & MBO, EO &  & 
\\
\"{B}\.{E} & (\"{B}\.{E}) & 3  & 
\begin{pspicture}(0,-0.2)(4.5,0.2)
\psline[linewidth=0.5pt]{->}(0,0)(4.5,0)
\psline[linewidth=0.5pt,linestyle=dashed](1,-0.2)(1,0.2)
\psline[linewidth=0.5pt,doubleline=true](0,-0.2)(0,0.2)
\psline[linewidth=1.2pt]{*-*}(0.2,0)( 0.5,0)
\psline[linewidth=1.2pt]{*-}(0.8,0)( 4.5,0)
\end{pspicture}  
 & BO, TWE &  & 
\\ \hline 
%%%%%%%%
B & (B) & 2  & 
\begin{pspicture}(0,-0.2)(4.5,0.2)
\psline[linewidth=0.5pt]{->}(0,0)(4.5,0)
\psline[linewidth=0.5pt,linestyle=dashed](1,-0.2)(1,0.2)
\psline[linewidth=0.5pt,doubleline=true](0,-0.2)(0,0.2)
\psline[linewidth=1.2pt]{*-*}(1.2,0)(2,0)
\end{pspicture}  
 & BO & \multirow{3}{*}{$<1$} & \multirow{6}{*}{ $>1$ } \\  
B$\rm _{E=0}$ & (B) & 2  & 
\begin{pspicture}(0,-0.2)(4.5,0.2)
\psline[linewidth=0.5pt]{->}(0,0)(4.5,0)
\psline[linewidth=0.5pt,linestyle=dashed](1,-0.2)(1,0.2)
\psline[linewidth=0.5pt,doubleline=true](0,-0.2)(0,0.2)
\psline[linewidth=1.2pt]{*-*}(1,0)(1,0)
\end{pspicture}  
 & MCO &  & \\  
\"{B} & (\"{B}) & 2  & 
\begin{pspicture}(0,-0.2)(4.5,0.2)
\psline[linewidth=0.5pt]{->}(0,0)(4.5,0)
\psline[linewidth=0.5pt,linestyle=dashed](1,-0.2)(1,0.2)
\psline[linewidth=0.5pt,doubleline=true](0,-0.2)(0,0.2)
\psline[linewidth=1.2pt]{*-*}(0.3,0)(0.7,0)
\end{pspicture}  
 & BO &  & \\ \cline{1-6}
%%%
E & (E) & 1  & 
\begin{pspicture}(0,-0.2)(4.5,0.2)
\psline[linewidth=0.5pt]{->}(0,0)(4.5,0)
\psline[linewidth=0.5pt,linestyle=dashed](1,-0.2)(1,0.2)
\psline[linewidth=0.5pt,doubleline=true](0,-0.2)(0,0.2)
\psline[linewidth=1.2pt]{*-}(1.2,0)( 4.5,0)
\end{pspicture}  
 & EO & \multirow{3}{*}{ $>1$ } & \\
BE & (BE) & 3  & 
\begin{pspicture}(0,-0.2)(4.5,0.2)
\psline[linewidth=0.5pt]{->}(0,0)(4.5,0)
\psline[linewidth=0.5pt,linestyle=dashed](1,-0.2)(1,0.2)
\psline[linewidth=0.5pt,doubleline=true](0,-0.2)(0,0.2)
\psline[linewidth=1.2pt]{*-*}(1.2,0)( 2,0)
\psline[linewidth=1.2pt]{*-}(2.6,0)( 4.5,0)
\end{pspicture}  
 & BO, EO &  & 
\\ 
\"{B}E & (\"{B}E) & 3  & 
\begin{pspicture}(0,-0.2)(4.5,0.2)
\psline[linewidth=0.5pt]{->}(0,0)(4.5,0)
\psline[linewidth=0.5pt,linestyle=dashed](1,-0.2)(1,0.2)
\psline[linewidth=0.5pt,doubleline=true](0,-0.2)(0,0.2)
\psline[linewidth=1.2pt]{*-*}(0.3,0)( 0.7,0)
\psline[linewidth=1.2pt]{*-}(2.6,0)( 4.5,0)
\end{pspicture}  
 & BO, EO &  & 
\\ \hline\hline
\end{tabular}
\caption{Orbit types of massive test particles: $\delta=1$. Thick lines represent the range of the orbits with thick dots as turning points. 
The horizon/pseudo-horizon is indicated by a vertical dashed line. The singularity at $x=r^2=0$ is shown as a solid line on the left. The orbit MCO is a circular orbit of a test particle with $E=0$ at the horizon/pseudo-horizon $x=1$. The types with a dot over a letter cross the horizon once, those without a dot do not cross $x=1$ and lie in the outer region, the types with two dots are concealed completely behind the horizon/pseudo-horizon. This table summarizes the discussion of sec.~\ref{sec:pot} and presents schematically the orbits shown in the effective potentials in figs.~\ref{fig:pots},~\ref{fig:potsjR} and~\ref{pot:jRcrit}.
\label{tab2}}
\end{center}
\end{table}

\subsubsection{Dynamics of massless test particles}~\label{sec:delta0}

Here we will see that planetary bound orbits are also possible when $\delta=0$, i.e., for massless particles. 

The variety of possibilities for the signs of the coefficients $a_i$ in equation~\eqref{reqn1_1} (or eq.~\eqref{reqn2}) given by~\eqref{req_coeff} reduces to a few cases when $\delta=0$. The coefficient $a_0$ stays negative. The coefficient $a_3$ is only positive now while the coefficient $a_2$ is only negative. Only the sign of the coefficient $a_1$ is not fixed. With the Descartes' rule of signs we infer that at most either $3$ (if $a_1>0$) or $1$ (if $a_1<0$) positive roots are possible.

For $\omega<1$ (underrotating case), when the potential barrier for $x\rightarrow \sqrt[3]{\omega}$ (VLS) is located behind the horizon, so that a test particle will necessary cross the degenerate horizon, the case of three positive zeros implies a bound orbit behind the horizon {\bf{BO}} and a two-world escape orbit {\bf{TWE}} or a many-world-bound orbit {\bf{MBO}} and an escape orbit {\bf{EO}}. In the case of one positive root only a two-world-escape orbit {\bf{TWE}} exists. 

For $\omega>1$ (overrotating case) the potential barrier will keep a test particle from crossing the pseudo-horizon, which allows for a planetary bound orbit {\bf{BO}} and an escape orbit {\bf{EO}} in the case of $3$ positive roots of $P(x)$~\eqref{reqn2}.  If $P(x)$ has one positive zero, only an escape orbit {\bf{EO}} exists. We summarize these results in the table~\ref{tab3}.

\begin{figure*}[th!]
\begin{center}
\subfigure[][$\omega=0.7, {K}= 1, {A}=0.1 \sqrt{K}$]{\label{pot1l}\includegraphics[width=7cm]{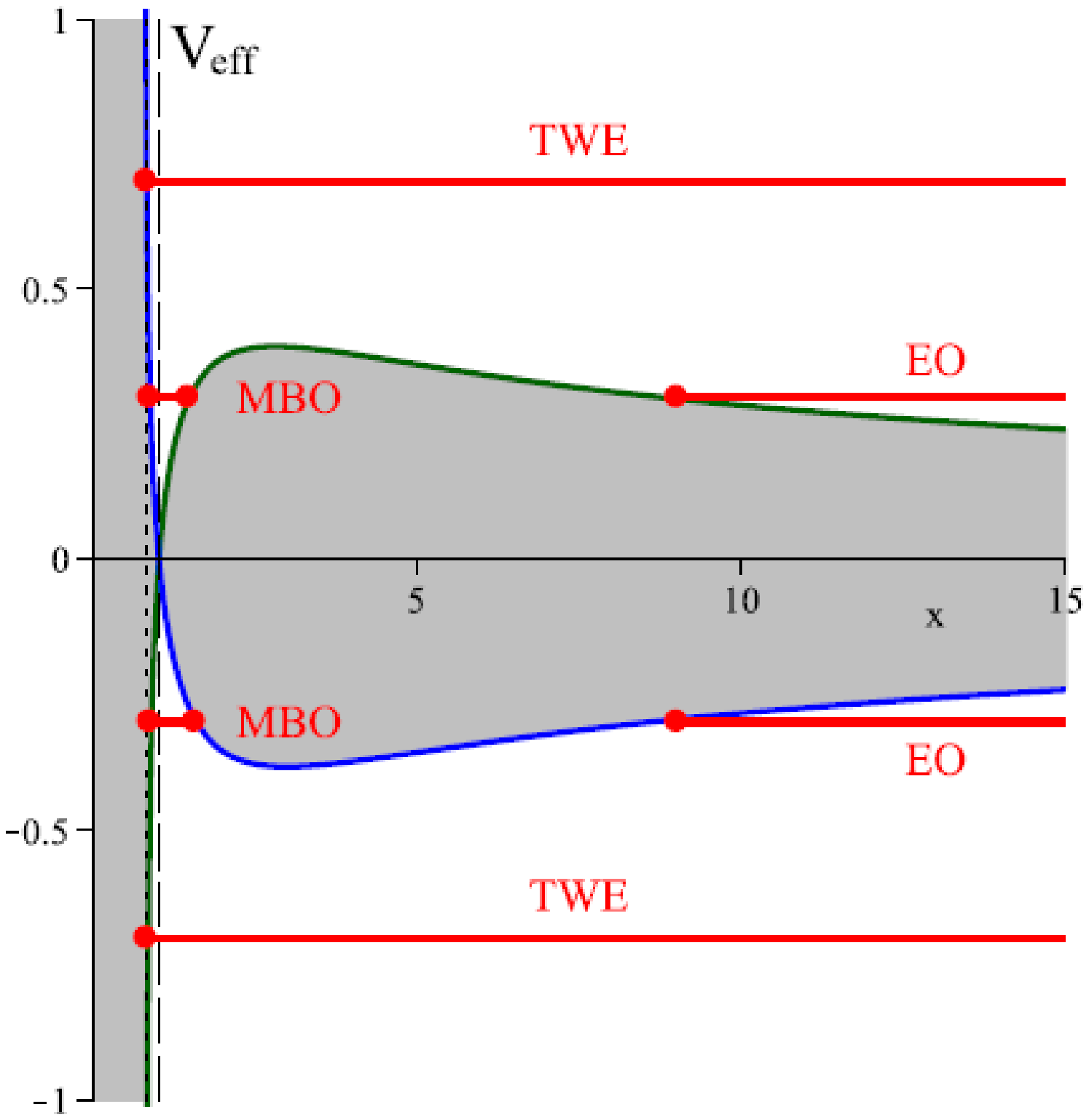}}
\subfigure[][$\omega=1.1, {K}= 36, {A}=0.1 \sqrt{K}$]{\label{pot2l}\includegraphics[width=7cm]{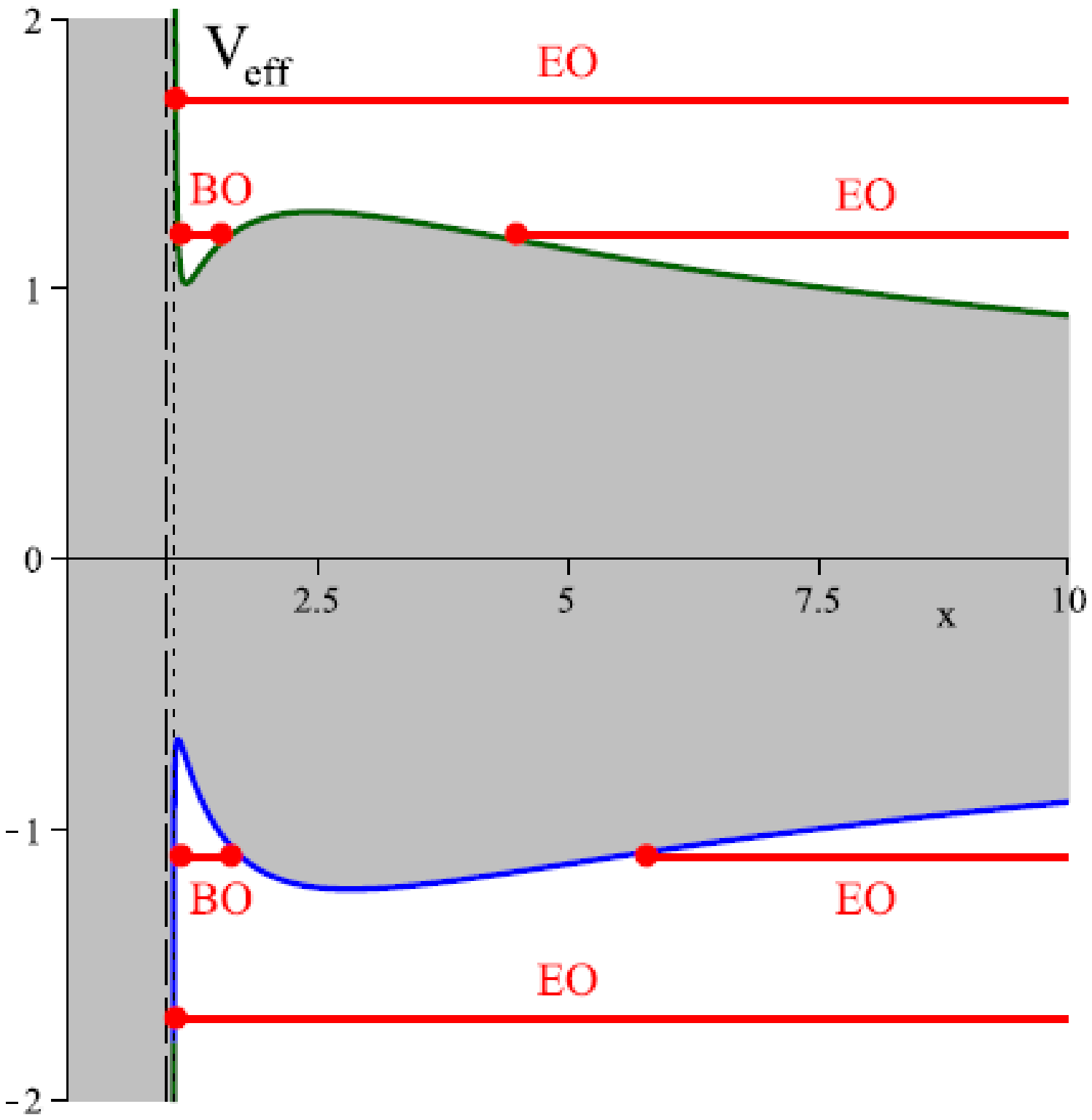}}
\end{center}
\caption{ Figures~\subref{pot1l} and~\subref{pot2l} show the effective potentials for $\omega<1$ (underrotating case) and $\omega>1$ (overrotating case), respectively, for $\boldsymbol{\delta=0}$. Both potentials tend to zero at infinity. The dashed line visualizes the horizon/pseudo-horizon at $x=1$ ($x=r^2$). The dotted line at $x=\sqrt[3]{\omega^2}$ (VLS) shows, where the effective potential~\eqref{rpot1} diverges (see the discussion for general $\delta$ in sec.~\ref{sec:pot}). The grey region marks the forbidden regions where the RHS of~\eqref{rpot} becomes negative. For $\omega<1$ a planetary bound orbit is not possible, but only a many-world-bound orbit MBO, as shown in fig.~\subref{pot1l}, since a test particle with $\omega<1$ will necessarily cross the degenerate horizon at $x=1$. This happens because the potential asymptotically approaches $x=\sqrt[3]{\omega^2}$, which is smaller than $1$ in this case, and thus lies behind the horizon. In the wedge of fig.~\subref{pot1l} for $E=0$ a circular orbit MCO at $x=1$ is possible. For $\omega>1$ both turning points of a bound orbit will lie in front of the pseudo-horizon, since in this case the value $x=\sqrt[3]{\omega^2}$, where the potential diverges, is larger than $1$ (see e.g. fig.~\subref{pot2l}). Thus, for $\omega>1$ planetary bound orbits for test particles with $\delta=0$ (massless) are possible.  \label{fig:potsl}}
\end{figure*}

\begin{figure*}[th!]
\begin{center}
\subfigure[][$\omega=0.7, {K}= 1, {A}=\sqrt{K}$]{\label{pot1ljR}\includegraphics[width=7cm]{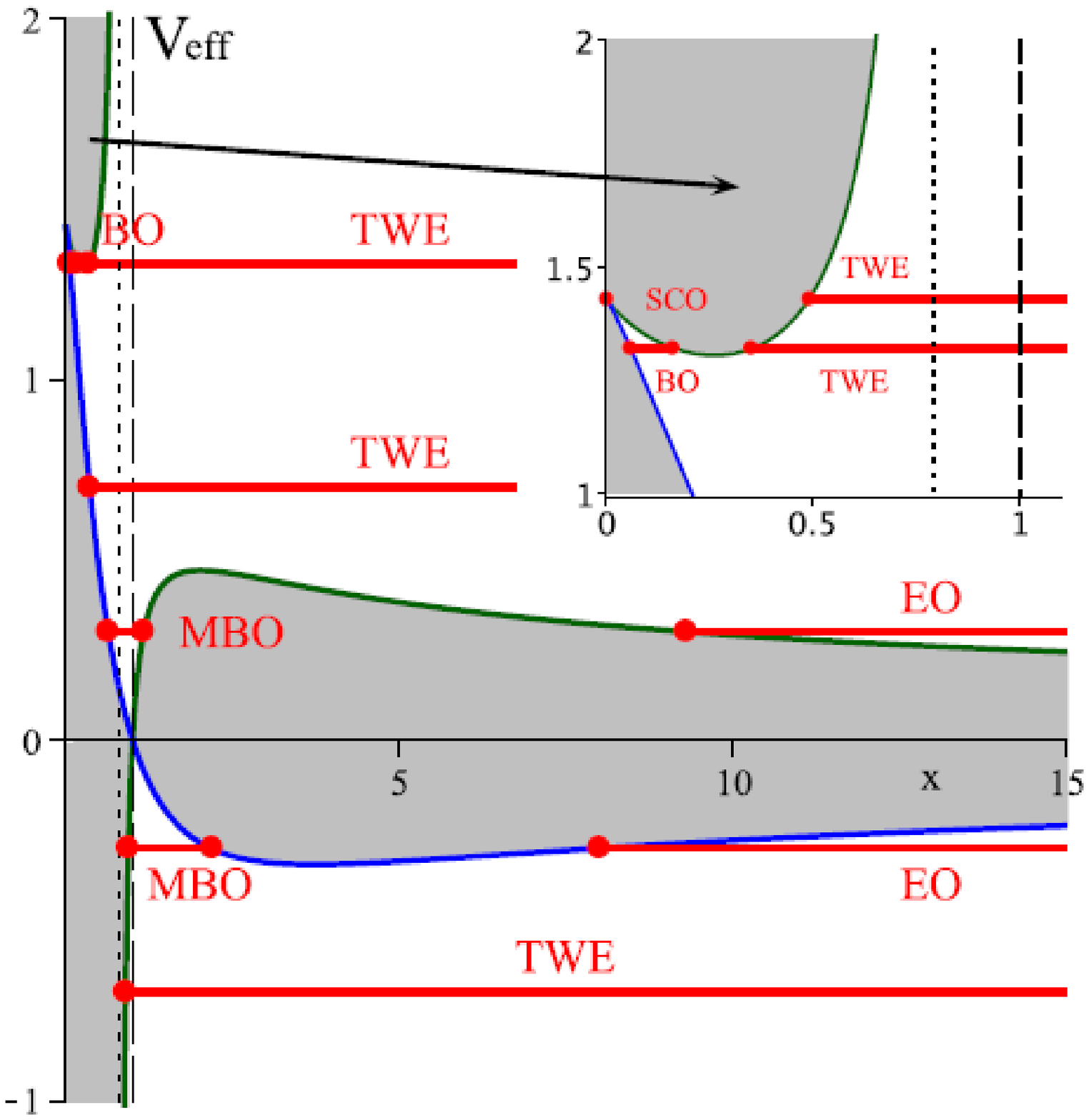}}
\subfigure[][$\omega=1.1, {K}= 10, {A}=0.95 \sqrt{K}$]{\label{pot2ljR}\includegraphics[width=7cm]{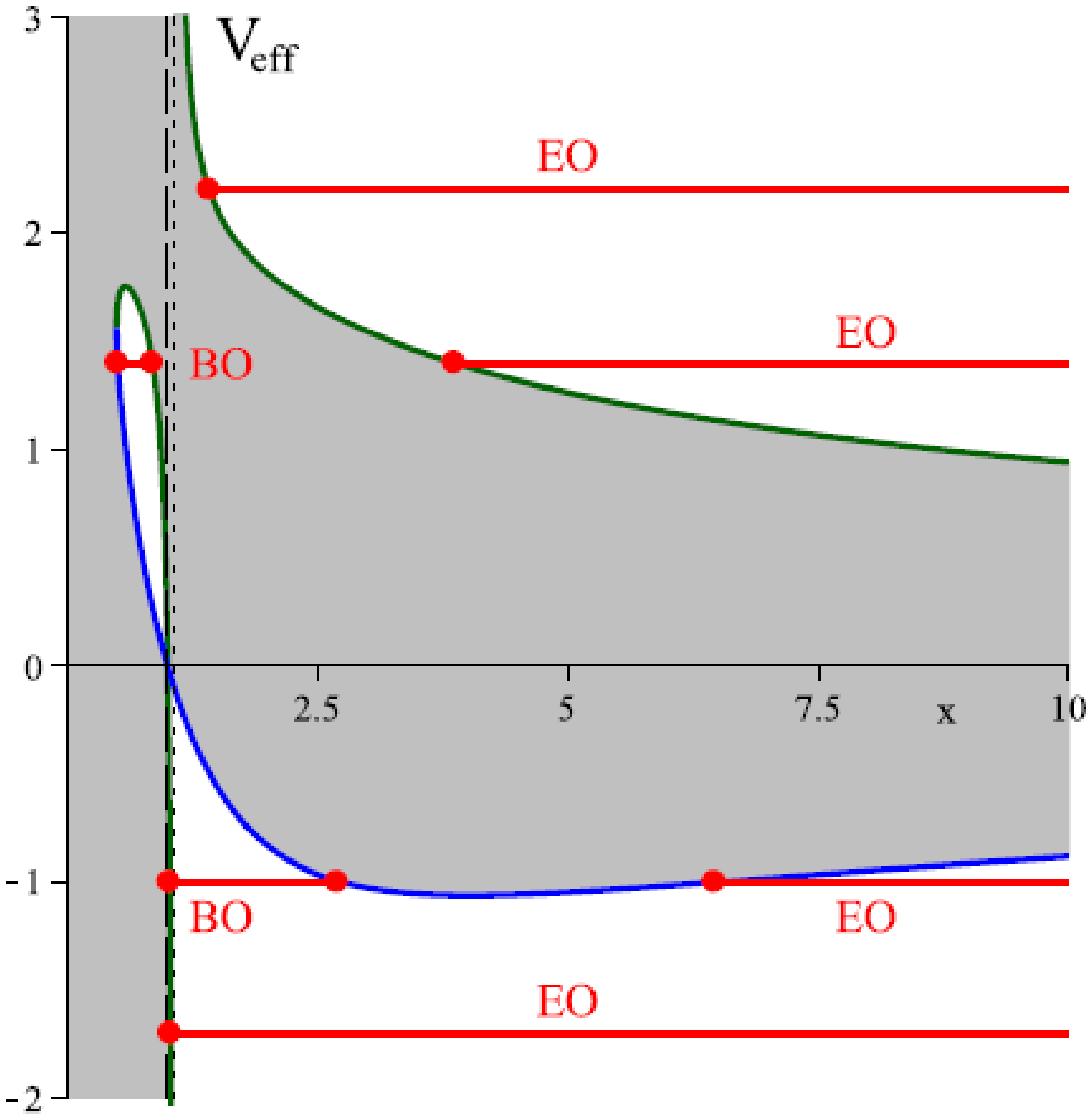}}
\subfigure[][$\omega=1.1, {K}= 10, {A}=-\sqrt{\frac{\omega^2-1}{\omega^2}{K}}\equiv {A}^{c}$]{\label{pot3ljR}\includegraphics[width=7cm]{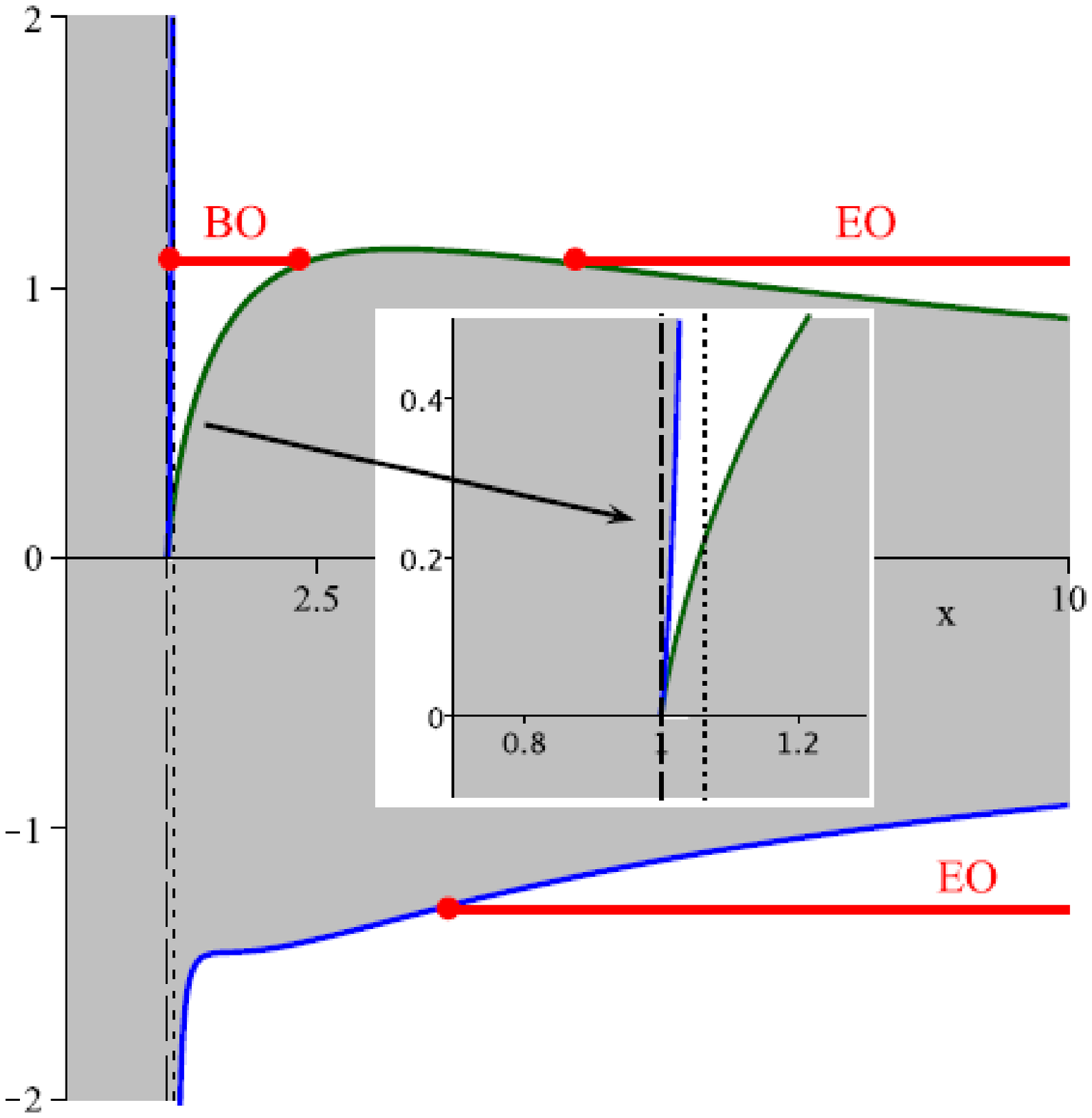}}
\subfigure[][$\omega=1.1, {K}= 10, {A}=\sqrt{K}$, ${K}>{K}^{c}$]{\label{pot4ljR}\includegraphics[width=7cm]{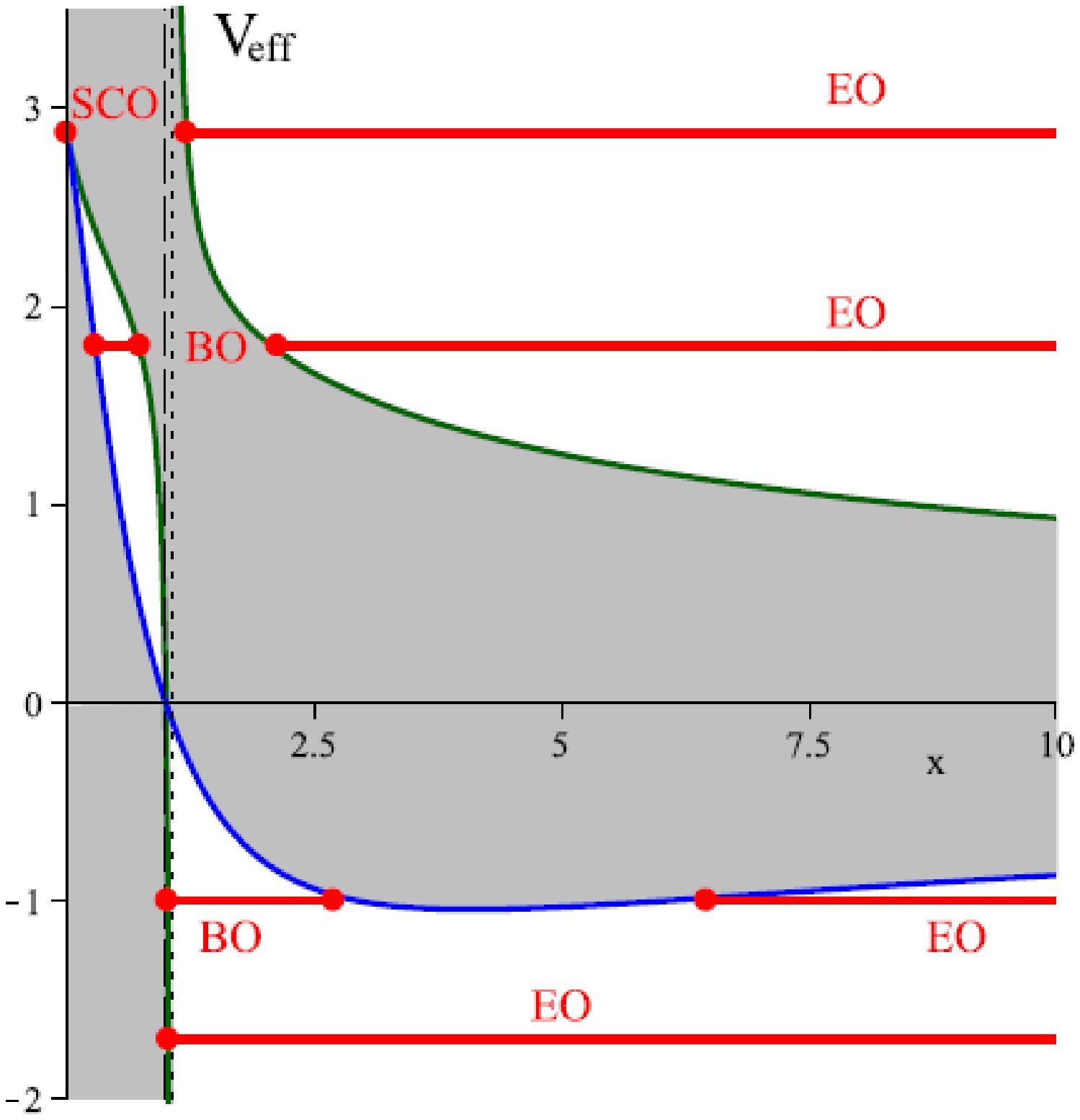}}
\end{center}
\caption{ Figures~\subref{pot1ljR}-\subref{pot3ljR} show effective potentials for large ${A}$ for massless test particles with $\boldsymbol{\delta=0}$. The potentials tend to zero at infinity. The dashed line visualizes the horizon/pseudo-horizon at $x=1$ ($x=r^2$). The dotted line at $x=\sqrt[3]{\omega^2}$ (VLS) shows where the effective potential~\eqref{rpot1} diverges (see the discussion for general $\delta$ in sec.~\ref{sec:pot}). The grey region marks the forbidden regions where the RHS of~\eqref{rpot} becomes negative. As in fig.~\ref{pot1l} for $\omega<1$ and small ${A}$ a planetary bound orbit does not exist. But a bound orbit behind the horizon exists similarly to fig.~\ref{pot2jR} for massive particles. For $\omega>1$ both planetary bound orbits and bound orbits hidden from a remote orbserver are possible (cp. figs.~\ref{pot3jR} and~\ref{pot4jR} for massive particles). Picture~\subref{pot3ljR} shows the effective potential for the critical value of ${A}$ from the equation~\eqref{jR_special} (cp. figure~\ref{pot:jRcrit}). Figure~\subref{pot4ljR} presents the potential for ${A}=\sqrt{{K}}$ where ${K}$ satisfies the inequality~\eqref{cond_j}. In this case as well as in the figure~\subref{pot1ljR} a singular solution SCO at $x=0$ is mathematically possible for $E=\frac{A}{\omega}$. For $E = 0$ a circular orbit MCO at $x = 1$ is possible in all the pictures. \label{fig:potsljR}}
\end{figure*}

The discussion of the effective potential properties from section~\ref{sec:pot} together with the results of the table~\ref{tab1} applies also here. In figures~\ref{fig:potsl} and~\ref{fig:potsljR} we show examples of the effective potential for massless test particles. At infinity it tends to zero. Planetary bound orbits for particles with $\delta=0$, present in the overrotating case, are shown in fig.~\ref{fig:potsl}\subref{pot2l} and fig.~\ref{pot3ljR}. This is another characteristic of the overrotating BMPV spacetime distinguishing it from the classical relativistic spacetimes. 

It is interesting to note, that because of the asymptotics of the potential at infinity the maximum for $\omega<1$ (underrotating case) in fig.~\ref{fig:potsl}\subref{pot1l} always exists, while for very large $\omega$, satisfying $\omega>1$ (overrotating case), neither a minimum nor a maximum may survive. In this case the potential resembles $\pm1/x$ curves, where the asymptotes are the $x$ axis and a vertical line at $x=\sqrt[3]{\omega}$ (VLS). In fig.~\ref{pot2l}, plotted for moderate values of $\omega$ and ${K}$, both a minimum and a maximum exist. For the same reason as before we notice that they always come in a pair for $\delta=0$.

In fig.~\ref{fig:potsljR} we choose large values of ${A}$. In this case for $\omega<1$ (underrotating case) bound orbits appear behind the horizon (fig.~\subref{pot1ljR}). For $\omega>1$ (overrotating case) in the fig.~\subref{pot2ljR} the value of ${A}$ satisfies the inequality~\eqref{cond_jR}. Then the $V^\pm_{\rm eff}$ parts of the effective potential~\eqref{rpot1} cross at $x=1$ and form a loop behind the pseudo-horizon, where bound orbits become possible. In fig.~\subref{pot3ljR} the value of ${A}$ coincides with the critical value~\eqref{jR_special}, where $x=1$ is additionally a discontinuity of the effective potential. In the wedge planetary bound orbits exist even for tiny absolute values of the energy, so that the pseudo-horizon might be approached very closely.

\renewcommand{\arraystretch}{1.5}
\begin{table}[t]
\begin{center}
\begin{tabular}{lcccl|c}\hline
type & region & + zeros & range of $x$ & orbit  & $\omega$ \\ \hline\hline
%%%
\.{E} & (\.{E}) & 1  & 
\begin{pspicture}(0,-0.2)(4.5,0.2)
\psline[linewidth=0.5pt]{->}(0,0)(4.5,0)
\psline[linewidth=0.5pt,linestyle=dashed](1,-0.2)(1,0.2)
\psline[linewidth=0.5pt,doubleline=true](0,-0.2)(0,0.2)
\psline[linewidth=1.2pt]{*-}(0.5,0)( 4.5,0)
\end{pspicture}  
 & TWE & \multirow{4}{*}{ $<1$ } \\  
\.{B}E & (\.{B}E) & 3  & 
\begin{pspicture}(0,-0.2)(4.5,0.2)
\psline[linewidth=0.5pt]{->}(0,0)(4.5,0)
\psline[linewidth=0.5pt,linestyle=dashed](1,-0.2)(1,0.2)
\psline[linewidth=0.5pt,doubleline=true](0,-0.2)(0,0.2)
\psline[linewidth=1.2pt]{*-*}(0.5,0)( 1.7,0)
\psline[linewidth=1.2pt]{*-}(2.2,0)( 4.5,0)
\end{pspicture}  
 & MBO, EO &   
\\  
\.{B}$\rm _{E=0}$ & (\.{B}E) & 2  & 
\begin{pspicture}(0,-0.2)(4.5,0.2)
\psline[linewidth=0.5pt]{->}(0,0)(4.5,0)
\psline[linewidth=0.5pt,linestyle=dashed](1,-0.2)(1,0.2)
\psline[linewidth=0.5pt,doubleline=true](0,-0.2)(0,0.2)
\psline[linewidth=1.2pt]{*-*}(1,0)( 1,0)
\end{pspicture}  
 & MCO &   
\\ 
\"{B}\.{E} & (\"{B}\.{E}) & 3  & 
\begin{pspicture}(0,-0.2)(4.5,0.2)
\psline[linewidth=0.5pt]{->}(0,0)(4.5,0)
\psline[linewidth=0.5pt,linestyle=dashed](1,-0.2)(1,0.2)
\psline[linewidth=0.5pt,doubleline=true](0,-0.2)(0,0.2)
\psline[linewidth=1.2pt]{*-*}(0.25,0)( 0.5,0)
\psline[linewidth=1.2pt]{*-}( 0.8,0)( 4.5,0)
\end{pspicture}  
 & BO, TWE &  
\\ 
\"{B}$\rm _0$\.{E} & (\"{B}\.{E}) & 3  & 
\begin{pspicture}(0,-0.2)(4.5,0.2)
\psline[linewidth=0.5pt]{->}(0,0)(4.5,0)
\psline[linewidth=0.5pt,linestyle=dashed](1,-0.2)(1,0.2)
\psline[linewidth=0.5pt,doubleline=true](0,-0.2)(0,0.2)
\psline[linewidth=1.2pt]{*-*}(0,0)( 0,0)
\psline[linewidth=1.2pt]{*-}( 0.8,0)( 4.5,0)
\end{pspicture}  
 & SCO, TWE &  
\\ \hline 
%%%%%%%%
%%%
E & (E) & 1  & 
\begin{pspicture}(0,-0.2)(4.5,0.2)
\psline[linewidth=0.5pt]{->}(0,0)(4.5,0)
\psline[linewidth=0.5pt,linestyle=dashed](1,-0.2)(1,0.2)
\psline[linewidth=0.5pt,doubleline=true](0,-0.2)(0,0.2)
\psline[linewidth=1.2pt]{*-}(1.2,0)( 4.5,0)
\end{pspicture}  
 & EO & \multirow{4}{*}{ $>1$ } \\  
BE & (BE) & 3  & 
\begin{pspicture}(0,-0.2)(4.5,0.2)
\psline[linewidth=0.5pt]{->}(0,0)(4.5,0)
\psline[linewidth=0.5pt,linestyle=dashed](1,-0.2)(1,0.2)
\psline[linewidth=0.5pt,doubleline=true](0,-0.2)(0,0.2)
\psline[linewidth=1.2pt]{*-*}(1.2,0)( 2,0)
\psline[linewidth=1.2pt]{*-}(2.6,0)( 4.5,0)
\end{pspicture}  
 & BO, EO &  
\\ 
B$\rm _{E=0}$ & (BE) & 2  & 
\begin{pspicture}(0,-0.2)(4.5,0.2)
\psline[linewidth=0.5pt]{->}(0,0)(4.5,0)
\psline[linewidth=0.5pt,linestyle=dashed](1,-0.2)(1,0.2)
\psline[linewidth=0.5pt,doubleline=true](0,-0.2)(0,0.2)
\psline[linewidth=1.2pt]{*-*}(1,0)( 1,0)
\end{pspicture}  
 & MCO &   
\\ 
\"{B}\.{E} & (\"{B}{E}) & 3  & 
\begin{pspicture}(0,-0.2)(4.5,0.2)
\psline[linewidth=0.5pt]{->}(0,0)(4.5,0)
\psline[linewidth=0.5pt,linestyle=dashed](1,-0.2)(1,0.2)
\psline[linewidth=0.5pt,doubleline=true](0,-0.2)(0,0.2)
\psline[linewidth=1.2pt]{*-*}( 0.25 ,0)(  0.75 ,0)
\psline[linewidth=1.2pt]{*-}(1.5,0)( 4.5,0)
\end{pspicture}  
 & BO, EO &  
\\ 
\"{B}$\rm _0$E & (\"{B}E) & 3  & 
\begin{pspicture}(0,-0.2)(4.5,0.2)
\psline[linewidth=0.5pt]{->}(0,0)(4.5,0)
\psline[linewidth=0.5pt,linestyle=dashed](1,-0.2)(1,0.2)
\psline[linewidth=0.5pt,doubleline=true](0,-0.2)(0,0.2)
\psline[linewidth=1.2pt]{*-*}(0,0)( 0,0)
\psline[linewidth=1.2pt]{*-}( 1.5,0)( 4.5,0)
\end{pspicture}  
 & SCO, EO & 
\\ \hline\hline
\end{tabular}
\caption{Orbit types of massless test particles: $\delta=0$. Thick lines represent the range of the orbits with thick dots as turning points. 
The horizon/pseudo-horizon is indicated by a vertical dashed line. The singularity at $x=r^2=0$ is shown as a solid line on the left. The types with a dot over a letter cross the horizon once, those without a dot do not cross $x=1$ and lie in the outer region, the types with two dots are concealed completely behind the horizon/pseudo-horizon. This table summarizes the discussion of sec.~\ref{sec:delta0} and presents schematically the orbits pictured in the effective potentials in the figs.~\ref{fig:potsl} and~\ref{fig:potsljR}. The orbit MCO is a circular orbit with a radius $x=1$ of a test particle with $E=0$. The type SCO is a singular solution located at $x=0$ for the parameters ${A}^2={K}$ and $E=\frac{{A}}{\omega}$.
\label{tab3}}
\end{center}
\end{table}

\subsubsection{Reaching the singularity.} 

The RHS of equation~\eqref{reqn1_1} or~\eqref{reqn2} must be non-negative to allow for physical motion for any test particle in principle. Setting $x=0$ in~\eqref{reqn2} for test particles which could reach the singularity we get 
\begin{equation}
P(x=0)=a_0=-({K}-{A}^2)-(\omega E - {A})^2 \geq 0 \ .
\end{equation}
The condition above is fulfilled if (recall the condition~\eqref{condj1} for ${K}$ and ${A}^2$)
\begin{equation}
{K}={A}^2 \quad \cup \quad E=\frac{{A}}{\omega} \label{cond_sing_omega} \ .
\end{equation}

From the previous discussion we know that for negative $\Delta_{\rm eff}$ in the effective potential~\eqref{rpot1} when $V^\pm_{\rm eff}$ becomes complex no physical motion is possible in general. This is the region to the left of the only positive zero of $\Delta_{\rm eff}$. The region to the right of this zero is generally allowed and the specific types of motion are finally determined by the non-negative RHS of the equation~\eqref{rpot} and corresponding conditions on the parameter $E$. Subsituting~\eqref{cond_sing_omega} into the expression for $\Delta_{\rm eff}$ in~\eqref{rpot1} we get
\begin{equation}
\Delta_{\rm eff} ({K}={A}^2) = x ( x^3 \delta + {A}^2 x^2 -\omega^2 \delta ) \label{Delta_sing_omega} \ .
\end{equation}

Analyzing the expression above we observe that for $\delta=1$ one positive zero always exists. Thus, between $x=0$ and the positive zero of~\eqref{Delta_sing_omega} the potential~\eqref{rpot1} has complex values and this region is forbidden in general. Hence a test particle with $\delta=1$ cannot reach the singularity in this case.

Consider now $\delta=0$. In this case all the roots of~\eqref{Delta_sing_omega} are equal to zero and a test particle can reach the singularity. Substituting~\eqref{cond_sing_omega} into $P(x)$ in~\eqref{reqn2} and solving $P(x)=0$ for $x$ we get the turning points: $x_1=x_2=0$ and $x_3=\omega^2$. Here $x_1=x_2$ indicates a singular solution at $x=0$ and $x_3$ is a turning point for a two world escape orbit of a massless test particle with ${K}={A}^2$ and $E=\frac{{A}}{\omega}$. The singular solution is indicated in the figures~\ref{pot1ljR} and~\ref{pot4ljR} (SCO). It is not a physical solution, but completes the set of all mathematically possible cases.

\subsubsection{Features of motion for $\boldsymbol{\omega=1}$}\label{sec:omega1}

Let us now consider the critical case where $\omega=1$. 
In this case the area of the surface $x=1$ vanishes, while the VLS also resides
at $x=1$.
We address the critical case separately,
since as we will see the features of the potential~\eqref{rpot1} change dramatically
in this case.
As in the overrotating case test particles travelling on geodesics
may not cross the surface $x=1$ to enter the region $x<1$.
However, the surface $x=1$ itself is reached by most types of geodesics.

For $\omega=1$ the polynomial $P(x)$ in the equation~\eqref{reqn2} takes the form 
\begin{equation}
P(x) = 4 (x-1)(a x^2 + b x + c) = 4 (x-1) P_1(x) \label{reqn2_omega1} \ ,
\end{equation}
with the coefficients
\begin{eqnarray}
&& a = E^2-\delta  \ , \quad b =  \delta  - {K} + E^2 \ , \nonumber \\
&& c = {K} - {A}^2 + (E - {A})^2 \label{req_coeff_omega1} \ .
\end{eqnarray}

We observe that $x=1$ is now a zero of $P(x)$. With the condition~\eqref{condj1} the coefficient $c$ is non-negative. With the Descates rule of signs we conclude that $P_1(x)$ has at most 2 positive zeros for $a>0$ and $b<0$, none -- if $a>0$ and $b>0$ and one positive zero if $a<0$ both for positive or negative $b$. 

Thus, the previously seen many world bound orbits will now convert into two types. One type is an exterior orbit ($x\ge 1$) with $x=1$ the inner boundary of the motion,
while the other type is an interior orbit ($x\le 1$) with $x=1$ the outer boundary
of the motion. This also means that no planetary bound orbits are possible in this case. 
Also the previously present two world escape orbit changes, since $x=1$ cannot be traversed.
For this type of orbit $x=1$ is now also the inner boundary of the motion.
Note, that we have kept the notation MBO and TWO for these orbits to see their connection
with the underrotating and overrotating cases.

The effective potential~\eqref{rpot1} becomes
\begin{equation}
V^\pm_{{\rm eff}_1} = \frac{ {A} \pm \sqrt{ \Delta_{\rm eff} } }{x^2+x+1}  \,\,\, \text{with} \,\,\, \Delta_{\rm eff}={A}^2 + (x^3-1) ({K} + x \delta)  \ , \label{rpot1_omega1}
\end{equation}
where we have substituted $x=r^2$. This effective potential has no pole at $x=\sqrt[3]{\omega^2}\equiv 1$ as compared to the effective potential in~\eqref{rpot1} for the
values of $\omega$ in the underrotating and overrotating cases.

Since $x=1$ is not any more a point where the plus and minus parts of the potential intersect and become zero (under certain conditions as discussed in the section~\ref{sec:pot}), the conditions~\eqref{cond_Delta} for $\Delta_{\rm eff}$ or~\eqref{cond_jR} for ${A}$ as well as further conditions and conclusions there cannot be directly applied here. But since $x=1$ is always a root of $P(x)$, evaluation of $V^\pm_{{\rm eff}_1}$ at this point gives: 
\begin{eqnarray}
&& {A}>0 \quad \Rightarrow \quad V^+_{{\rm eff}_1}(x=1)=\frac{2}{3} {A} \quad \text{and} \quad V^-_{{\rm eff}_1}(x=1)=0 \ , \\
&& {A}<0 \quad \Rightarrow \quad V^+_{{\rm eff}_1}(x=1)=0 \quad \text{and} \quad V^-_{{\rm eff}_1}(x=1)=\frac{2}{3} {A} \ .
\end{eqnarray}

\begin{figure*}[th!]
\begin{center}
\subfigure[][${K}= 16, {A}=0.8 \sqrt{{K}} $]{\label{pot1_o1}\includegraphics[width=7cm]{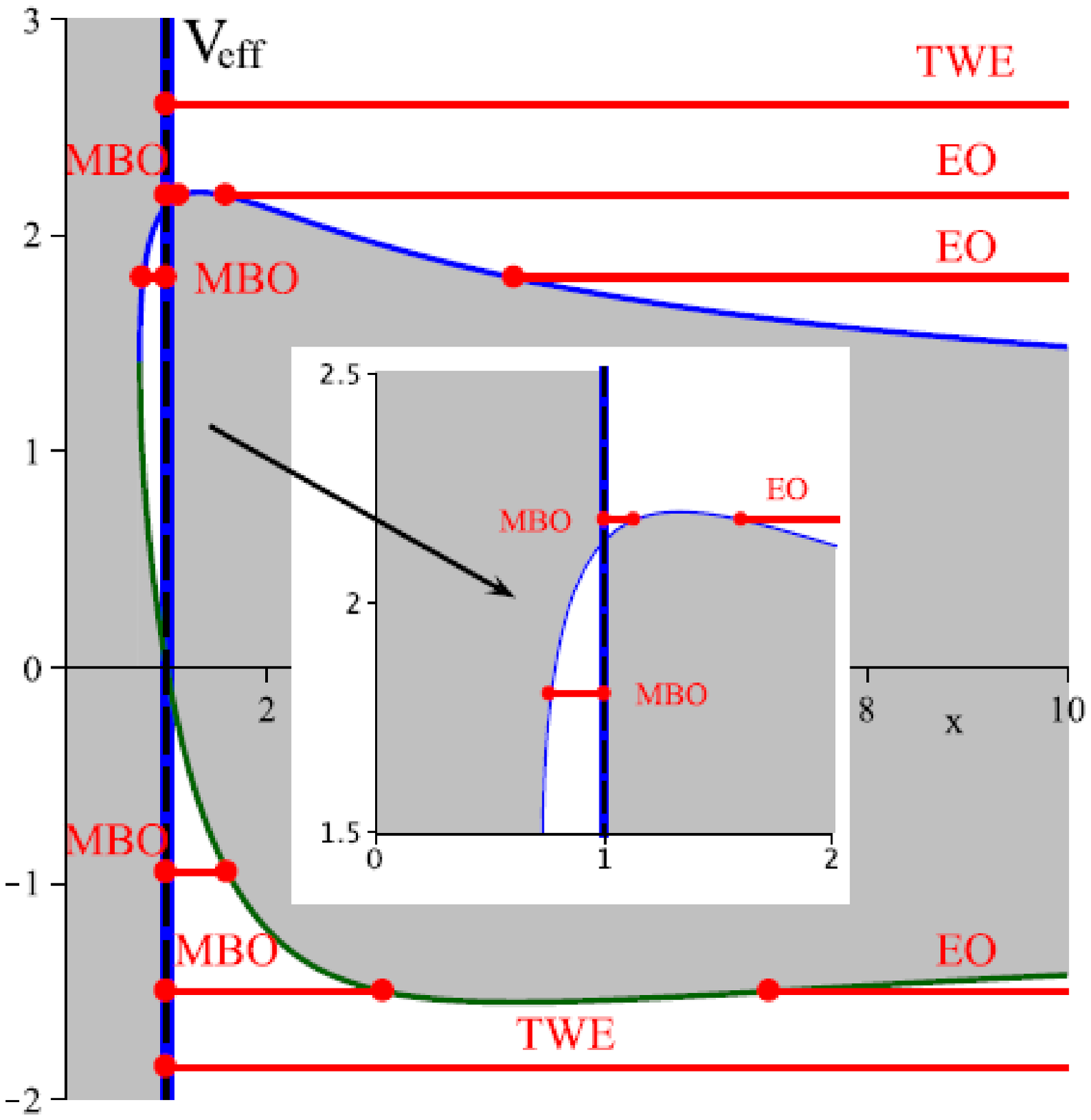}}
\subfigure[][${K}= 16, {A}= \sqrt{{K}} $]{\label{pot2_o1}\includegraphics[width=7cm]{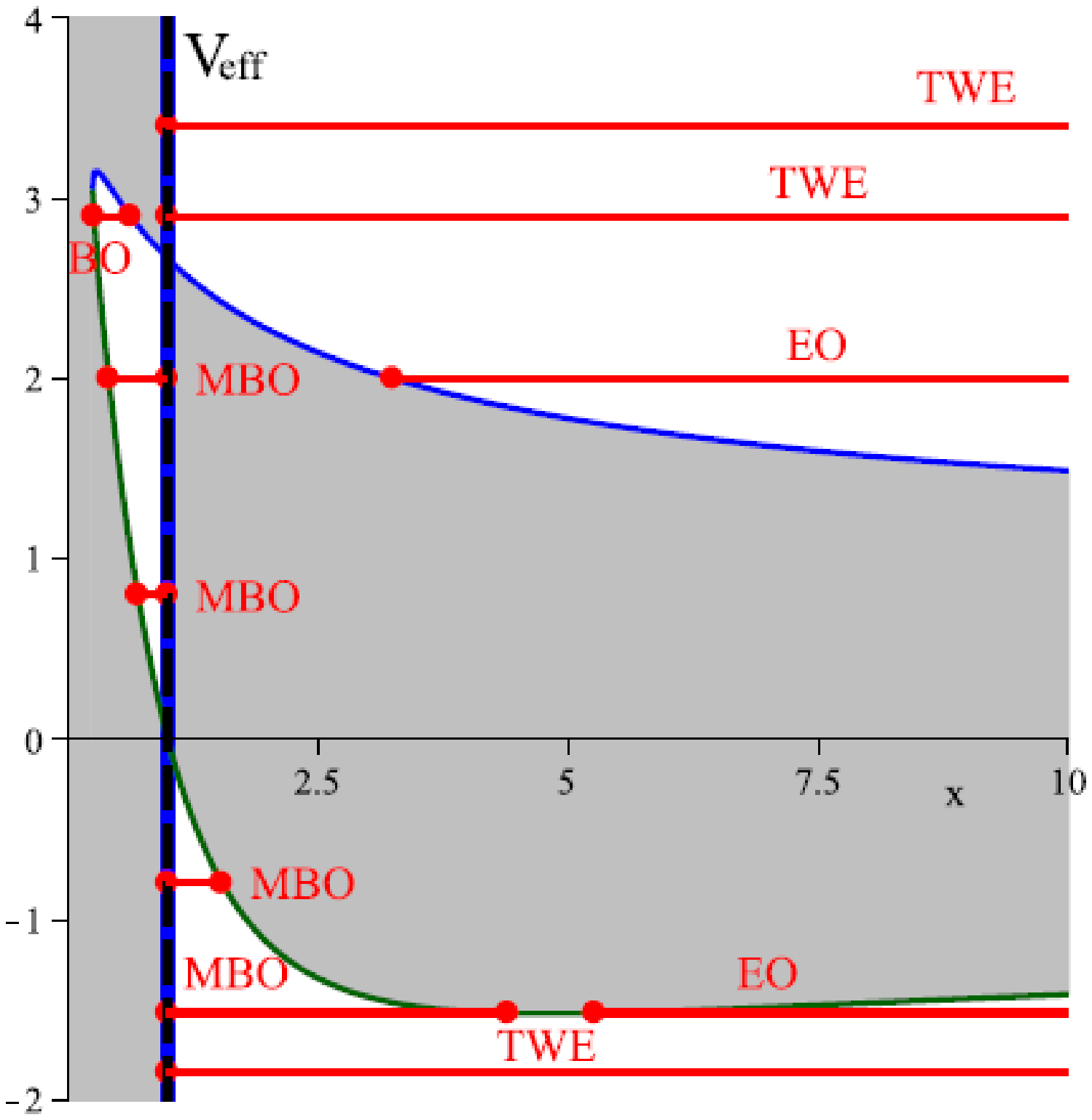}}
\end{center}
\caption{ Figures~\subref{pot1_o1} and~\subref{pot2_o1} show the effective potentials for $\omega=1$ (critical case) for massive test particles with $\boldsymbol{\delta=1}$. For $\omega=1$ no planetary bound orbits are possible even for massive particles,
since $x=1$ is always a zero of the polynomial $P(x)$ in~\eqref{reqn2_omega1}. 
For $E=0$ a circular orbit MCO at $x=1$ exists. \label{fig:pot_omega1}}
\end{figure*}

\begin{figure*}[th!]
\begin{center}
\subfigure[][${K}= 16, {A}=0.8 \sqrt{{K}} $]{\label{potl1_o1}\includegraphics[width=7cm]{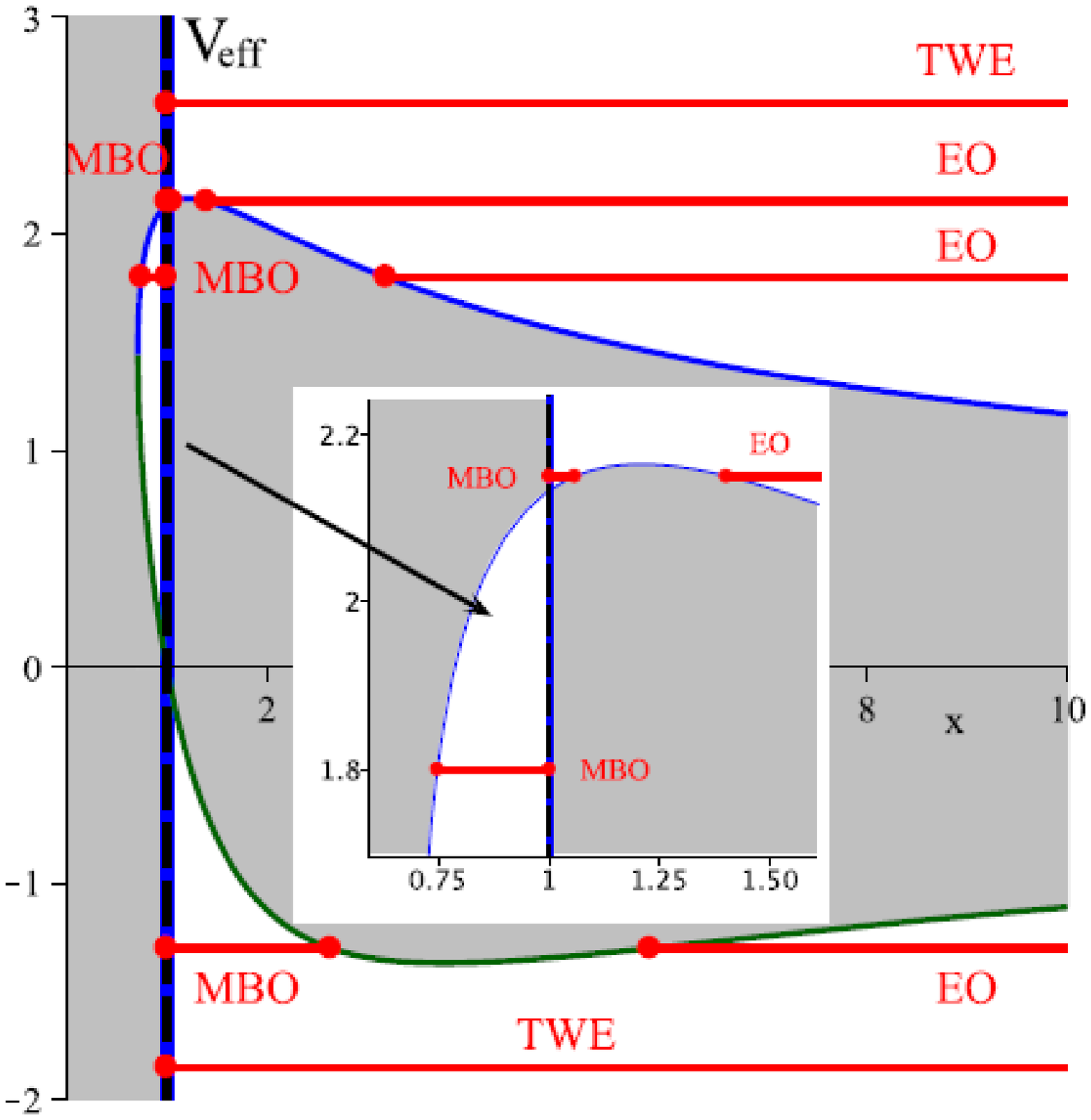}}
\subfigure[][${K}= 16, {A}= \sqrt{{K}} $]{\label{potl2_o1}\includegraphics[width=7cm]{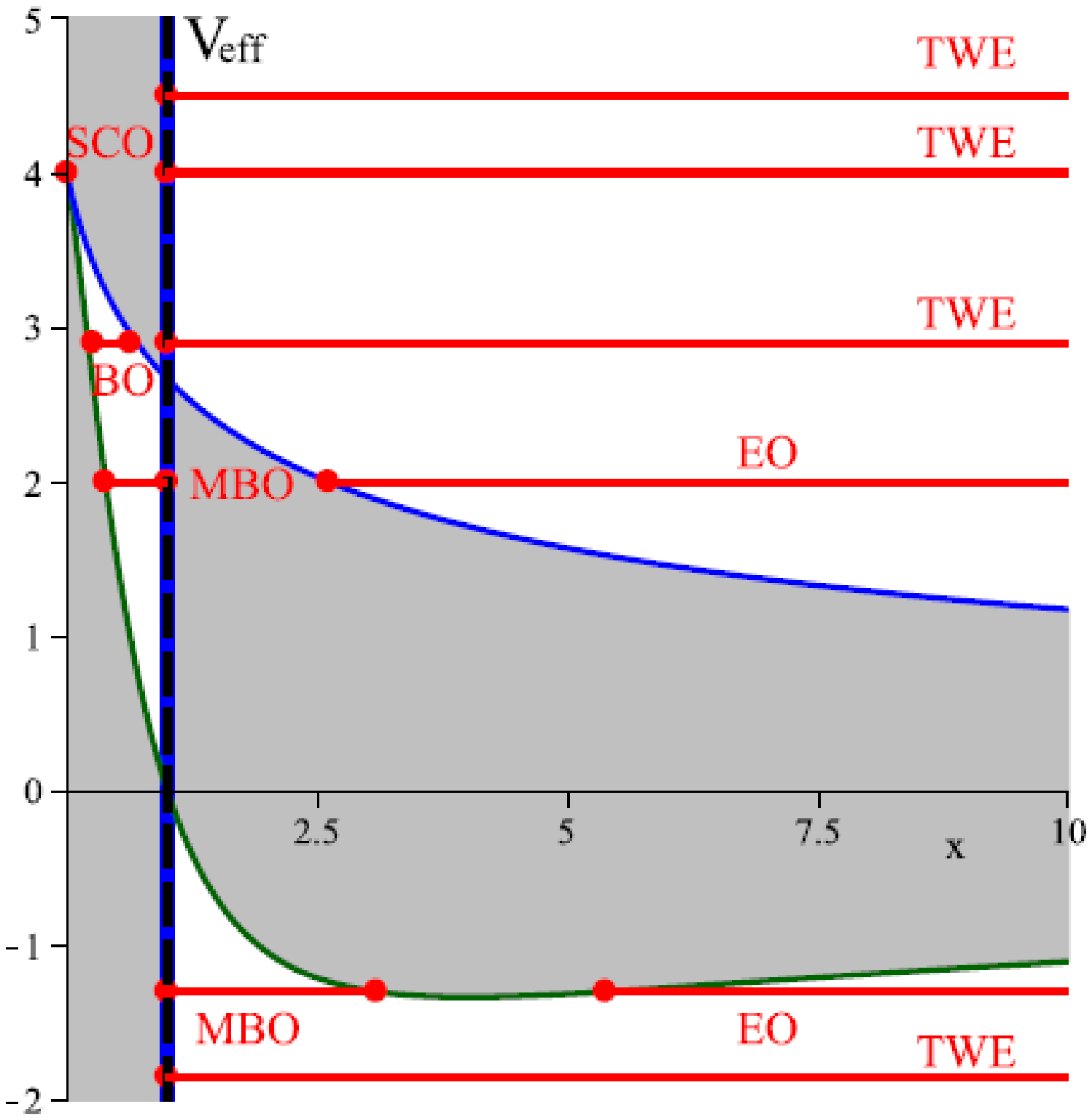}}
\end{center}
\caption{ Figures~\subref{potl1_o1} and~\subref{potl2_o1} show the effective potentials for $\omega=1$ (critical case) for massless test particles with $\boldsymbol{\delta=0}$. For $\omega=1$ no planetary bound orbits are possible,
since $x=1$ is always a zero of the polynomial $P(x)$ in~\eqref{reqn2_omega1}. 
The type SCO in the plot~\subref{potl2_o1} is a singular solution at $x=0$ with $E={A}$. For $E=0$ a circular orbit MCO at $x=1$ exists in both plots. \label{fig:potl_omega1}}
\end{figure*}

Consider again ${A}>0$ since its negative counterpart just mirrors the potential w.r.t.~the $x$-axis. We observe that the plus $V^+_{{\rm eff}_1}$ and minus $V^-_{{\rm eff}_1}$ parts of the potential~\eqref{rpot1_omega1} are identically zero at $x=1$ only if ${A}=0$. If ${A} \neq 0$ then only the minus part of the potential is zero at that point, while the plus part has a non-negative value. Thus, for ${A} \neq 0$ the two parts of the potential cross behind the $x=1$-line, forming an additional allowed region between the singularity and the surface $x=1$. We show a few examples of the effective potential in the fig.~\ref{fig:pot_omega1} for massive test particles and in the fig.~\ref{fig:potl_omega1} for massless test particles.

Analyzing the potentials in the figs.~\ref{fig:pot_omega1} and~\ref{fig:potl_omega1} we observe that the bound orbits are either located behind the surface $x=1$ 
or have one turning point exactly at $x=1$. A two world escape orbit always reaches $x=1$ 
and only an escape orbit has a turning point at a finite distancs from $x=1$. 

Consider in detail ${A}=0$, when the effective potential is symmetric w.r.t.~the $x$-axis and at $x=1$ both parts of the effective potential vanish. For $E=0$, the polynomial $P(x)$ has three zeros in the form: $x_{1,2}=1$ and $x_3=-\frac{{K}}{\delta}<0$. Thus, for $E=0$ and ${A}=0$ a circular orbit at $x=1$ for a massive test particle exists. For $\delta=0$ and $E=0$ the polynomial $P(x)$ has a double zero at $x=1$. Thus, also massless test particles with ${A}=0$, $E=0$ can be on a circular orbit at $x=1$.

In the tables~\ref{tab4} and~\ref{tab5} we list the types of orbits for $\delta=1$ and $\delta=0$ schematically.

\renewcommand{\arraystretch}{1.5}
\begin{table}[t]
\begin{center}
\begin{tabular}{lcccl|c}\hline
type & region & + zeros & range of $x$ & orbit  & $|E|$ \\ \hline\hline
%%%
\.{B}$\rm _1$ & (\.{B}$\rm _1$) & 2  & 
\begin{pspicture}(0,-0.2)(4.5,0.2)
\psline[linewidth=0.5pt]{->}(0,0)(4.5,0)
\psline[linewidth=0.5pt,linestyle=dashed](1,-0.2)(1,0.2)
\psline[linewidth=0.5pt,doubleline=true](0,-0.2)(0,0.2)
\psline[linewidth=1.2pt]{*-*}(0.5,0)( 1.0,0)
\end{pspicture}  
 & MBO & \multirow{3}{*}{ $<1$ } \\  
$\rm _1$\.{B} & ($\rm _1$\.{B}) & 2  & 
\begin{pspicture}(0,-0.2)(4.5,0.2)
\psline[linewidth=0.5pt]{->}(0,0)(4.5,0)
\psline[linewidth=0.5pt,linestyle=dashed](1,-0.2)(1,0.2)
\psline[linewidth=0.5pt,doubleline=true](0,-0.2)(0,0.2)
\psline[linewidth=1.2pt]{*-*}(1.5,0)( 1.0,0)
\end{pspicture}  
 & MBO &   
\\ 
\.{B}$\rm _{E=0}$ & ($\rm _1$\.{B}) & 2  & 
\begin{pspicture}(0,-0.2)(4.5,0.2)
\psline[linewidth=0.5pt]{->}(0,0)(4.5,0)
\psline[linewidth=0.5pt,linestyle=dashed](1,-0.2)(1,0.2)
\psline[linewidth=0.5pt,doubleline=true](0,-0.2)(0,0.2)
\psline[linewidth=1.2pt]{*-*}(1,0)( 1.0,0)
\end{pspicture}  
 & MCO &   
\\
 \hline 
%%%%%%%%
%%%
\.{B}$\rm _1$E & (\.{B}$\rm _1$E) & 3  & 
\begin{pspicture}(0,-0.2)(4.5,0.2)
\psline[linewidth=0.5pt]{->}(0,0)(4.5,0)
\psline[linewidth=0.5pt,linestyle=dashed](1,-0.2)(1,0.2)
\psline[linewidth=0.5pt,doubleline=true](0,-0.2)(0,0.2)
\psline[linewidth=1.2pt]{*-*}(0.5,0)( 1.0,0)
\psline[linewidth=1.2pt]{*-}(1.3,0)( 4.5,0)
\end{pspicture}  
 & MBO, EO & \multirow{4}{*}{ $>1$ } \\  
$\rm _1$\.{B}E & ($\rm _1$\.{B}E) & 3  & 
\begin{pspicture}(0,-0.2)(4.5,0.2)
\psline[linewidth=0.5pt]{->}(0,0)(4.5,0)
\psline[linewidth=0.5pt,linestyle=dashed](1,-0.2)(1,0.2)
\psline[linewidth=0.5pt,doubleline=true](0,-0.2)(0,0.2)
\psline[linewidth=1.2pt]{*-*}(1,0)( 1.5,0)
\psline[linewidth=1.2pt]{*-}(1.8,0)( 4.5,0)
\end{pspicture}  
 & MBO, EO & \\
\"{B} $\rm _1$\.{E} & (\"{B} $\rm _1$\.{E}) & 3  & 
\begin{pspicture}(0,-0.2)(4.5,0.2)
\psline[linewidth=0.5pt]{->}(0,0)(4.5,0)
\psline[linewidth=0.5pt,linestyle=dashed](1,-0.2)(1,0.2)
\psline[linewidth=0.5pt,doubleline=true](0,-0.2)(0,0.2)
\psline[linewidth=1.2pt]{*-*}(0.3,0)( 0.6,0)
\psline[linewidth=1.2pt]{*-}(1,0)( 4.5,0)
\end{pspicture}  
 & BO, TWE &
\\ 
$\rm _1$\.{E} & ($\rm _1$\.{E}) & 1  & 
\begin{pspicture}(0,-0.2)(4.5,0.2)
\psline[linewidth=0.5pt]{->}(0,0)(4.5,0)
\psline[linewidth=0.5pt,linestyle=dashed](1,-0.2)(1,0.2)
\psline[linewidth=0.5pt,doubleline=true](0,-0.2)(0,0.2)
\psline[linewidth=1.2pt]{*-}(1,0)( 4.5,0)
\end{pspicture}  
 & TWE &
\\
\hline\hline
\end{tabular}
\caption{Orbit types for $\omega=1$ (critical case) for massive test particles with $\delta=1$. Thick lines represent the range of the orbits with thick dots as turning points. The surface $x=1$ is indicated by a vertical dashed line. The singularity at $x=r^2=0$ is shown by a solid line at the left. The types with a dot over a letter reach $x=1$, those without a dot do not reach $x=1$ and lie in the outer region, the types with two dots are completely inside the inner region. The index $\rm _1$ on the left or right of a letter indicates the position of the surface $x=1$. This table summarizes the discussion of sec.~\ref{sec:omega1} and presents schematically the orbits pictured in the effective potentials in the fig.~\ref{fig:pot_omega1}. The orbit MCO is a circular orbit of a test particle with $E=0$ and ${A}=0$ with a radius $x=1$. 
\label{tab4}}
\end{center}
\end{table}

\paragraph{Reaching the singularity when $\omega=1$.} 

The singularity is located at $x=0$. Setting $x=0$ in the effective potential~\eqref{rpot1_omega1} we obtain:
\begin{equation}
V^\pm_{{\rm eff}_1} (x=0) = {A} \pm \sqrt{{A}^2-{K}}  \ . \label{rpot1_s1}
\end{equation}
Taking into account the condition~\eqref{condj1} the expression above makes sense if ${K}={A}^2$. In this case 
\begin{equation}
V^\pm_{{\rm eff}_1} (x=0) = {A} \ . \label{rpot1_s2}
\end{equation}

Set now ${K}={A}^2$ and $E={A}$ in $P_1(x)$ in~\eqref{reqn2_omega1} and calculate its roots. They read
\begin{equation}
x_1=0 \quad \text{and} \quad x_2=\frac{\delta}{\delta-{K}} \label{x12_s1} \ .
\end{equation}

Consider $\delta=1$. If ${K}<1$ then $x_2>1$. Then the three roots of the polynomial $P(x)$ are: $0$, $1$ and $x_2>1$. Keeping in mind the form and properties of the effective potential we conclude that the singularity is located in the forbidden grey region. If ${K}>1$ then $x_2<0$ and there are two non-negative roots of $P(x)$: $1$ and $x_1=0$. Again, the form and properties of the effective potential tell us that the singularity is located in the forbidden grey region and $x=1$ is the only physically relevant solution being a boundary point of a two world escape orbit. Thus, a massive test particle cannot reach the singularity.

Consider $\delta=0$. In this case $x_2=0$. A massless test particle with ${K}={A}^2$ and $E={A}$ 
%can indeed fall into the singularity. T
may mathematically be on a circular orbit at $x=0$. This corresponds to the peak of the allowed white part of the effective potential behind the horizon in fig.~\ref{potl2_o1}.

\renewcommand{\arraystretch}{1.5}
\begin{table}[t]
\begin{center}
\begin{tabular}{lcccl}\hline
type & region & + zeros & range of $x$ & orbit  \\ \hline\hline
%%%
\.{B}$\rm _1$E & (\.{B}$\rm _1$E) & 3  & 
\begin{pspicture}(0,-0.2)(4.5,0.2)
\psline[linewidth=0.5pt]{->}(0,0)(4.5,0)
\psline[linewidth=0.5pt,linestyle=dashed](1,-0.2)(1,0.2)
\psline[linewidth=0.5pt,doubleline=true](0,-0.2)(0,0.2)
\psline[linewidth=1.2pt]{*-*}(0.5,0)( 1.0,0)
\psline[linewidth=1.2pt]{*-}(1.3,0)( 4.5,0)
\end{pspicture}  
 & MBO, EO  \\  
$\rm _1$\.{B}E & ($\rm _1$\.{B}E) & 3  &
\begin{pspicture}(0,-0.2)(4.5,0.2)
\psline[linewidth=0.5pt]{->}(0,0)(4.5,0)
\psline[linewidth=0.5pt,linestyle=dashed](1,-0.2)(1,0.2)
\psline[linewidth=0.5pt,doubleline=true](0,-0.2)(0,0.2)
\psline[linewidth=1.2pt]{*-*}(1,0)( 1.5,0)
\psline[linewidth=1.2pt]{*-}(1.8,0)( 4.5,0)
\end{pspicture}  
 & MBO, EO  \\
\.{B}$\rm _{E=0}$ & ($\rm _1$\.{B}E) & 2  &
\begin{pspicture}(0,-0.2)(4.5,0.2)
\psline[linewidth=0.5pt]{->}(0,0)(4.5,0)
\psline[linewidth=0.5pt,linestyle=dashed](1,-0.2)(1,0.2)
\psline[linewidth=0.5pt,doubleline=true](0,-0.2)(0,0.2)
\psline[linewidth=1.2pt]{*-*}(1,0)( 1,0)
\end{pspicture}  
 & MCO  \\
\"{B} $\rm _1$\.{E}  & (\"{B} $\rm _1$\.{E}) & 3  &
\begin{pspicture}(0,-0.2)(4.5,0.2)
\psline[linewidth=0.5pt]{->}(0,0)(4.5,0)
\psline[linewidth=0.5pt,linestyle=dashed](1,-0.2)(1,0.2)
\psline[linewidth=0.5pt,doubleline=true](0,-0.2)(0,0.2)
\psline[linewidth=1.2pt]{*-*}(0.3,0)( 0.6,0)
\psline[linewidth=1.2pt]{*-}(1,0)( 4.5,0)
\end{pspicture}  
 & BO, TWE 
\\ 
\"{B}$\rm _0$ $\rm _1$\.{E} & (\"{B} $\rm _1$\.{E}) & 3  &
\begin{pspicture}(0,-0.2)(4.5,0.2)
\psline[linewidth=0.5pt]{->}(0,0)(4.5,0)
\psline[linewidth=0.5pt,linestyle=dashed](1,-0.2)(1,0.2)
\psline[linewidth=0.5pt,doubleline=true](0,-0.2)(0,0.2)
\psline[linewidth=1.2pt]{*-*}(0,0)( 0,0)
\psline[linewidth=1.2pt]{*-}(1,0)( 4.5,0)
\end{pspicture}  
 & SCO, TWE 
\\
$\rm _1$\.{E} & ($\rm _1$\.{E}) & 1  &
\begin{pspicture}(0,-0.2)(4.5,0.2)
\psline[linewidth=0.5pt]{->}(0,0)(4.5,0)
\psline[linewidth=0.5pt,linestyle=dashed](1,-0.2)(1,0.2)
\psline[linewidth=0.5pt,doubleline=true](0,-0.2)(0,0.2)
\psline[linewidth=1.2pt]{*-}(1,0)( 4.5,0)
\end{pspicture}  
 & TWE 
\\
\hline\hline
\end{tabular}
\caption{Orbit types for $\omega=1$ (critical case) for massless test particles with $\delta=0$. Thick lines represent the range of the orbits with thick dots as turning points. The surface $x=1$ is indicated by a vertical dashed line. The singularity at $x=r^2=0$ is shown by a solid line at the left. The types with a dot over a letter reach $x=1$, those without a dot do not reach $x=1$ and lie in the outer region, the types with two dots are completely inside the inner region. The index $\rm _1$ on the left or right of a letter indicates the position of the surface $x=1$. 
This table summarizes the discussion of sec.~\ref{sec:omega1} and presents schematically the orbits shown in the effective potentials in fig.~\ref{fig:potl_omega1}. The orbit MCO is a circular orbit of a test particle with $E=0$ and ${A}=0$ with a radius $x=1$. The singular solution SCO at $x=0$ for ${K}={A}^2$ and $E={A}$ completes the set of all mathematically possible cases, but is of no physical relevance.
\label{tab5}}
\end{center}
\end{table}

\subsubsection{The ${K}$-$E$ diagrams}~\label{sec:diag}

To supplement the study on the influence of the parameters of the spacetime and a test particle itself on the test particle's dynamics which was done in the previous subsections in terms of the effective potentials, we use here the method of double zeros approach to polish our knowledge.

For this we consider the resultant $\mathcal{R}$ of the two equations
\begin{equation}
P(x)=0 \quad \text{and} \quad P^\prime(x) = 0 \ ,
\end{equation}
where $P(x)$ is the polynomial in~\eqref{reqn2} and $P^\prime(x)$ is the derivative of 
$P(x)$ w.r.t.~$x$. The resultant is an algebraic function of the form $\mathcal{R}=(E^2-\delta)P(\omega,\frac{{A}}{\sqrt{{K}}},{K},E,\delta)$ with a long polynomial $P(\omega,\frac{{A}}{\sqrt{{K}}},{K},E,\delta)$ of $\omega$, the ratio $\frac{{A}}{\sqrt{{K}}}$, ${K}$, $E$ and $\delta$, and we do not give it here. Instead, we visualize it giving $\omega$ some value and letting ${K}$ and $E$ vary. The ratio ${A}/\sqrt{{K}}$ is also fixed for a single plot. We use here the notation of regions introduced in table~\ref{tab2} for massive test particles and table~\ref{tab3} for massless test particles.

\paragraph{Massive test particles.}

\begin{figure*}[th!]
\begin{center}
\subfigure[][$\omega= 0.7, {A}=0.1 \sqrt{{K}} $]{\label{jE_w_lt1_1}\includegraphics[width=6cm]{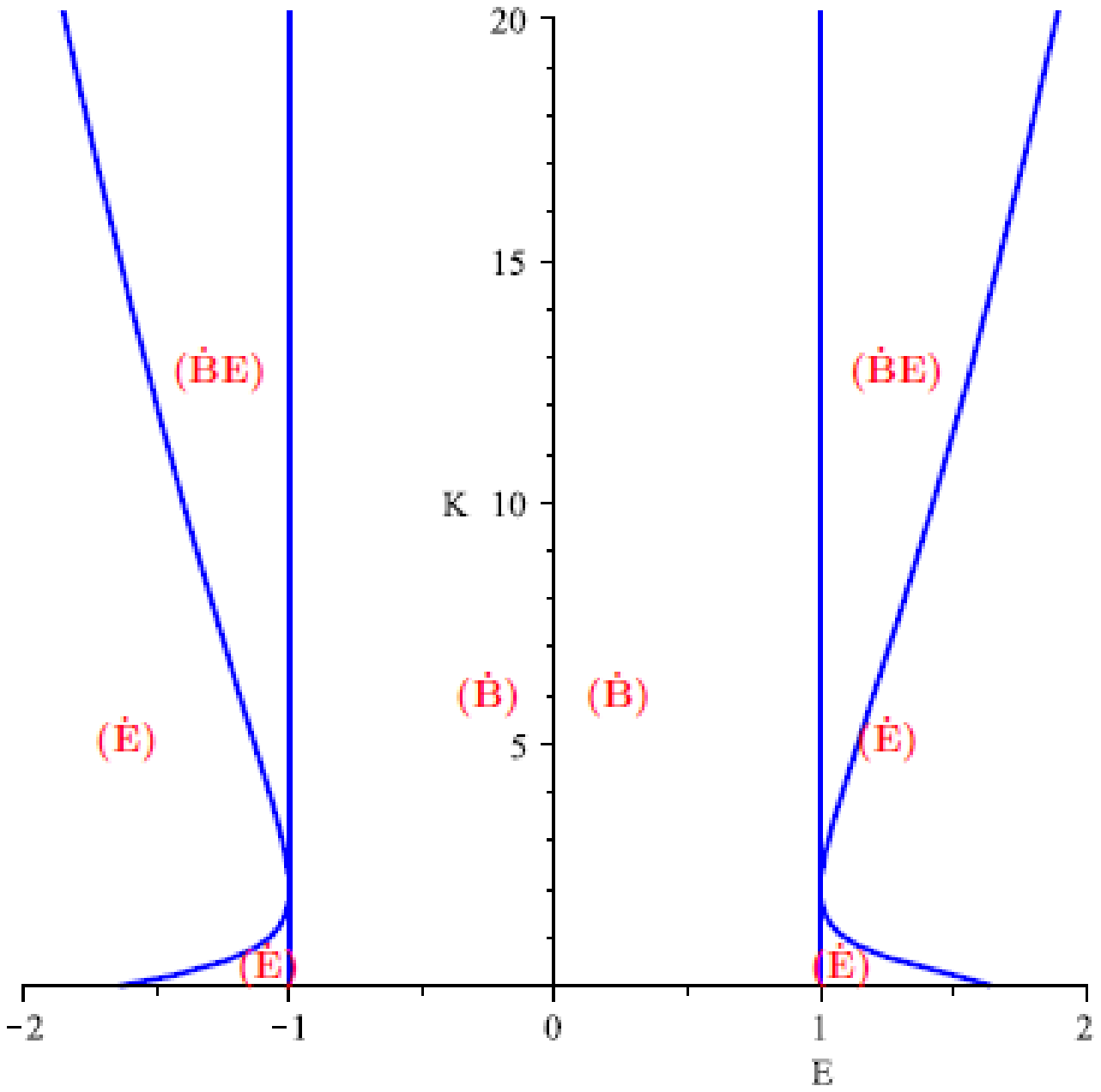}}
\subfigure[][$\omega= 0.7, {A}=\sqrt{{K}} $]{\label{jE_w_lt1_2}\includegraphics[width=6cm]{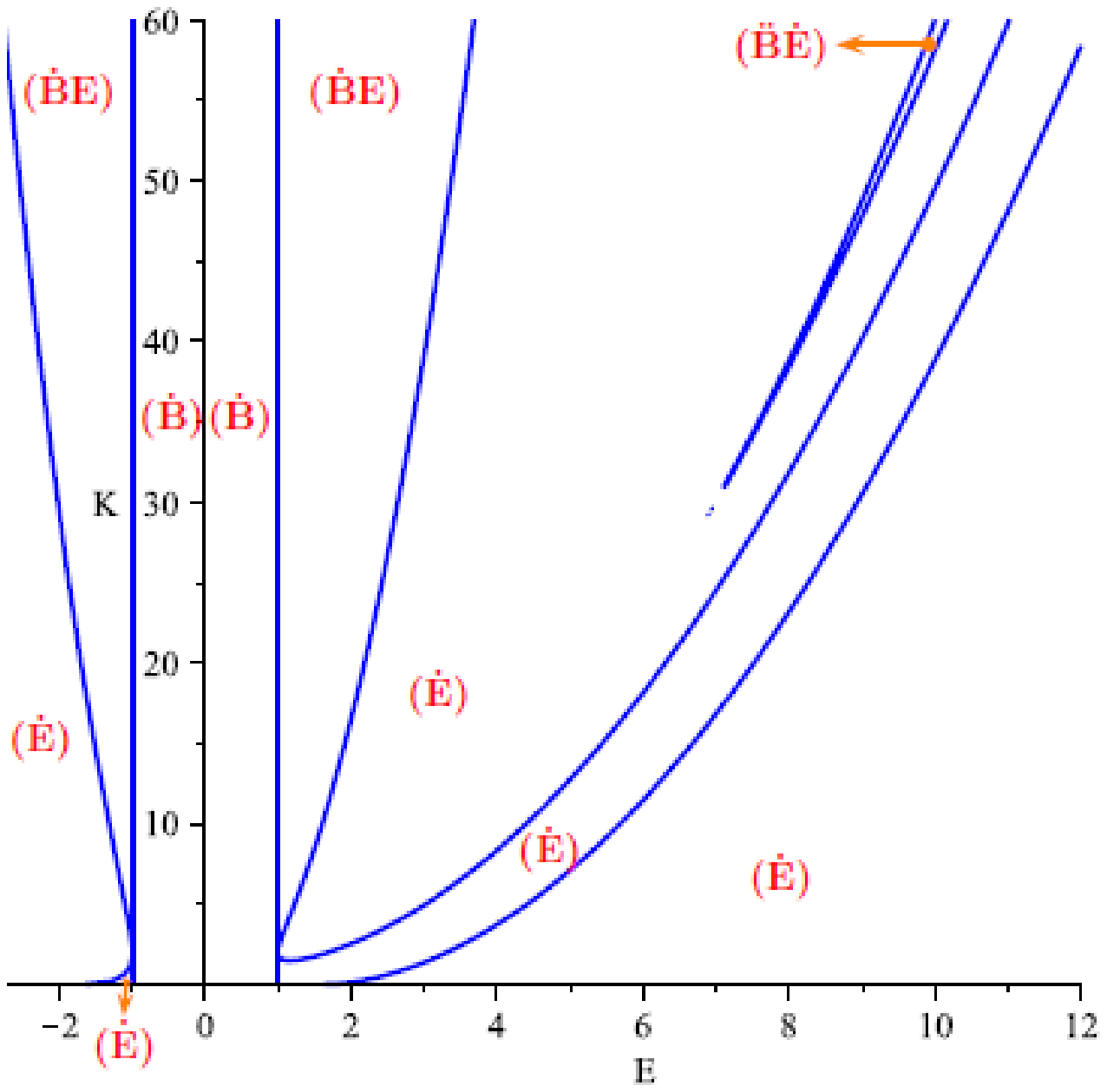}}
\end{center}
\caption{ Figures~\subref{jE_w_lt1_1} and~\subref{jE_w_lt1_2} show the ${K}$-$E$ diagram for $\omega=0.7$ ($\omega<1$, i.e., underrotating case) for massive test particles with $\boldsymbol{\delta=1}$. The regions are in accordance with the table~\ref{tab2}. See also the description in the section~\ref{sec:diag}. \label{fig:jEwlt1}}
\end{figure*}

Consider first fig.~\ref{fig:jEwlt1} for $\omega<1$ (underrotating case) and massive test particles. In the pictures~\subref{jE_w_lt1_1} and~\subref{jE_w_lt1_2} we choose $\omega=0.7$. Having plotted the diagrams for other values of $\omega$ smaller than $1$ we do not see any qualitative difference between the plots. In this case it is the ratio ${A}/\sqrt{{K}}$ which influences the form of the diagram essentially. Looking at~\ref{jE_w_lt1_1} where ${A}=0.1 \sqrt{{K}}$ we observe that the possible orbits in the region (\.{B}) are many world bound orbits for $|E|<1$. For $|E|>1$ region (\.{E}) with two world escape orbits, and region (\.{B}E) containing many world bound orbit and escape orbit are possible. 

This diagram generalizes the effective potential in fig.~\ref{pot1} for higher ${K}$ values. For ${A}=\sqrt{{K}}$ in fig.~\ref{jE_w_lt1_2} for $\omega=0.7$ a new region (\"{B}\.{E}) with bound orbits hidden behind the horizon and a two world escape orbit appear. This happens for large ${K}$ values. Such regions we have seen in the effective potential for $\omega=0.9$ in fig.~\ref{pot2jR}. We note that the regions (\.{E}) separated from each other by a blue line indicating the presence of double roots in $P(x)$ have one positive zero but either 2 negative zeros or 2 complex conjugate, which are however not relevant for the physical motion.

\begin{figure*}[th!]
\begin{center}
\subfigure[][$\omega= 1.1, {A}=0.1 \sqrt{{K}} $]{\label{jE_w_gt1_1}\includegraphics[width=6cm]{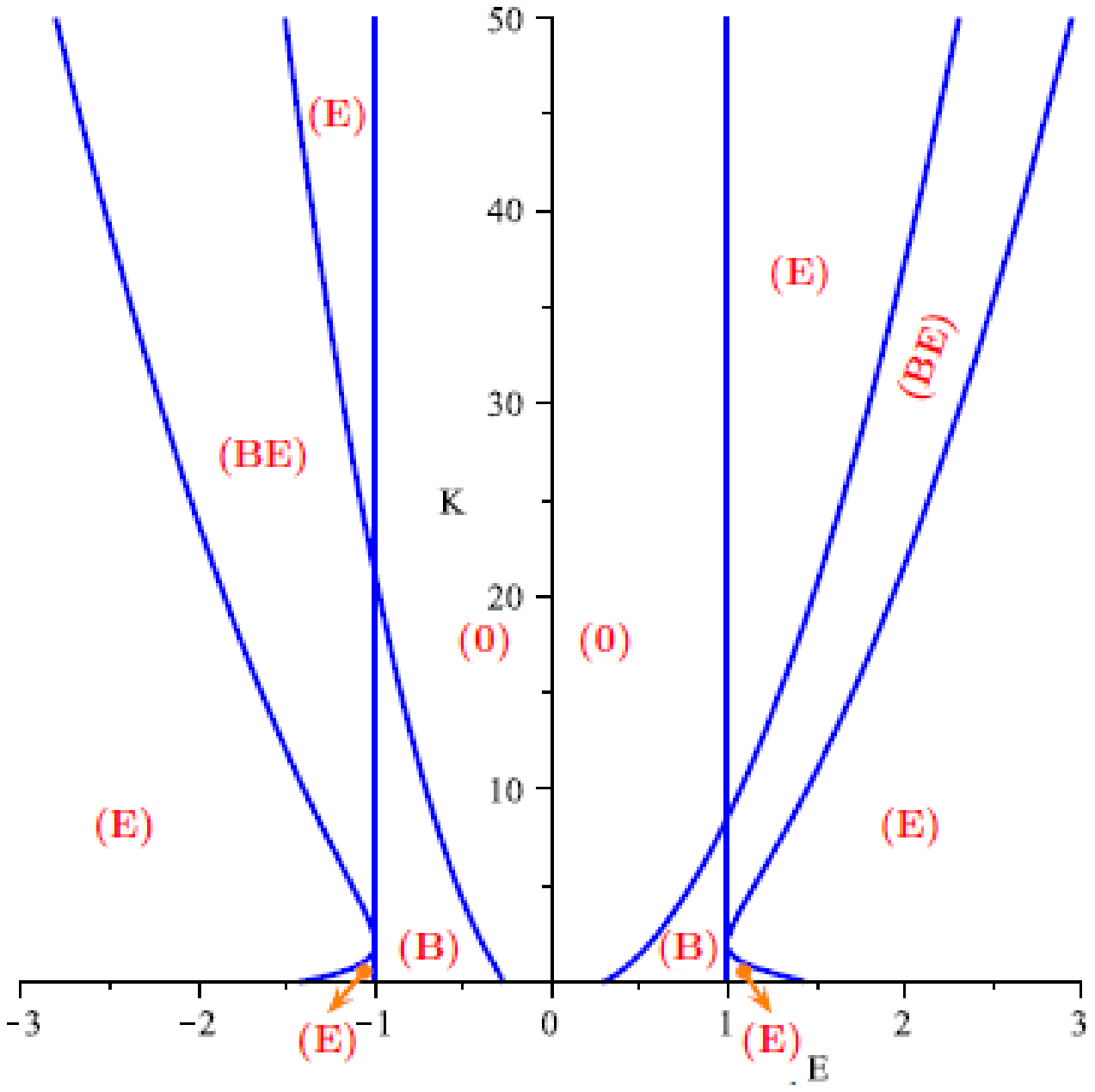}}
\subfigure[][$\omega= 2.1, {A}=0.1 \sqrt{{K}} $]{\label{jE_w_gt1_4}\includegraphics[width=9.5cm]{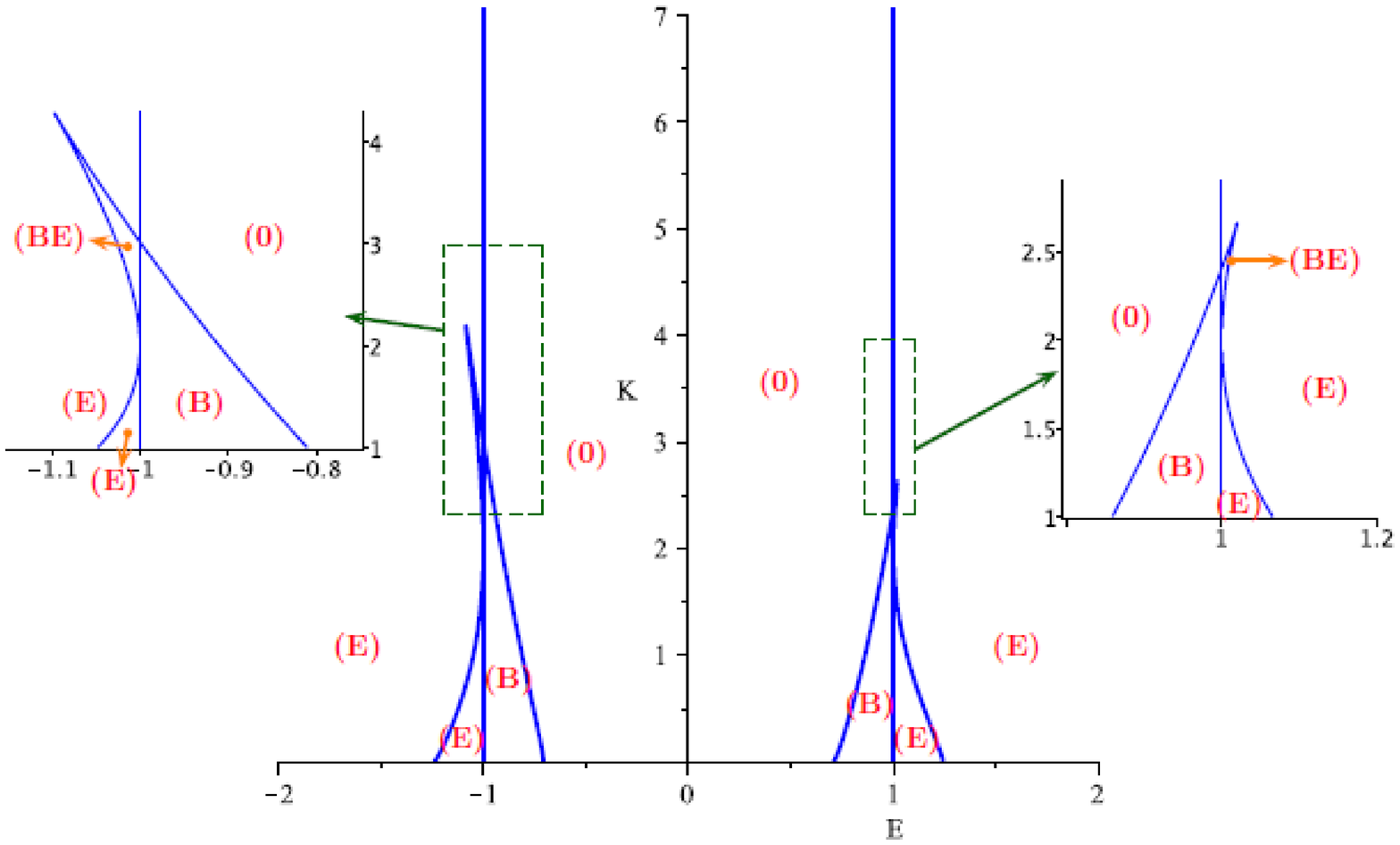}}
%%%
\subfigure[][$\omega= 1.1, {A}=0.59 \sqrt{{K}} $]{\label{jE_w_gt1_2}\includegraphics[width=7.7cm]{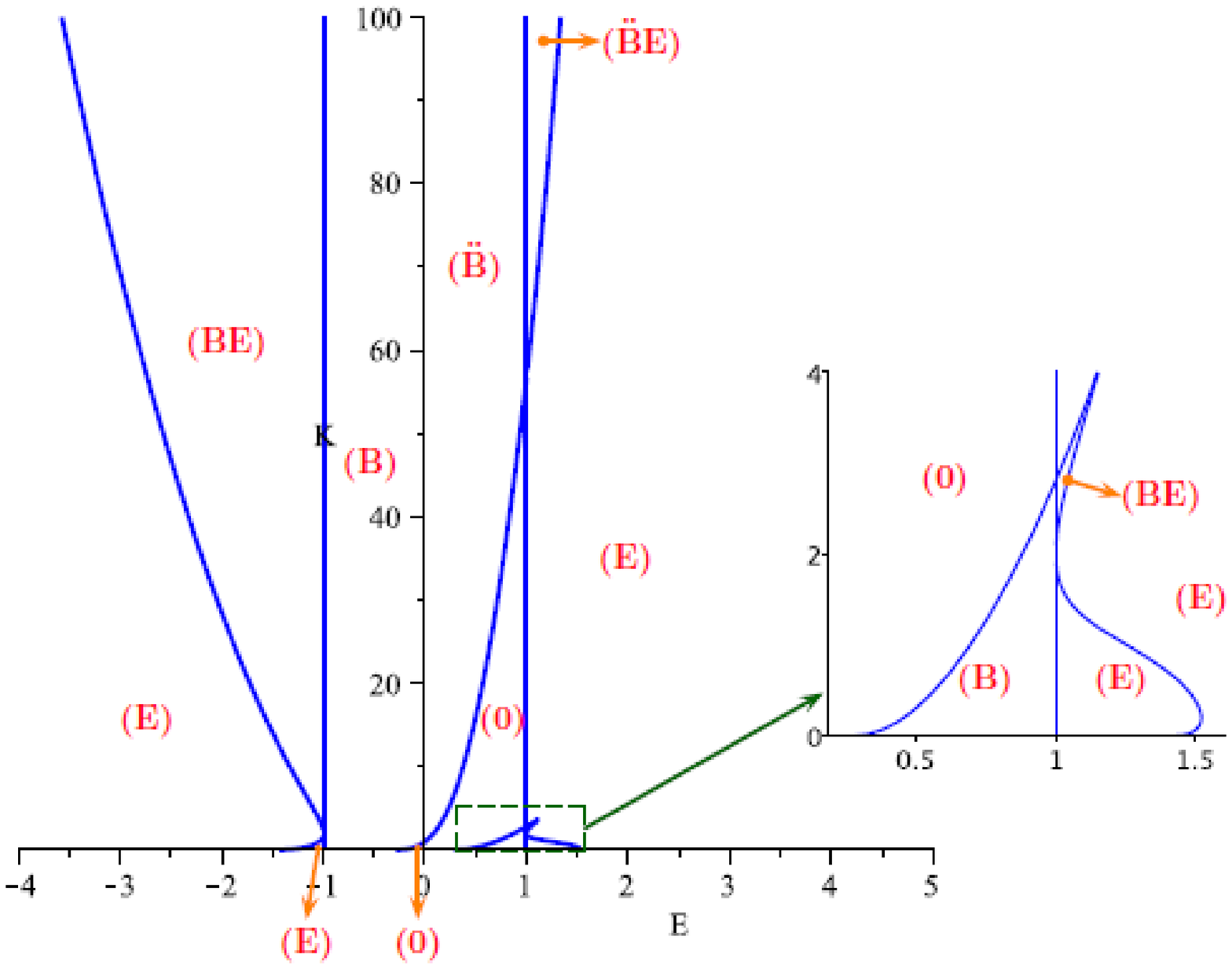}}
\subfigure[][$\omega= 2.1, {A}=0.59 \sqrt{{K}} $]{\label{jE_w_gt1_5}\includegraphics[width=6cm]{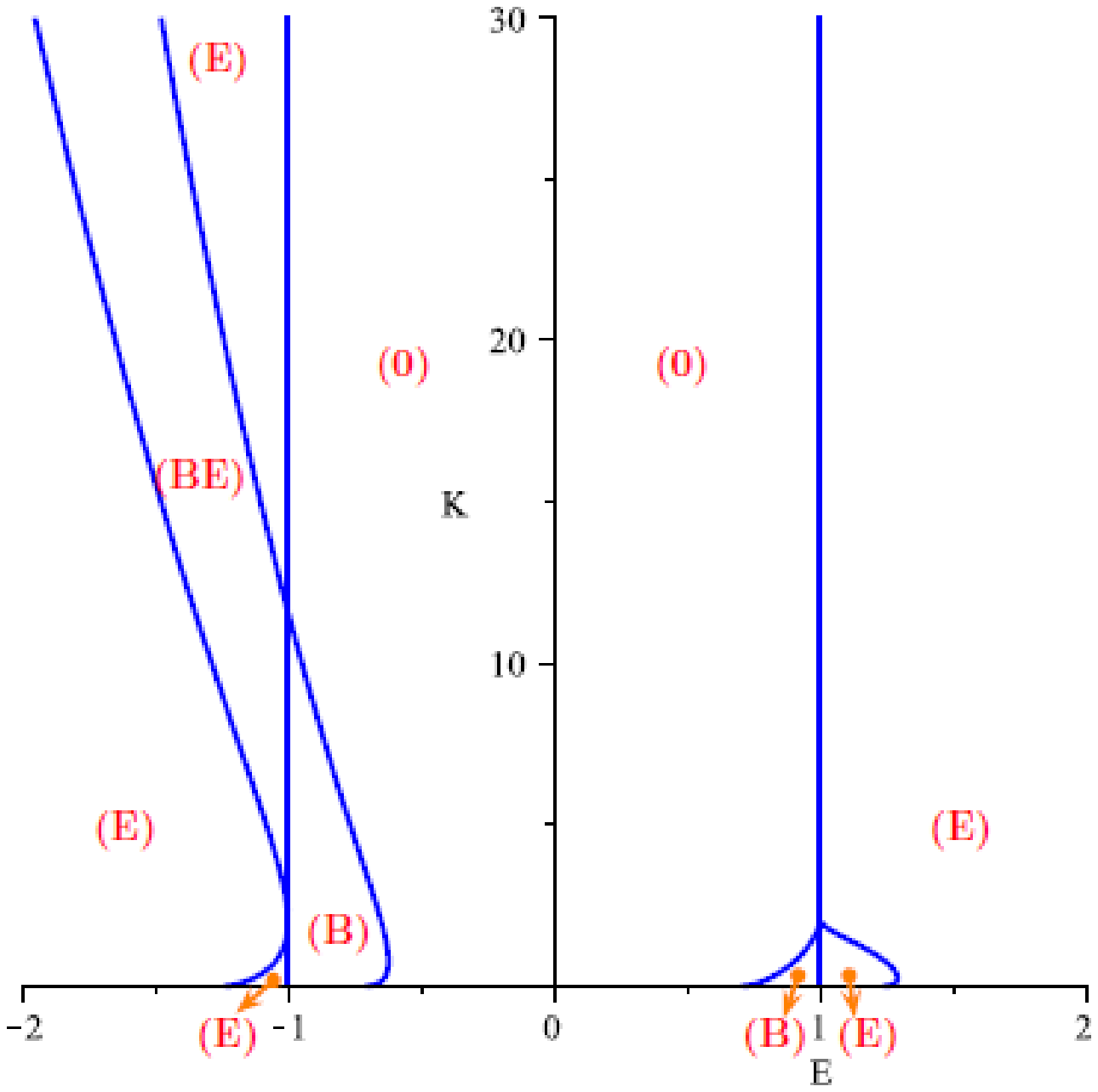}}
%%%
\subfigure[][$\omega= 1.1, {A}=\sqrt{{K}} $]{\label{jE_w_gt1_3}\includegraphics[width=6cm]{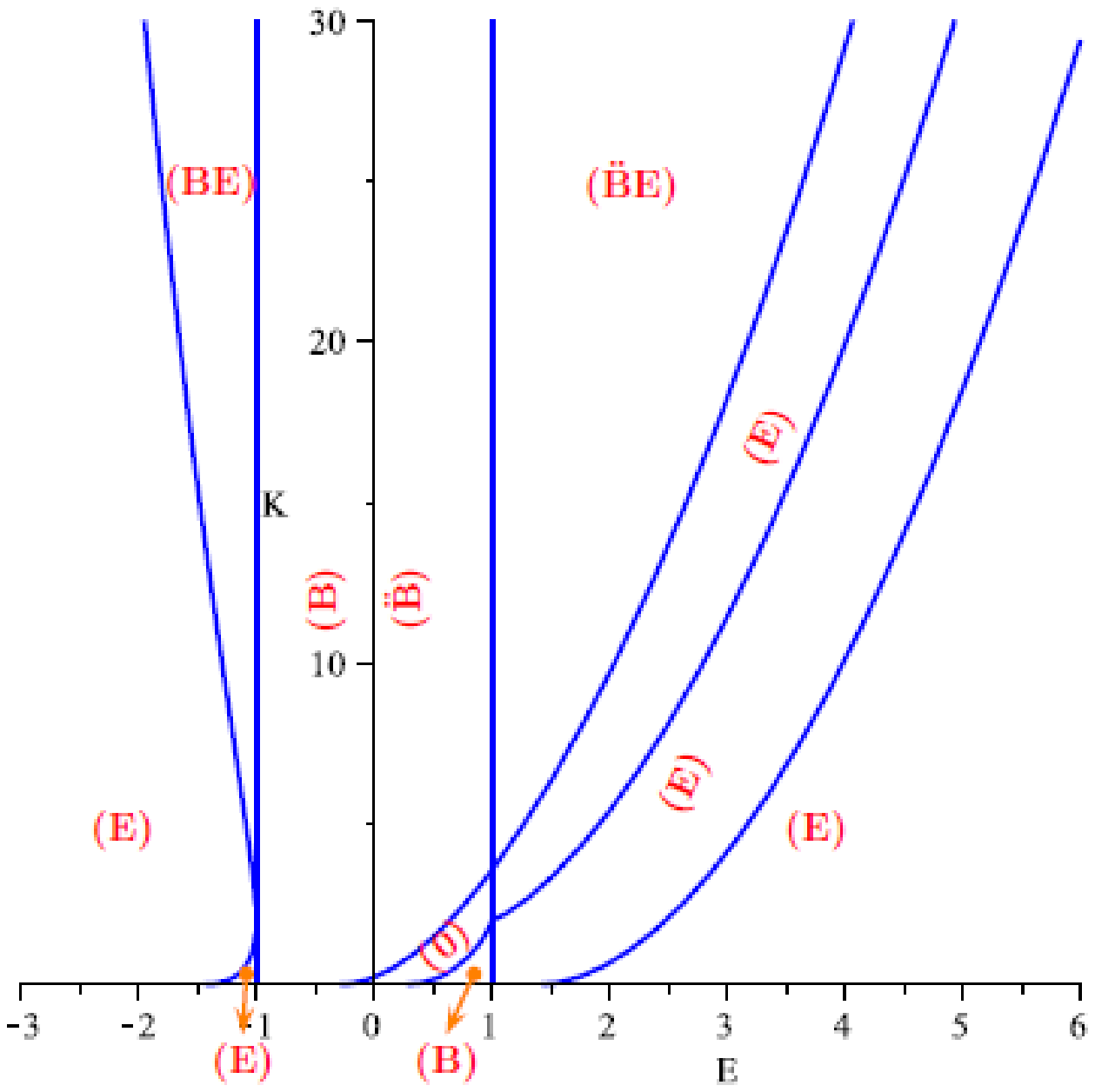}}
\subfigure[][$\omega= 2.1, {A}=\sqrt{{K}} $]{\label{jE_w_gt1_6}\includegraphics[width=6cm]{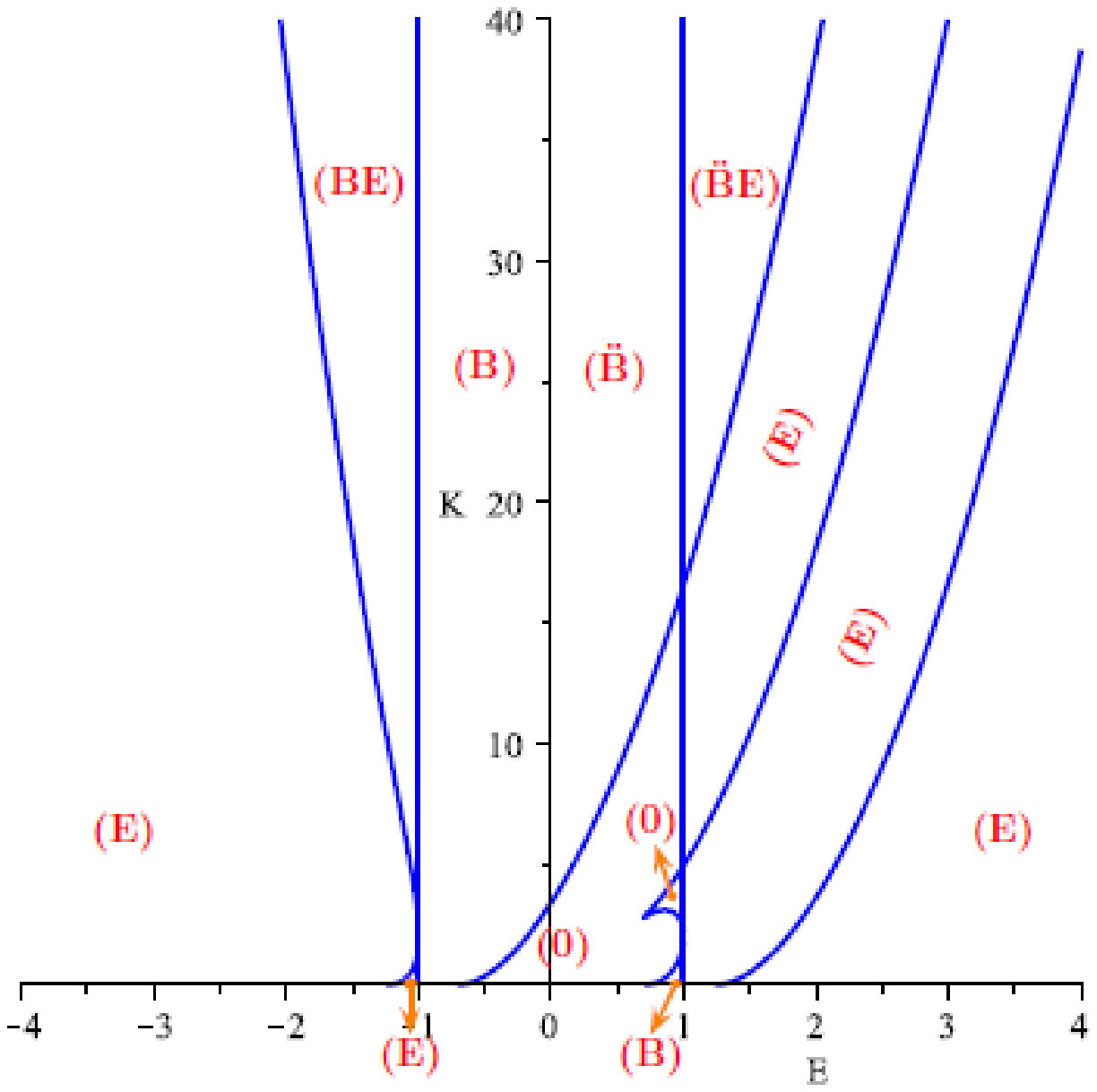}}
\end{center}
\caption{ Figures~\subref{jE_w_gt1_1},\subref{jE_w_gt1_2},\subref{jE_w_gt1_3} show the ${K}$-$E$ diagram for $\omega=1.1$ ($\omega>1$, i.e., overrotating case) and growing ${A}$ for massive test particles with $\boldsymbol{\delta=1}$. Figures~\subref{jE_w_gt1_4},\subref{jE_w_gt1_5},\subref{jE_w_gt1_6} -- for $\omega=2.1$. The regions are in accordance with the table~\ref{tab2}. There are no real positive roots in the region $(0)$. See also the discussion in the section~\ref{sec:diag}. \label{fig:jEwgt2}}
\end{figure*}

In fig.~\ref{fig:jEwgt2} we show the ${K}$-$E$ diagrams for $\omega>1$ (overrotating case), namely for $\omega=1.1$ in the first column and $\omega=2.1$ in the second column. Compare figures~\subref{jE_w_gt1_1} and~\subref{jE_w_gt1_4} for small ${A}$. In the region (B) planetary bound orbits BO for $|E|<1$ are possible. These orbits can be found in the potential plots~\ref{pot3} and~\subref{pot4}. In the region (0) no orbits exist: it corresponds to the forbidden grey regions in the figures~\ref{pot3},~\subref{pot4} for $|E|<1$. Here the value of ${A}$ is smaller than the critical value ${{A}^{c}}^2$ defined in~\eqref{cond_jR} for which a loop with bound orbits in the inner region forms (see also discussion in the section~\ref{sec:pot}).

Regions (E) contain escape orbits. They are separated in the pictures since the number of negative or complex zeros, irrelevant for physical motion, varies there. In the region (BE) for larger ${K}$ values and $|E|>1$ both planetary bound and escape orbits exist. From these diagrams we can infer that for growing $\omega$ the region (BE) becomes smaller. It disappears for large $\omega$. For very large $\omega$ also the region (B) does not exist and only escape orbits in the region (E) are left. In this case the potential is reminiscent of $\pm 1/x$ curves (we have already observed this for massless test particles in the section~\ref{sec:delta0}).

Continuing with the analysis of fig.~\ref{fig:jEwgt2} we compare figures~\subref{jE_w_gt1_2} and~\subref{jE_w_gt1_5} for ${A}=0.59 \sqrt{{K}}$. In the plot~\ref{jE_w_gt1_2} for smaller $\omega$ new regions form. For $0<E<1$ this is a region (\"{B}) with bound orbits behind the pseudo-horizon like the one in the potentials~\ref{pot3jR} or~\ref{pot4jR}. For $-1<E<0$ the region ({B}) grows and the region (0) with no motion becomes smaller compared to the plot~\subref{jE_w_gt1_1}. Also for $E>1$ a new region (\"{B}E) with bound orbits behind the pseudo-horizon and an escape orbit appears. The region (BE) with planetary bound and escape orbits is still there, which is best seen in the inlay of the picture~\ref{jE_w_gt1_2}. On the contrary this region disappears for $\omega=2.1$ in the plot~\ref{jE_w_gt1_5}, and no new region appears here.  

For the maximal ${A}$ value in the plots~\ref{jE_w_gt1_3} and~\subref{jE_w_gt1_6} the regions with inner bound orbits (\"{B}) and (\"{B}E) dominate in both plots for $E>0$ and the region (BE) is present only for negative $E$. Planetary bound orbits for positive energies are possible in the plots~\ref{jE_w_gt1_3} and~\subref{jE_w_gt1_6} only for $0<E<1$ and small ${K}$ in the region (B). For large $\omega$ and ${A}=\sqrt{{K}}$ the region (B) for $0<E<1$ disappears. The described behaviour is inverted of course for negative ${A}$: in this case the region (BE) is on the side with positive energies and the regions with inner bound orbits (\"{B}) and (\"{B}E) are on the negative energy side. An example of the effective potential for ${A}=-\sqrt{{K}}$ is shown in fig.~\ref{pot4jR}. Here planetary bound orbits and escape orbits located in the regions (B) and (BE) exist for positive $E$.

\paragraph{Massless test particles.}

As we know from table~\ref{tab3} the variety of trajectory types for massless test particles is not that rich as for massive particles. But photons can still move on a bound trajectory. In the following diagrams we will see for which ${A}/\sqrt{{K}}$ and $E$ that is possible. 

\begin{figure*}[th!]
\begin{center}
\subfigure[][$\omega= 0.7, {A}=0.1 \sqrt{{K}} $]{\label{jEl_w_lt1_1}\includegraphics[width=6cm]{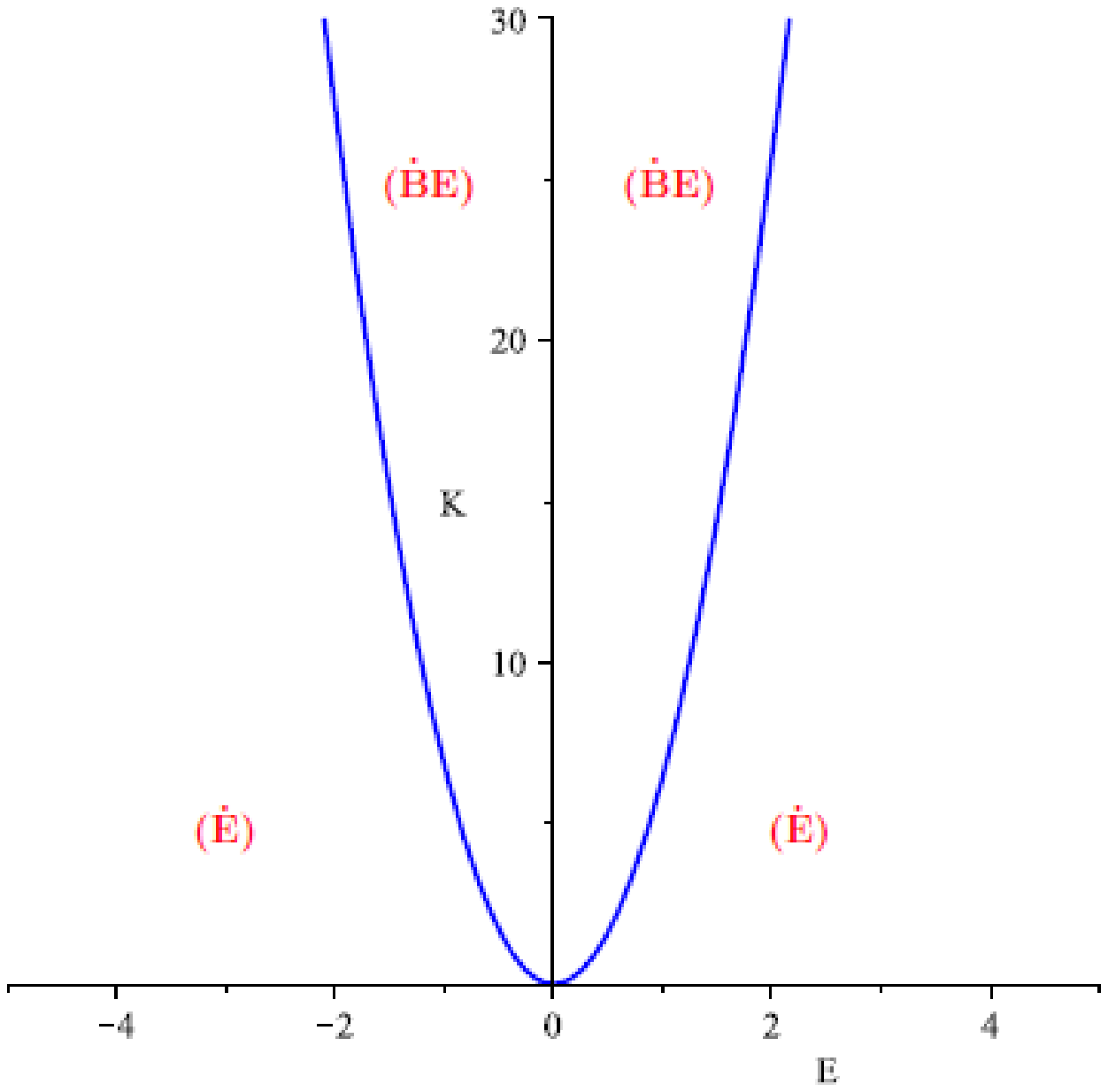}}
\subfigure[][$\omega= 0.7, {A}= \sqrt{{K}} $]{\label{jEl_w_lt1_2}\includegraphics[width=6cm]{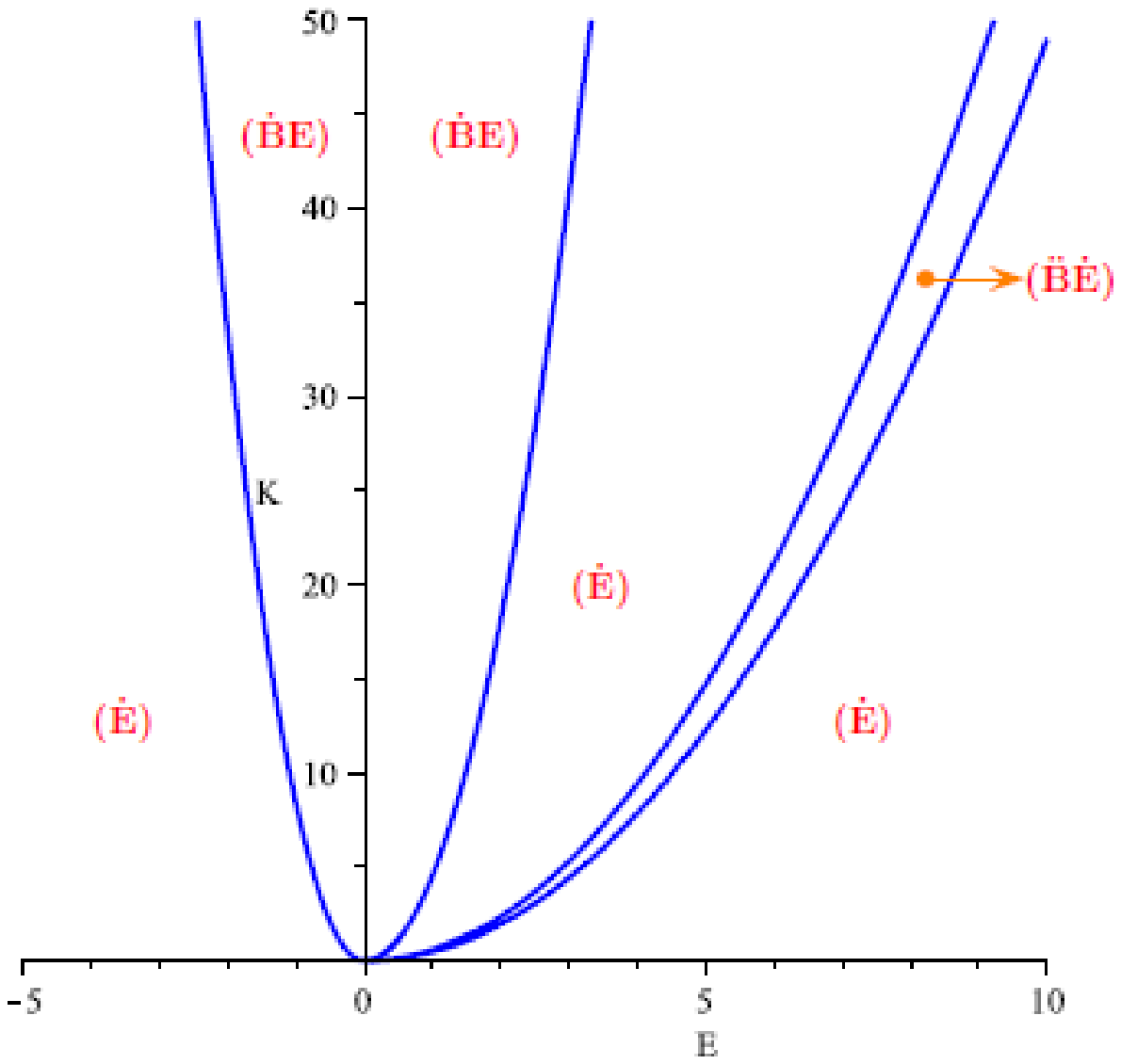}}
\end{center}
\caption{ Figures~\subref{jEl_w_lt1_1} and~\subref{jEl_w_lt1_2} show the ${K}$-$E$ diagram for $\omega=0.7$ ($\omega<1$, i.e., underrotating case) for massless test particles with $\boldsymbol{\delta=0}$. The regions in the diagrams are in accordance with table~\ref{tab3}. Since for $\omega<1$ the effective potential~\eqref{rpot1} has a minimum and a maximum the many world bound orbits and escape orbits corresponding to the region (\.BE) exist for positive and negative energies. For large $\frac{{A}}{\sqrt{{K}}}$ ratio (provided the inequality~\eqref{condj1} is fulfilled) a new region (\"B\.E) appears. Here inner bound orbits behind the horizon and two world escape orbits are possible. In the region (\.E) in both diagrams two world bound orbits exist. See also the description in the section~\ref{sec:diag}. \label{fig:jElwlt1}}
\end{figure*}

In fig.~\ref{fig:jElwlt1} we present two ${K}$-$E$ diagrams for $\omega=0.7$ (underrotating case) and growing $\frac{{A}}{\sqrt{{K}}}$ ratio. In the figure~\subref{jEl_w_lt1_1} we have $\left. i \right)$ the region (\.BE) with many world bound and escape orbits, and $\left. ii \right)$ the region (\.E) with two world escape orbits. A typical effective potential for this diagram is shown in fig.~\ref{pot1l}. From the discussion in the section~\ref{sec:pot} we know that inner bound orbits hidden from external observers are possible. This happens for high $\frac{{A}}{\sqrt{{K}}}$ ratios and corresponds to a new region (\"B\.E) in the figure~\subref{jEl_w_lt1_2} with inner bound orbits and two world escape orbits. Such orbits can be found for example in the potential~\ref{pot1ljR} for $\omega=0.7$ and ${A}=\sqrt{{K}}$.

\begin{figure*}[th!]
\begin{center}
\subfigure[][$\omega= 1.1, {A}=0.1  \sqrt{{K}} $, $\frac{{A}}{ \sqrt{{K}}}<\frac{{A}^{c}}{\sqrt{{K}}}$]{\label{jEl_w_gt1_1}\includegraphics[width=6cm]{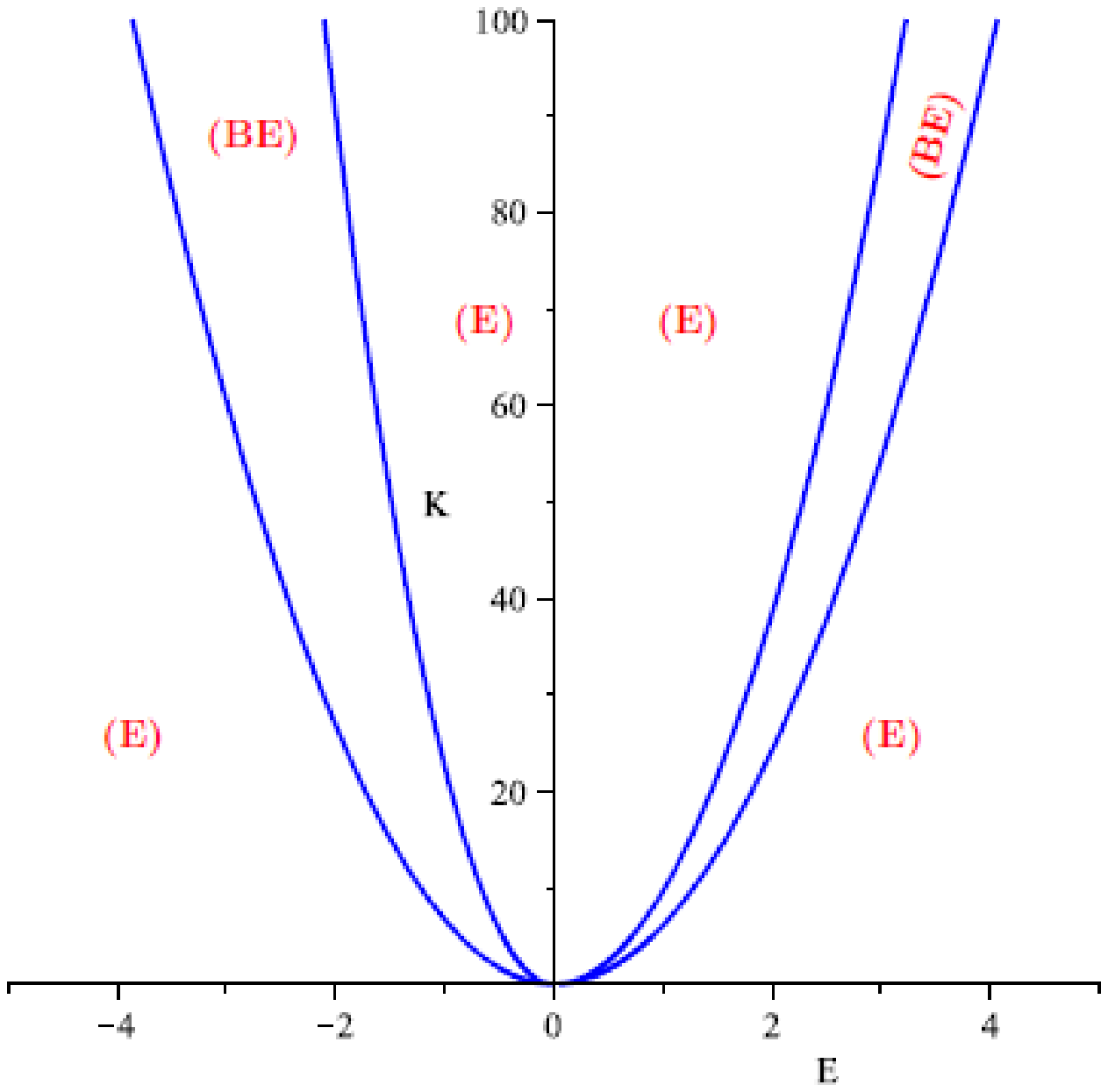}}
\subfigure[][$\omega= 1.1$, $\frac{{A}}{\sqrt{{K}}}=\frac{{A}^{c}}{ \sqrt{{K}}}=\sqrt{\frac{\omega^2-1}{\omega^2}}$]{\label{jEl_w_gt1_3}\includegraphics[width=6cm]{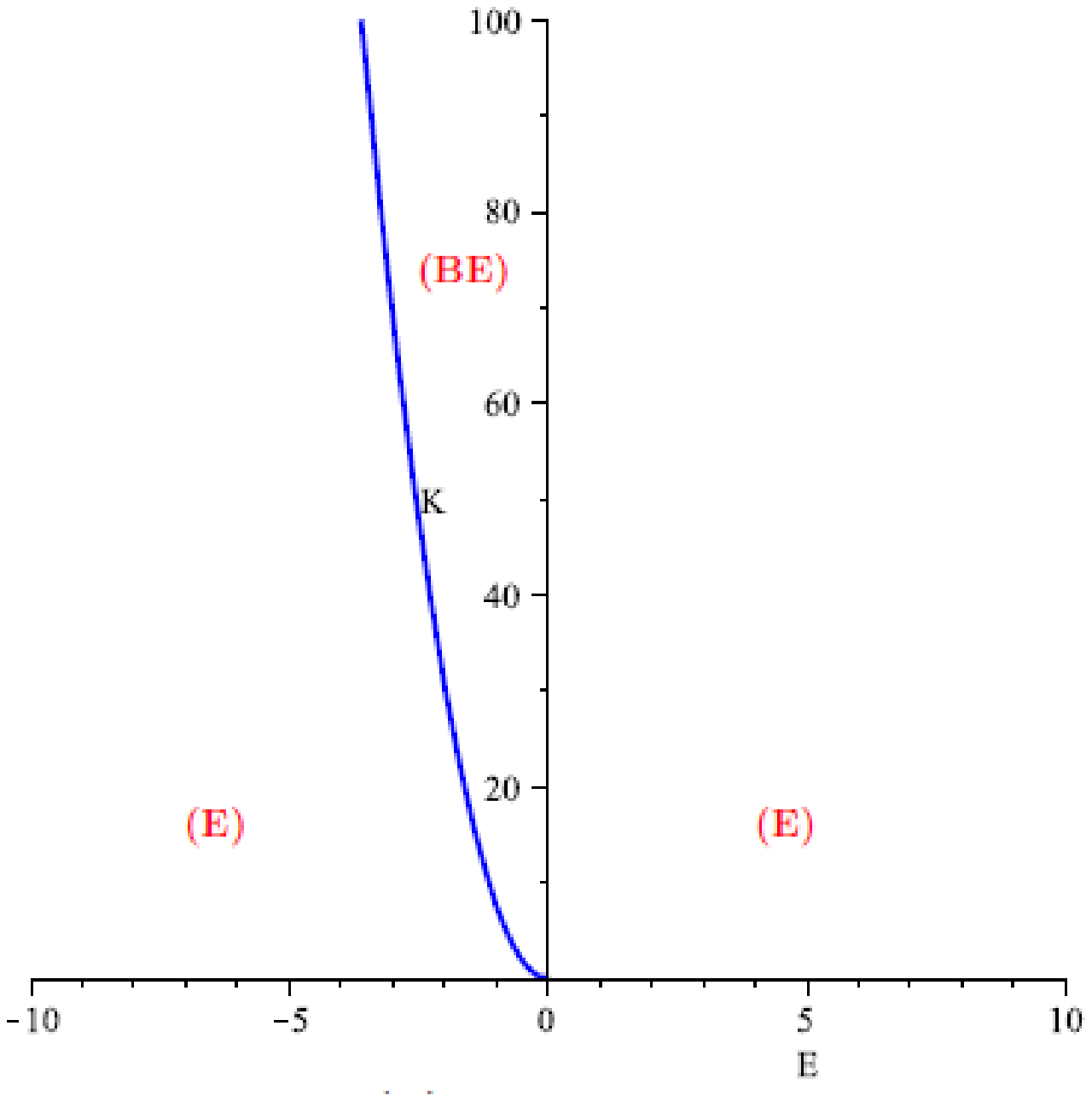}}
\subfigure[][$\omega= 1.1, {A}=  \sqrt{{K}}$, $\frac{{A}}{ \sqrt{{K}}}>\frac{{A}^{c}}{ \sqrt{{K}}}$]{\label{jEl_w_gt1_2}\includegraphics[width=6cm]{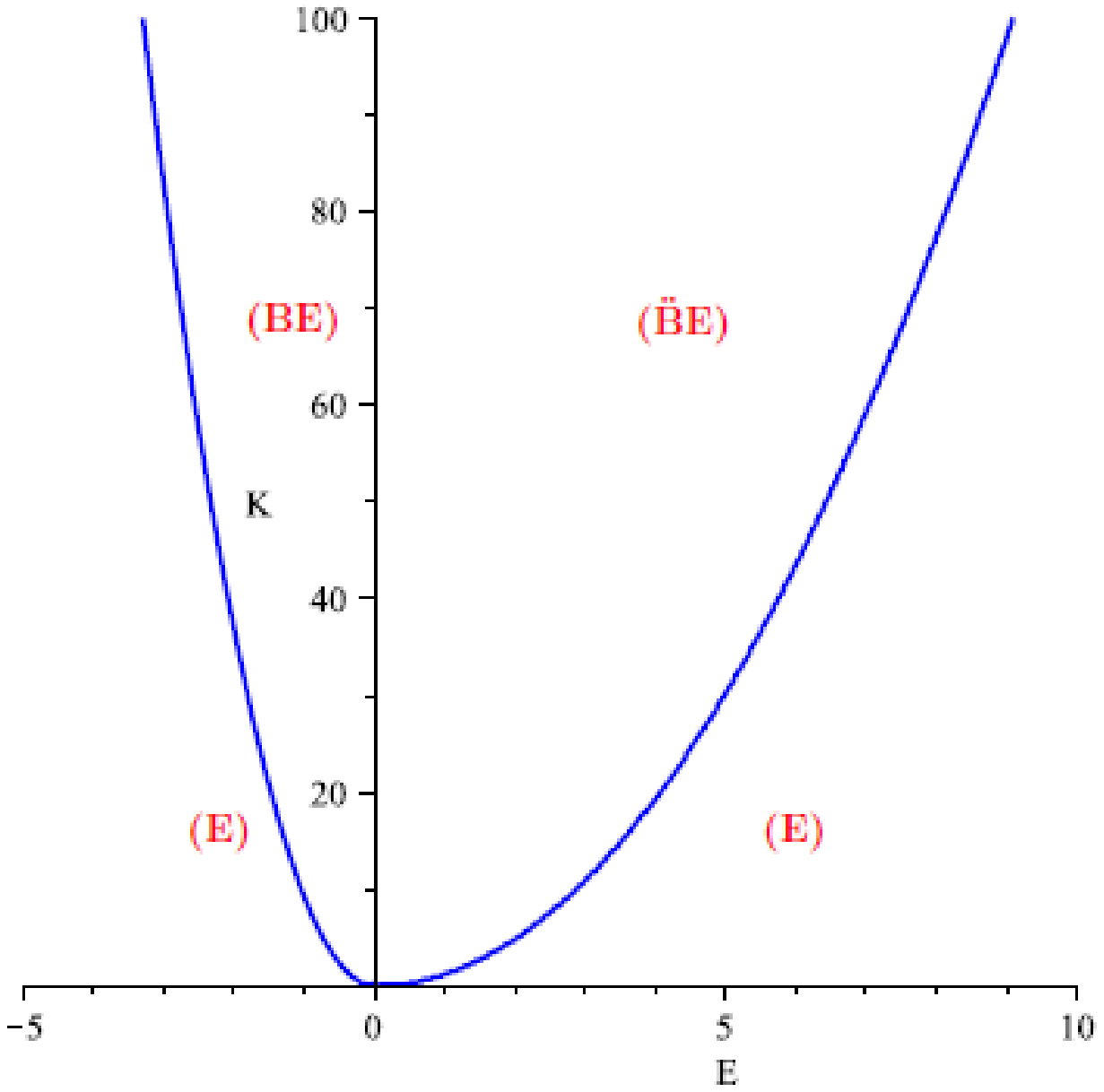}}
\subfigure[][$\omega= 2.1, {A}= 0.5  \sqrt{{K}}$, $\frac{{A}}{ \sqrt{{K}}}<\frac{{A}^{c}}{ \sqrt{{K}}}$]{\label{jEl_w_gt1_4}\includegraphics[width=6cm]{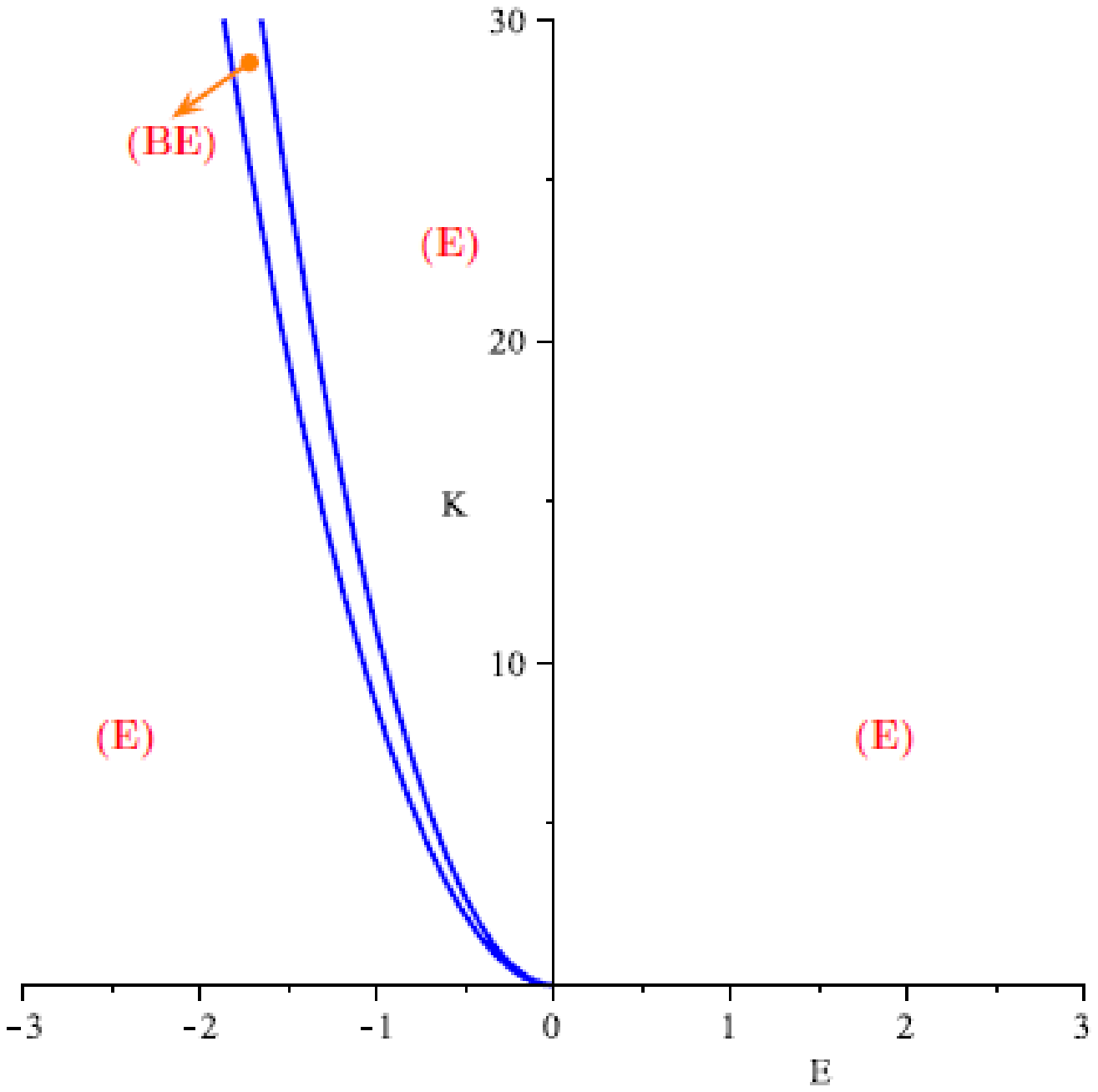}}
\end{center}
\caption{ Figures~\subref{jEl_w_gt1_1},~\subref{jEl_w_gt1_3} and~\subref{jEl_w_gt1_2} show the ${K}$-$E$ diagram for $\omega=1.1$  and figure~\subref{jEl_w_gt1_4} for $\omega=2.1$ ($\omega>1$, i.e., overrotating case) for massless test particles with $\boldsymbol{\delta=0}$. The regions in the diagrams are in accordance with table~\ref{tab3}. For small $\omega$ and ${A}<{A}^{c}$ defined by~\eqref{cond_jR} as in the diagram~\subref{jEl_w_gt1_1} planetary bound orbits and escape orbits exist both for positive and negative energies in the region (BE). In the region (E) escape orbits are found. For the critical value ${A}={A}^{c}$ the (BE) region is located only in the negative energies part and the ${K}$-axis where $E=0$ corresponds to a circular MCO orbit at $r^2=x=1$  (diagram~\subref{jEl_w_gt1_3}). If ${A}>{A}^{c}$ a new region (\"{B}E) with inner bound and escape orbits forms (diagram~\subref{jEl_w_gt1_2}). For larger $\omega$ and ${A}<{A}^{c}$ the (BE) region survives only for negative energies and for positive energies only escape orbits in the region (E) exist (diagram~\subref{jEl_w_gt1_4}). The cases when ${A}\geq {A}^{c}$ stay qualitatively unchanged. For very large $\omega$ also for negative energies only the region (E) can be found. The diagrams for ${A}\geq {A}^{c}$ are of the type~\subref{jEl_w_gt1_3} and~\subref{jEl_w_gt1_2}. See also the description in the section~\ref{sec:diag}. 
\label{fig:jElwgt2}}
\end{figure*}

Consider now $\omega>1$ (overrotating case) and fig.~\ref{fig:jElwgt2}. In the plot~\subref{jEl_w_gt1_1} where the ratio $\frac{{A}}{\sqrt{{K}}}$ is smaller (for positive ratio) than the critical value $\frac{{A}^{c}}{\sqrt{{K}}}=\pm\sqrt{\frac{\omega^2-1}{\omega^2}}$ from~\eqref{jR_special}, the region (BE) with planetary bound and escape orbits exists both for positive and negative energies. This corresponds to the effective potential in the plot~\ref{pot2l}. The second possible region (E) contains escape orbits. When the ratio $\frac{{A}}{\sqrt{{K}}}$ is equal to the critical value $\pm\sqrt{\frac{\omega^2-1}{\omega^2}}$ the region (BE) for positive $\frac{{A}}{\sqrt{{K}}}$ disappears for positive energies. Thus, for $E>0$ only region (E) exists as it is shown in the plot~\subref{jEl_w_gt1_3}. Regions (BE) and (E) are still there for $E<0$ and there is only one blue line for negative energies separating these regions. This is inverted for negative $\frac{{A}^{c}}{\sqrt{{K}}}$. For $E=0$ the polynomial $P(x)$ has 2 positive zeros equal to $1$ all over the ${K}$-axis. This corresponds to the MCO orbit in table~\ref{tab3}. 

In fig.~\ref{pot3ljR} we show an effective potential for $\frac{{A}}{\sqrt{{K}}}=-\sqrt{\frac{\omega^2-1}{\omega^2}}$. Let the ratio $\frac{{A}}{\sqrt{{K}}}$ be larger than $\frac{{A}^{c}}{\sqrt{{K}}}=\sqrt{\frac{\omega^2-1}{\omega^2}}$, choosing the positive sign (for the negative sign of the ratio the picture is inverted). Then a new region (\"BE) for positive energies appears. This region grows for increasing $\frac{{A}}{\sqrt{{K}}}$ and reaches a maximal size for ${A}=\sqrt{{K}}$ as it is shown in the diagram~\subref{jEl_w_gt1_2}. An example of the effective potential for ${A} =0.95 \sqrt{{K}} > {A}^{c}$ is shown in fig.~\ref{pot2ljR}, and ${A} = \sqrt{{K}} > {A}^{c}$ is presented in fig.~\ref{pot4ljR}. In the region (\"BE) in the diagram~\subref{jEl_w_gt1_2} inner bound and escape orbits exist. For $E<0$ outer bound and escape orbits are still present.

For a bit larger  value of $\omega$, e.g. $\omega=2.1$, and ${A}<{A}^{c}$ from equation~\eqref{jR_special} (we choose positive ${A}$) only the region (E) survives and the region (BE) for positive energies does not exist, since the effective potential has no minimum and no maximum for $E>0$ any longer. These exist only for negative energies. In this case the ${K}$-$E$ diagram has a form like in fig.~\ref{jEl_w_gt1_4}. For the critical value ${A}={A}^{c}$ and for ${A}>{A}^{c}$ the diagrams look similar to the pictures~\ref{jEl_w_gt1_3} and~\ref{jEl_w_gt1_2}, respectively.

When further increasing $\omega$ and for non-critical (and positive) ${A}$ the minimum and maximum in the effective potential like in fig.~\ref{pot2l} do not exist any longer also for negative energies, and the potential is reminiscent of $\pm \frac{1}{x}$ curves as we already know. Both sides of the ${K}$-$E$ diagram for positive and negative energies contain only the (E) regions. For the value of the ratio $\frac{{A}}{\sqrt{{K}}} > \sqrt{\frac{\omega^2-1}{\omega^2}}$ (larger than the critical value), the regions with inner bound orbit and escape orbit (\"BE) for positive energies and bound and escape orbits (BE) for negative energies form. Here again the diagrams are of type~\ref{jEl_w_gt1_2} for $\frac{{A}}{\sqrt{{K}}}>\frac{{A}^{c}}{\sqrt{{K}}}$ and of type~\ref{jEl_w_gt1_3} for $\frac{{A}}{\sqrt{{K}}}=\frac{{A}^{c}}{\sqrt{{K}}}$, where the (BE) region for negative energies exist.

\paragraph{${K}$-$E$ diagrams for $\omega=1$.}

\begin{figure*}[th!]
\begin{center}
\subfigure[][$j_R=0.3 \sqrt{{K}}$]{\label{jE_w_1}\includegraphics[width=6cm]{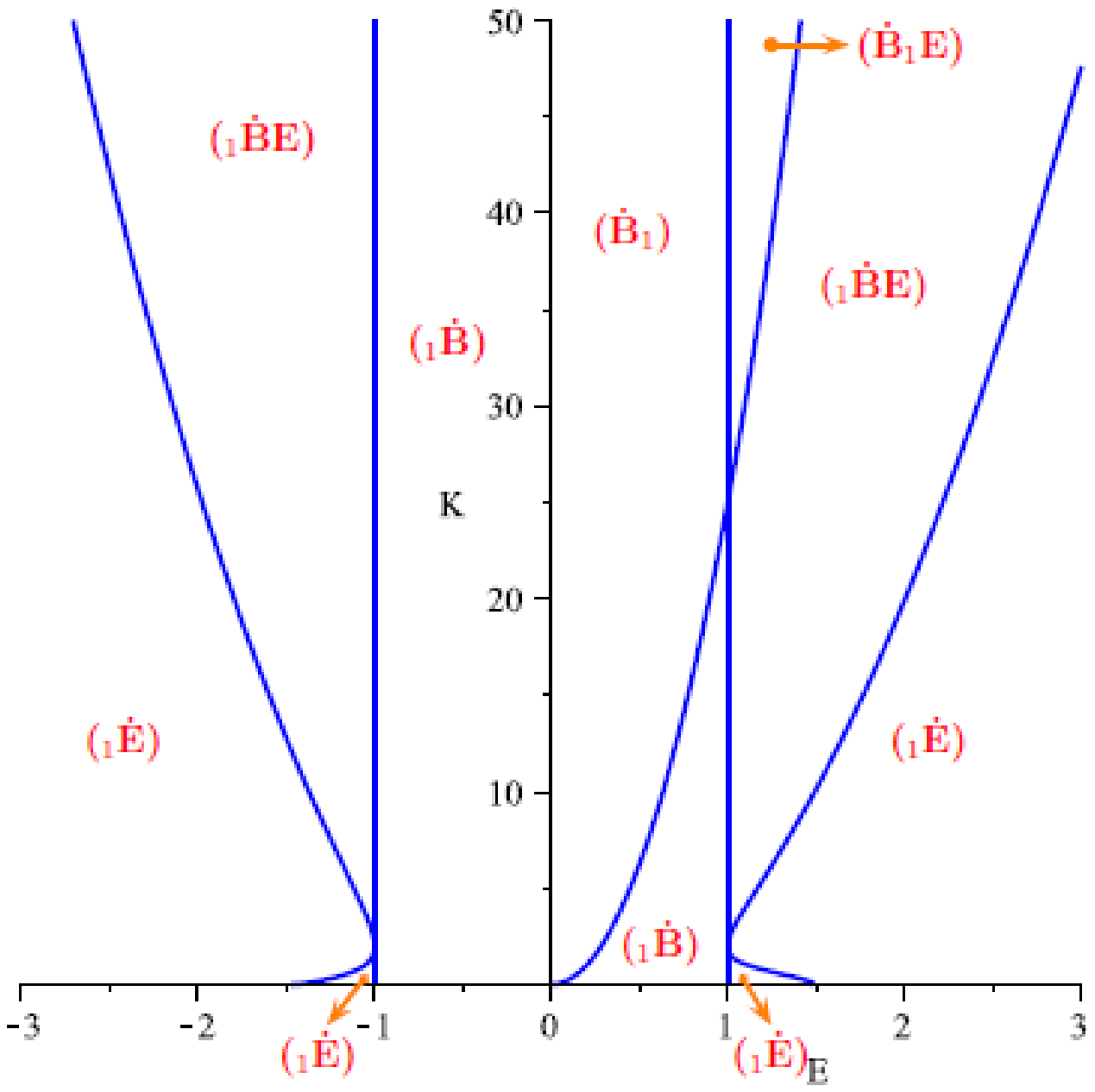}}
\subfigure[][$j_R= \sqrt{{K}}$]{\label{jE_w_2}\includegraphics[width=7.3cm]{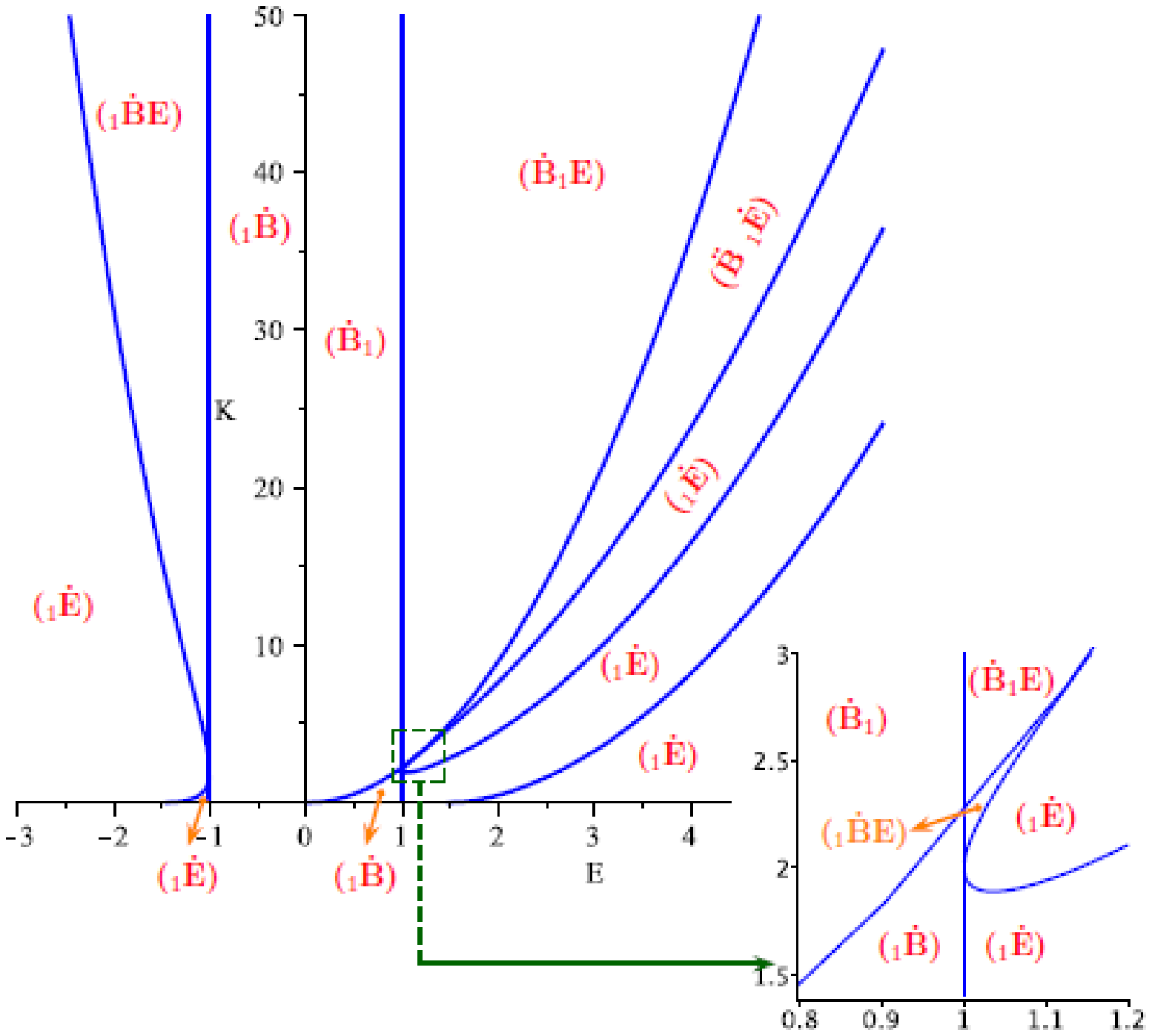}}
\end{center}
\caption{Figures~\subref{jE_w_1} and~\subref{jE_w_2} show the ${K}$-$E$ diagram for $\omega=1.0$ (critical case) for ${A}= 0.3 \sqrt{{K}}$ and the maximal value ${A}=\sqrt{{K}}$ respectively. The regions are in accordance with table~\ref{tab4}. Here $\delta=1$. See also the discussion in the section~\ref{sec:diag} and the diagrams in fig.~\ref{fig:jElw1} for massless test particles. 
\label{fig:jEw1}}
\end{figure*}

\begin{figure*}[th!]
\begin{center}
\subfigure[][$j_R=0.3 \sqrt{{K}}$]{\label{jEl_w_1}\includegraphics[width=6cm]{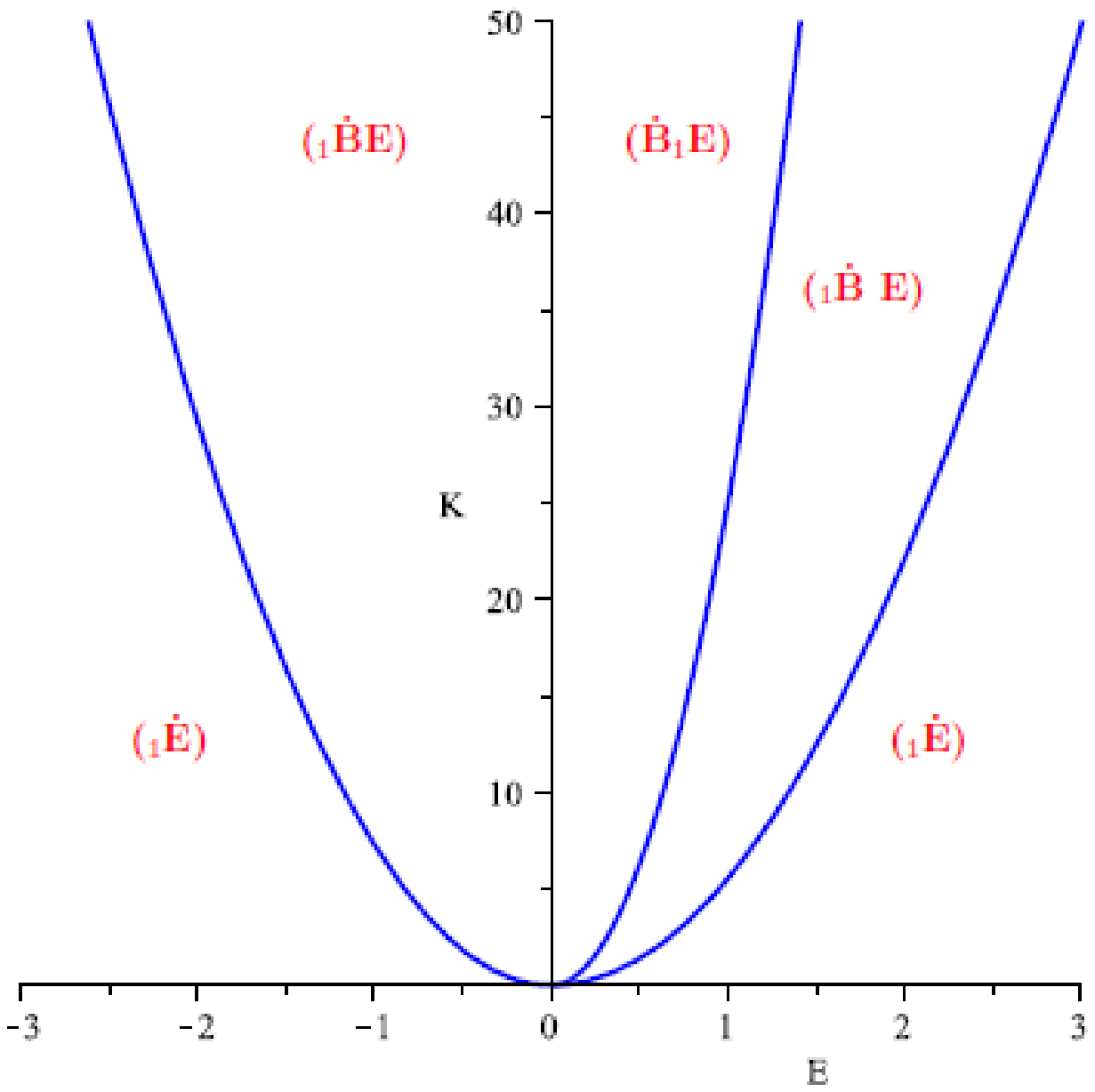}}
\subfigure[][$j_R=\sqrt{{K}}$]{\label{jEl_w_2}\includegraphics[width=6cm]{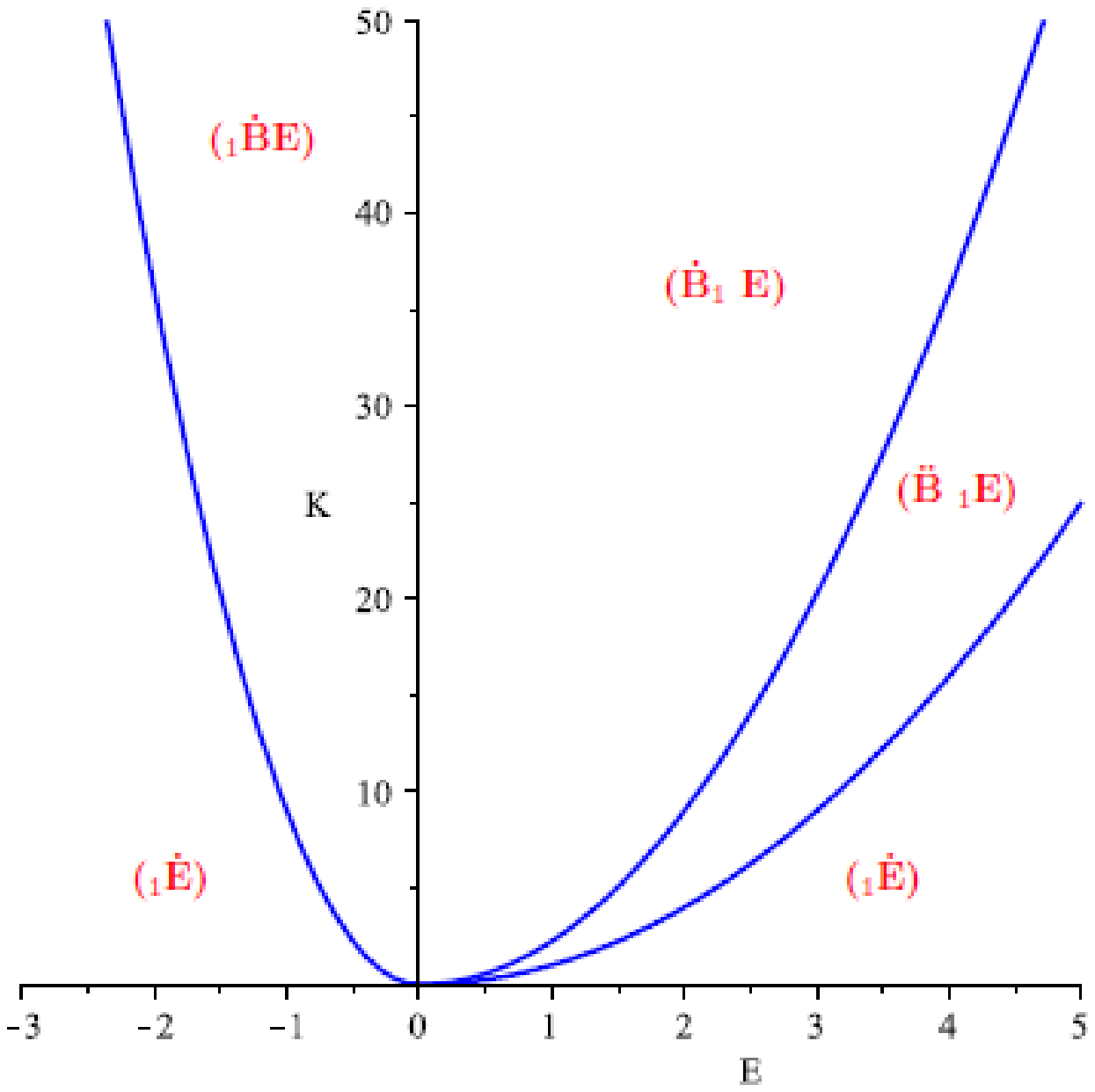}}
\end{center}
\caption{Figures~\subref{jEl_w_1} and~\subref{jEl_w_2} show the ${K}$-$E$ diagram for $\omega=1.0$ (critical case) for ${A}= 0.3 \sqrt{{K}}$ and the maximal value ${A}=\sqrt{{K}}$ respectively. The regions are in accordance with table~\ref{tab5}. Here $\delta=0$. See also the discussion in the section~\ref{sec:diag} and the diagrams in fig.~\ref{fig:jEw1} for massive test particles. 
\label{fig:jElw1}}
\end{figure*}

Consider the critical case $\omega=1$. We have already studied the properties of the motion in the section~\ref{sec:omega1}. 
A feature of this critical $\omega$-value is that most orbit types reach
the surface $x=1$. 
But no orbits can pass the surface $x=1$ from the outer region
to reach smaller values of $x$, or pass the surface
from the inner region to reach larger values of $x$. 
Thus $x=1$ presents a boundary for the geodesic motion.
Fig.~\ref{fig:pot_omega1} and~\ref{fig:potl_omega1} present effective potentials for massive test particles and massless test particles for large (till maximal) values of ${A}$. Tables~\ref{tab4} and~\ref{tab5} show all possible orbit types for $\delta=1$ and $\delta=0$. Here we present the ${K}$-$E$ diagrams for massive (figure~\ref{fig:jEw1}) and massless test particles (figure~\ref{fig:jElw1}). 

Consider first massive test particles. The diagram~\ref{jE_w_1} is plotted for ${A}=0.3 \sqrt{{K}}$. Here all orbits have as turning point $x=1$. For increasing ${A}$, as for example in the plot~\ref{jE_w_2} for ${A}=\sqrt{{K}}$, a region \"B$_1$\.E with a bound orbit behind the horizon and a two world escape orbit with a turning point at $x=1$ appear. Since we choose ${A}>0$ the many world bound orbits of the region (\.B$_1$) and the many world bound and escape orbits of the region (\.B$_1$E) exist only for positive energies. This corresponds to the white allowed region in the potential~\ref{fig:pot_omega1}.

In fig.~\ref{fig:jElw1} we show the diagrams for massless test particles again for ${A}=0.3 \sqrt{{K}}$ in the plot~\subref{jEl_w_1} and for ${A} = \sqrt{{K}}$ in the plot~\subref{jEl_w_2}. All orbit types from the schematical representation in table~\ref{tab5} can be found there. The region (\"B$_1$\.E) with an inner bound orbit and a two world escape orbit exist also here (diagram~\subref{jEl_w_2}). But contrary to the diagrams for massive test particles in the figure~\ref{jE_w_2}, it also exists for very small ${K}$ and $E$.  

\subsection{The $\varphi$-equation}~\label{sec:varphi}

The $\varphi$--equation~\eqref{varphieqn1} consists of $r$-- and $\vartheta$--dependent parts
\begin{equation}
d\varphi = \frac{\omega E}{r^2-1} d\tau - \frac{\Phi}{\sin^2\vartheta} d\tau = \frac{\omega E}{r^2-1} \frac{rdr}{\sqrt{R}} - \frac{\Phi}{\sin^2\vartheta} \frac{d\vartheta}{\sqrt{\Theta}}  \ . \label{varphieqn2}
\end{equation}

With the notations
\begin{eqnarray}
&& I_r= \frac{1}{r^2-1} \frac{rdr}{\sqrt{R}} \label{Ir} \ , \\
&& I^\varphi_\vartheta =  \frac{1}{\sin^2\vartheta} \frac{d\vartheta}{\sqrt{\Theta}}                            \label{Ivartheta} \,
\end{eqnarray}
we integrate the equation~\eqref{varphieqn2} getting
\begin{equation}
\varphi - \varphi_0 = \omega E \int^r_{r_0} I_r - \Phi \int^\vartheta_{\vartheta_0} I^\varphi_\vartheta \ . \label{varphieqn3}
\end{equation}

Consider first the radial differential $I_r$. We make the substitution $r^2=x=\frac{1}{a_3}(y-\frac{a_2}{3})$ as in the section~\ref{sec:radial}:
\begin{equation}
I_r= \frac{a_3}{y-p} \frac{dy}{\sqrt{P_3(y)}} \label{Ir2} \ .
\end{equation}

Substituting $y=\wp(v)$ from~\eqref{soly} where
\begin{equation}
v=v(\tau)=\tau - \tau^\prime \, \label{t_v}
\end{equation}
and $\tau^\prime$ is given by~\eqref{tauprime}, we get:
\begin{equation}
I_r = \frac{a_3}{(\wp(v)-p)} dv  \ , \label{Ir3}
\end{equation}
where $p =   a_3 +\frac{a_2}{3}$.

The integration~\eqref{Ir3} reads
\begin{equation}
\int^v_{v_0} I_r = a_3 I_1   \ , 
\end{equation}
with $I_1$ given by~\cite{Markush,Kagramanova:2010bk,Grunau:2010gd,Hackmann:2010zz,Kagramanova:2012hw}
\begin{equation}
I_1 = \int^v_{v_0} \frac{1}{\wp(v)-p} = \frac{1}{\wp^\prime(v_{p})}
\Biggl( 2\zeta(v_{p})(v-v_{ 0 }) + \ln\frac{\sigma(v-v_{p})}{\sigma(v_0 - v_{p})}
- \ln\frac{\sigma(v + v_{p})}{\sigma(v_{ 0 } + v_{p})} \Biggr) \ , \label{I1} 
\end{equation}
with $\wp(v_{p})=p$, $v(\tau)$ given by~\eqref{t_v} and $v_0 = v(0)$.

Consider now the angular part $I^\varphi_\vartheta$.

Like in the section~\ref{sec:beta} we make the substitution $\xi=\cos^2\vartheta$:
\begin{equation}
I^\varphi_\vartheta =  \frac{1}{2(\xi-1)} \frac{d\xi}{\sqrt{\Theta_\xi}}  \label{Ivartheta2} \ ,
\end{equation}
where $\Theta_\xi$ is defined in~\eqref{xieqn1}. 

With the substitution $u=\frac{2b_2\xi+b_1}{\sqrt{D_\xi}}$, where the coefficients $b_i$ and the discriminant $D_\xi$ are given by~\eqref{b_coeffs} and~\eqref{Dxi}, the integration of $I^\varphi_\vartheta$ is given by an elementary function:
\begin{equation}
\int^\xi_{\xi_0} I^\varphi_\vartheta =  \frac{1}{|{A}-{B}|} \arctan{ \frac{1-u\beta}{\sqrt{1-u^2}\sqrt{\beta^2-1}} } \Bigl|^{\xi(\tau)}_{\xi_{0} }    \ , \label{Ivartheta3}
\end{equation}
where 
\begin{equation}
 \beta = \frac{-{K}+{A}{B}}{\sqrt{D_\xi}} \ , \quad \beta^2-1 =  \frac{{K}({A}-{B})^2}{D_\xi} \geq 0  \ , \label{Ivartheta4}
\end{equation}

Finally, the integration of the $\varphi$--equation yields:
\begin{equation}
\varphi(\tau) =  \varphi_0 + \omega E a_3 I_1 -   \frac{\Phi}{|{A}-{B}|} \arctan{ \frac{1-u\beta}{\sqrt{1-u^2}\sqrt{\beta^2-1}} } \Bigl|^{\xi(\tau)}_{\xi_{0} }   \ , \label{varphieqn4}
\end{equation}
with $I_1$ given by~\eqref{I1}.

$\varphi(\tau)$ is a function of $\tau$ since $\xi(\tau)=\cos^2\vartheta(\tau)$ (equation~\eqref{betaeqn5}) is a function of $\tau$.

\subsection{The $\psi$-equation}~\label{sec:psi}

The $\psi$--equation~\eqref{psieqn1} consists of $r$-- and $\vartheta$ dependent parts similarly to the $\varphi$--equation in the section~\ref{sec:varphi}
\begin{equation}
d\psi = - \frac{\omega E}{r^2-1} d\tau - \frac{\Psi}{\cos^2\vartheta} d\tau = - \frac{\omega E}{r^2-1} \frac{rdr}{\sqrt{R}} - \frac{\Psi}{\cos^2\vartheta} \frac{d\vartheta}{\sqrt{\Theta}}  \ . \label{psieqn2}
\end{equation}

With the same substitutions as in the section~\ref{sec:varphi} for the $\varphi$--equation we can write down the expression for the coordinate $\psi$:
\begin{equation}
\psi (\tau)= \psi_0 - \omega E a_3 I_1 +   \frac{\Psi}{|{A}+{B}|} \arctan{ \frac{1-u\beta_1}{\sqrt{1-u^2}\sqrt{\beta^2_1-1}} }\Bigl|^{\xi(\tau)}_{\xi_{0} }   \ , \label{psieqn3}
\end{equation}
with $I_1$ given by~\eqref{I1} and 
\begin{equation}
 \beta_1 = \frac{{K}+{A}{B}}{\sqrt{D_\xi}} \ , \quad \beta_1^2-1 =  \frac{{K}({A}+{B})^2}{D_\xi} \geq 0  \ , \label{psieq:beta1}
\end{equation}

$\psi(\tau)$ is a function of $\tau$ since $\xi(\tau)=\cos^2\vartheta(\tau)$ (equation~\eqref{betaeqn5}) is a function of $\tau$.

\subsection{The $t$-equation}~\label{sec:time}

We replace $d\tau$ in~\eqref{teqn1} by~\eqref{reqn1} and make the substitution $r^2=\frac{1}{a_3}(y-\frac{a_2}{3})$ as in the section~\ref{sec:radial}. Next we apply the partial fraction decomposition and substitute as in the section~\ref{sec:varphi} $y=\wp(v)$ 
\begin{equation}
dt = \left( \frac{E}{a_3} \left(2  a_3 - \frac{a_2}{3}\right)  + \frac{E}{a_3} \wp(v) + \frac{  a_3 (3  E - \omega {A})}{\wp(v)-p} + \frac{a_3^2 E ( 1 - \omega^2 )}{(\wp(v)-p)^2}  \right) dv  \ , \label{teqn4}
\end{equation}
where $p =   a_3 +\frac{a_2}{3}$ and again $v=v(\tau)=\tau - \tau^\prime$ as defined by~\eqref{t_v} with $\tau^\prime$ given by~\eqref{tauprime}.

The integration of~\eqref{teqn4} reads
\begin{equation}
t (\tau) = t_0 + \frac{E}{a_3} \left(2  a_3 - \frac{a_2}{3}\right)  (v-v_0) - \frac{E}{a_3} (\zeta(v) - \zeta(v_0)) +   a_3 (3  E - \omega A) I_1 + a_3^2 E ( 1 - \omega^2 ) I_2  \ , \label{teqn5}
\end{equation}
where $I_2$ is given by~\cite{Markush,Kagramanova:2010bk,Grunau:2010gd,Hackmann:2010zz,Kagramanova:2012hw}
\begin{eqnarray}
&& I_2 = \int^v_{v_0} \frac{1}{(\wp(v)-p)^2} = -\frac{\wp^{\prime\prime}(v_p)}{(\wp^{\prime}(v_p))^2} I_1 \nonumber \\ 
&& \quad  - \frac{1}{(\wp^{\prime}(v_p))^2} \left( 2\wp(v_p)(v-v_0) + 2(\zeta(v)-\zeta(v_0)) + \frac{\wp^{\prime}(v)}{\wp(v)-\wp(v_p)}-\frac{\wp^{\prime}(v_0)}{\wp(v_0)-\wp(v_p)}   \right) \label{I2}
\end{eqnarray}
and $I_1$ by~\eqref{I1} with $\wp(v_p)=p$, $v(\tau)$ given by~\eqref{t_v} and $v_0 = v(0)$.

\subsection{Causality}~\label{sec:ctc}

The equation~\eqref{teqn1} can be written in the form
\begin{equation}
\frac{dt}{d\tau} = -\frac{(\omega^2 - x^3)}{(x-1)^2}\left( E - V^t \right) \label{teqn1_Vtime} \ , 
\end{equation}
with $x=r^2$ and the potential $V^t$
\begin{equation}
V^t=-\frac{\omega A (x-1)}{\omega^2 - x^3} \label{Vtime} \ .
\end{equation}
Note the factor $\Delta_\omega = 1 - \frac{\omega^2}{x^3}$ in these expressions
and recall that $\Delta_\omega = 0$ represents the VLS.

\begin{figure*}[th!]
\begin{center}
\subfigure[][$\omega=0.7, {K}= 1$, ${A}=0.1 \sqrt{{K}}$]{\label{ctcpot1}\includegraphics[width=7cm]{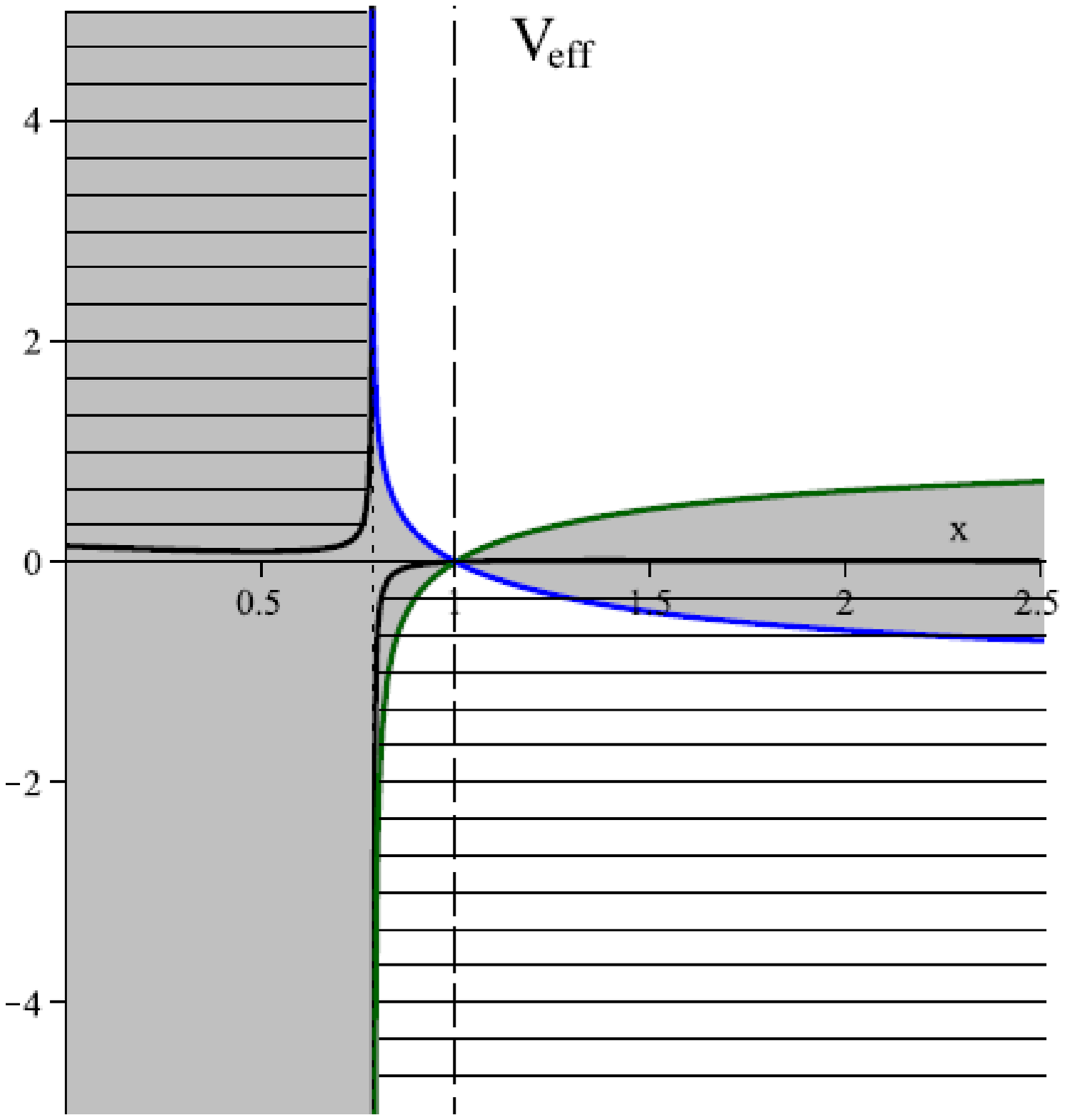}}
\subfigure[][$\omega=1.1, {K}= 10$, ${A}=-0.1  \sqrt{{K}}$; $|{A}|<|{A}^c|$]{\label{ctcpot3}\includegraphics[width=7cm]{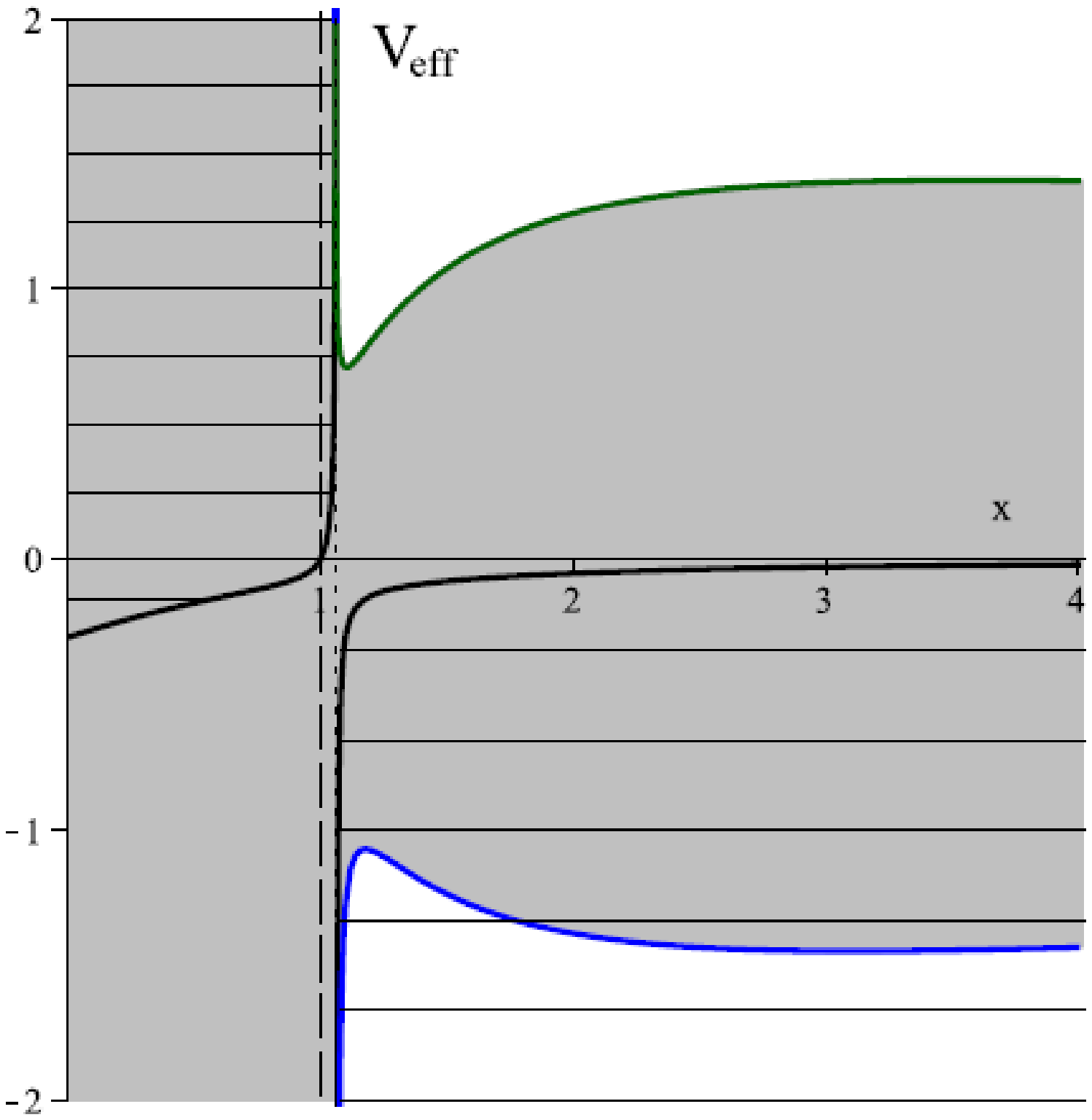}}
\end{center}
\caption{We here supplement the potentials in the figs.~\ref{pot1} and~\ref{pot3} for massive test particles with additional information on the direction of the time flow. The grey region marks as before the forbidden regions. The black solid line is the time potential~\eqref{Vtime}. The dashed regions denote a negative time flow with $\frac{dt}{d\tau}<0$. \label{fig:ctcpots}}
\end{figure*}

\begin{figure*}[th!]
\begin{center}
\subfigure[][$\omega=0.9, {K}= 36$, ${A}=\sqrt{{K}}$]{\label{ctcpot2jR}\includegraphics[width=7cm]{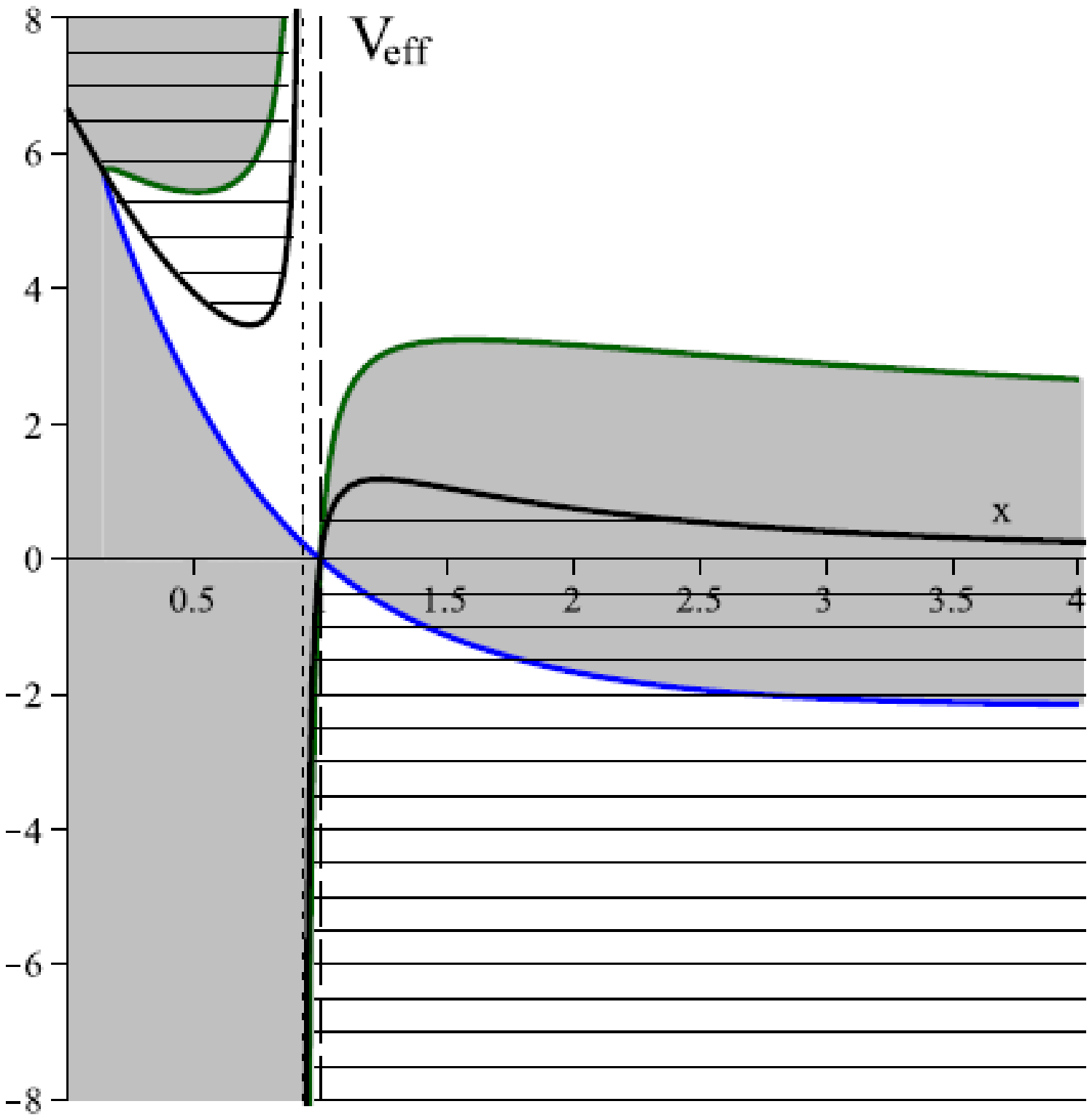}}
\subfigure[][$\omega=1.1, {K}= 10$, ${A}= \sqrt{{K}}$, $|{A}|>|{A}^c|$]{\label{ctcpot3jR}\includegraphics[width=7cm]{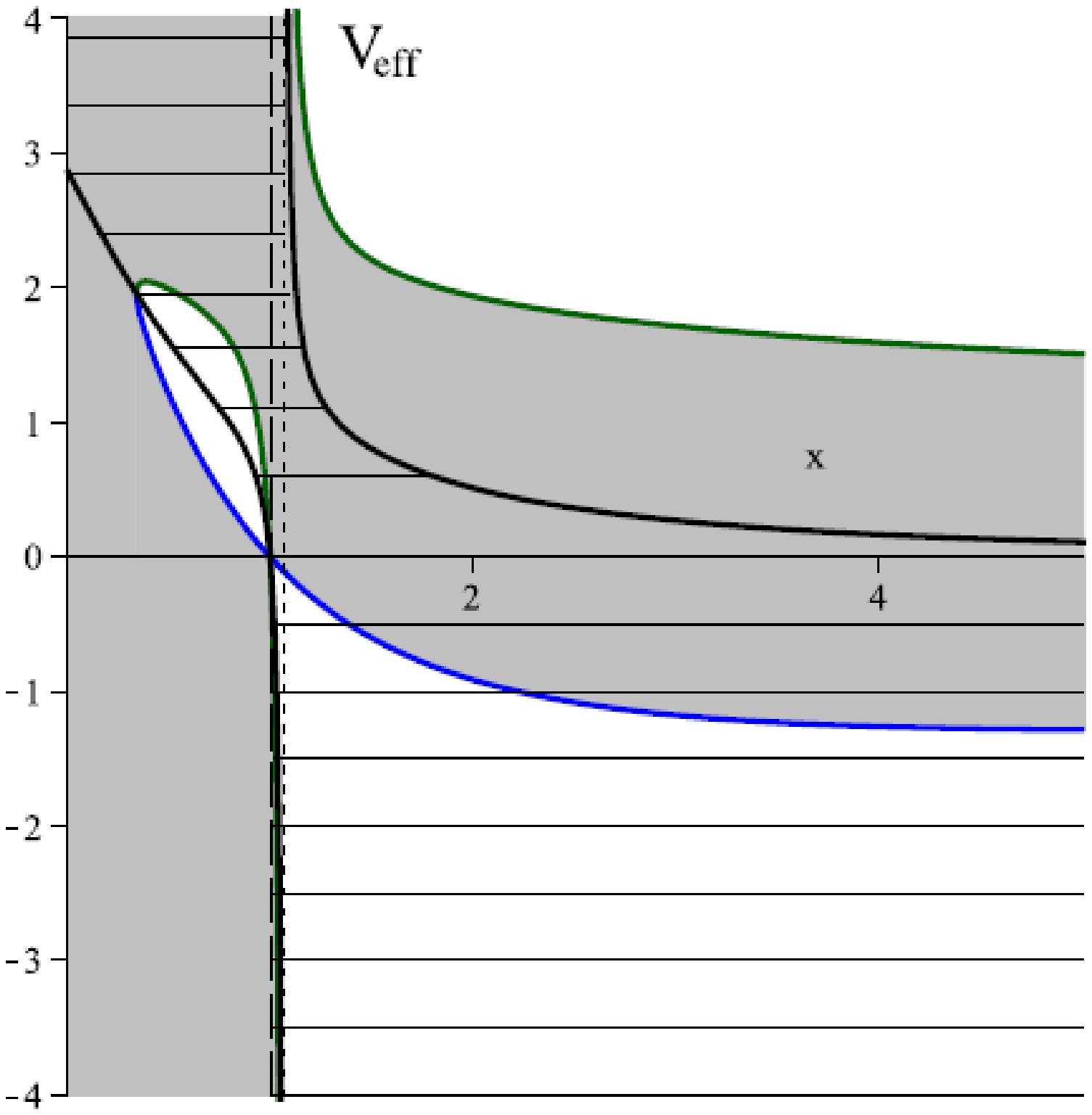}}
\end{center}
\caption{We here supplement the potentials in the figs.~\ref{pot2jR} and~\ref{pot3jR} for massive test particles with additional information on the direction of the time flow. The grey region marks as before the forbidden regions. The black solid line is the time potential~\eqref{Vtime}. The dashed regions denote a negative time flow with $\frac{dt}{d\tau}<0$. \label{fig:ctcpots2} }
\end{figure*}

\begin{figure*}[th!]
\begin{center}
\subfigure[][$\omega=0.9, {K}= 36$, ${A}=-\sqrt{{K}}$]{\label{ctcpotadd1}\includegraphics[width=7cm]{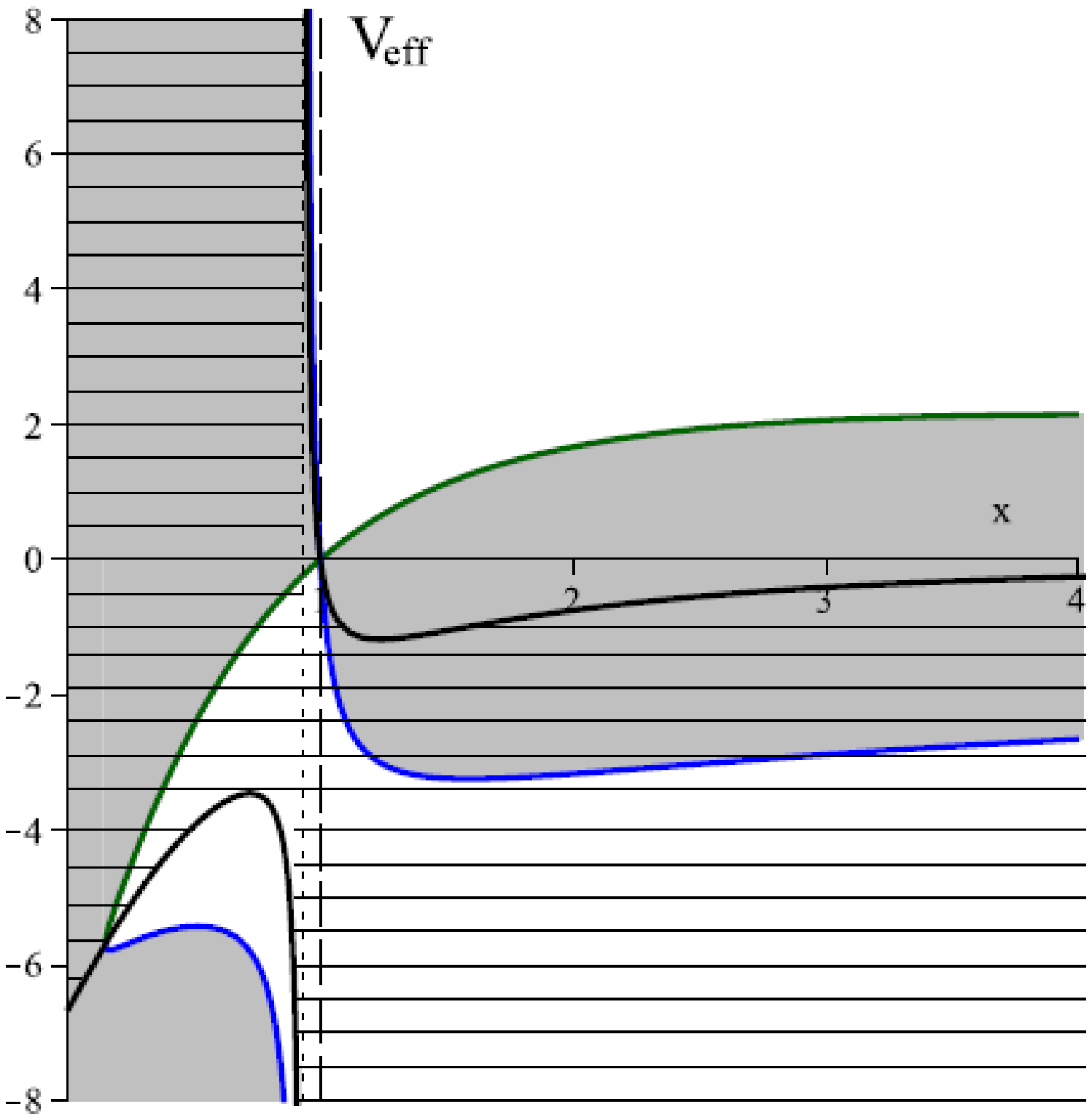}}
\subfigure[][$\omega=1.1, {K}= 10$, ${A}= 0.1 \sqrt{{K}}$]{\label{ctcpotadd2}\includegraphics[width=7cm]{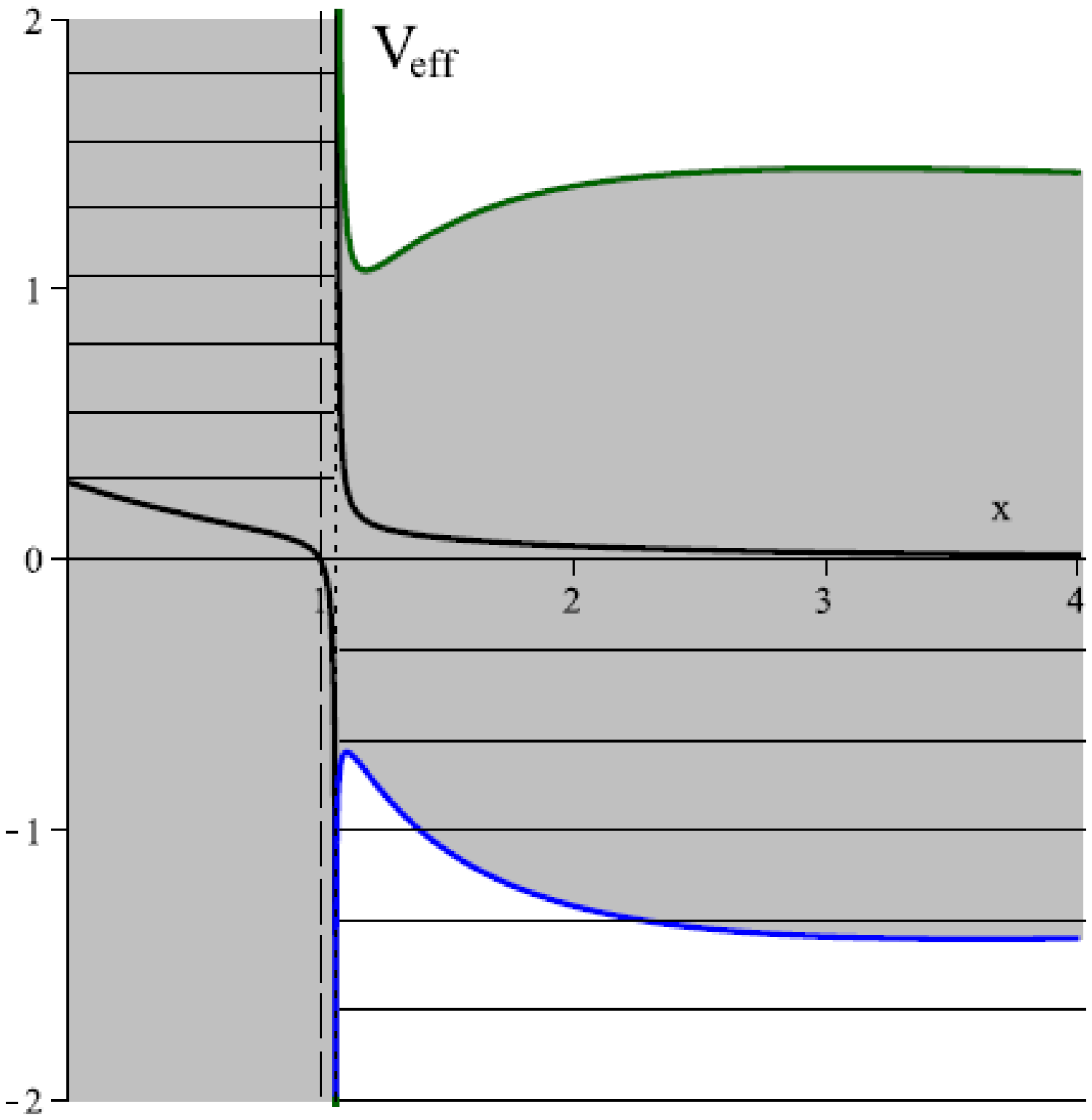}}
\end{center}
\caption{In the plot~\subref{ctcpotadd1} the value of $A$ is chosen to be opposite to the figure~\ref{ctcpot2jR} for comparison. In the plot~\subref{ctcpotadd2} the value of $A$ is chosen to be opposite to the figure~\ref{ctcpot3}. The grey region marks as before the forbidden regions. The black solid line is the time potential~\eqref{Vtime}. The dashed regions denote a negative time flow with $\frac{dt}{d\tau}<0$. \label{fig:ctcpots3} }
\end{figure*}

In the figs.~\ref{fig:ctcpots},~\ref{fig:ctcpots2} and~\ref{fig:ctcpots3} the black solid line denotes the time potential~\eqref{Vtime}. It presents the border of the dashed region. In the dashed regions for positive and negative values of energy the direction of time flow in~\eqref{teqn1_Vtime} changes, i.e. $\frac{dt}{d\tau}$ becomes negative there. 
%This implies the existence of antiparticles in this region. 
The grey region, as before, denotes the forbidden regions for motion.

\section{Orbits}~\label{section:orbits}

To visualize the geodesics we use the cartesian coordinates $(X,Y,Z,W)$ in the form:
\begin{eqnarray}
&&X=r\sin\vartheta\cos\varphi \ , \, Y=r\sin\vartheta\sin\varphi \ , \nonumber \\
&&Z=r\cos\vartheta\cos\psi \ , \, W=r\cos\vartheta\sin\psi \label{XYZW} \ ,
\end{eqnarray}
where $r\in[0,\infty) \ , \vartheta \in [0, \frac{\pi}{2}] \ , \varphi \in [0, 2 \pi) \ , \psi \in [0, 2 \pi)$.

\subsection{$\vartheta=\frac{\pi}{2}$}~\label{section:2dorbits}

We first consider motion in the plane $\theta=\pi/2$. Then only motion w.r.t.~the angle 
$\varphi$ is present.
From the function $\Theta$ in the equation~\eqref{varthetaeqn2} follows that in this case $\Psi=0$ and ${K}=\Phi^2$. This implies ${A}=-\Phi$, ${B}=\Phi$ and ${A}=\pm\sqrt{{K}}$.

For $\vartheta=\frac{\pi}{2}$ the $\varphi$--equation~\eqref{varphieqn1} consists only of the $r$--dependent part and a constant:
\begin{equation}
d\varphi = \left( \frac{\omega E}{r^2-1} - \Phi \right) d\tau =  \left( \frac{\omega E}{r^2-1} - \Phi \right) \frac{rdr}{\sqrt{R}}  \ . \label{varphieqn2_1}
\end{equation}

Next we carry out the same substitutions as in the section~\ref{sec:varphi}. Integration of the equation~\eqref{varphieqn2_1} then yields
\begin{equation}
\varphi (\tau) = \varphi_0 + \int^y_{y_0} \left( \omega E \frac{a_3}{y-p} -\Phi \right) \frac{dy}{\sqrt{P_3(y)}} = \varphi_0 + \omega E a_3 I_1 - \Phi (v-v_{0})  \ , \label{varphieqn3_1}
\end{equation}
where as in the section~\ref{sec:varphi} $p =   a_3 +\frac{a_2}{3}$, $I_1$ is given by~\eqref{I1} and $v_0 = v(0)$ for $v(\tau)=\tau-\tau^\prime$ from the equation~\eqref{t_v}.

From the $\varphi$-equation~\eqref{varphieqn1} we observe that the left hand side vanishes at
 \begin{equation}
x\equiv r^2= 1+\frac{\omega E}{\Phi}
\ . \label{turn} \end{equation}
This means that the angular direction of the test particle motion will be changed when arriving at this point. 
Such an effect is usually known to occur in the presence of an ergosphere as for example in the Kerr or Kerr-Newman spacetimes, where a counterrotating orbit will be forced to corotate with the black hole spacetime. But contrary to those spacetimes, the BMPV spacetime does not possess an ergoregion,
since its horizon angular velocity vanishes. In the following we will call this surface the {\sl turnaround boundary}.

\begin{figure*}[th!]
\begin{center}
\subfigure[][MBO, $E=0.8 \ , {A}=\sqrt{{K}}$]{\label{orb1}\includegraphics[width=6cm]{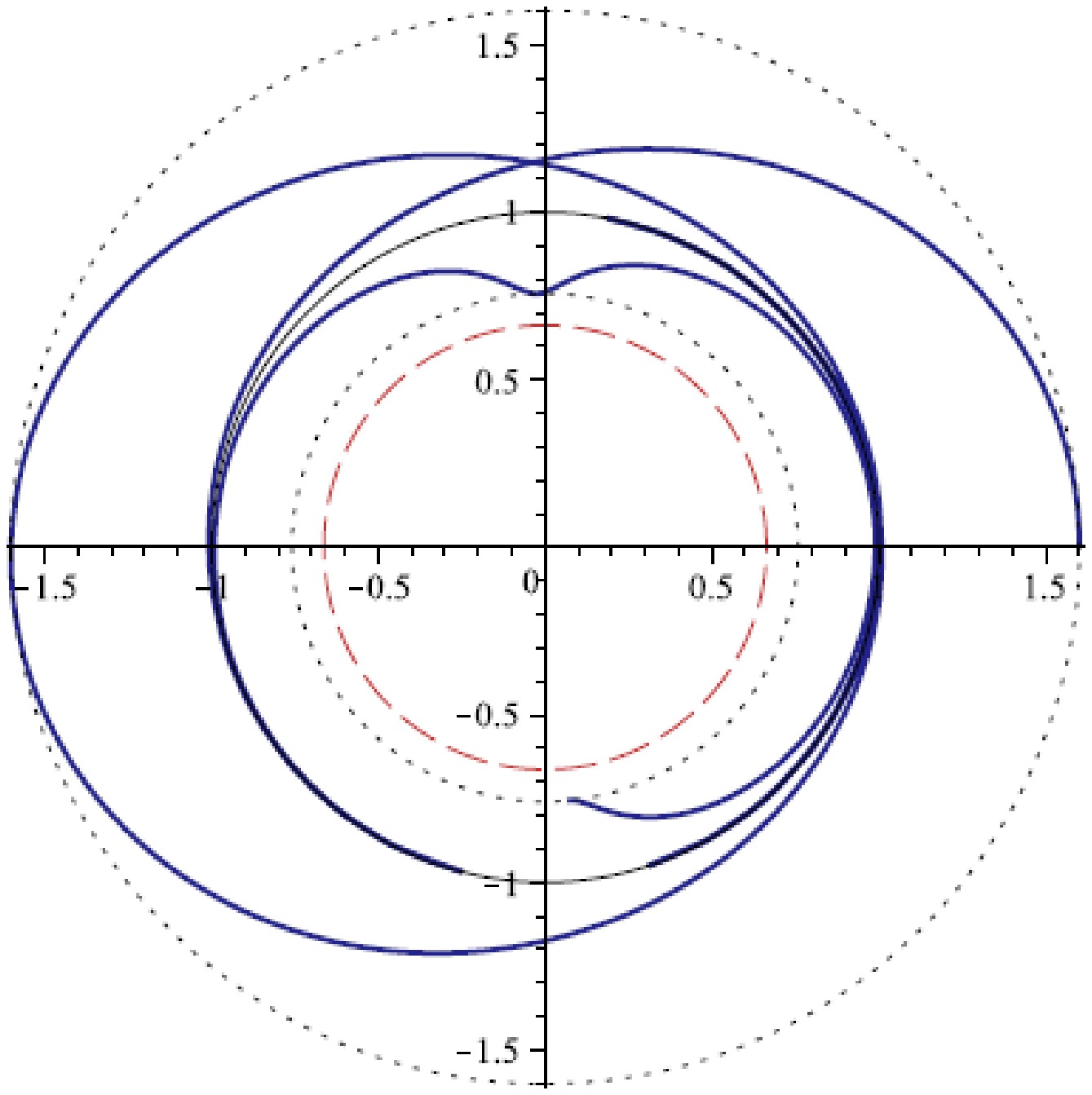}}
\subfigure[][MBO, $E=0.15 \ , {A}=\sqrt{{K}}$]{\label{orb2}\includegraphics[width=6cm]{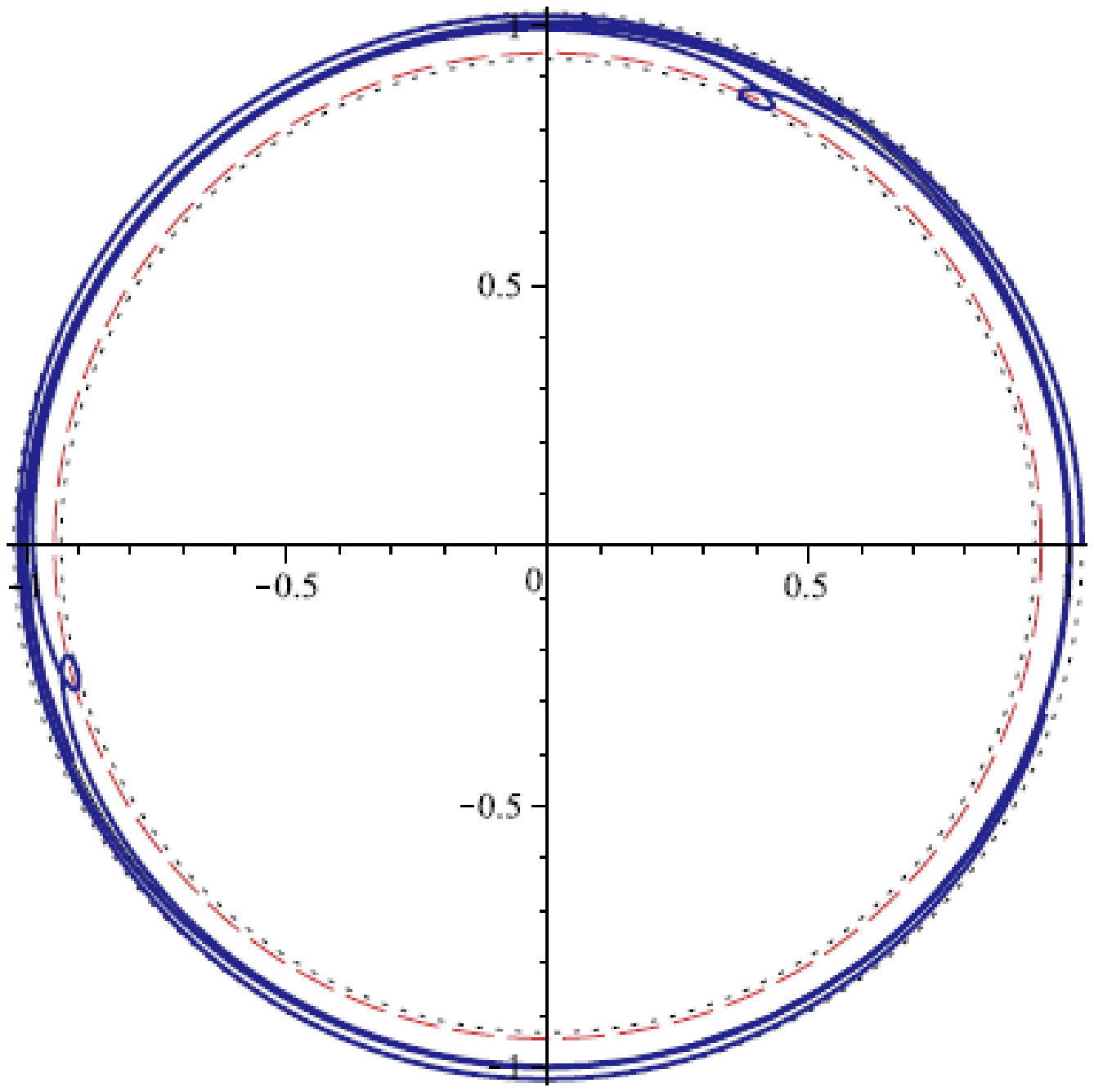}}
\subfigure[][MBO, $E=0.8 \ , {A}=-\sqrt{{K}}$]{\label{orb11}\includegraphics[width=6cm]{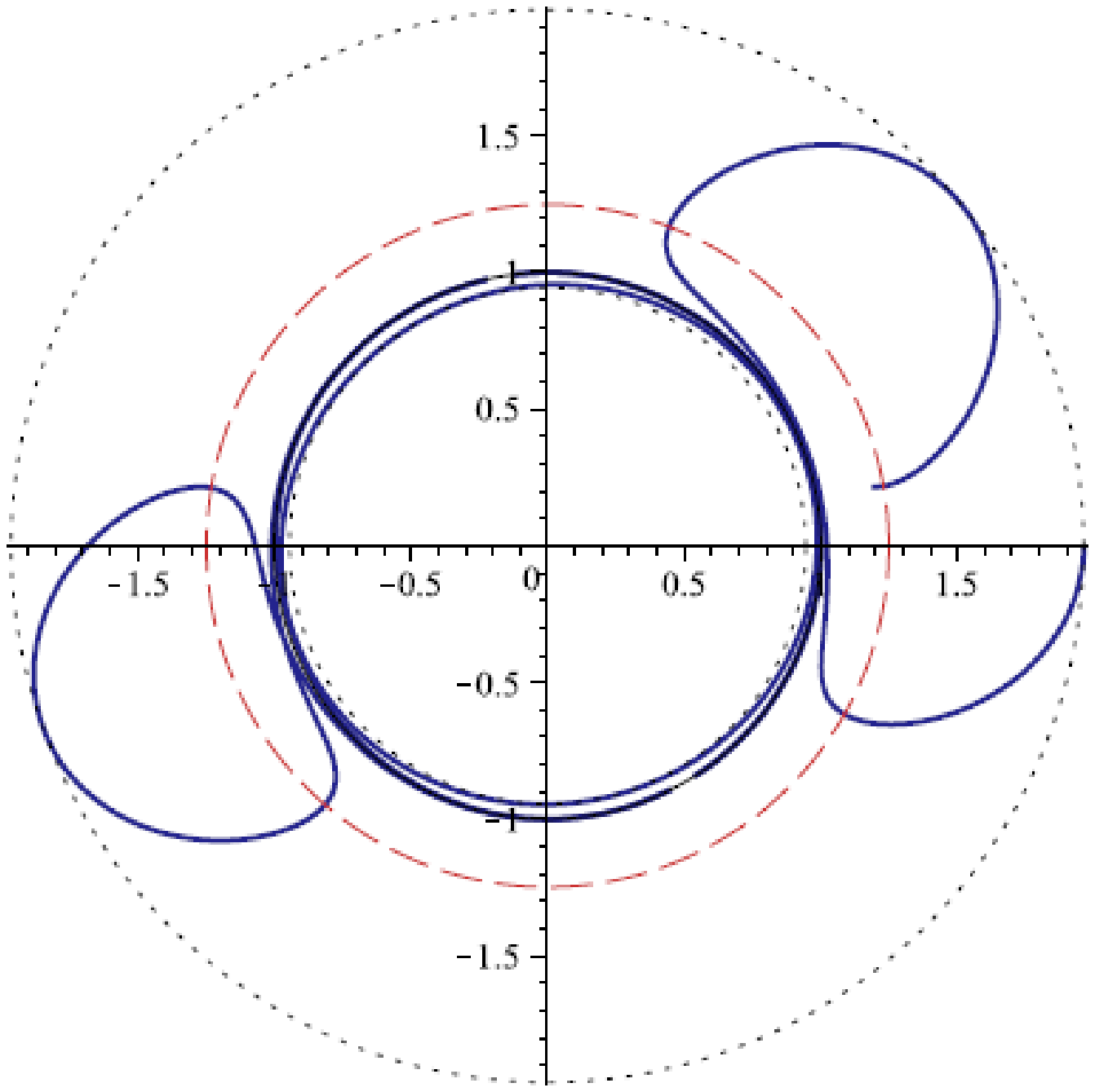}}
\subfigure[][TWE, $E=1.02 \ , {A}=-\sqrt{{K}}$]{\label{orb22}\includegraphics[width=6cm]{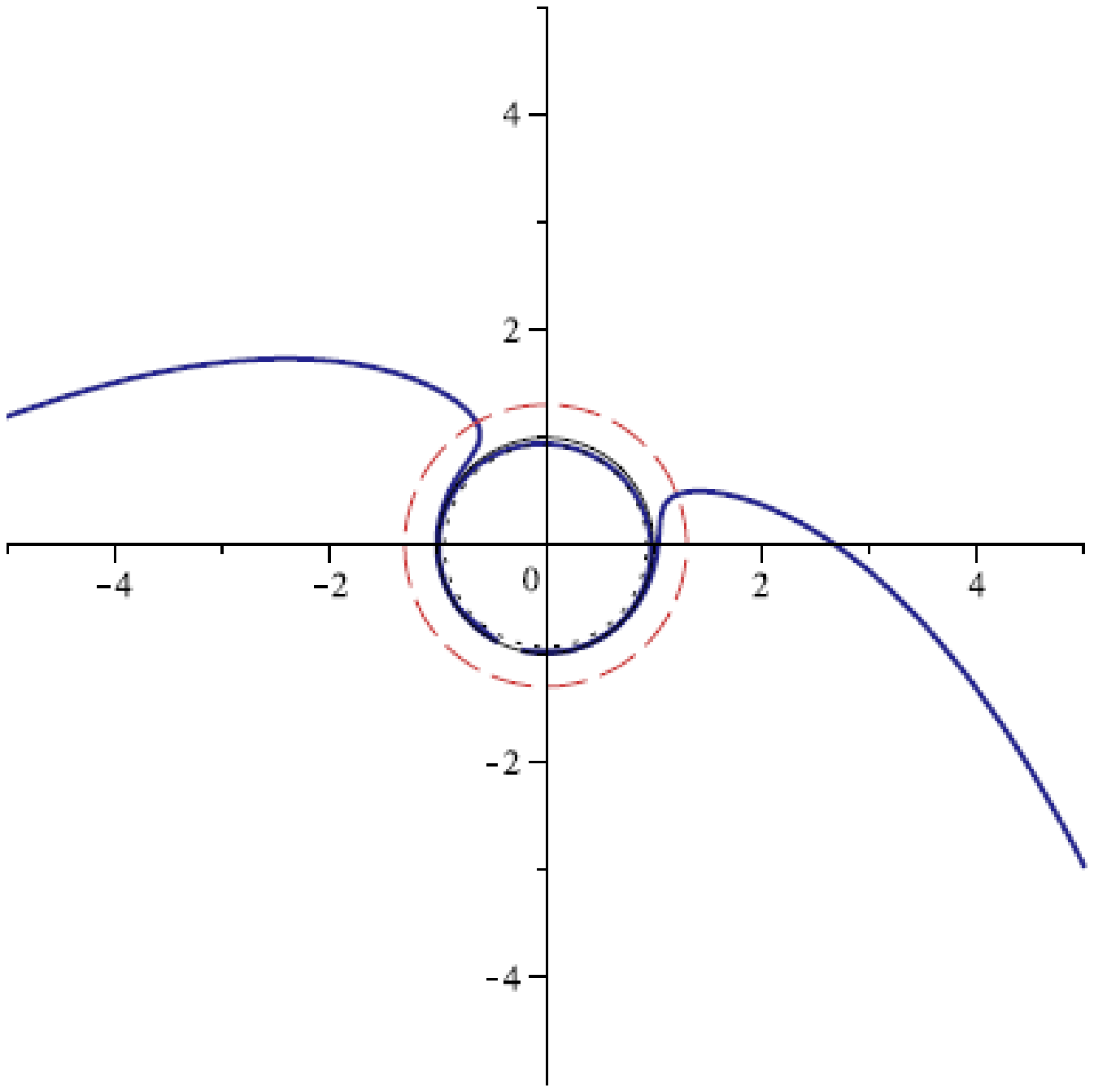}}
\end{center}
\caption{$X-Y$ plot for $\omega=0.7$ (underrotating case), ${K}=1$. In the two plots~\subref{orb1} and~\subref{orb2} $\Phi=-1$ and in the plots~\subref{orb11} and~\subref{orb22} $\Phi=1$. The solid circle marks the horizon at $x=r^2=1$, the two dotted circles mark the minimal and maximal values of the radial coordinate. The dashed circle is located at
the `turnaround boundary' $x\equiv r^2= 1+\frac{\omega E}{\Phi}$, signalling vanishing $\frac{d\varphi}{d\tau}$. The orbit~\subref{orb1} is located outside that circle.  \label{fig1:orb}}
\end{figure*}

\begin{figure*}[th!]
\begin{center}
\subfigure[][MBO, ${K}=0.1  \ , E=0.4 \ , {A}=-\sqrt{{K}} \ , \Phi=\sqrt{0.1}$]{\label{orb111}\includegraphics[width=6cm]{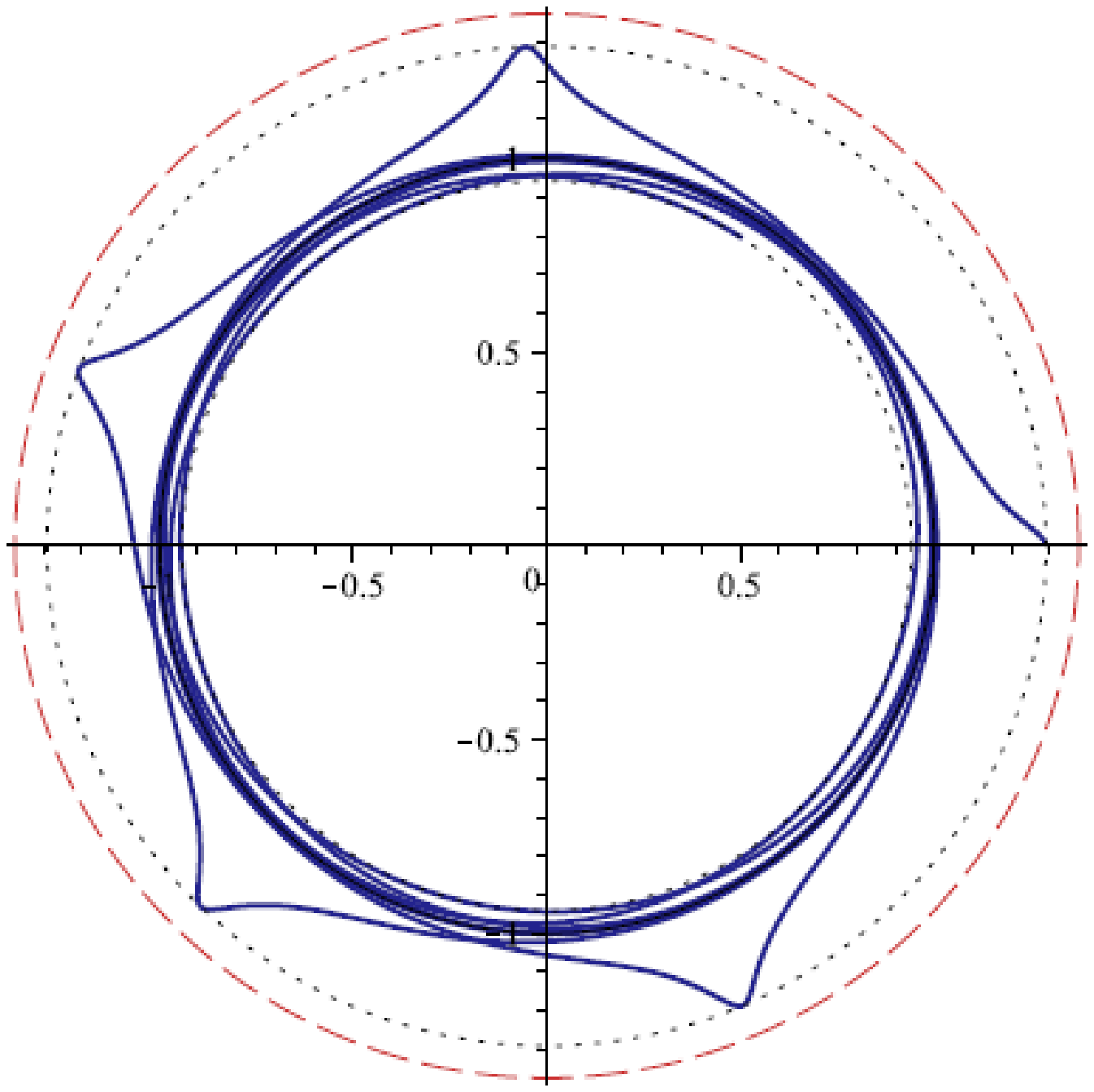}}
\subfigure[][EO, ${K}=6 \ , E=1.14083 \ , {A}=-\sqrt{{K}} \ , \Phi=\sqrt{6}$]{\label{orb112}\includegraphics[width=6cm]{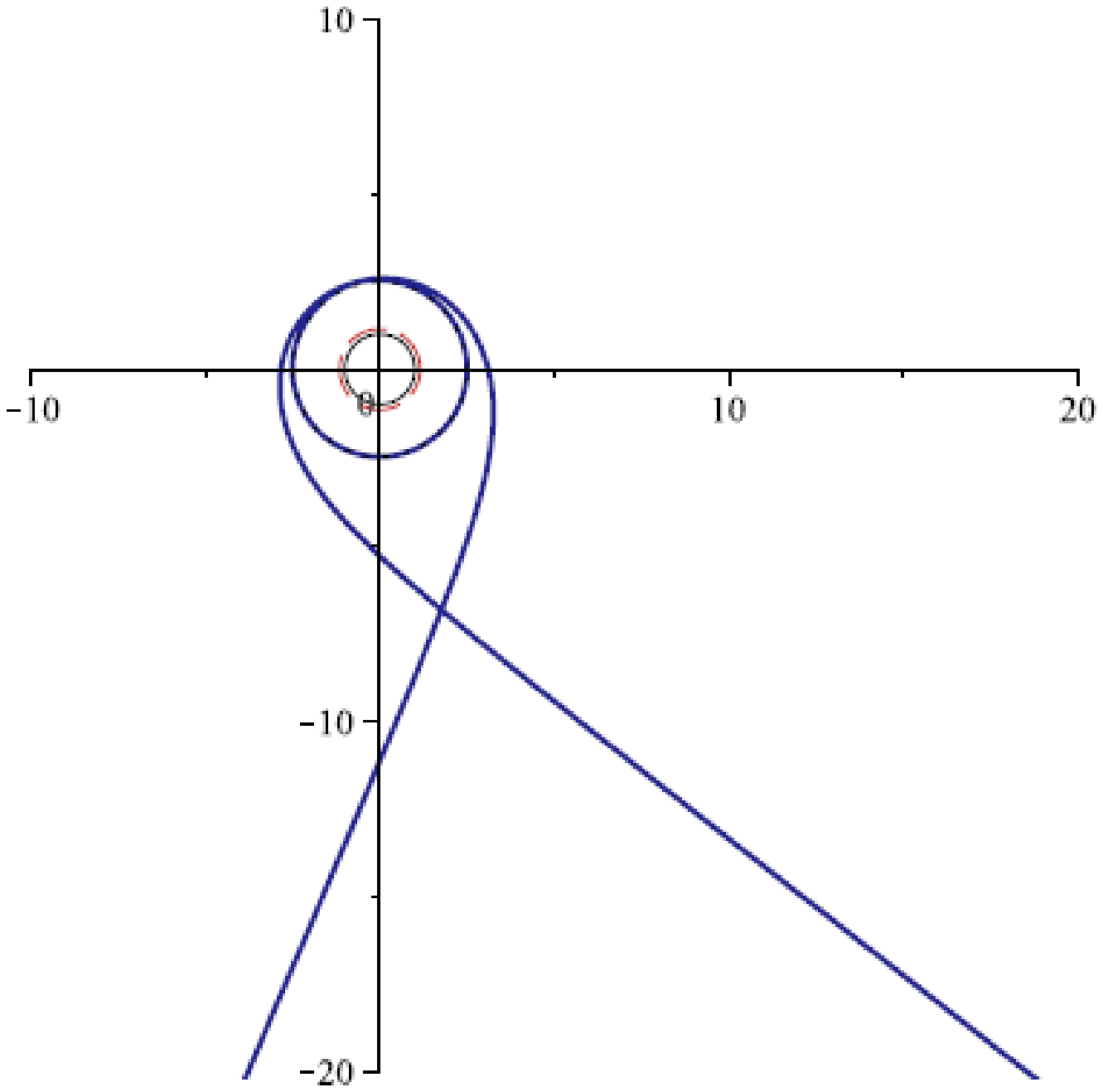}}
\subfigure[][TWE, ${K}=6 \ , E=1.14084 \ , {A}=-\sqrt{{K}} \ , \Phi=\sqrt{6}$]{\label{orb120}\includegraphics[width=6.5cm]{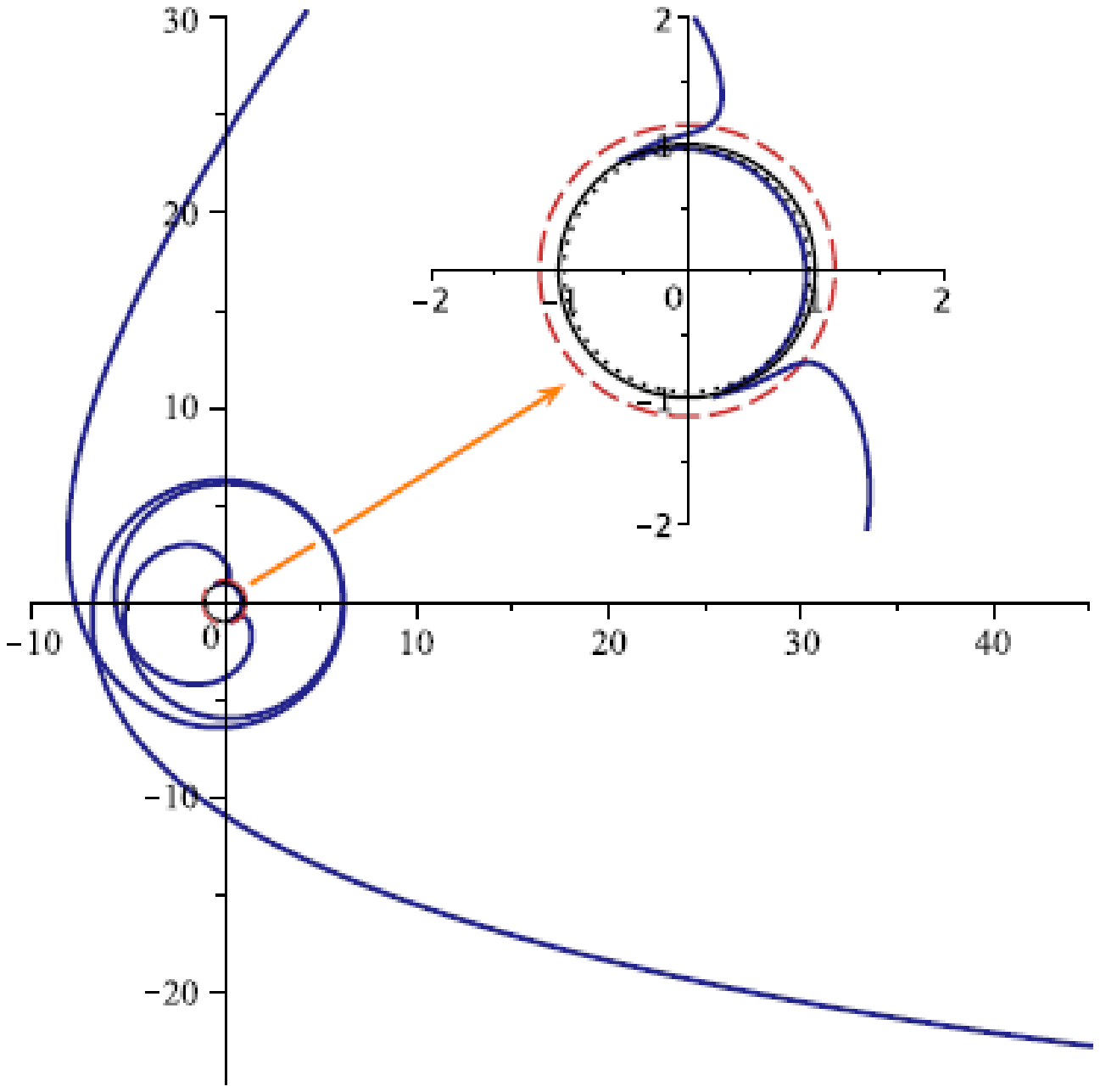}}
\end{center}
\caption{$X-Y$ plot for $\omega=0.7$ (underrotating case). The solid circle marks the horizon at $x=r^2=1$, the two dotted circles mark the minimal and maximal values of the radial coordinate. The dashed circle is located at the `turnaround boundary' $x\equiv r^2= 1+\frac{\omega E}{\Phi}$, signalling vanishing $\frac{d\varphi}{d\tau}$. The orbit~\subref{orb111} is located inside that circle.  \label{fig2:orb}}
\end{figure*}

\begin{figure*}[th!]
\begin{center}
\subfigure[][MBO, $E=3.24252 \ , {A}=\sqrt{{K}}$]{\label{orb7}\includegraphics[width=6cm]{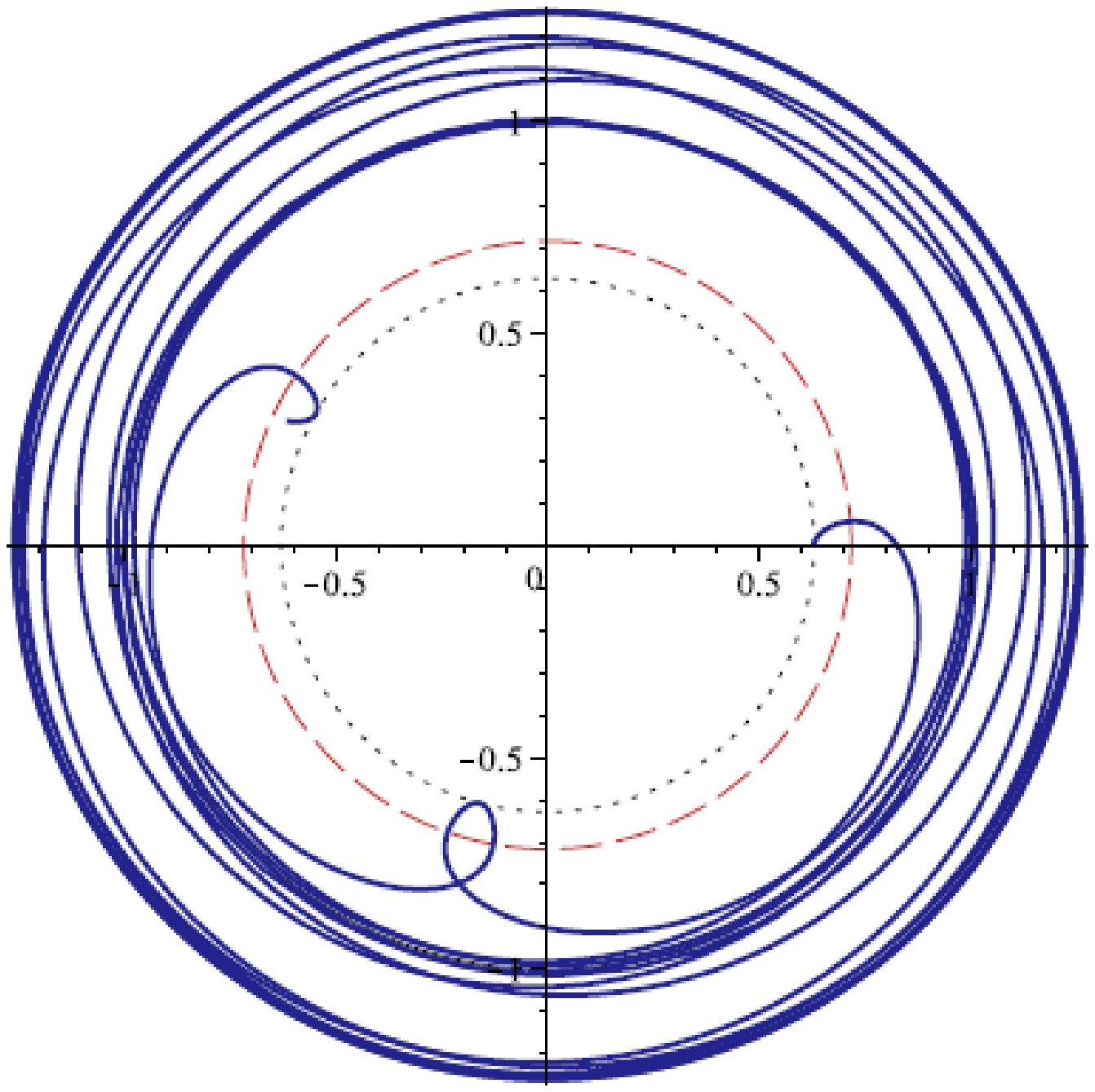}}
\subfigure[][EO, $E=3.24252 \ , {A}=\sqrt{{K}}$]{\label{orb71}\includegraphics[width=6cm]{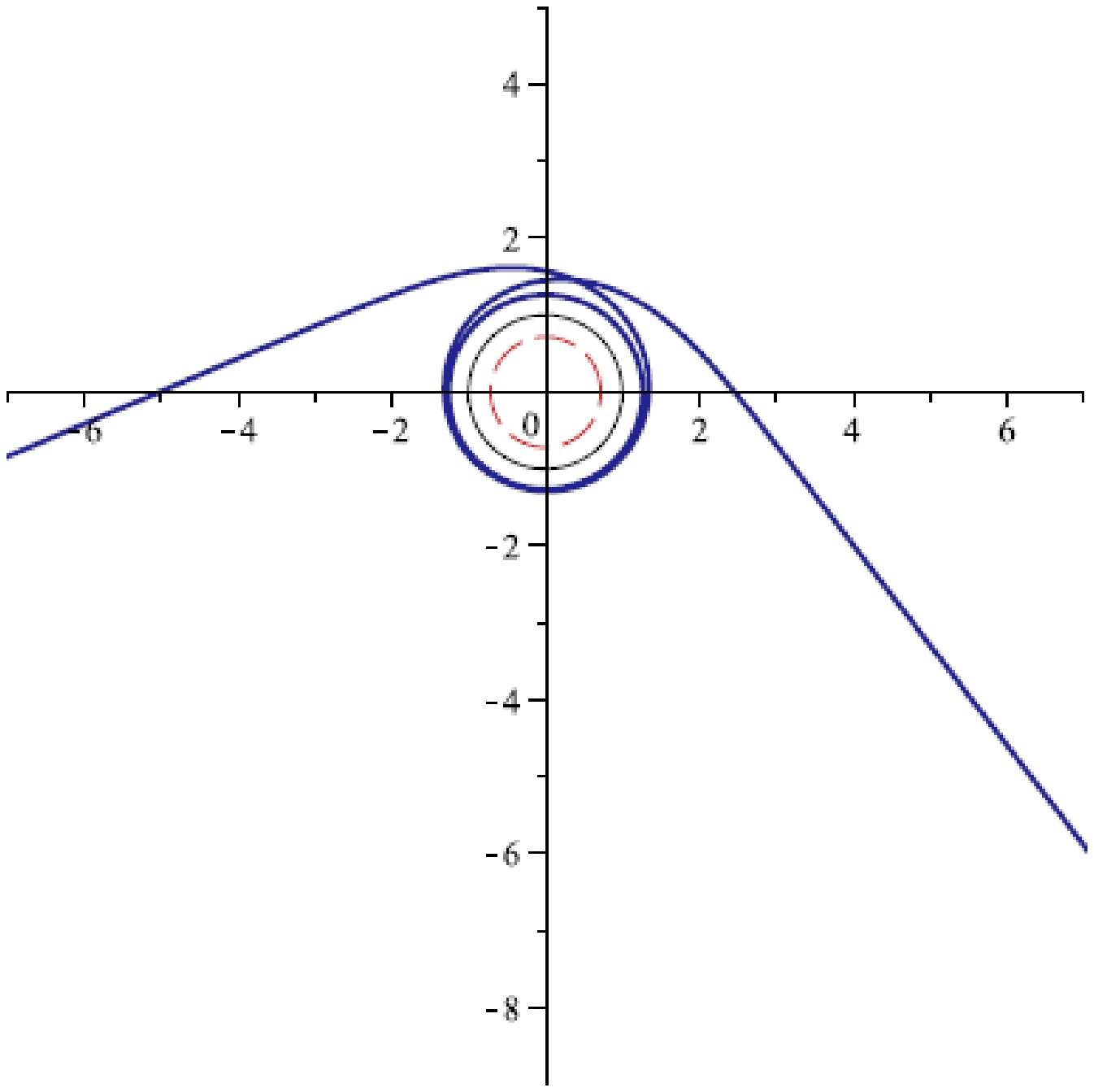}}
\subfigure[][MBO, $E=2.14606 \ , {A}=-\sqrt{{K}}$]{\label{orb8}\includegraphics[width=6cm]{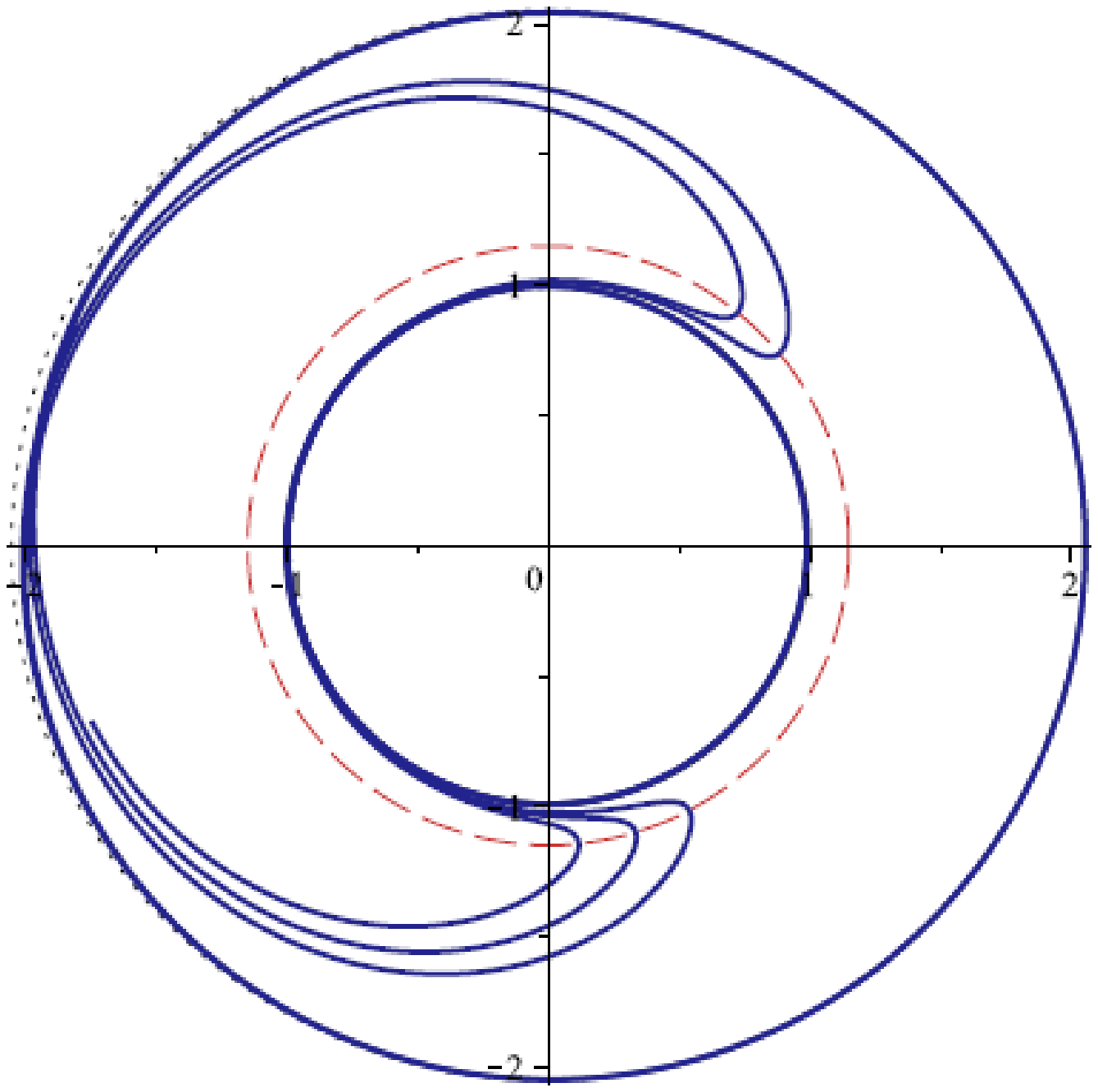}}
\subfigure[][EO, $E=2.14606 \ , {A}=-\sqrt{{K}}$]{\label{orb81}\includegraphics[width=6cm]{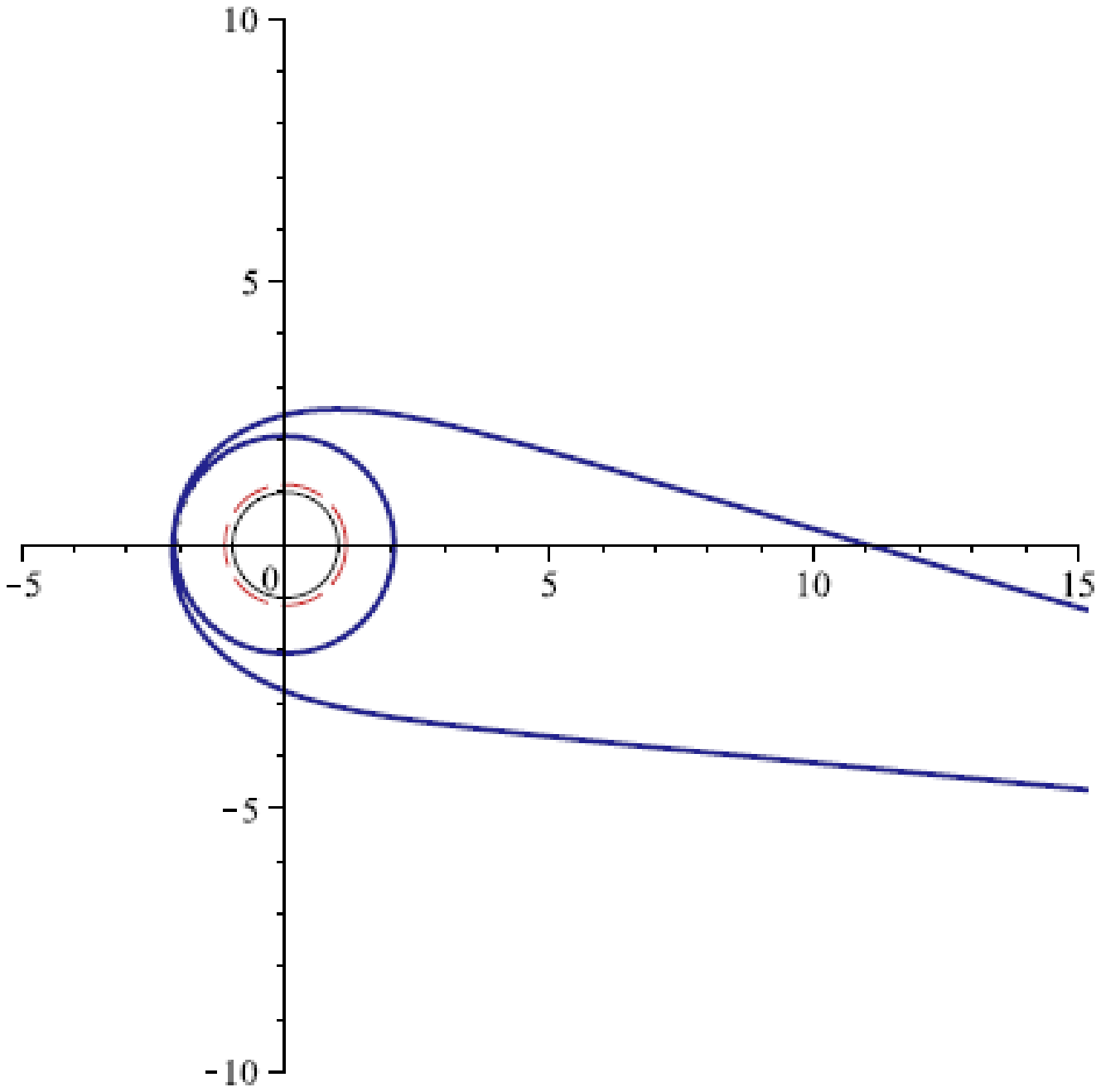}}
\end{center}
\caption{$X-Y$ plot for $\omega=0.9$ (underrotating case), ${K}=36$. In the two plots~\subref{orb7} and~\subref{orb71} $\Phi=-6$ and in the plots~\subref{orb8} and~\subref{orb81} $\Phi=6$. The solid circle marks the horizon at $x=r^2=1$, the two dotted circles mark the minimal and maximal values of the radial coordinate. The dashed circle is located at the `turnaround boundary' $x\equiv r^2= 1+\frac{\omega E}{\Phi}$, signalling vanishing $\frac{d\varphi}{d\tau}$. The orbits~\subref{orb71} and~\subref{orb81} are located outside that circle. \label{fig3:orb}}
\end{figure*}

In the following we show in the figs.~\ref{fig1:orb},~\ref{fig2:orb} and~\ref{fig3:orb}
 two-dimensional $X-Y$ plots for the underrotating case, $\omega<1$. 

The first three orbits in the figure~\ref{fig1:orb} are many-world-bound orbits, and the orbit in the fourth figure is a two-world escape orbit. The orbits~\ref{orb1} and~\ref{orb2} correspond to the potential~\ref{pot1jR} with $A=\sqrt{K}$. The orbits~\ref{orb11} and~\ref{orb22} have opposite $A$ value. We see that in the plots~\ref{orb2},~\ref{orb11},~\ref{orb22} the orbits cross the dashed circle corresponding to the `turnaround boundary', where the test particle changes its angular direction of motion. In the plot~\ref{orb1} the `turnaround boundary'  has no influence on the orbit. 

In the figure~\ref{fig2:orb} we show orbits for the same value of $\omega$. In the fig.~\ref{orb111} the many-world-bound orbit is located inside the `turnaround boundary' and the escape orbit~\ref{orb112}--outside. The two-world-escape orbit~\ref{orb120}  experiences the influence of the `turnaround boundary', similar to the orbit in the picture~\ref{orb22}.

In the fig.~\ref{fig3:orb} we show $X-Y$ orbits for $\omega=0.9$. The many-world bound orbits~\ref{orb7} and~\ref{orb8}, plotted for different values of the energy $E$, are angularly deflected at the `turnaround boundary', while the escape orbits~\ref{orb71} and~\ref{orb81} remain far away from  the `turnaround boundary'.

\begin{figure*}[th!]
\begin{center}
\subfigure[][BO, $E=1.3 \ , {A}=\sqrt{{K}}$]{\label{orb3}\includegraphics[width=6cm]{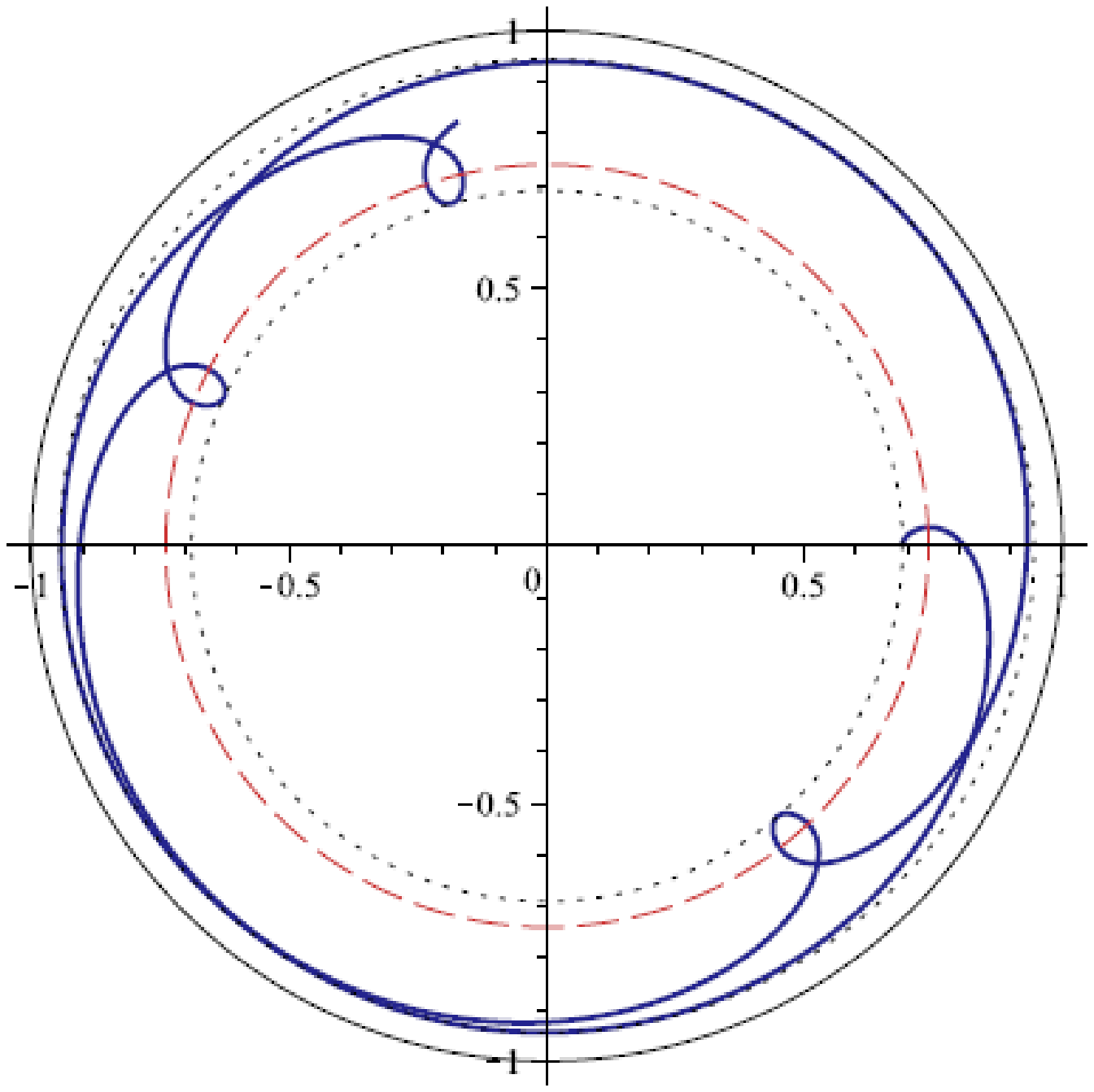}}
\subfigure[][EO, $E=1.3 \ , {A}=\sqrt{{K}}$]{\label{orb31}\includegraphics[width=6cm]{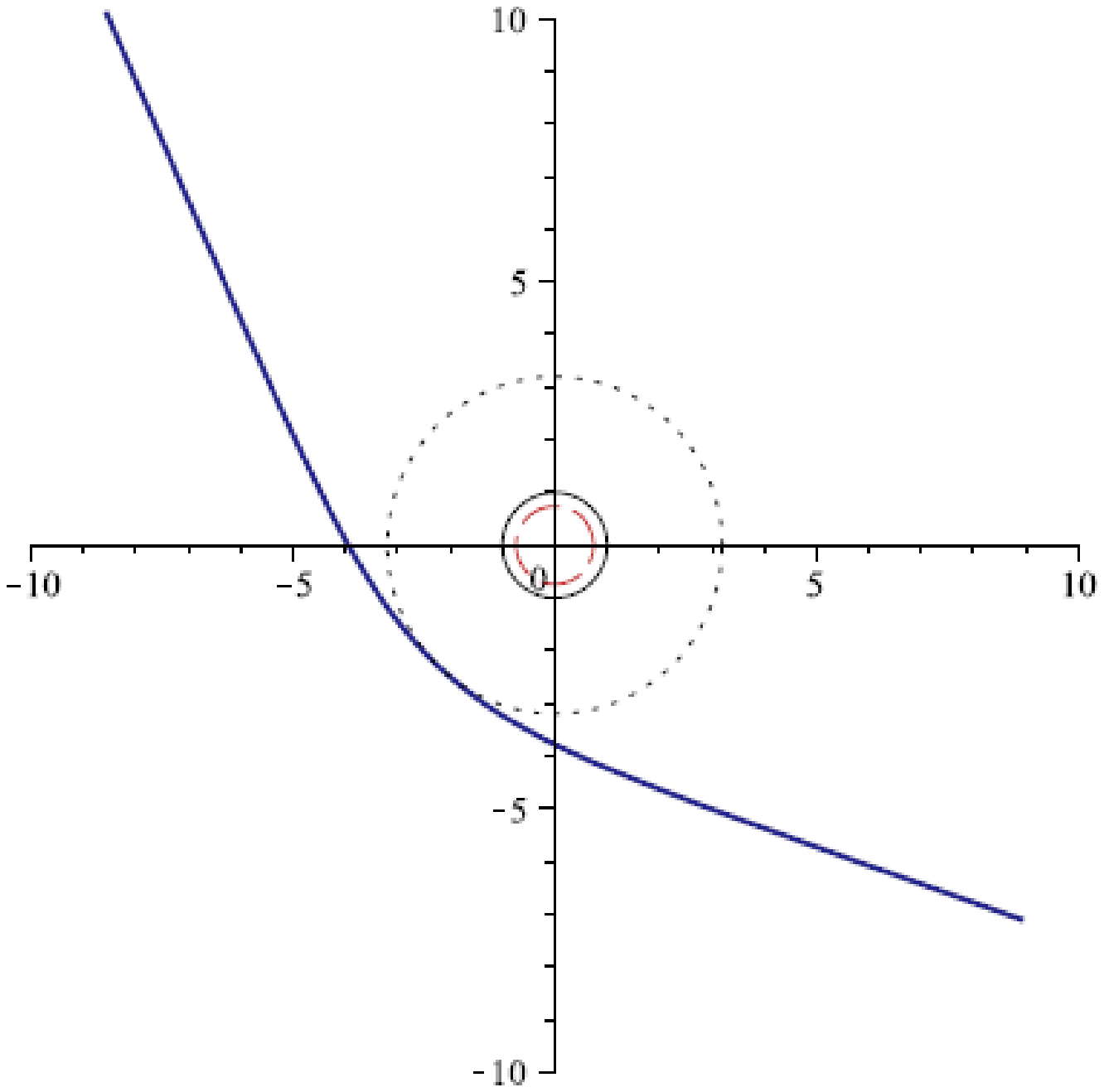}}
\subfigure[][BO, $E=1.28706 \ , {A}=-\sqrt{{K}}$]{\label{orb4}\includegraphics[width=6cm]{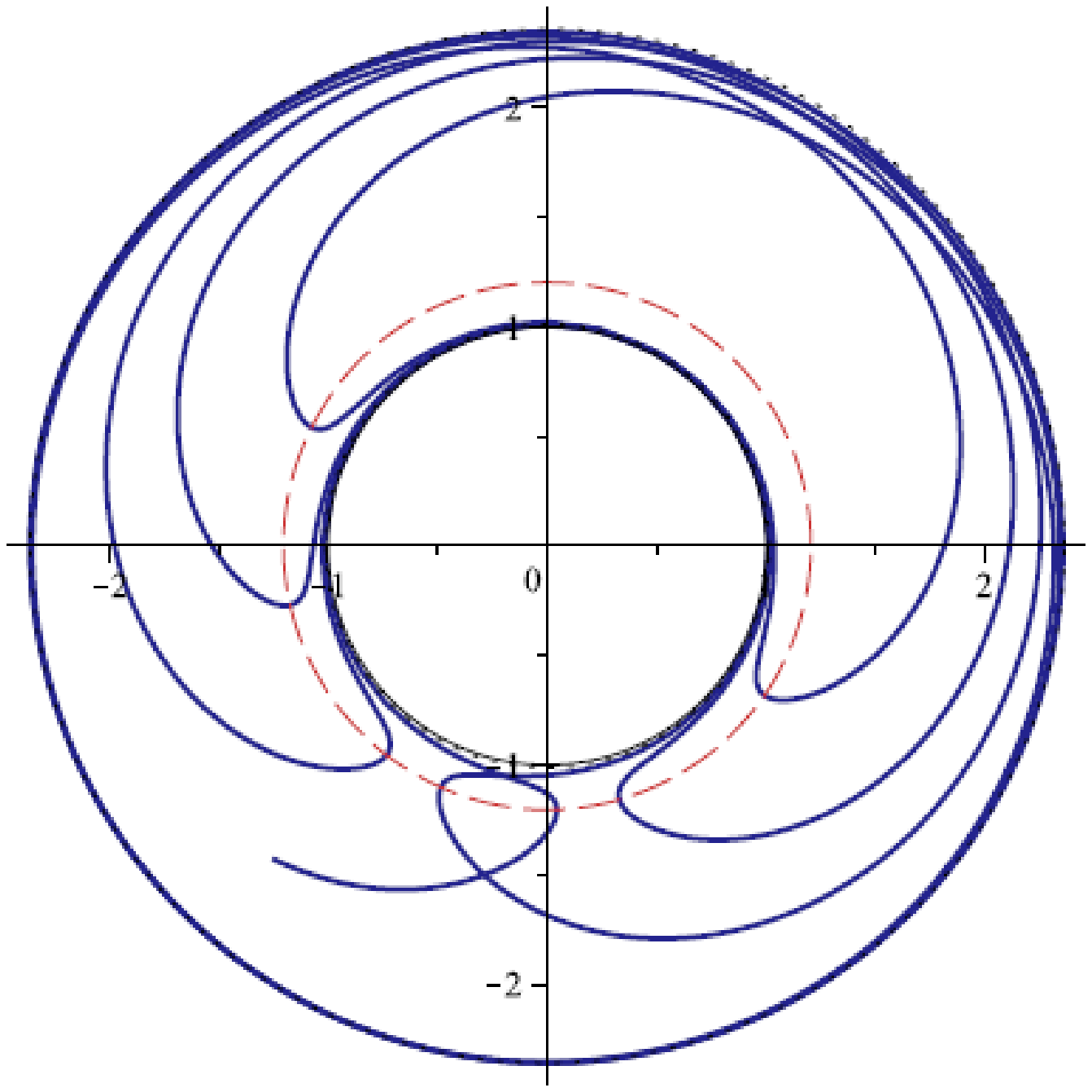}}
\subfigure[][EO, $E=1.28706 \ , {A}=-\sqrt{{K}}$]{\label{orb41}\includegraphics[width=6cm]{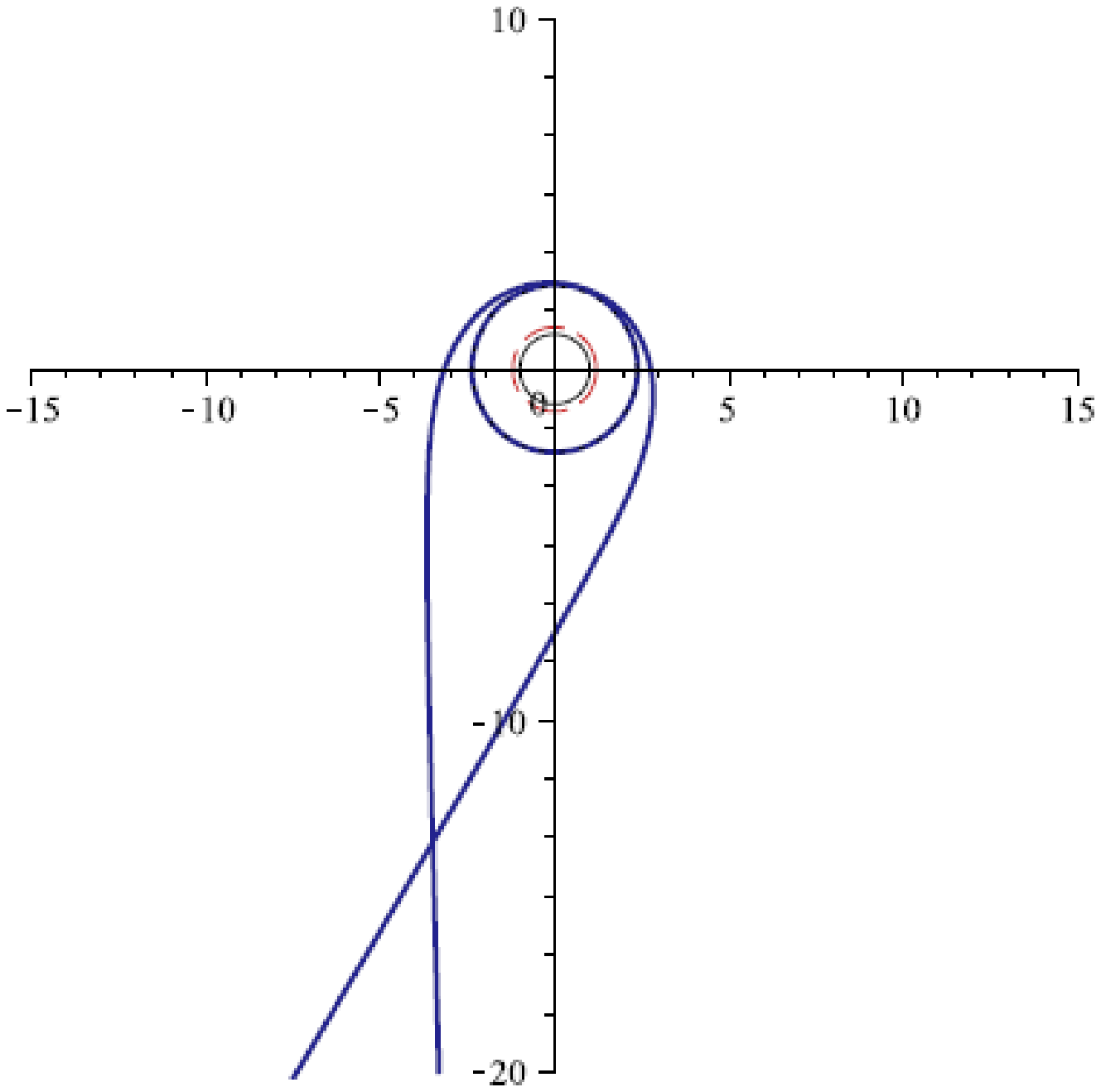}}
\end{center}
\caption{$X-Y$ plot for $\omega=1.1$ (overrotating case), ${K}=10$. In the two plots~\subref{orb3} and~\subref{orb31} $\Phi=-\sqrt{10}$ and in the plots~\subref{orb4} and~\subref{orb41} $\Phi=\sqrt{10}$. The solid circle marks the pseudo-horizon at $x=r^2=1$, the two dotted circles mark the minimal and maximal values of the radial coordinate. The orbit~\subref{orb3} is located inside the peudo-horizon. The dashed circle is located at the `turnaround boundary' $x\equiv r^2= 1+\frac{\omega E}{\Phi}$, signalling vanishing $\frac{d\varphi}{d\tau}$.  The orbits~\subref{orb31} and~\subref{orb41} are located outside that circle. \label{fig4:orb}}
\end{figure*}

\begin{figure*}[th!]
\begin{center}
\subfigure[][BO, $E=0.25 \ , {A}=-\sqrt{{K}}$]{\label{orb5}\includegraphics[width=6cm]{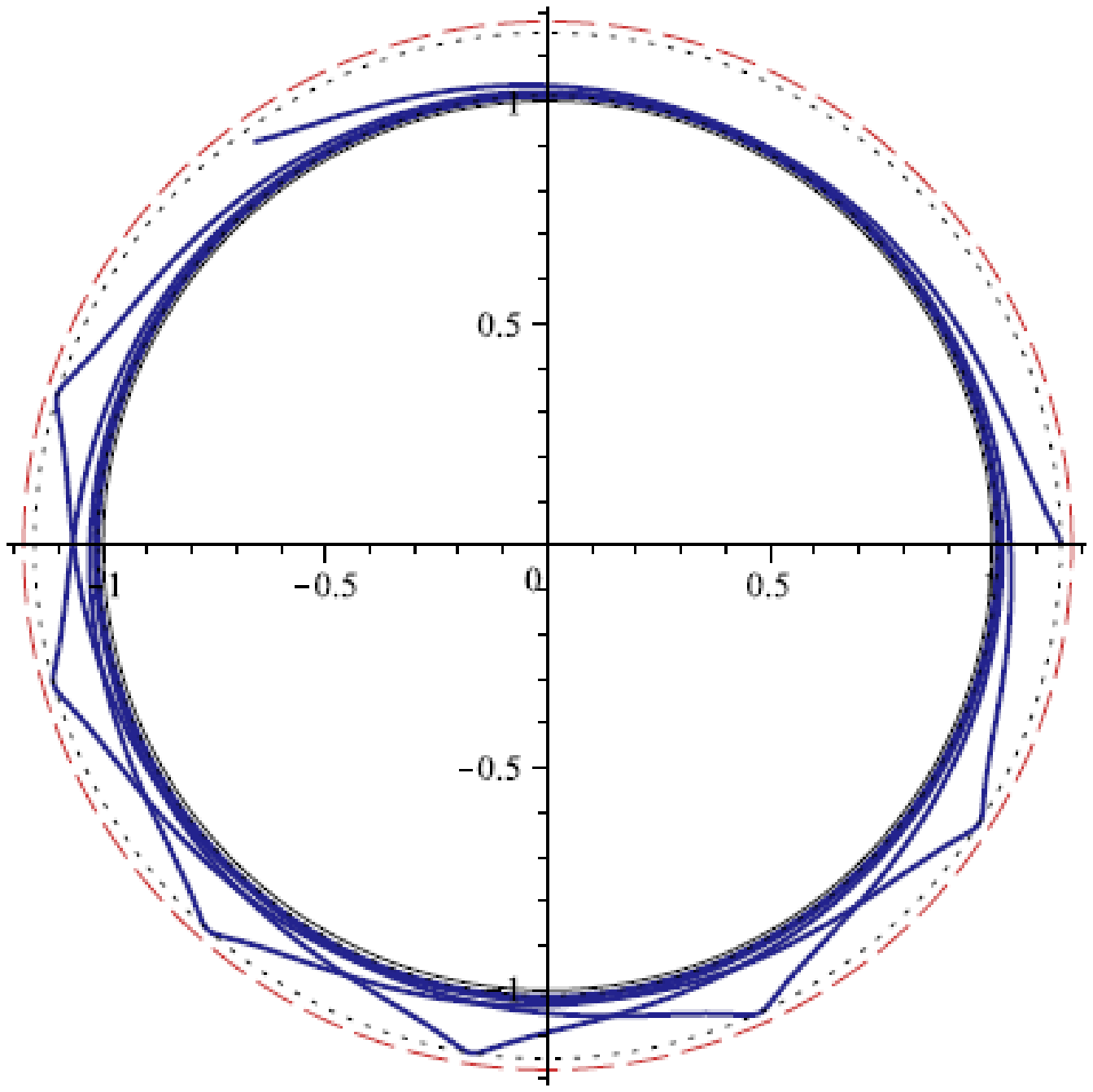}}
\subfigure[][EO, $E=1.01 \ , {A}=-\sqrt{{K}}$]{\label{orb51}\includegraphics[width=6cm]{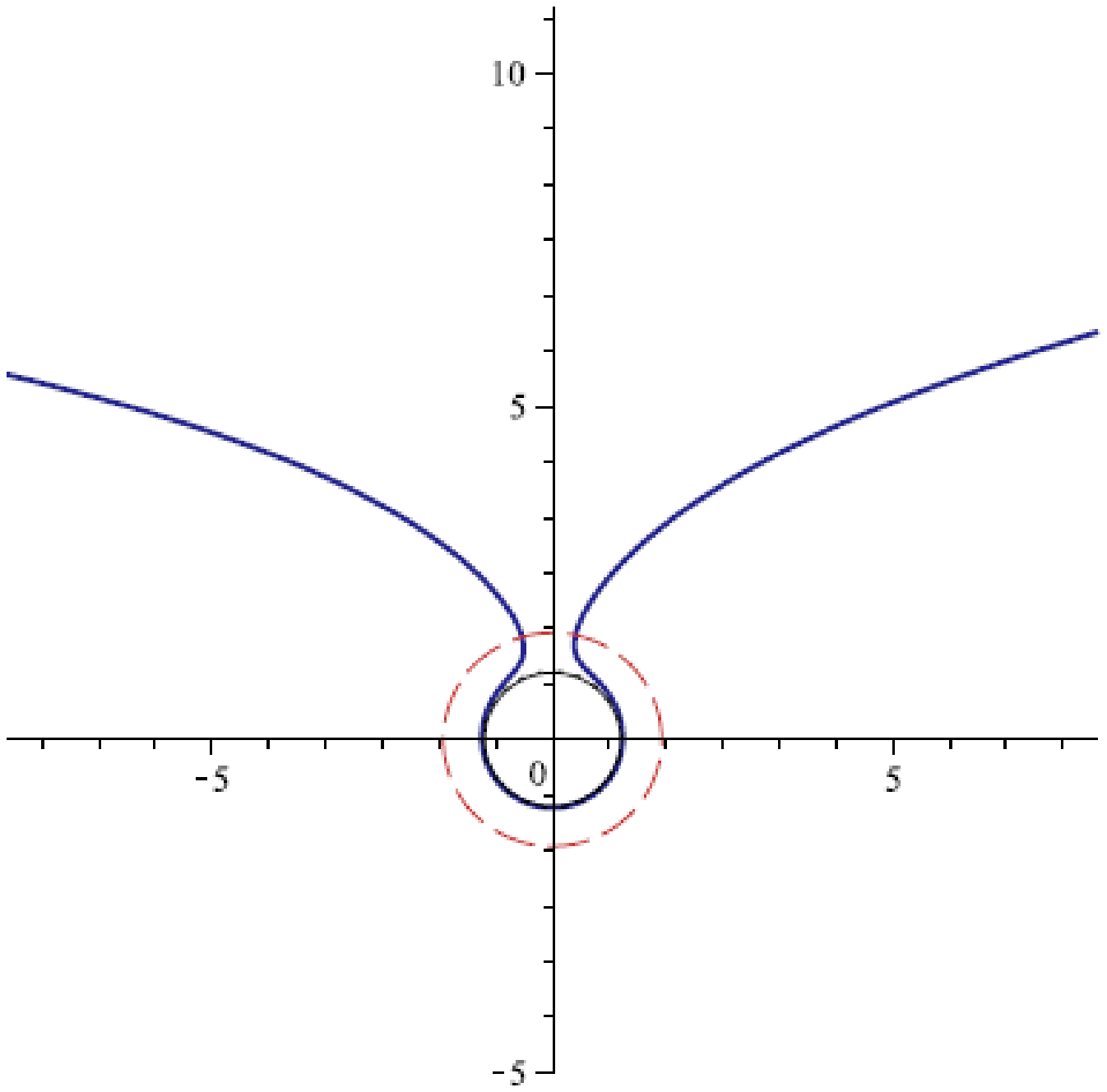}}
\end{center}
\caption{$X-Y$ plot for $\omega=1.1$ (overrotating case), ${K}=0.5$. In the plots~\subref{orb5} and~\subref{orb51} $\Phi=\sqrt{0.5}$. The solid circle marks the pseudo-horizon at $x=r^2=1$, the two dotted circles mark the minimal and maximal values of the radial coordinate. The dashed circle is located at the `turnaround boundary' $x\equiv r^2= 1+\frac{\omega E}{\Phi}$, signalling vanishing $\frac{d\varphi}{d\tau}$. The orbit~\subref{orb5} is located inside this circle. 
%The orbit~\subref{orb51} crosses that circle. 
\label{fig5:orb}}
\end{figure*}

\begin{figure*}[th!]
\begin{center}
\subfigure[][BO, $E=1.15 \ , {A}=\sqrt{{K}}$]{\label{orb9}\includegraphics[width=6cm]{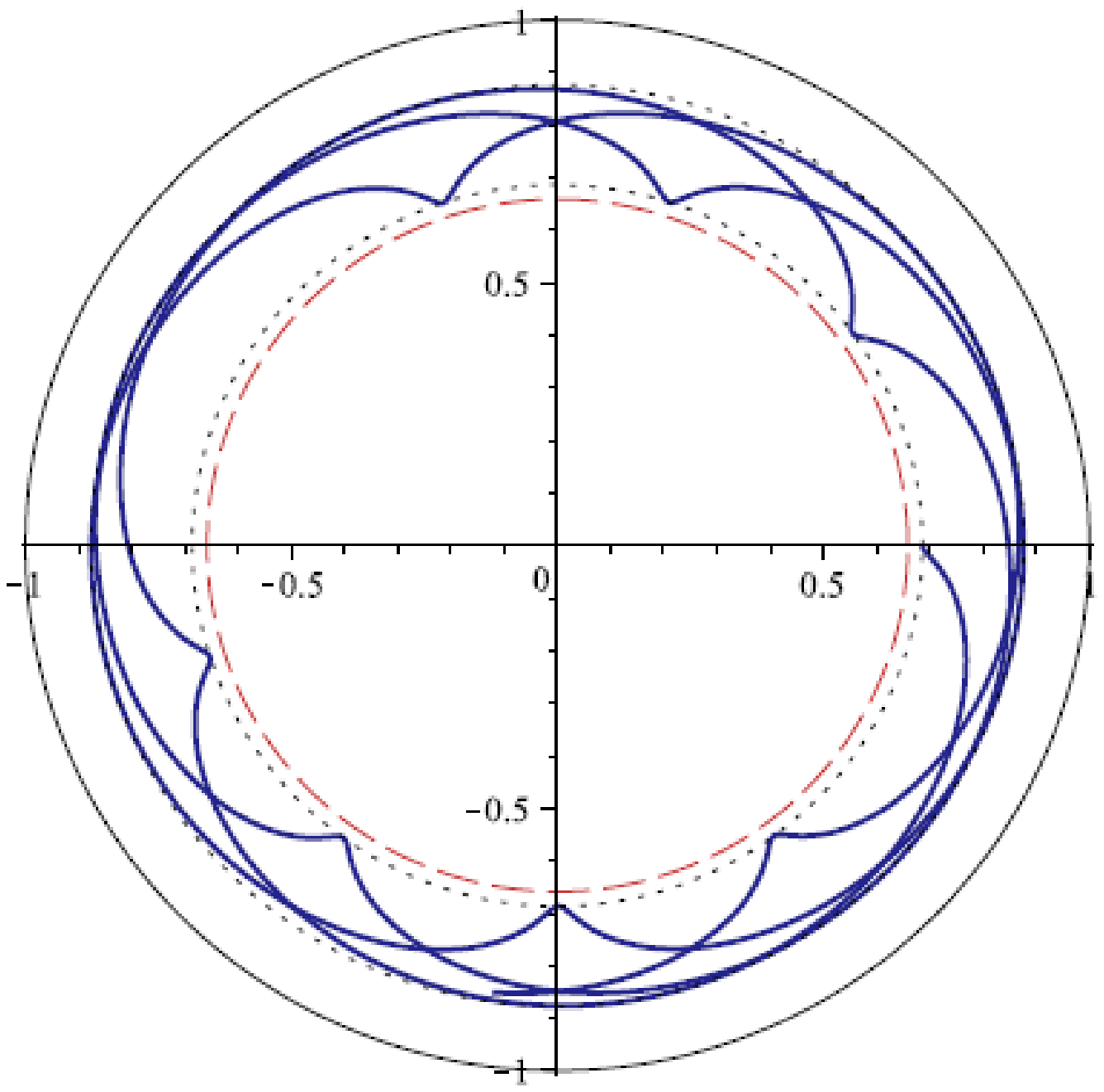}}
\subfigure[][EO, $E=1.15 \ , {A}=\sqrt{{K}}$]{\label{orb91}\includegraphics[width=6cm]{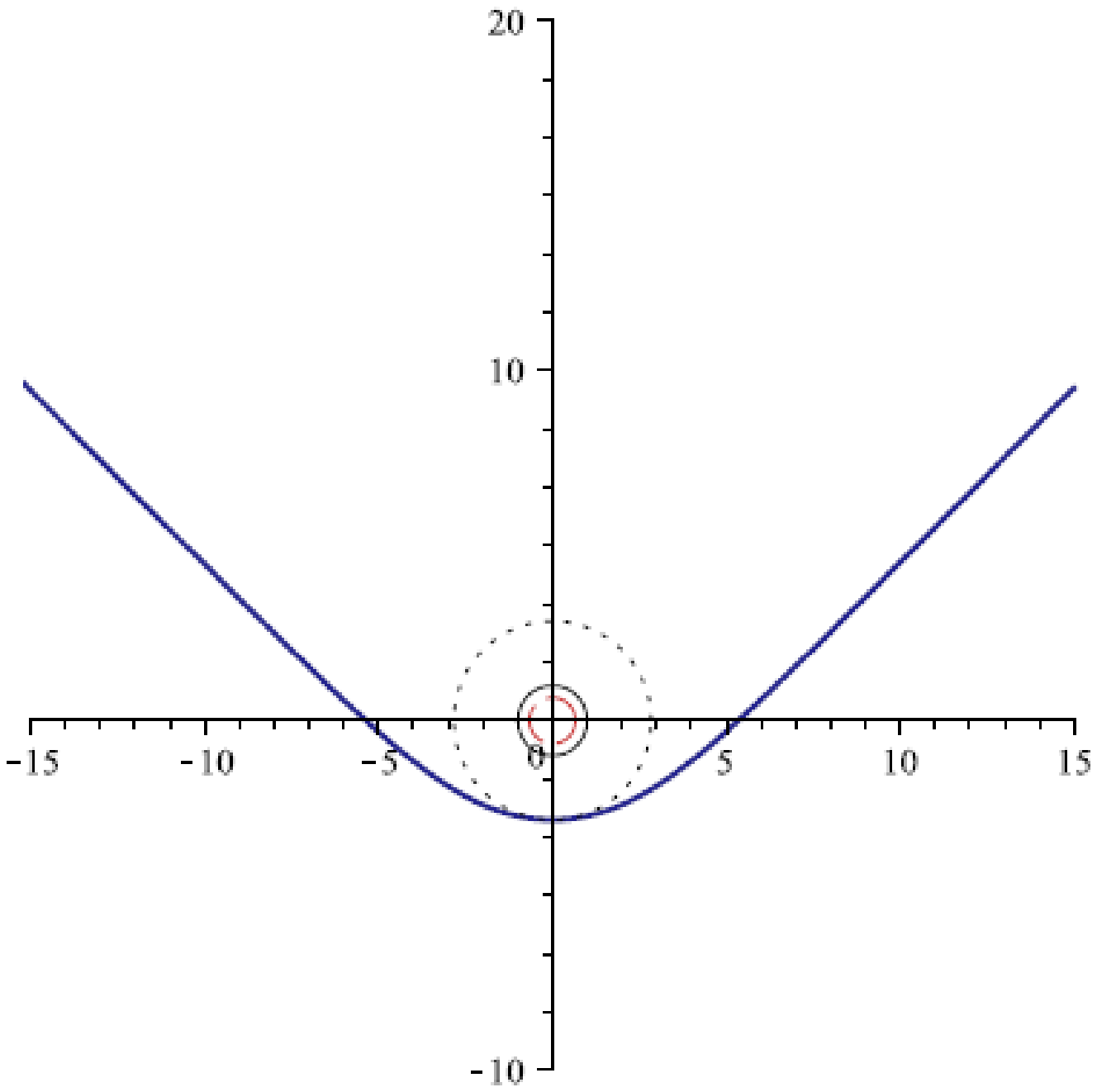}}
\end{center}
\caption{$X-Y$ plot for $\omega=1.1$ (overrotating case), ${K}=5 \ , \Phi=-\sqrt{5}$. The solid circle marks the pseudo-horizon at $x=r^2=1$, the two dotted circles mark the minimal and maximal values of the radial coordinate. The dashed circle is located at the `turnaround boundary' $x\equiv r^2= 1+\frac{\omega E}{\Phi}$, signalling vanishing $\frac{d\varphi}{d\tau}$. The orbits are located outside this circle. \label{fig6:orb}}
\end{figure*}

\begin{figure*}[th!]
\begin{center}
\subfigure[][BO, $E=6.0687 \ , {A}=-\sqrt{{K}}$]{\label{orb6}\includegraphics[width=6cm]{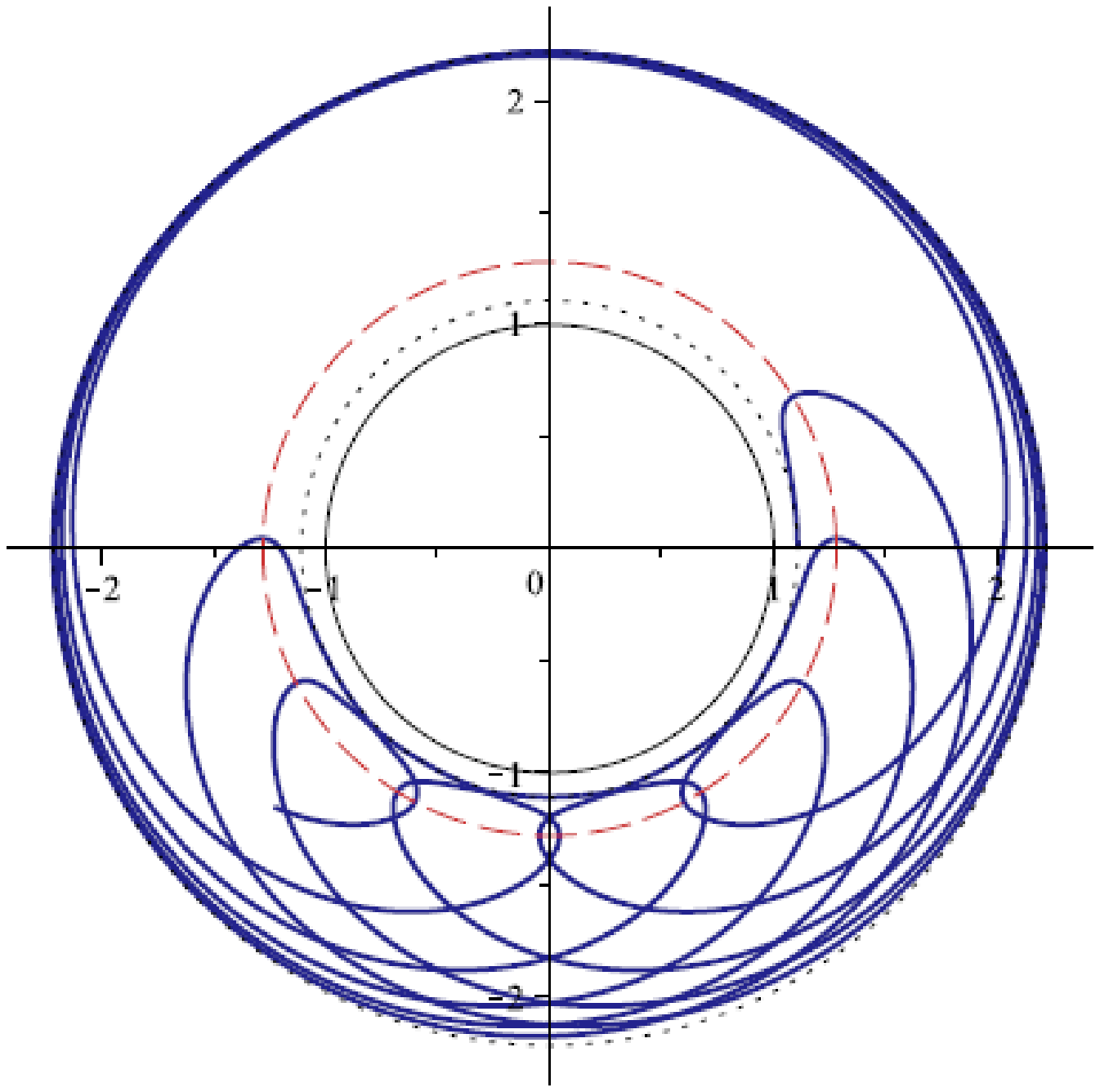}}
\subfigure[][EO, $E=6.0687 \ , {A}=-\sqrt{{K}}$]{\label{orb61}\includegraphics[width=6cm]{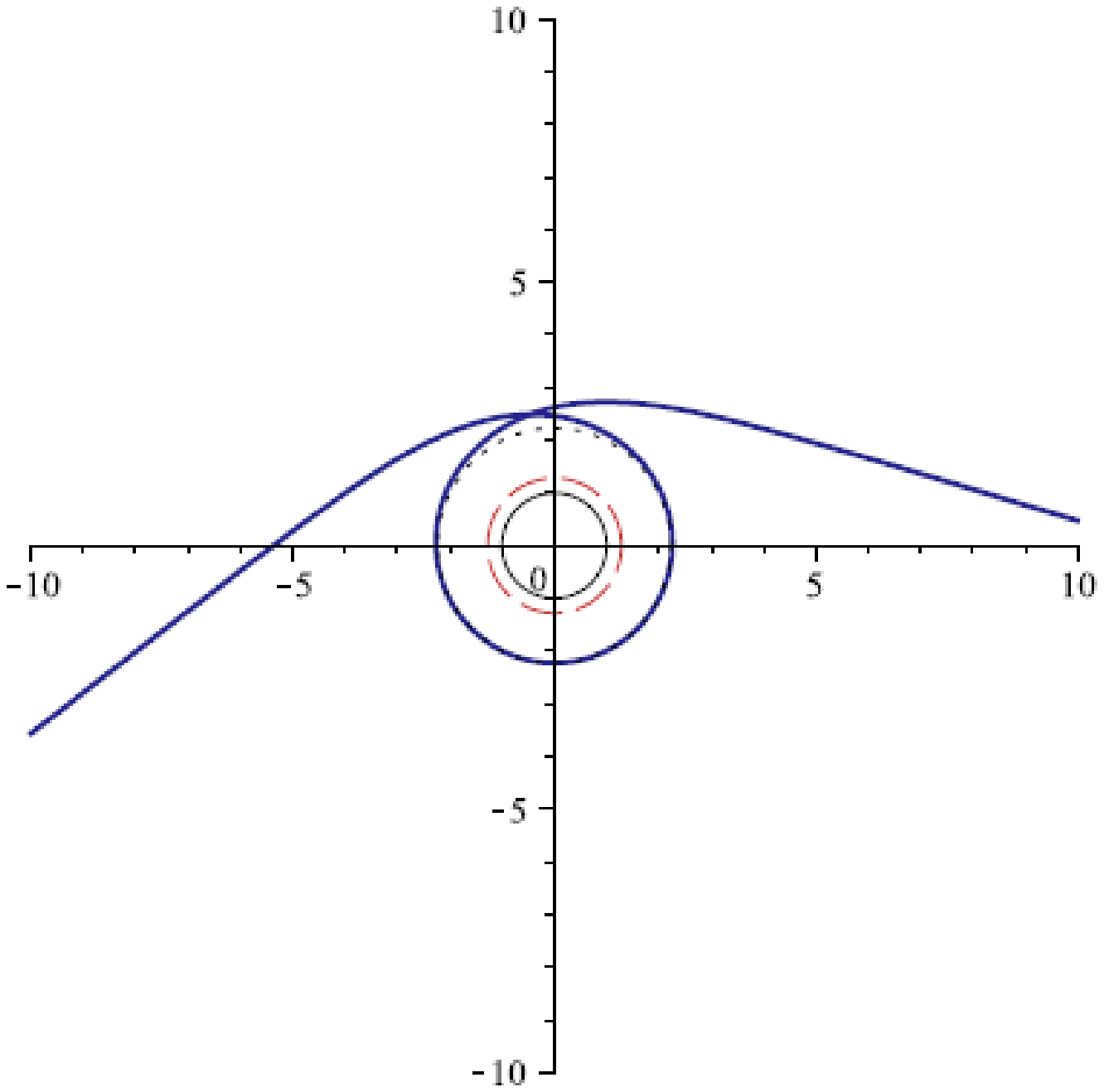}}
\end{center}
\caption{$X-Y$ plot for $\omega=2.1$ (overrotating case), ${K}=400$. In the plots~\subref{orb6} and~\subref{orb61} $\Phi=20$. The solid circle marks the pseudo-horizon at $x=r^2=1$, the two dotted circles mark the minimal and maximal values of the radial coordinate. The dashed circle is located at the `turnaround boundary' $x\equiv r^2= 1+\frac{\omega E}{\Phi}$, signalling vanishing $\frac{d\varphi}{d\tau}$. The orbit~\subref{orb61} is located outside that circle. \label{fig7:orb}}
\end{figure*}

\begin{figure*}[th!]
\begin{center}
\subfigure[][BO, $E=0.65 \ , {A}=-\sqrt{{K}}$]{\label{orb10}\includegraphics[width=6cm]{orbit10.eps}}
\subfigure[][EO, $E=1.001 \ , {A}=-\sqrt{{K}}$]{\label{orb101}\includegraphics[width=6cm]{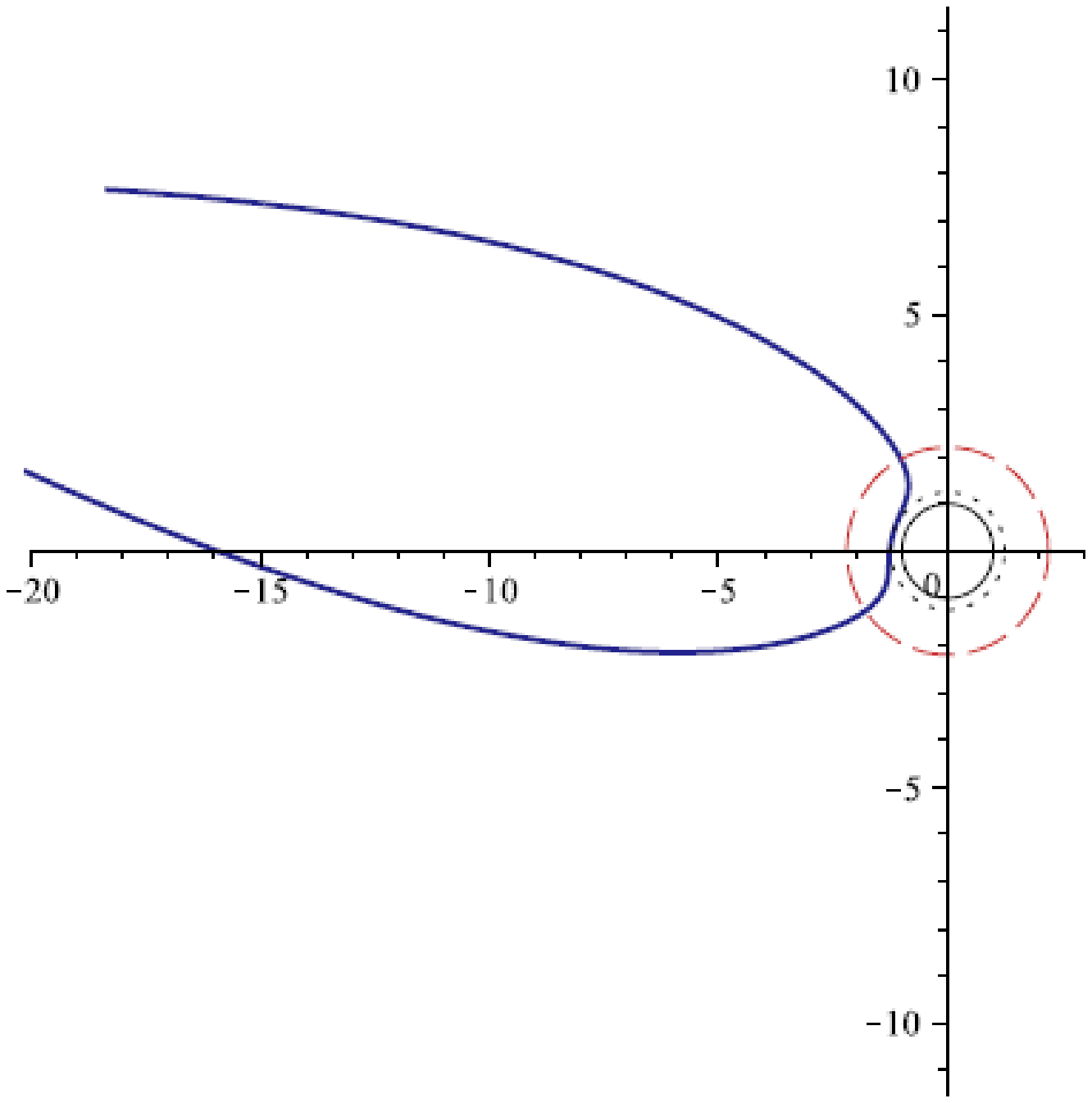}}
\end{center}
\caption{$X-Y$ plot for $\omega=2.1$ (overrotating case), ${K}=0.3$. In the plots~\subref{orb10} and~\subref{orb101} $\Phi=\sqrt{0.3}$. The solid circle marks the pseudo-horizon at $x=r^2=1$, the two dotted circles mark the minimal and maximal values of the radial coordinate. The dashed circle is located at the `turnaround boundary' $x\equiv r^2= 1+\frac{\omega E}{\Phi}$, signalling vanishing $\frac{d\varphi}{d\tau}$. The orbit~\subref{orb10} is located inside that circle. \label{fig8:orb}}
\end{figure*}

We now turn to orbits in the overrotating case, $\omega>1$.
In the figs.~\ref{fig4:orb},~\ref{fig5:orb} and~\ref{fig6:orb} we show trajectories for $\omega=1.1$ and in the figs.~\ref{fig7:orb} and~\ref{fig8:orb}  trajectories for $\omega=2.1$ and varying values of the separation constant $K$ or angular momentum $\Phi$. For $\omega>1$ there are bound and escape orbits possible as we know from the previous chapters. 

In the figs.~\ref{orb3} and~\ref{orb4} bound orbits deflected at the `turnaround boundary' are shown. The escape orbits~\ref{orb31} and~\ref{orb41} do not get influenced by the `turnaround boundary'.
In the plot~\ref{orb5} a bound orbit is located inside the `turnaround boundary', while the escape orbit~\ref{orb51} will be angularly deflected there.
In the fig.~\ref{orb9} the bound orbit is behind the pseudo-horizon and lies beyond the `turnaround boundary'. The escape orbit~\ref{orb91} for the same value of energy is of general hyperbolic type.

The figs.~\ref{orb6} and~\ref{orb61} for $\omega=2.1$ show a bound orbit influenced by the `turnaround boundary' and an escape orbit. In the orbit~\ref{orb10} a bound orbit in the form of a Christmas star is located inside the `turnaround boundary', while the particle in the escape orbit~\ref{orb101} will change its angular direction of motion at the `turnaround boundary'.

\subsection{Three dimensional orbits}~\label{section:3dorbits}

In this section we visualize three dimensional geodesics. 
The three dimensional orbits represent in this case 
a three-dimensional projection of the general four-dimensional orbits. 
This projection explains the sometimes not very smooth looking 
parts of the orbits in the figures below.

In the fig.~\ref{fig3d:orb1} we show a many-world bound orbit~\subref{3dorb1} and an escape orbit~\subref{3orb11} for the underrotating case with $\omega=0.7$. Because of the choice of the coordinates there is a divergence at the horizon. The motion is continued on the inner side of the horizon. 

\begin{figure*}[th!]
\begin{center}
\subfigure[][MBO, $E=0.8$]{\label{3dorb1}\includegraphics[width=6cm]{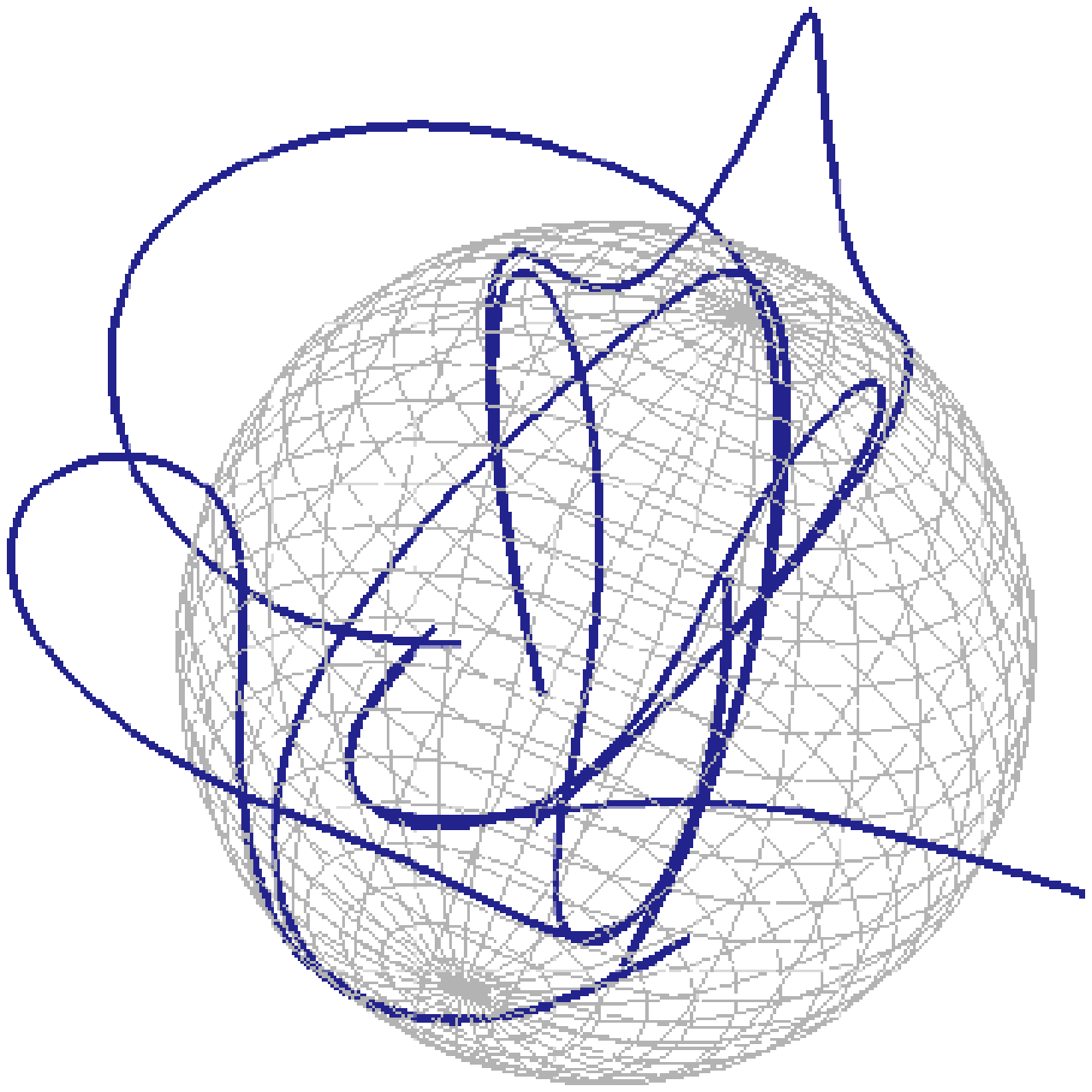}}
\subfigure[][TWE, $E=1.001$]{\label{3orb11}\includegraphics[width=6cm]{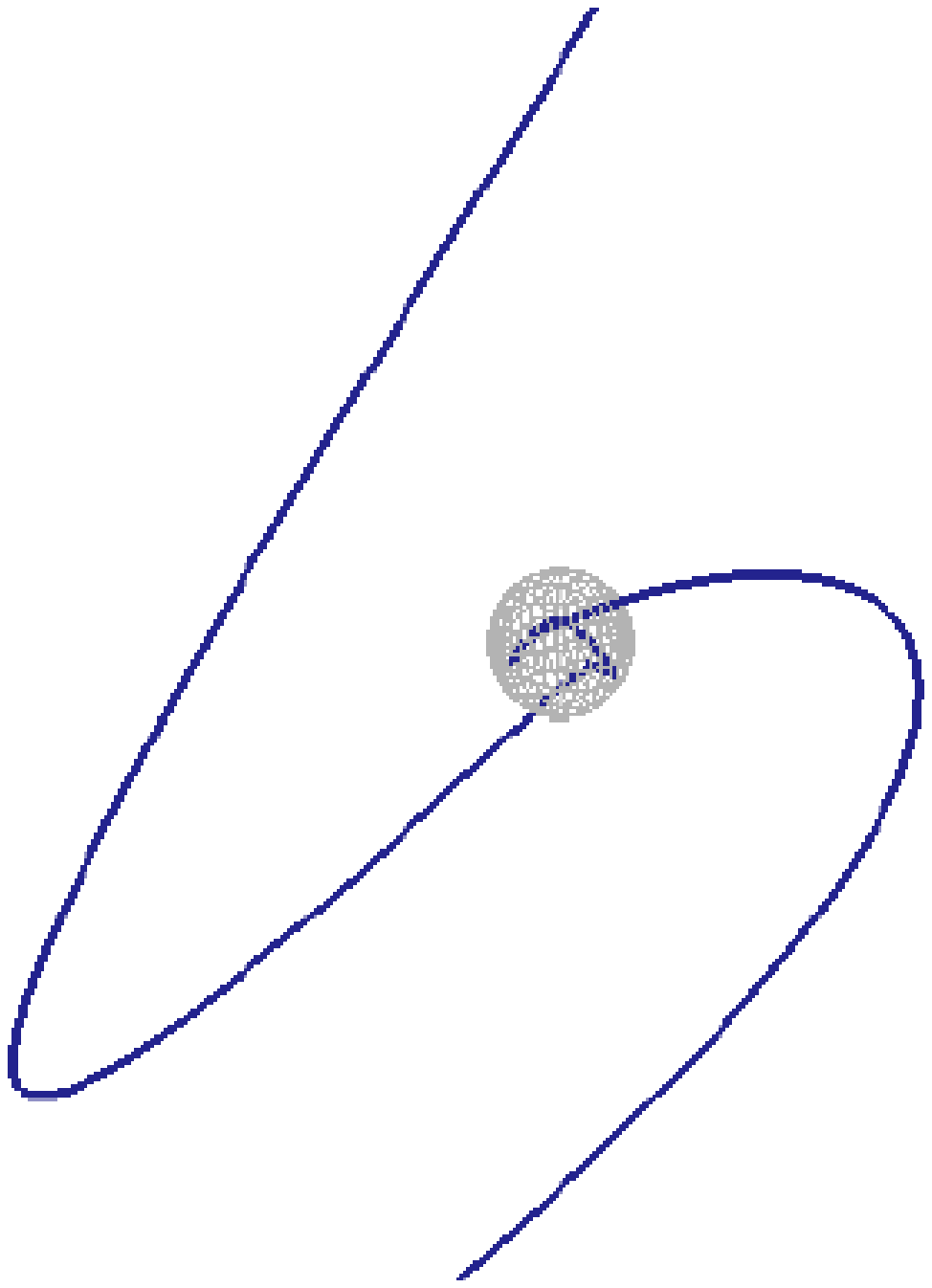}}
\end{center}
\caption{$X-Y-Z$ plot for a massive test particle with ${A}={B}=0.1\sqrt{{K}} \ , {K}=1$ in an underrotating spacetime with $\omega=0.7$. The sphere represents the horizon. \label{fig3d:orb1}}
\end{figure*}

\begin{figure*}[th!]
\begin{center}
\subfigure[][BO, $E=1.4$]{\label{3dorb2}\includegraphics[width=6cm]{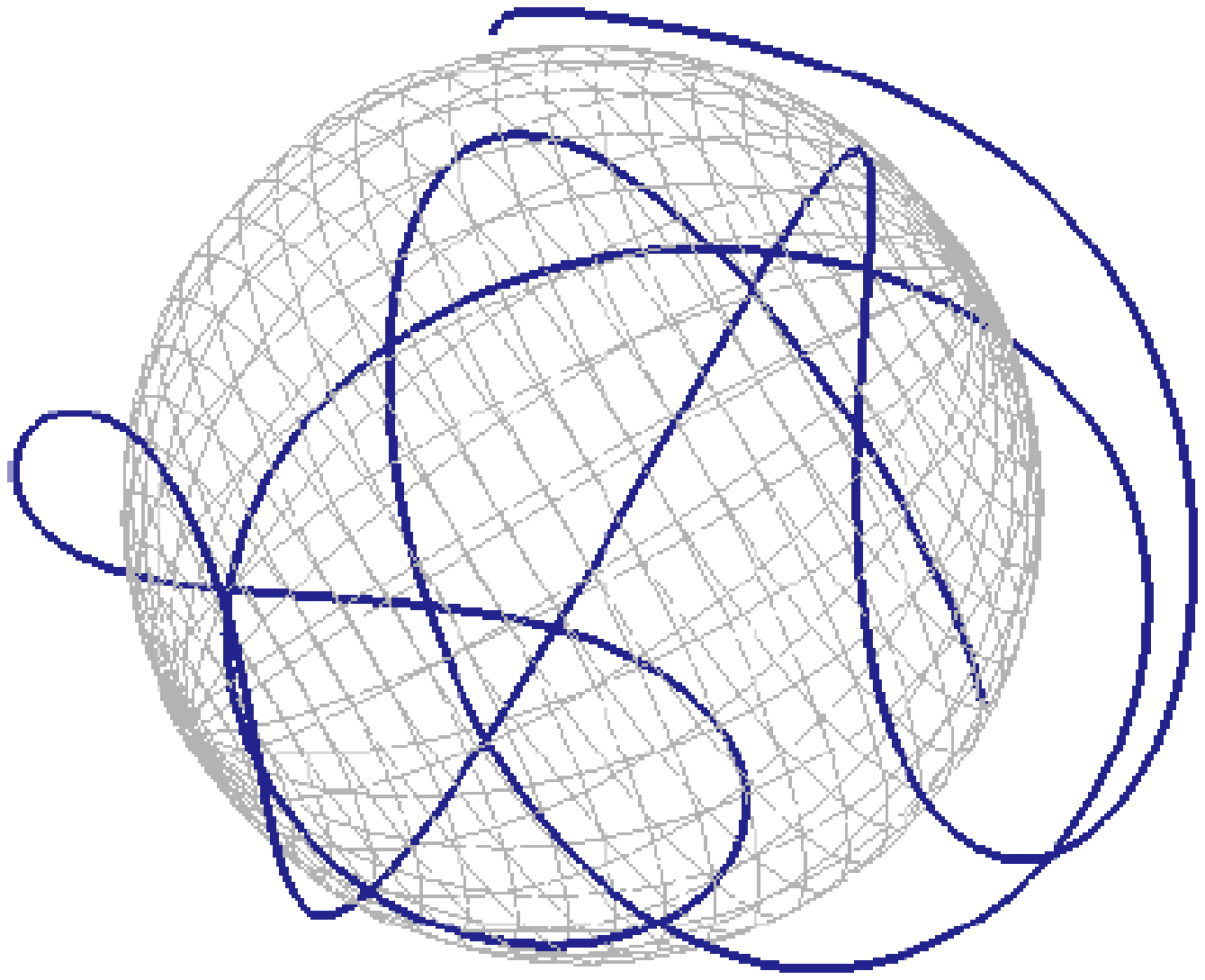}}
\subfigure[][EO, $E=1.4$]{\label{3orb21}\includegraphics[width=6cm]{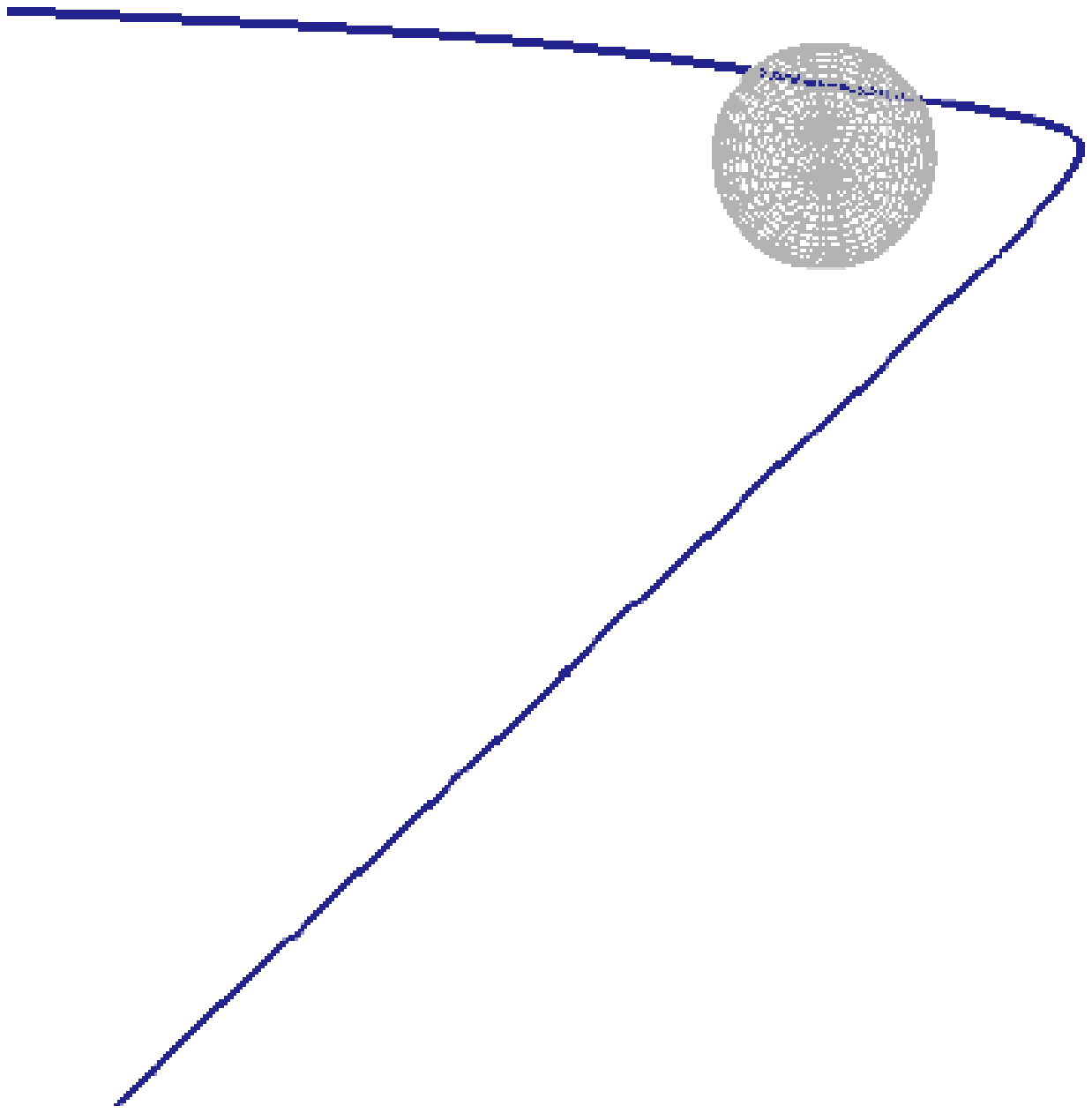}}
\subfigure[][BO, $E=1.1$]{\label{3dorb3}\includegraphics[width=6cm]{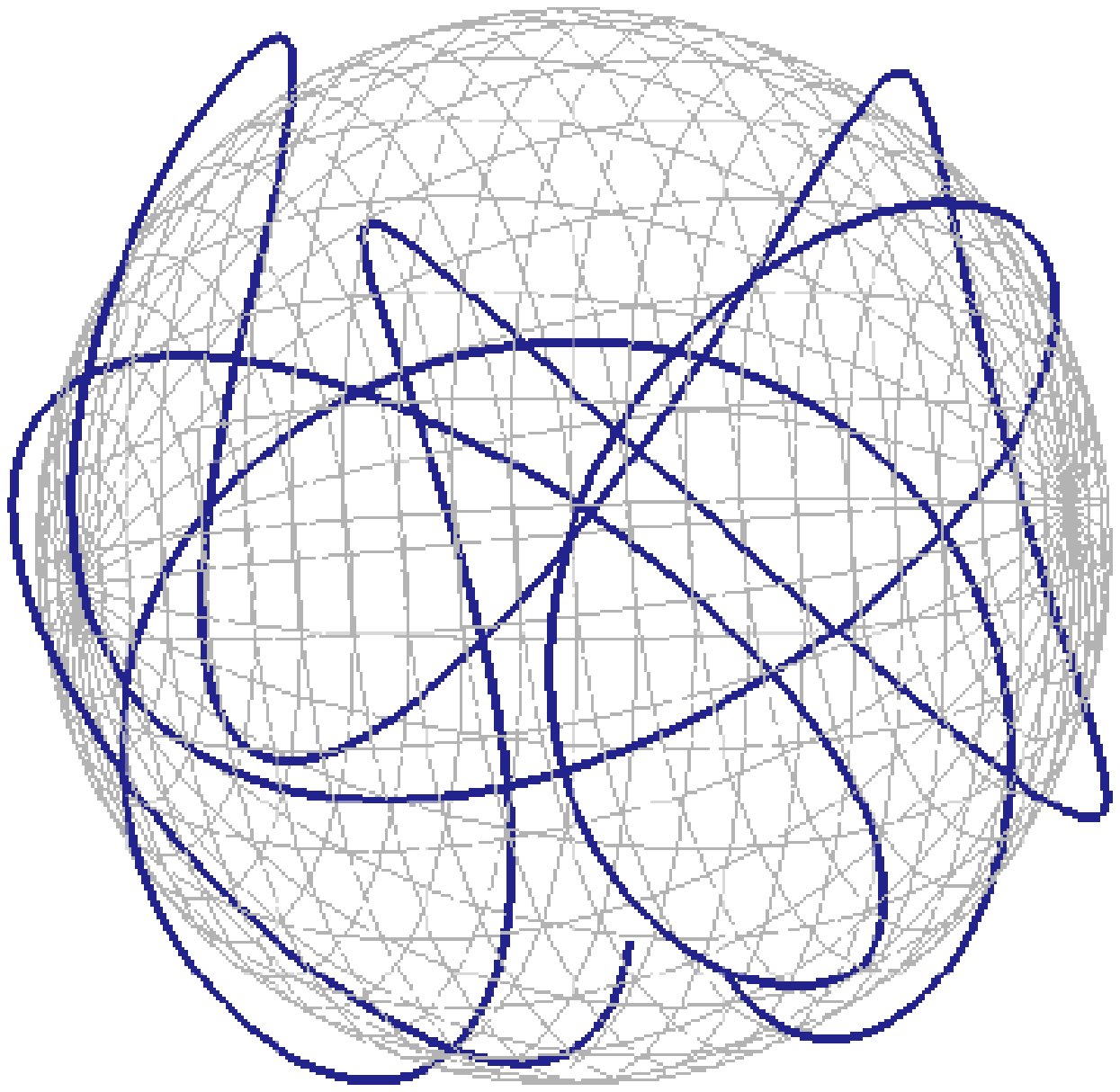}}
\subfigure[][EO, $E=1.44752$]{\label{3dorb31}\includegraphics[width=6cm]{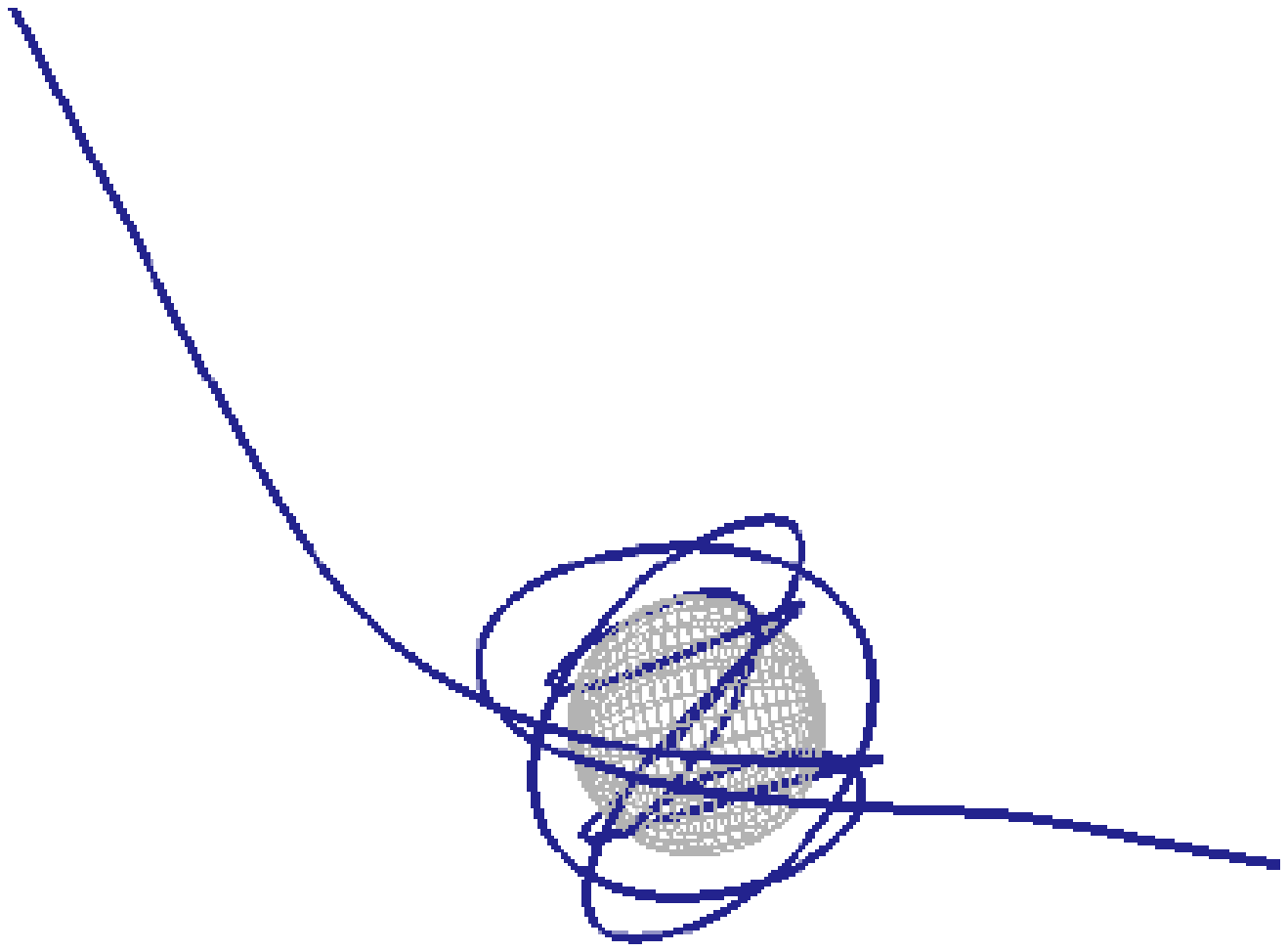}}
\end{center}
\caption{$X-Y-Z$ plot for a massive test particle with ${A}=0.1\sqrt{{K}} \ , {B}=0.7\sqrt{{K}} \ , {K}=10$ in an overrotating spacetime with $\omega=1.1$. The sphere represents the pseudo-horizon. \label{fig3d:orb2}}
\end{figure*}

In the fig.~\ref{fig3d:orb2} we show trajectories for the overrotating case with $\omega=1.1$. In the figs.~\ref{3dorb2} and~\ref{3dorb3} the orbits are bound and possess different values of the energy $E$. In the fig.~\ref{3dorb3} the energy value is close to the minimum of the effective potential. Fig.~\ref{3orb21} shows a corresponding escape orbit for the bound orbit~\ref{3dorb2}, while the orbit~\ref{3dorb31} has an almost critical value of the energy $E$ corresponding to the maximum of the effective potential.

\begin{figure*}[th!]
\begin{center}
\subfigure[][inner BO, $E=4$]{\label{3dorb4}\includegraphics[width=6cm]{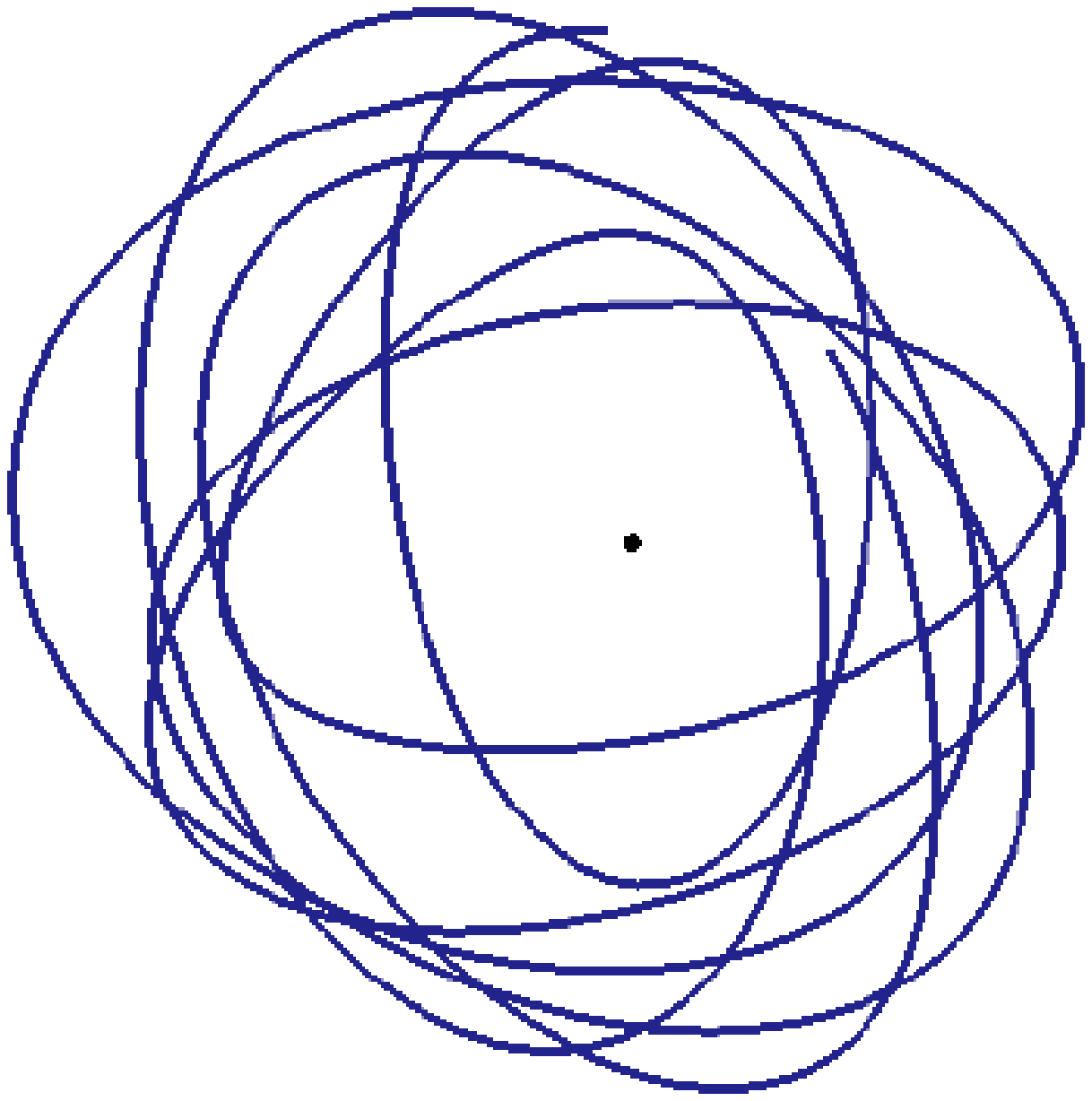}}
\subfigure[][EO, $E=4$]{\label{3orb51}\includegraphics[width=6cm]{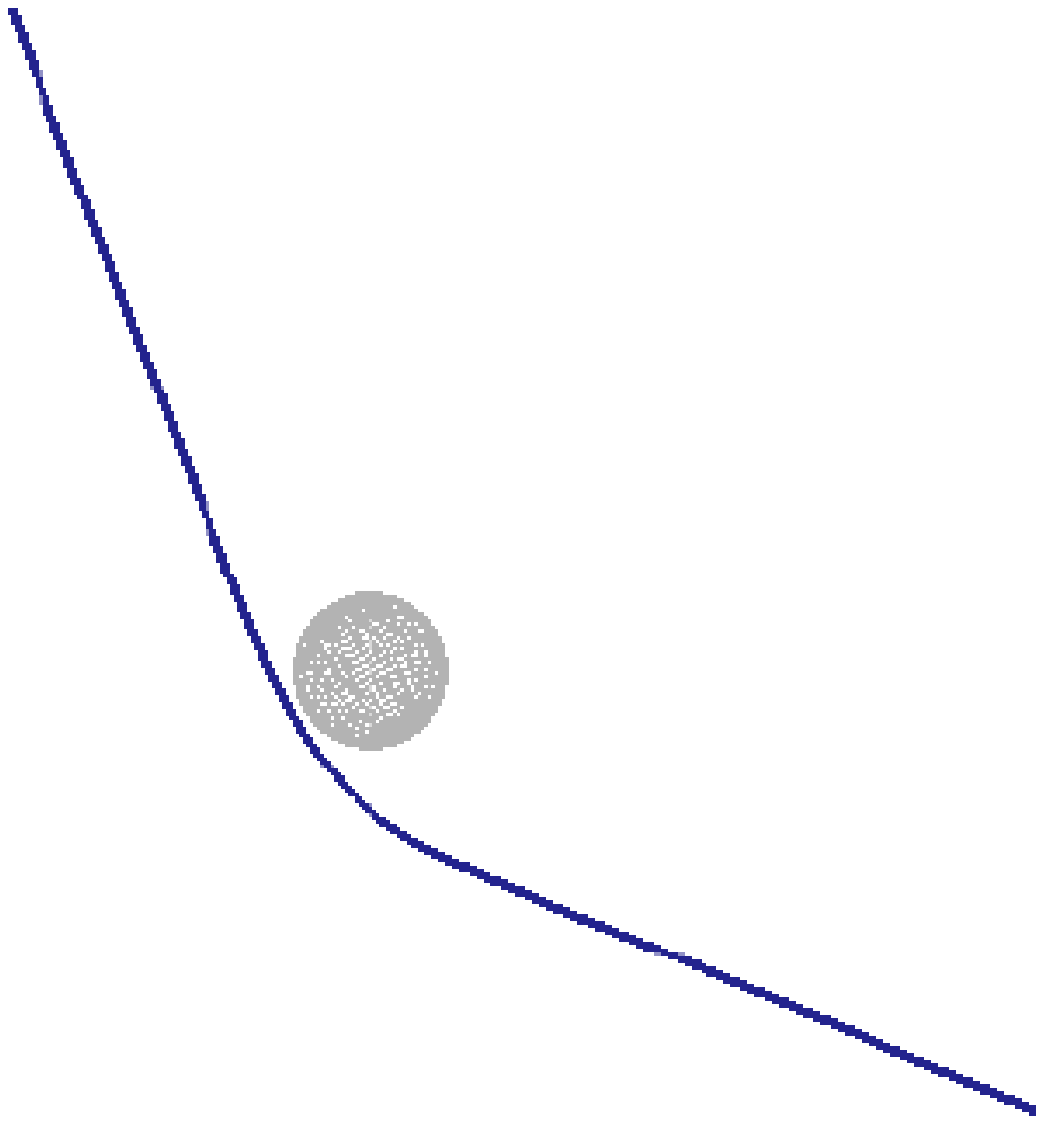}}
\end{center}
\caption{$X-Y-Z$ plot for a massive test particle with ${A}=0.9\sqrt{{K}} \ , {B}=0.2\sqrt{{K}} \ , {K}=100$ in an overrotating spacetime with $\omega=1.1$. The sphere represents the pseudo-horizon. The orbit~\subref{3dorb4} is between the singularity and the pseudo-horizon, it is hidden from a remote observer. The black dot in the plot denotes the singularity at the origin of the coordinates. \label{fig3d:orb3}}
\end{figure*}

The trajectories in the fig.~\ref{fig3d:orb3} are a bound orbit behind the pseudo-horizon~\ref{3dorb4} and an escape orbit~\ref{3orb51} for the same value of the energy.

\section{Conclusions}

In this paper we have discussed the orbits of neutral test particles in the BMPV spacetime.
We have solved the full set of geodesic equations analytically in terms of the Weierstrass' functions.
We have analyzed in detail the effective potential of the radial equation, and presented
a complete classification of the possible orbits in this spacetime.
We have also addressed the causal properties of the BMPV spacetime. 
Our results are in full accordance with previous more qualitative discussions
\cite{Gibbons:1999uv,Herdeiro:2000ap,Herdeiro:2002ft,Cvetic:2005zi}.

\begin{itemize}
\item
In the underrotating case, 
when the rotation parameter $\omega <1$, the BMPV spacetime describes 
supersymmetric black holes. Here the velocity of light surface,
which forms the boundary inside which causality violation can occur,
is hidden behind the horizon, 
%which is located at $x=1$.
located at $x=1$.

The possible types of orbits in this black hole spacetime
are classified in table~\ref{tab2} and table~\ref{tab3}.
There exist no planetary type orbits in the outer spacetime.
But there are many world bound orbits. 
Moreover, bound orbits are found in the interior region $x<1$
for massive and massless particles.
\item
In the overcritical case $\omega >1$ the BMPV spacetime represents in its
exterior region $x \ge 1$ a repulson. 
No geodesics can cross the pseudo-horizon located at $x=1$. 
The exterior region is geodesically complete.
The velocity of light surface is, however, outside the pseudo-horizon.
Therefore this spacetime represents a naked time machine.

The possible types of orbits in this overrotating spacetime
are also classified in table~\ref{tab2} and table~\ref{tab3}.
The outer BMPV spacetime now allows for planetary bound orbits
for particles and light.
But there are also bound orbits in the interior region
for massive and massless particles.

\item
In the critical case $\omega=1$ the surface $x=1$ 
has vanishing area and coincides
with the velocity of light surface. 
As in the repulson case, no geodesics can cross
this surface to reach the interior region $x<1$. 
However, most types of orbits reach
the surface $x=1$.

The possible types of orbits in this critical spacetime
are classified in table~\ref{tab4} and table~\ref{tab5}.
Only escape orbits in the outer region and bound
orbits in the interior region do not reach the surface $x=1$.
%as a boundary point. 
This holds for massive particles and for light.
\end{itemize}

To illustrate the analytical solutions we have presented various types of trajectories 
in the plane $\theta=\pi/2$, 
and subsequently performed a three-dimensional projection for some
selected orbits.
An interesting effect present in many of the orbits is the change of
the angular direction at the `turnaround boundary', where the 
derivative of the azimuthal angle w.r.t.~the radial coordinate 
vanishes. This effect arises although there is no ergosphere
in the spacetime. The resulting orbits have rather intriguing
shapes, differing from the ones of the Kerr, Kerr-Newman or
Myers-Perry spacetimes.

In should be interesting to next address the motion of charged test particles
in the BMPV spacetime.
This should yield the full analytical solution of the set of equations 
presented and discussed by Herdeiro \cite{Herdeiro:2000ap,Herdeiro:2002ft}.
An analogous analysis as the one given here
should yield the complete classification of the possible types of orbits.

Since the BMPV solution may be considered
as a subset of the more general family of solutions
found by Chong, Cvetic, L\"u and Pope \cite{Chong:2005hr},
it will be interesting to extend the present study to this 
set of solutions.
They include two unequal rotation parameters and 
describe also non-extremal solutions \cite{Chong:2005hr}.
Moreover a cosmological constant is present.
Particularly interesting should be the analysis of 
the included set of
supersymmetric black holes and topological solitons.

Moreover, an extension of the present work to the intriguing set of
solutions of gauged supergravities in four, five and seven dimensions,
discussed by
Cvetic, Gibbons, L\"u and Pope \cite{Cvetic:2005zi},
appears interesting.
However, in analytical studies of the geodesics 
of such extended sets of solutions
it may be necessary to employ more advanced mathematical tools
based on hyperelliptic functions
\cite{Hackmann:2008zza,Hackmann:2008tu,Enolski:2010if}.

On the other hand, the BMPV spacetime can also be viewed as a special
case of a more general family of solutions, where the Chern-Simons coupling 
constant, multiplying the $F^2 A$ term in the action, 
is a free parameter \cite{Gauntlett:1998fz}.
When this free parameter assumes a particular value, the
BMPV solutions of minimal supergravity are obtained.
For other values of this coupling constant new surprising phenomena
occur. For instance, the resulting black hole solutions need no longer
be uniquely specified in terms of their global charges
\cite{Kunz:2005ei} or an infinite sequence of extremal
radially excited rotating black holes arises
\cite{Blazquez-Salcedo:2013muz}.
The study of the geodesics in these spacetime may also 
reveal some surprises.

\bigskip

{\bf Acknowledgments.} We gratefully acknowledge discussions with Saskia Grunau, Burkhard Kleihaus and Eugen Radu, and support by the Deutsche Forschungsgemeinschaft (DFG), in particular, within the framework of the DFG Research Training group 1620 {\it Models of gravity}.

\bibliographystyle{unsrt}

\end{document}